\documentclass[%
 reprint,
superscriptaddress,
 amsmath,amssymb,
 prd, 
]{revtex4-1}

\usepackage{float}
\usepackage{graphicx}% Include figure files
\usepackage{lipsum}% http://ctan.org/pkg/lipsum
\usepackage{dcolumn}% Align table columns on decimal point
\usepackage{bm}% bold math
\usepackage{times}
\usepackage{multirow}
\usepackage{epstopdf}
\usepackage{color}
\usepackage[normalem]{ulem}
\definecolor{cmntblue}{rgb}{0.0, 0.58, 0.71}
\definecolor{cmntgreen}{rgb}{0.0, 0.42, 0.24}
\definecolor{cmntbrown}{rgb}{0.6, 0.3, 0.2}

\usepackage[draft, inline, nomargin]{fixme}
\fxsetup{theme=color, mode=multiuser}
\FXRegisterAuthor{bl}{1}{\color{blue}BL}  
\FXRegisterAuthor{dj}{2}{\color{red}DJ}
\FXRegisterAuthor{nb}{3}{\color{cmntblue}NB}
\FXRegisterAuthor{pm}{4}{\color{cmntgreen}PM}
\FXRegisterAuthor{kh}{5}{\color{green}KH}
\FXRegisterAuthor{pts}{6}{\color{cmntbrown}PTS}
\newcommand{\LS}{liquid scintillator (LS) \renewcommand{\LS}{LS}}

\newcommand{\nuebar}{\ensuremath{\overline{\nu}_{e}} }
\newcommand{\uFive}{$^{235}$U}
\newcommand{\uEight}{$^{238}$U}
\newcommand{\pNine}{$^{239}$Pu}
\newcommand{\pOne}{$^{241}$Pu}
\newcommand{\NIST}{National Institute of Standards and Technology (NIST) \renewcommand{\NIST}{NIST}}
\newcommand{\NBSR}{National Bureau of Standards Reactor (NBSR) \renewcommand{\NBSR}{NBSR}}
\newcommand{\INL}{Idaho National Laboratory (INL) \renewcommand{\INL}{INL}}
\newcommand{\ATR}{Advanced Test Reactor (ATR) \renewcommand{\ATR}{ATR}}
\newcommand{\ORNL}{Oak Ridge National Laboratory (ORNL) \renewcommand{\ORNL}{ORNL}}
\newcommand{\HFIR}{High Flux Isotope Reactor (HFIR) \renewcommand{\HFIR}{HFIR}}
\newcommand{\LLNL}{Lawrence Livermore National Laboratory (LLNL) \renewcommand{\LLNL}{LLNL}}

\begin{document}

\preprint{APS/123-QED}

\title{Improved Short-Baseline Neutrino Oscillation Search and  Energy Spectrum Measurement with the PROSPECT Experiment at HFIR} 
%\title{Observation of antineutrinos from ${}^{235}{\rm U}$ fission by the PROSPECT experiment at HFIR}% Force line breaks with \\
%\input{June2018AuthorListRevised.tex}

%\affiliation{%
%  Authors' institution and/or address\\
%  This line break forced with \textbackslash\textbackslash
% }%

\affiliation{Brookhaven National Laboratory, Upton, NY, USA}
\affiliation{Department of Physics, Drexel University, Philadelphia, PA, USA}
\affiliation{George W.\,Woodruff School of Mechanical Engineering, Georgia Institute of Technology, Atlanta, GA USA}
\affiliation{Department of Physics \& Astronomy, University of Hawaii, Honolulu, HI, USA}
\affiliation{Department of Physics, Illinois Institute of Technology, Chicago, IL, USA}
\affiliation{Nuclear and Chemical Sciences Division, Lawrence Livermore National Laboratory, Livermore, CA, USA}
\affiliation{Department of Physics, Le Moyne College, Syracuse, NY, USA}
\affiliation{National Institute of Standards and Technology, Gaithersburg, MD, USA}
\affiliation{High Flux Isotope Reactor, Oak Ridge National Laboratory, Oak Ridge, TN, USA}
\affiliation{Physics Division, Oak Ridge National Laboratory, Oak Ridge, TN, USA}
\affiliation{Department of Physics, Temple University, Philadelphia, PA, USA}
\affiliation{Department of Physics and Astronomy, University of Tennessee, Knoxville, TN, USA}
\affiliation{Institute for Quantum Computing and Department of Physics and Astronomy, University of Waterloo, Waterloo, ON, Canada}
\affiliation{Department of Physics, University of Wisconsin, Madison, Madison, WI, USA}
\affiliation{Wright Laboratory, Department of Physics, Yale University, New Haven, CT, USA}

\author{M.\,Andriamirado}
\affiliation{Department of Physics, Illinois Institute of Technology, Chicago, IL, USA}
\author{A.\,B.\,Balantekin}
\affiliation{Department of Physics, University of Wisconsin, Madison, Madison, WI, USA}
\author{H.\,R.\,Band}
\affiliation{Wright Laboratory, Department of Physics, Yale University, New Haven, CT, USA}
\author{C.\,D.\,Bass}
\affiliation{Department of Physics, Le Moyne College, Syracuse, NY, USA}
\author{D.\,E.\,Bergeron}
\affiliation{National Institute of Standards and Technology, Gaithersburg, MD, USA}
\author{D.\,Berish}
\affiliation{Department of Physics, Temple University, Philadelphia, PA, USA}
\author{N.\,S.\,Bowden}
\affiliation{Nuclear and Chemical Sciences Division, Lawrence Livermore National Laboratory, Livermore, CA, USA}
\author{J.\,P.\,Brodsky}
\affiliation{Nuclear and Chemical Sciences Division, Lawrence Livermore National Laboratory, Livermore, CA, USA}
\author{C.\,D.\,Bryan}
\affiliation{High Flux Isotope Reactor, Oak Ridge National Laboratory, Oak Ridge, TN, USA}
\author{T.\,Classen}
\affiliation{Nuclear and Chemical Sciences Division, Lawrence Livermore National Laboratory, Livermore, CA, USA}
\author{A.\,J.\,Conant}
\affiliation{George W.\,Woodruff School of Mechanical Engineering, Georgia Institute of Technology, Atlanta, GA USA}
\author{G.\,Deichert}
\affiliation{High Flux Isotope Reactor, Oak Ridge National Laboratory, Oak Ridge, TN, USA}
\author{M.\,V.\,Diwan}
\affiliation{Brookhaven National Laboratory, Upton, NY, USA}
\author{M.\,J.\,Dolinski}\affiliation{Department of Physics, Drexel University, Philadelphia, PA, USA}
\author{A.\,Erickson}
\affiliation{George W.\,Woodruff School of Mechanical Engineering, Georgia Institute of Technology, Atlanta, GA USA}
\author{B.\,T.\,Foust}
\affiliation{Wright Laboratory, Department of Physics, Yale University, New Haven, CT, USA}
\author{J.\,K.\,Gaison}
\affiliation{Wright Laboratory, Department of Physics, Yale University, New Haven, CT, USA}
\author{A.\,Galindo-Uribarri}\affiliation{Physics Division, Oak Ridge National Laboratory, Oak Ridge, TN, USA} \affiliation{Department of Physics and Astronomy, University of Tennessee, Knoxville, TN, USA}
\author{C.\,E.\,Gilbert}\affiliation{Physics Division, Oak Ridge National Laboratory, Oak Ridge, TN, USA} \affiliation{Department of Physics and Astronomy, University of Tennessee, Knoxville, TN, USA}
\author{B.\,W.\,Goddard}\affiliation{Department of Physics, Drexel University, Philadelphia, PA, USA}
\author{B.\,T.\,Hackett}\affiliation{Physics Division, Oak Ridge National Laboratory, Oak Ridge, TN, USA} \affiliation{Department of Physics and Astronomy, University of Tennessee, Knoxville, TN, USA}
\author{S.\,Hans}\affiliation{Brookhaven National Laboratory, Upton, NY, USA}
\author{A.\,B.\,Hansell}
\affiliation{Department of Physics, Temple University, Philadelphia, PA, USA}
\author{K.\,M.\,Heeger}
\affiliation{Wright Laboratory, Department of Physics, Yale University, New Haven, CT, USA}
%\author{B.\,Heffron}\affiliation{Physics Division, Oak Ridge National Laboratory, Oak Ridge, TN, USA} \affiliation{Department of Physics and Astronomy, University of Tennessee, Knoxville, TN, USA}
\author{D.\,E.\,Jaffe}
\affiliation{Brookhaven National Laboratory, Upton, NY, USA}
\author{X.\,Ji}
\affiliation{Brookhaven National Laboratory, Upton, NY, USA}
\author{D.\,C.\,Jones}
\affiliation{Department of Physics, Temple University, Philadelphia, PA, USA}
\author{O.\,Kyzylova}\affiliation{Department of Physics, Drexel University, Philadelphia, PA, USA}
\author{C.\,E.\,Lane}\affiliation{Department of Physics, Drexel University, Philadelphia, PA, USA}
\author{T.\,J.\,Langford}
\affiliation{Wright Laboratory, Department of Physics, Yale University, New Haven, CT, USA}
\author{J.\,LaRosa}
\affiliation{National Institute of Standards and Technology, Gaithersburg, MD, USA}
\author{B.\,R.\,Littlejohn}
\affiliation{Department of Physics, Illinois Institute of Technology, Chicago, IL, USA}
\author{X.\,Lu}\affiliation{Physics Division, Oak Ridge National Laboratory, Oak Ridge, TN, USA} \affiliation{Department of Physics and Astronomy, University of Tennessee, Knoxville, TN, USA}
\author{J.\,Maricic}\affiliation{Department of Physics \& Astronomy, University of Hawaii, Honolulu, HI, USA}
\author{M.\,P.\,Mendenhall}\affiliation{Nuclear and Chemical Sciences Division, Lawrence Livermore National Laboratory, Livermore, CA, USA}
\author{A.\,M.\,Meyer}\affiliation{Department of Physics \& Astronomy, University of Hawaii, Honolulu, HI, USA}
\author{R.\,Milincic}\affiliation{Department of Physics \& Astronomy, University of Hawaii, Honolulu, HI, USA}
\author{I.\,Mitchell}\affiliation{Department of Physics \& Astronomy, University of Hawaii, Honolulu, HI, USA}
\author{P.\,E.\,Mueller}\affiliation{Physics Division, Oak Ridge National Laboratory, Oak Ridge, TN, USA} 
\author{H.\,P.\,Mumm}
\affiliation{National Institute of Standards and Technology, Gaithersburg, MD, USA}
\author{J.\,Napolitano}
\affiliation{Department of Physics, Temple University, Philadelphia, PA, USA}
\author{C.\,Nave}\affiliation{Department of Physics, Drexel University, Philadelphia, PA, USA}
\author{R.\,Neilson}\affiliation{Department of Physics, Drexel University, Philadelphia, PA, USA}
\author{J.\,A.\,Nikkel}
\affiliation{Wright Laboratory, Department of Physics, Yale University, New Haven, CT, USA}
\author{D.\,Norcini}
\affiliation{Wright Laboratory, Department of Physics, Yale University, New Haven, CT, USA}
\author{S.\,Nour}
\affiliation{National Institute of Standards and Technology, Gaithersburg, MD, USA}
\author{J.\,L.\,Palomino}
\affiliation{Department of Physics, Illinois Institute of Technology, Chicago, IL, USA}
\author{D.\,A.\,Pushin}\affiliation{Institute for Quantum Computing and Department of Physics and Astronomy, University of Waterloo, Waterloo, ON, Canada}
\author{X.\,Qian}
\affiliation{Brookhaven National Laboratory, Upton, NY, USA}
\author{E.\,Romero-Romero}\affiliation{Physics Division, Oak Ridge National Laboratory, Oak Ridge, TN, USA} \affiliation{Department of Physics and Astronomy, University of Tennessee, Knoxville, TN, USA}
\author{R.\,Rosero}
\affiliation{Brookhaven National Laboratory, Upton, NY, USA}
\author{P.\,T.\,Surukuchi}
\affiliation{Wright Laboratory, Department of Physics, Yale University, New Haven, CT, USA}
\author{M.\,A.\,Tyra}
\affiliation{National Institute of Standards and Technology, Gaithersburg, MD, USA}
\author{R.\,L.\,Varner}\affiliation{Physics Division, Oak Ridge National Laboratory, Oak Ridge, TN, USA} 
\author{D.\,Venegas-Vargas}\affiliation{Physics Division, Oak Ridge National Laboratory, Oak Ridge, TN, USA} \affiliation{Department of Physics and Astronomy, University of Tennessee, Knoxville, TN, USA}
\author{P.\,B.\,Weatherly}\affiliation{Department of Physics, Drexel University, Philadelphia, PA, USA}
\author{C.\,White}
\affiliation{Department of Physics, Illinois Institute of Technology, Chicago, IL, USA}
\author{J.\,Wilhelmi}
\affiliation{Wright Laboratory, Department of Physics, Yale University, New Haven, CT, USA}
\author{A.\,Woolverton}\affiliation{Institute for Quantum Computing and Department of Physics and Astronomy, University of Waterloo, Waterloo, ON, Canada}
\author{M.\,Yeh}
\affiliation{Brookhaven National Laboratory, Upton, NY, USA}
\author{A.\,Zhang}
\affiliation{Brookhaven National Laboratory, Upton, NY, USA}
\author{C.\,Zhang}
\affiliation{Brookhaven National Laboratory, Upton, NY, USA}
\author{X.\,Zhang}
\affiliation{Nuclear and Chemical Sciences Division, Lawrence Livermore National Laboratory, Livermore, CA, USA}

\collaboration{PROSPECT Collaboration}%\noaffiliation
\email{prospect.collaboration@gmail.com}

%  \homepage{http://www.Second.institution.edu/~Charlie.Author}
% \affiliation{
%  Second institution and/or address\\
%  This line break forced% with \\
% }%
% \affiliation{
%  Third institution, the second for Charlie Author
% \affiliation{%
%  Authors' institution and/or address\\
%  This line break forced with \textbackslash\textbackslash
% }%

% \collaboration{CLEO Collaboration}%\noaffiliation

\date{\today}% It is always \today, today,

\begin{abstract}
We present a detailed report on sterile neutrino oscillation and \uFive~\nuebar{} energy spectrum measurement results from the PROSPECT experiment at the highly enriched High Flux Isotope Reactor (HFIR) at Oak Ridge National Laboratory.  
In 96 calendar days of data taken at an average baseline distance of 7.9~m from the center of the 85~MW HFIR core, the PROSPECT detector has observed more than 50,000 interactions of \nuebar{} produced in beta decays of \uFive{} fission products.  
New limits on the oscillation of \nuebar{} to light sterile neutrinos have been set by comparing the detected energy spectra of ten reactor-detector baselines between 6.7 and 9.2\,meters.  
Measured differences in energy spectra between baselines show no statistically significant indication of \nuebar{} to sterile neutrino oscillation and disfavor the Reactor Antineutrino Anomaly best-fit point at the 2.5$\sigma$ confidence level.  
The reported \uFive{} \nuebar{} energy spectrum measurement shows excellent agreement with energy spectrum models generated via conversion of the measured \uFive{} beta spectrum, with a $\chi^2$/DOF of 31/31.  
PROSPECT is able to disfavor at 2.4$\sigma$ confidence level the hypothesis that \uFive{} \nuebar{} are solely responsible for  spectrum discrepancies between model and data obtained at commercial reactor cores.  
A data-model deviation in PROSPECT similar to that observed by commercial core experiments is preferred with respect to no observed deviation, at a 2.2$\sigma$ confidence level.  

%in predicted and measured energy spectra from commercial reactor cores.
%Despite deployment on the earth's surface, the PROSPECT detector 
%To set new limits on short-baseline reactor antineutrino disappearance, energy spectra at ten different reactor-detector baselines are compared to search for evidence of baseline-dependent variations in 

%new limits on sterile neutrino oscillations 

%\begin{description}
%\end{description}
\end{abstract}

\maketitle

%\tableofcontents

%%%%%%%%%%%%%%%%%%%%%%%%%%%%%%%INTRO
\section{Introduction}
\label{sec:intro}

Neutrinos arguably remain the least well-understood fundamental particles in the Standard Model: their absolute masses are only constrained within a few orders of magnitude, properties of their right-handed versions and differences between matter and antimatter versions are undetermined, and many of their flavor mixing parameters remain uncertain at the 10\% level or greater~\cite{pdg_2018}.  
Further improvement in understanding of these properties requires new high-precision measurements using high-intensity neutrino sources.  % with either well-understood modes of neutrino production, highly controlled detector systematics, or a combination of the two.  
Nuclear reactors are the highest intensity artificial neutrino sources on Earth, producing MeV-scale energy electron-type antineutrinos ($\overline{\nu}_e$) predominantly via $\beta$-particle decay of neutron-rich fission daughter products of \uFive{}, \uEight{}, \pNine{}, and \pOne{}~\cite{vogel_review}.  
%A single 1~GW$_{th}$ commercial reactor core will produce more neutrinos in one minute ($\sim$10$^{21}$) than the combined neutrino output of the NuMI and BNB beamlines at Fermilab in 2018~\cite{beam_power,grant_dar} .%Fermilab flux numbers, DAR neutrino/proton numbers.  

These prodigious reactor \nuebar emissions have been used in past experiments to substantially expand our understanding of neutrino properties.  
Early reactor \nuebar experiments provided the first direct evidence of the particle's existence~\cite{cowan1956} and measured its rate of charged and neutral current interaction~\cite{bib:reines_cc,bib:reines_nc,bib:reines_escat}.  
More recently, the KamLAND experiment used fluxes from many reactors at 180~km average distance to measure distortion of the predicted reactor \nuebar energy spectrum due to \nuebar disappearance~\cite{KamLAND_rate, KamLAND_shape}, confirming large-amplitude lepton flavor mixing as the solution to the long-standing `solar neutrino problem'~\cite{bib:solar}. 

Subsequently, three reactor neutrino experiments at km-scale baselines -- Daya Bay, Double Chooz, and RENO -- also measured substantial \nuebar disappearance and energy spectrum distortion~\cite{bib:prl_rate,bib:prl_shape,bib:reno,bib:reno_shape,bib:dc}.  
% which were demonstrated to be dependent on the \nuebar energy, ($E_{nu}$), and on the distance between \nuebar production and detection (baseline), $L$.  
These results produced the first confirmation of a non-zero value for the neutrino mixing parameter $\theta_{13}$, opening the door to future accelerator-based measurements of leptonic CP violation and determination of the ordering of the three Standard Model neutrino masses~\cite{if_white}.  
Reactor neutrino experiments provide leading or competitive precision in the determination of three of the six parameters describing Standard Model neutrino mixing: $\theta_{13}$, $\Delta m^2_{21}$, and $|\Delta m^2_{31}|$~\cite{pdg_2018}.  
The observed discrepancies between measured and modelled reactor \nuebar fluxes~\cite{bib:mueller2011} has motivated new experiments and analyses that focus on probing the active-sterile mixing parameters $\Delta m^2_{41}$ and $\theta_{14}$~\cite{giunti_review}.  

As these measurements have improved in precision, they have also enabled a more detailed understanding of reactors as a source of \nuebar.    
The production of $\nuebar$ in a nuclear reactor core per second at time $t$, given in terms of neutrinos per unit energy, can be described as follows:
\begin{equation}\label{eq:abs}
\frac{d\phi(E_{\nu},t)}{dE_{\nu}} = \frac{W_{\mathrm{th}}(t)}{\overline{E}(t)}\sum_{i=1}^4f_{i}(t)s_{i}(E_{\nu}),
\end{equation}
where $W_{\mathrm{th}}$ is the core's thermal power output,
$f_i$ and $s_{i}(E_{\nu}$) are respectively the fission fraction and antineutrino flux from isotope $i$,
and $\overline{E}(t) = \sum_{i}f_{i}(t)e_{i}$ is the average energy release per fission, with individual fission isotope energy releases $e_i$.
Some of these inputs are computed directly from measurements of the core or its fuel: $W_{\mathrm{th}}$ is derived from real-time in-reactor measurements~\cite{bib:cpc_reactor}, while $f_i$ are determined by reactor simulations benchmarked to measurements of spent fuel  content~\cite{bib:science2010,bib:ff}.  
Other inputs are based on theoretical calculations.  
The energy released per fission $\overline{E}$ is primarily dependent on mass differences between fission isotopes and their products, and can be calculated with relatively little uncertainty based on existing nuclear databases~\cite{bib:fr_ma}.  
On the other hand, calculations of $s_{i}(E_{\nu})$ suffer from a variety of systematic uncertainties.  
The favored method performs conversion of measured aggregate $\beta$-particle spectra from each fission isotope~\cite{bib:ILL_1,bib:ILL_2,bib:ILL_3} into \nuebar spectra using energy conservation and various spectrum shape assumptions and corrections~\cite{bib:huber,bib:haag}.  
These spectrum inputs have sizable systematic uncertainties; nonetheless, this method serves as the standard method for calculating $s_{i}(E_{\nu})$.  
An alternate summation method calculates $s_{i}(E_{\nu})$ by adding the \nuebar produced by each fission daughter using their evaluated nuclear data (e.g. fission yields and $\beta$ decay properties)~\cite{bib:mueller2011,bib:fallot}; here, uncertainty is contributed by the incomplete and sometimes inaccurate nature of the inputs~\cite{bib:fallot,bib:dwyer,sonzongi2,tas_rb,tas_lots}. 
%Rather than relying on these uncertain methods, 
To gain further insight into the potential deficiencies of these methods, recent reactor \nuebar measurements have been used to directly determine \nuebar production by reactors and individual fission isotopes.  

Direct \nuebar measurements are usually reported in terms of the inverse beta decay (IBD) yield and energy spectrum per fission, $\sigma_{i}(E_{\nu}) = \sigma_{IBD}(E_{\nu}) s_{i}(E_{\nu})$, where $\sigma_{IBD}$ is the well-known cross-section for the inverse beta decay interaction used for detection in most reactor \nuebar experiments, 
\nuebar~+~$p$~$\rightarrow$~$e^+$~+~$n$~\cite{Vogel:1999zy}.  
Using results from cores of differing fuel composition, direct determination of isotopic IBD yields and spectra now approaches or exceeds the precision of the theoretically-calculated counterparts.  %., while energy spectrum measurements have recently begun to approach the precision of theoretical values.  
The IBD yield for \uFive{} has been measured to better than 1.5\% precision via historical measurements at highly \uFive{} enriched reactor cores~\cite{Giunti}, while measurements from commercial cores during periods of differing fuel content at Daya Bay~\cite{bib:prl_evol} produce better than 2.5\% and 6\% precision in \uFive{} and \pNine{} yields, respectively~\cite{bib:prl_235239}.  
Global fits of all IBD yield measurements produce 1.5\%, 14\%, and 4.5\% precision in production by \uFive{}, \uEight{}, and \pNine{}, respectively~\cite{giunti_diagnose}.  
Daya Bay has also provided measurements of IBD energy spectra from \uFive{} and \pNine{} fission below 9~MeV, with precision better than 5\% and 12\% over most of the relevant energy range~\cite{bib:prl_235239}.  
The PROSPECT experiment has recently performed the first high-statistics measurement of IBD  energy spectra at a highly \uFive{} enriched reactor, with precision better than 10\% between 1-6~MeV~\cite{prospect_spec}.  
These measurements confirm differing rates and energies of \nuebar production for the different fission isotopes, and provide improved justifications for and demonstration of capabilities in monitoring the status, power, and fuel content of nuclear reactors using their \nuebar emissions~\cite{hubermon1,hubermon2,bernmon1,bernmon2,Bernstein:2019hix}.  

Comparison of theoretical conversion predictions and direct \nuebar flux and spectrum measurements yields numerous inconsistencies.  
An overall deficit in measured IBD yields with respect to predictions of approximately 6\% is observed~\cite{AnomalyWhite,bib:mention2011}; this deficit is referred to throughout the rest of this paper as the `Reactor Antineutrino Anomaly.'
In addition, the size of this discrepancy is partially dependent on the fuel content of the reactors producing the observed \nuebar~\cite{bib:prl_evol}.  
Measured IBD energy spectra from numerous experiments are found to be in clear disagreement with conversion-based predictions~\cite{dc_bump,bib:reno_shape,bib:prl_reactor,bib:neos}.  
Recently improved summation models predict a smaller IBD yield excess and correct dependence of IBD yields with fuel content, but also cannot reproduce the measured IBD spectrum per fission~\cite{bib:fallot2}.  
%\todo{BLAH}: sentence about improved summation predictions.  
%The level of agreement between measured and \empha{ab initio} spectrum predictions is dependent upon input nuclear datasets and shape assumptions~\cite{bib:dwyer,sonzongi2,hayen_initio}, while absolute flux systematics for this prediction are too large for a meaningful comparison to the measured flux. 
%If these discrepancies are taken seriously, then they must be 
These discrepancies are indicative of a lack of understanding of neutrino production in nuclear reactor cores and/or their fundamental properties. % of neutrinos.  

%- Reactor flux anomaly in wide variety of experiments
%- Reactor evolution anomaly in LEU -- anomaly is largest when 235 fission fractions are highest.  
%- Reactor spectrum anomaly in a variety of experiments.  

As observed in previous experiments, reactor \nuebar undergo flavor transformations, or oscillation, as they travel from source to detection point, a quantum mechanical phenomenon resulting from the interacting flavor states being a superposition of underlying mass eigenstates~\cite{osc1,osc2,osc3}. 
When only one neutrino mass difference is involved, this oscillation probability $P_{dis} $can be described as
\begin{equation}\label{eq:osc}
\begin{aligned}
P_{\rm{dis}} = \sin^22\theta \sin^2 \left(1.27 \Delta m^2({\rm eV}^2) \frac{L({\rm m})}{E_\nu ({\rm MeV})}\right),
\end{aligned}
\end{equation}
where $\Delta m^2$ is the squared mass difference, $\theta$ is the mixing angle between the mass and flavor states, and $E_{\nu}$ and $L$ are the energy and travel distance (baseline) of the neutrino, respectively.  
For a reactor \nuebar experiment detecting neutrinos via IBD, this transformation manifests itself as a deficit in detection rates that varies with baseline and neutrino energy.  
According to Eq.~\ref{eq:osc}, a mass splitting on the order of 1~eV$^2$ or larger will manifest as an observed average deficit in energy-integrated reactor \nuebar detection rates with respect to predictions for reactor experiments with $L$ of order 10~m and larger~\cite{bib:mention2011}.  
This mass splitting is much larger than those associated with the three Standard Model neutrinos~\cite{pdg_2018}, requiring the existence of new neutrino mass states; to maintain consistency with existing collider physics measurements, these new states must be `sterile,' or incapable of interacting via the weak force~\cite{AnomalyWhite}.  
Demonstration of the existence and properties of such a particle would have far-reaching implications in particle physics and cosmology. %Citations here...?  

To unambiguously investigate whether these neutrino propagation effects contribute to the observed discrepancies between reactor \nuebar measurements and predictions, experiments must directly probe the baseline and neutrino energy dependence of reactor \nuebar signals.  Any deviation from 1/r$^2$ behaviour as a function baseline and energy would indicate an oscillation effect and provide the ability to infer the parameters describing such oscillation.  
This investigation can be performed using \nuebar energy spectrum measurements at multiple short ($\mathcal{O}$(10~m)) reactor-detector baselines~\cite{VSBL}.  
Historical and more recent measurements of short-baseline IBD energy spectrum ratios have excluded large regions of sterile oscillation parameter space~\cite{bib:Bugey3_osc,bib:neos,danss_osc,stereo_2018}.  
Using 33 days of reactor-on data-taking, the PROSPECT experiment has recently placed limits on sterile neutrino oscillations through relative comparison of measured IBD spectra between multiple baselines within a single stationary detector~\cite{prospect_osc}.  

The observed deviation between measured and predicted IBD energy spectra, as well as the dependence of measured IBD yield deficits on reactor fuel content, cannot be caused by neutrino oscillations, and are likely the result of incorrect modelling of the \nuebar flux~\cite{hayes_shoulder,hayesEvol}.  
New, more precise \nuebar measurements from reactors of differing fuel content will enable further study of the nature of this mis-modelling~\cite{vogel_review,surukuchi_flux} and facilitate improved understanding of \nuebar production by fission daughters.  
Of particular importance is understanding whether or not \nuebar spectrum and flux predictions are similarly incorrect for all four primary fission isotopes.  
Inconsistencies present in individual isotopes could direct additional scrutiny towards specific fission $\beta$ spectrum measurements~\cite{bib:ILL_calib}, corrections to be applied during the beta-conversion process~\cite{neutrons_shoulder,hayes_shape,hayen_initio}, or nuclear data for  fission daughters~\cite{tas_nb,tas_lots}.  
Considering isotopic IBD yields, global fits currently favor incorrect prediction of only \uFive{}~and \uEight~\nuebar fluxes, but inclusion of sterile neutrino oscillation effects clouds this picture~\cite{giunti_evol,giunti_diagnose}.  
For isotopic IBD energy spectra, Daya Bay and RENO both appear to show disagreement between measurement and prediction for \uFive{}, but they cannot presently determine whether other isotopes exhibit similar discrepancies~\cite{bib:prl_235239,reno_evol}.  
Meanwhile, the first PROSPECT measurement of the pure \uFive{}~IBD energy spectrum at the highly-enriched HFIR reactor core is consistent with Daya Bay's \uFive{} result, while also slightly disfavoring \uFive{} as being the sole isotope exhibiting a spectrum discrepant with its prediction~\cite{prospect_spec}.  

This paper will present new results from the PROSPECT experiment using an enlarged dataset including 96 (73) days of reactor-on (-off) data.  
Improved sterile neutrino oscillation search results and an improved measurement of the reactor \nuebar spectrum produced by \uFive{} fission will be described, in addition to reviewing in detail how the inputs and systematic uncertainties for these two different analyses are determined.  
Section~\ref{sec:exp} will describe the experimental layout and detector design.  
Sections~\ref{sec:calib} and~\ref{sec:ereco} will describe the detector calibrations and subsequent event and physics metric reconstruction, respectively.  
Section~\ref{sec:select} will then describe the selection and Monte Carlo-based modelling of IBD candidates, with background to this selection described in Section~\ref{sec:bkg}; selected IBD candidate datasets are then described in Section~\ref{sec:signal}.  
New oscillation and \uFive{} \nuebar energy spectrum measurements will be presented in Sections~\ref{sec:osc} and~\ref{sec:spec}, respectively, with concluding remarks given in Section~\ref{sec:summary}.

%%%%%%%%%%%%%%%%%%%%%%%%%%%%%%%EXPERIMENTAL 
\section{Experiment Description}
\label{sec:exp}

\begin{figure*}[hptb]
\includegraphics[trim = 0.0cm 0.0cm 0.0cm 0.0cm, clip=true, 
width=0.8\textwidth]{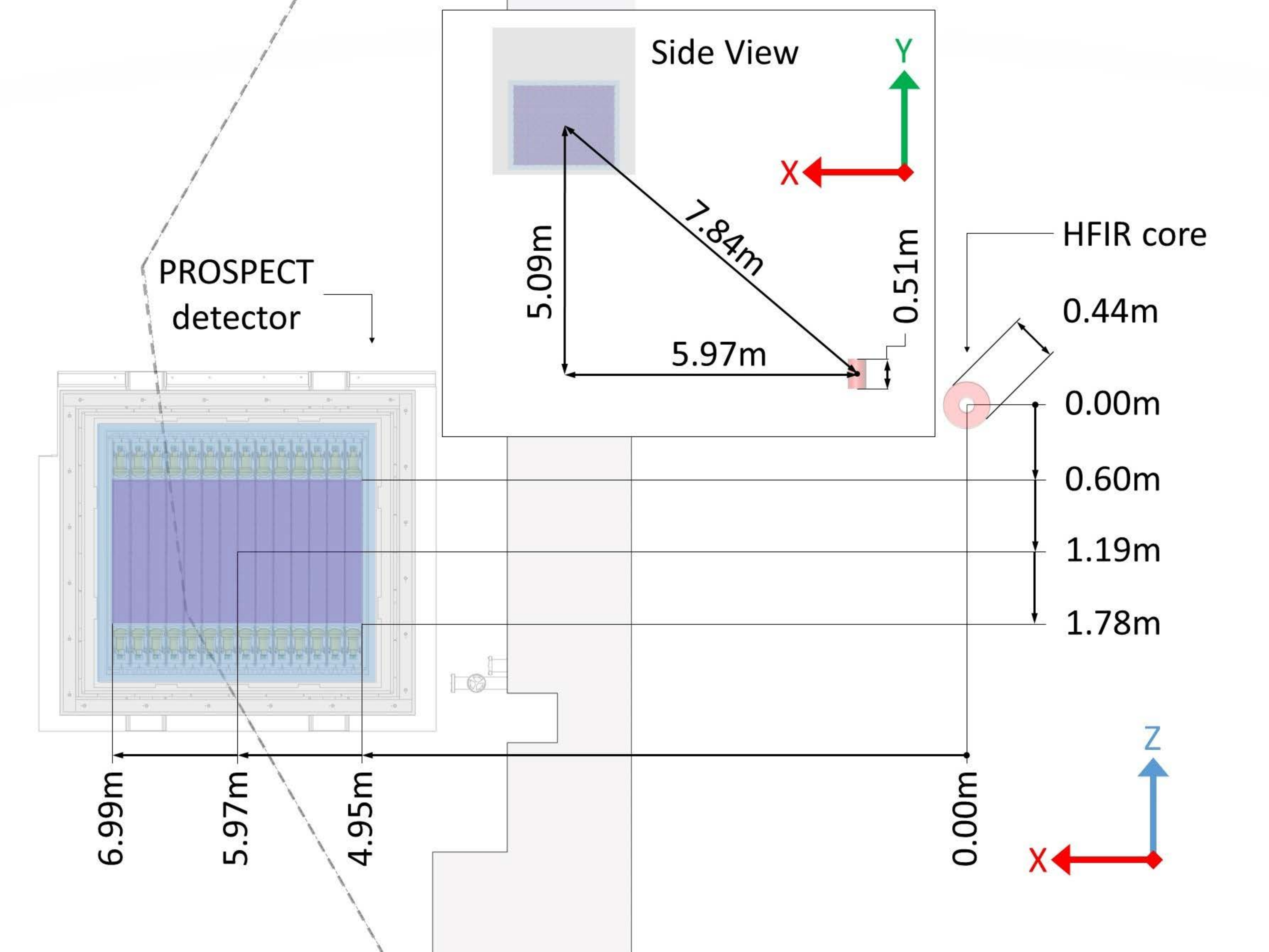}
\caption{A top and side view of the PROSPECT experimental layout in the HFIR building.  The HFIR reactor and PROSPECT detector package are illustrated, along with the reactor pool containment wall (gray) and the concrete monolith located underneath much of the detector package (dashed line).  Horizonal and vertical distances from the reactor core center to various detector locations are shown, as well as coordinate axes used to describe the orientation of the reactor and detector.  The floor on which the detector sits is parallel to the $x$-$z$ plane, and the zenith is in the +$y$ direction.}
\label{fig:layout}
\end{figure*}

The PROSPECT experiment is located at the High Flux Isotope Reactor (HFIR) facility at Oak Ridge National Laboratory in Oak Ridge, Tennessee.  
Among other factors, the high power and compact core of the highly \uFive{} enriched HFIR research reactor, the availability of unoccupied near-reactor floor space~\cite{prospect_reactor}, and the status of HFIR as a DOE user facility make it a favorable site for the PROSPECT detector.  
To probe disappearance caused by the existence of an eV-scale sterile neutrino, the PROSPECT detector must be located in close proximity to the HFIR core and without substantial overburden, necessitating an IBD measurement in an intrinsically  high-background environment.  
Moreover, demonstration of the $L/E_{\nu}$ nature of this disappearance requires the reconstruction of neutrino interaction locations and energies within the PROSPECT detector target~\cite{VSBL}.  
These requirements served as the primary drivers behind the PROSPECT experimental layout and detector design.  
A detailed description of these aspects of PROSPECT are provided in Ref.~\cite{prospect_nim}.  
The aspects of the experimental layout and detector design most relevant to performing a sterile neutrino search and absolute \nuebar{} spectrum measurement with PROSPECT are outlined below.  

\subsection{Experimental Layout}

%HFIR Core
The HFIR reactor is located at an elevation of roughly 250 meters above sea level in the HFIR building.  
The HFIR core contains two concentric cylindrical rings of \uFive{} fuel plates (93\% enrichment) with an outer diameter of 0.435~m and height of 0.508~m.  
The fuel is surrounded by an aluminum cladding and structural environment, which is in turn surrounded by a thick concentric cylindrical beryllium reflector.  
Each reactor cycle starts with fresh fuel and lasts for $\sim$24 days running at a nominal thermal power of 85~MW$_{th}$.  
The reactor core and pressure vessel are operated inside a large water pool whose surface is nominally eight meters above the midplane of the core.  
To enable more direct access to the reactor vessel during reactor-off operations, reactor pool water levels are occasionally reduced by 5~m for time periods no longer than a few days.  
Spent fuel elements are stored in an adjacent water pool $\mathcal{O}$(10~m) from the reactor core.  
A more detailed description of the HFIR core and facility is given in Ref.~\cite{conant_thesis}.  

The PROSPECT detector is located one floor above the HFIR core in a ground-level hallway running along the outer side of the pool walls, as illustrated in Figure~\ref{fig:layout}.  
The detector package, consisting of inner detector, liquid containment vessels, $\gamma$-ray and neutron shielding, and detector movement elements, is partially sited above a thick concrete monolith that significantly attenuates $\gamma$-ray and neutron backgrounds associated with Neutron scattering experiments located one floor below.   The operational cycles of these instruments is a source of non-negligible time-varying $\gamma$-ray backgrounds.  Lead walls built between the detector package and the reactor pool wall provide targeted  shielding of $\gamma$ radiation emanating from the reactor environment and unused neutron beam tubes.

The PROSPECT inner detector, which serves as the antineutrino target, is illustrated in Figure~\ref{fig:layout}, including size and orientation with respect to the HFIR core.  
The $x$, $y$, and $z$ coordinate system used to describe the orientation of detector and reactor throughout this work are also indicated in Figure~\ref{fig:layout}, with the center of the inner detector serving as the system's origin.  
The PROSPECT inner detector approximates a rectangular prism with dimensions of 2.045~m long (in $x$), 1.607~m tall (in $y$), and 1.176~m wide (in $z$).  
The inner detector center is displaced from the center of the reactor core by -1.19~m in $z$ and +5.09~m in $y$.  
The distance from the front-most (back-most) midpoint edge of the inner detector to the reactor center is 6.65~m (9.22~m), respectively, with a core-detector center-to-center distance of 7.93~m~\cite{conant_thesis}.  
Distances between the inner detector edges and the detector package exterior range from 40~cm on the detector sides and bottom to 100~cm on top of the detector. 
Detector distance from the reactor was determined with respect to a reference location on the detector package exterior with $\pm$10~cm estimated precision using  HFIR facility mechanical drawings and a measuring tape.  
Negligible additional baseline uncertainty is contributed from the knowledge of relative inner detector placement with respect to this detector-external reference point.  

\subsection{Antineutrino Detection Strategy}

\begin{figure*}[bthp!]
\includegraphics[trim = 4.3cm 23.0cm 6.3cm 6.5cm, clip=true, 
width=0.48\textwidth]{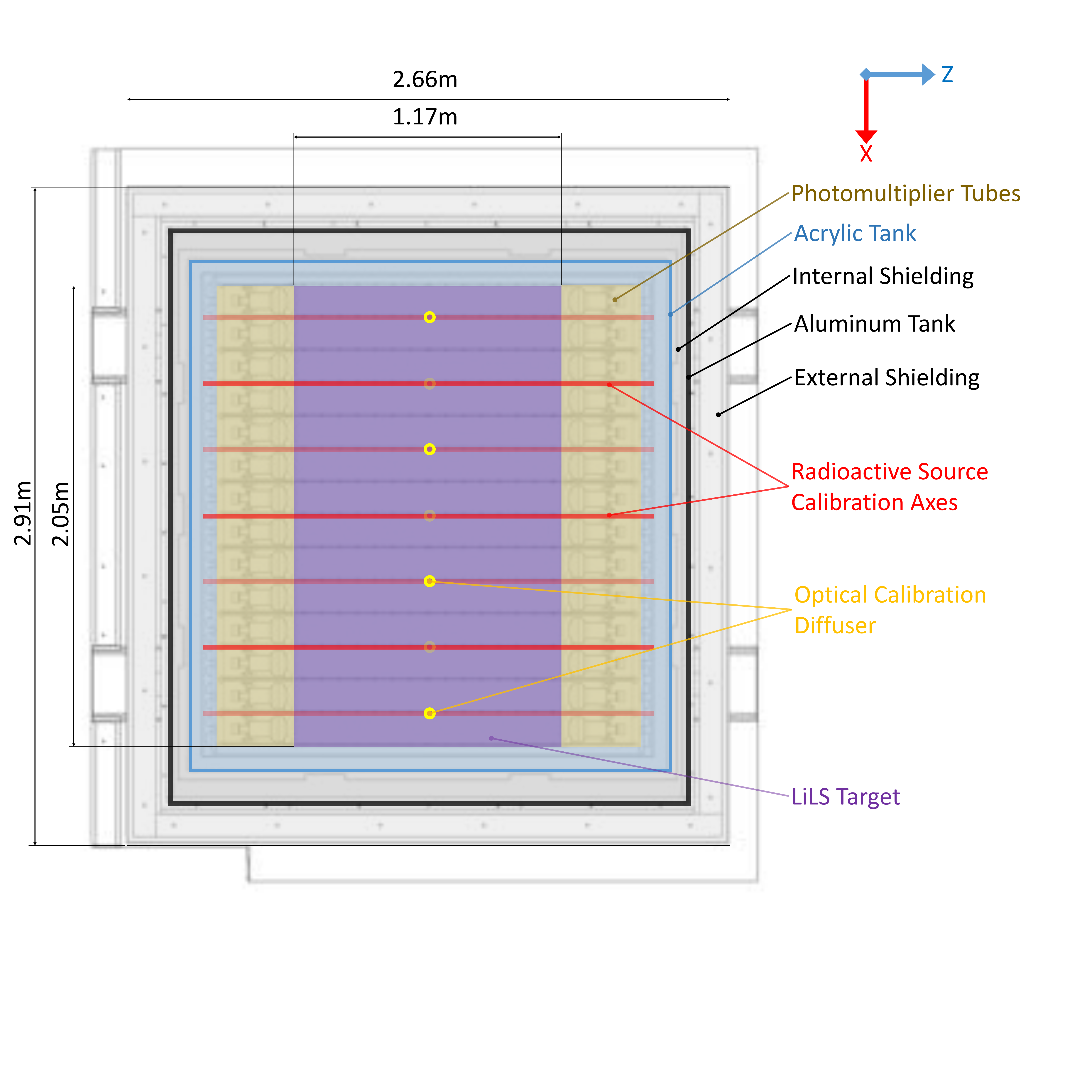}
\includegraphics[trim = 4.3cm 23.0cm 6.3cm 6.5cm, clip=true, 
width=0.48\textwidth]{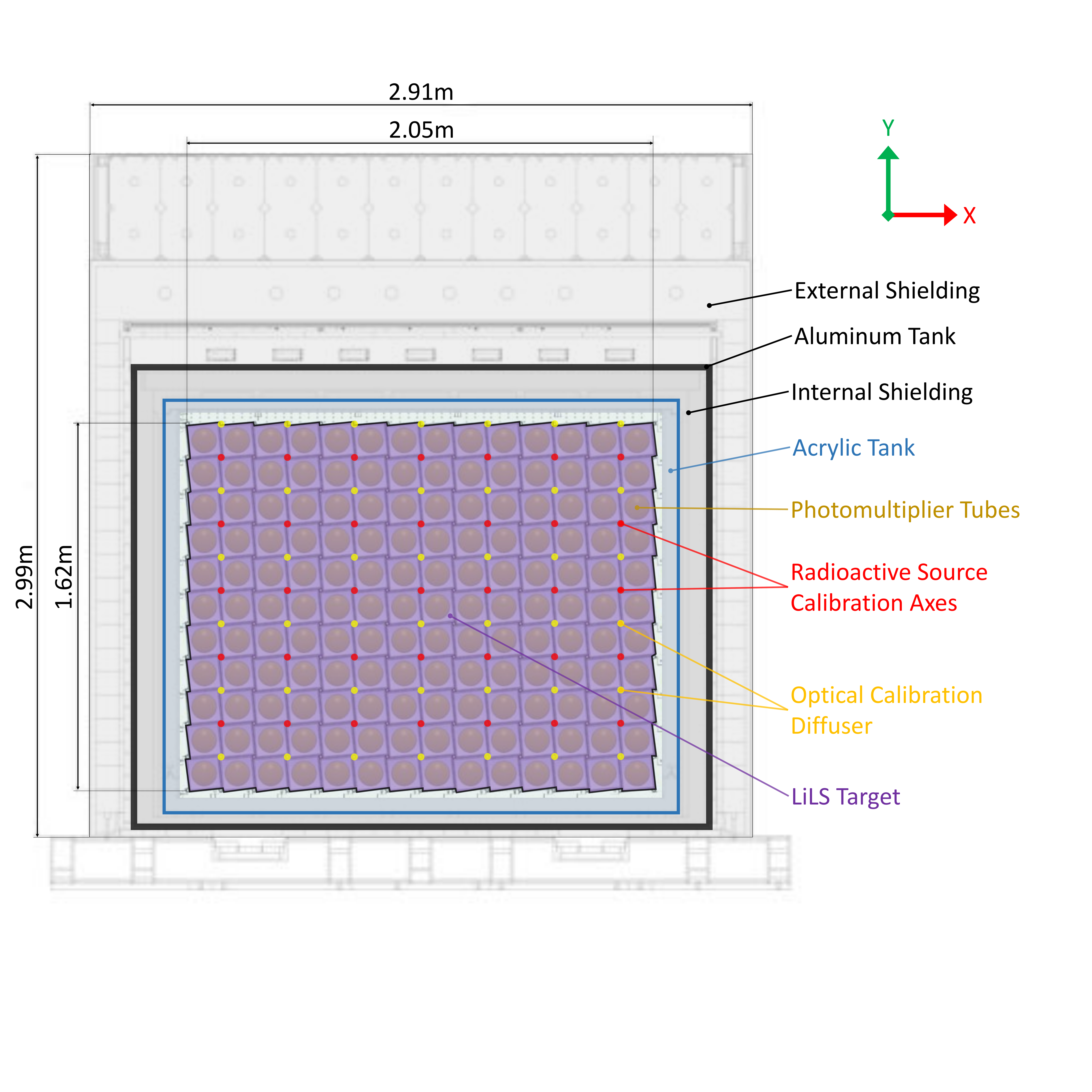}
\caption{A side- (right) and top-view (left) diagram demonstrating primary PROSPECT detector design features.  The coordinate axes are also shown.}
\label{fig:detector}
\end{figure*}

PROSPECT detects IBD \nuebar{} interactions with hydrogen nuclei in liquid organic scintillator comprising most of the volume of the inner detector.  
The IBD signal consists of time- and position-correlated energy depositions produced by an IBD positron and the capture of the IBD neutron on $^6$Li doped into the liquid scintillator.  
The IBD positron produces a signal with low ionization density and extended (tens of centimeters) topology due to the production of positron annihilation $\gamma$-rays. 
The energy deposited by the IBD positron, $E_p$, is related to the energy of the incoming $\overline{\nu}_e$: 
\begin{equation}\label{eq:IBD}
E_p = E_{\nu} - 0.78~MeV - E_n,
\end{equation}
with outgoing IBD neutron kinetic energy, $E_n$, of order 10~keV or less.  
The IBD neutron preferentially captures on $^6$Li within a few tens of centimeters and roughly 50~$\mu$s, producing $^3$H and $^4$He ions with 4.78~MeV of total kinetic energy due to the mass difference between parent and daughter nuclei.  
These products generate a compact ($\mu$m-range) mono-energetic energy deposit with high ionization density.  
Ionization signals from the liquid scintillator produce visible light, which can be collected and converted to an electronic signal by photomultiplier tubes.  
The PROSPECT inner detector is designed to capture the unique energy, energy density, spatial, and temporal signatures specific to IBD interaction products.  

%\todo{Describe PSD and role in detection strategy near here; reference \ref{fig:waveform}}

\subsection{PROSPECT Inner Detector}

The PROSPECT inner detector is pictured in Figure~\ref{fig:detector}.  
It contains four tons of pulse shape discriminating (PSD) liquid scintillator loaded to a mass fraction of 0.08\% with $^6$Li~\cite{prospect_p50,prospect_ls}.  
This scintillator is sub-divided into 154 14.5\,cm\,$\times$14.5\,cm\,$\times$117.6\,cm optically isolated segments of rectangular cross-section by an optical grid composed of thin (1.18 mm) specularly reflecting panels held together by white, diffusely reflecting, hollow 3D-printed polylactic acid (PLA) plastic support rods.
%an array of thin (1.18~mm) specularly reflecting panels and white diffusely reflecting hollow PLA plastic support rods~\cite{prospect_grid}.  
The grid permits the liquid scintillator to fill the entire volume of all segments.  
One subset of hollow support rod axes are instrumented to allow the use of removable radioactive calibration sources, while another is equipped with stationary optical sources, as indicated in Figure~\ref{fig:detector}.  
Un-instrumented axes are filled with square acrylic rods.  
The inactive materials composing the optical grid and calibration sub-systems comprise 3.5\% of the mass of the scintillator contained in the 154 active segments.  

The long axis of each segment is oriented along $z$, running parallel to the front reactor-facing side of the detector, and is bounded on either end by a mineral oil-filled acrylic box containing a 5'' photomultiplier tube (PMT), a magnetic shield, a light reflector, and a support structure.  
240 housings contain one Hamamatsu R6594 PMT, while 68 segments on the inner detector top and side edges contain one ET 9372KB PMT.  
Individual PMT housings are mechanically connected to one another and to acrylic supports running along the inner detector $z$ axis outside the outer rows of segments; this rigid support structure ensures mechanical integrity of the inner detector and maintains consistent target and segment dimensions.  
To achieve better dimensional uniformity, during detector dry assembly, segment dimensions were measured to mm-scale precision with metrological surveys.  
The inner detector and support structure are contained within a rectangular acrylic vessel under continuously flushed nitrogen cover gas inside a water-filled aluminum tank providing secondary containment of detector liquids.  
Ultrasonic sensors above two corners of the detector target monitor the scintillator liquid level above the top row of detector segments to sub-millimeter precision.  
Additional sensors monitor temperatures in the scintillator and surrounding detector regions,  as well as humidity and pressure in the cover gas region.  

%%%%%%%%%%%%%%%%%%%%%%%%%%%%%%

%Michael's DAQ sub-section
\subsection{Readout, Triggering, Data Acquisition, and Storage}
\label{subsec:daq}

\begin{figure}[hptb]
\includegraphics[trim = 0.0cm 0.15cm 0.0cm 0.25cm, clip=true, 
width=0.48\textwidth]{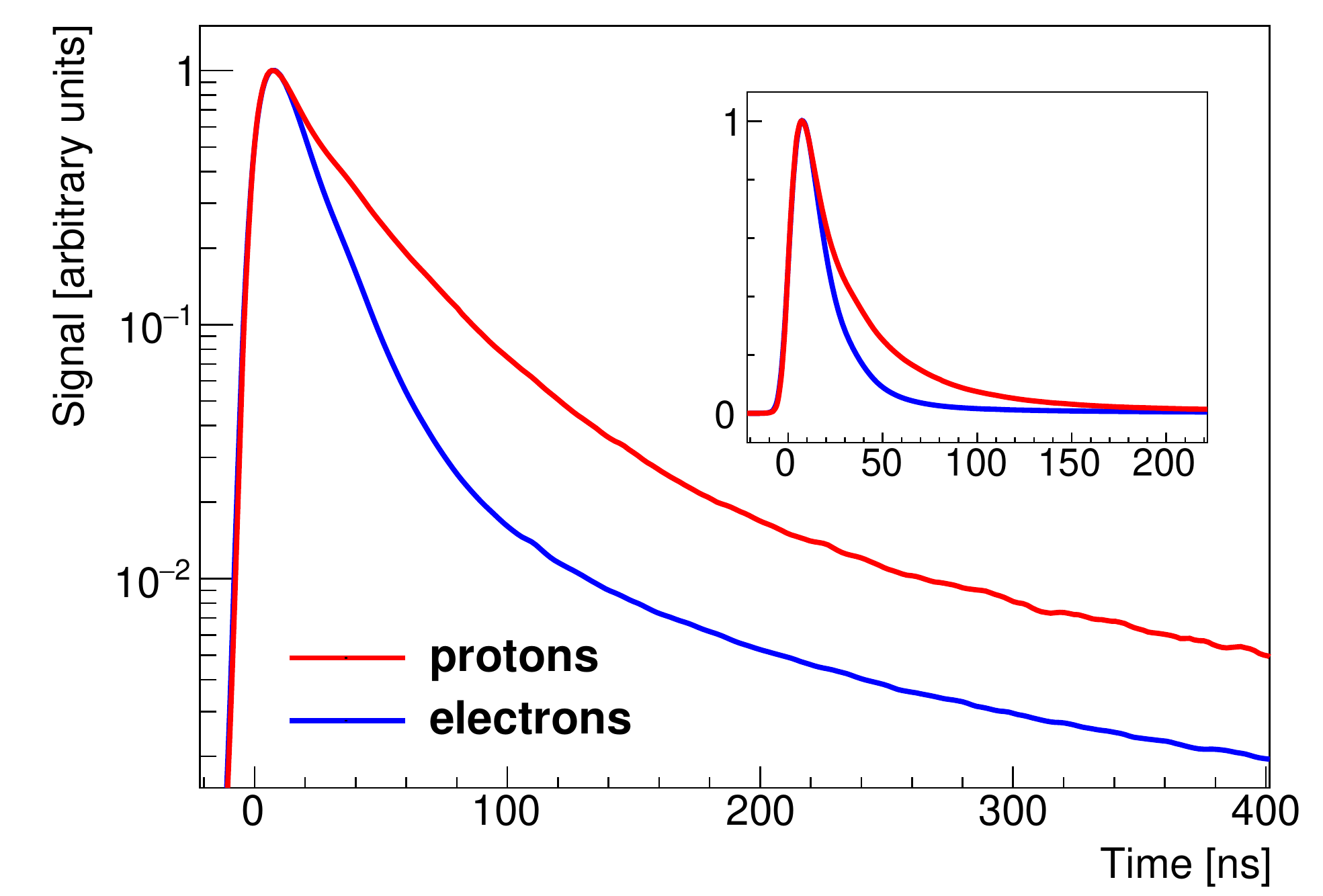}
\caption{Averaged waveforms typical for low-energy-density (electron) tracks (lower, blue), and high-energy-density (proton recoil) tracks (upper, red). Electron-like pulse shapes are independent of energy, while recoil pulses have varying proportions of the longer-time tail component (asymptotically approaching the electron track shape in the high energy, minimum-ionizing-density limit). The inset panel shows the same waveforms on a linear scale.}
\label{fig:waveform}
\end{figure}

Scintillation light produced in a detector segment via interaction of ionizing particles is efficiently transported by the reflecting walls to the corresponding PMTs, whose analog responses are individually processed by CAEN V1725 250 MHz 14-bit waveform digitizer (WFD) modules\footnote{Certain trade names and company products are mentioned in the text or identified in illustrations in order to adequately specify the experimental procedure and equipment used. In no case does such identification imply recommendation or endorsement, nor does it imply that the products are necessarily the best available for the purpose.}.  
The shape of digitized waveforms are primarily determined by the timing characteristics
    of the PROSPECT scintillator, photon transit time dispersion in the segment,
    photoelectron transit times in the PMTs,
    and impedance mismatch in the connections and cabling en route to the detector-external digitizing electronics.  
Scintillator light output is from a combination of processes characterized by different exponential decay times.
Ionization density affects the relative fractions of these processes,
    causing differences in pulse shape between light and heavy ionizing particles.
Lower-ionization-density events such as electron tracks are dominated by a 16\,ns scintillator decay time, while a 38\,ns component increasingly affects  higher-ionization-density ({\em e.g.}, proton recoil) events, with a small contribution from a 225\,ns tail.
Averaged PROSPECT waveforms representative of electron and proton interactions are illustrated in Figure~\ref{fig:waveform}.

The PROSPECT detector implements a trigger configuration and zero-suppression scheme that enables unbiased readout of all energy depositions above $\sim$200~keV in energy despite operating in a challenging background environment.  
A data acquisition (DAQ) trigger starts with a pair of PMTs on the same segment producing a signal 50 ADC channels (~5 photoelectrons) above baseline within 16\,ns of each other.
The pairwise logical AND for every segment is combined via a logical OR operation at the WFD board level (up to 8 pairs). 
A resulting logic output signal from each of 21 WFD boards is further combined via logical OR by a Phillips Scientific 757D Fan-in/Fan-out NIM module,
    modified by the manufacturer for 32-in-to-32-out operation.
This logical OR of all segment pairs defines the DAQ global acquisition trigger, which is fanned out to the acquisition trigger input of each WFD.
The global trigger rate is $\sim 5\cdot10^3/$s when the HFIR reactor is not operating (`reactor-off') and $\sim 2\cdot10^4/$s when it is on (`reactor on'), with the latter dominated by $\gamma$-ray backgrounds related to the reactor's operation.  

On receipt of the global trigger signal, the WFD records a 592~ns (148-sample) data sequence for each channel,
    including $\sim$200~ns preceding the trigger signal.
New events arriving within 592~ns of the initial trigger do not re-trigger the DAQ,
    resulting in a deadtime after each trigger during which light pulses may be
    recorded in the waveform but are truncated at the end of the sampling sequence ---
    an $\mathcal{O}$(1\%) deadtime effect, depending on total trigger rate.

\begin{figure}[hptb]
\includegraphics[trim = 0.0cm 0.20cm 0.0cm 0.25cm, clip=true, 
width=0.49\textwidth]{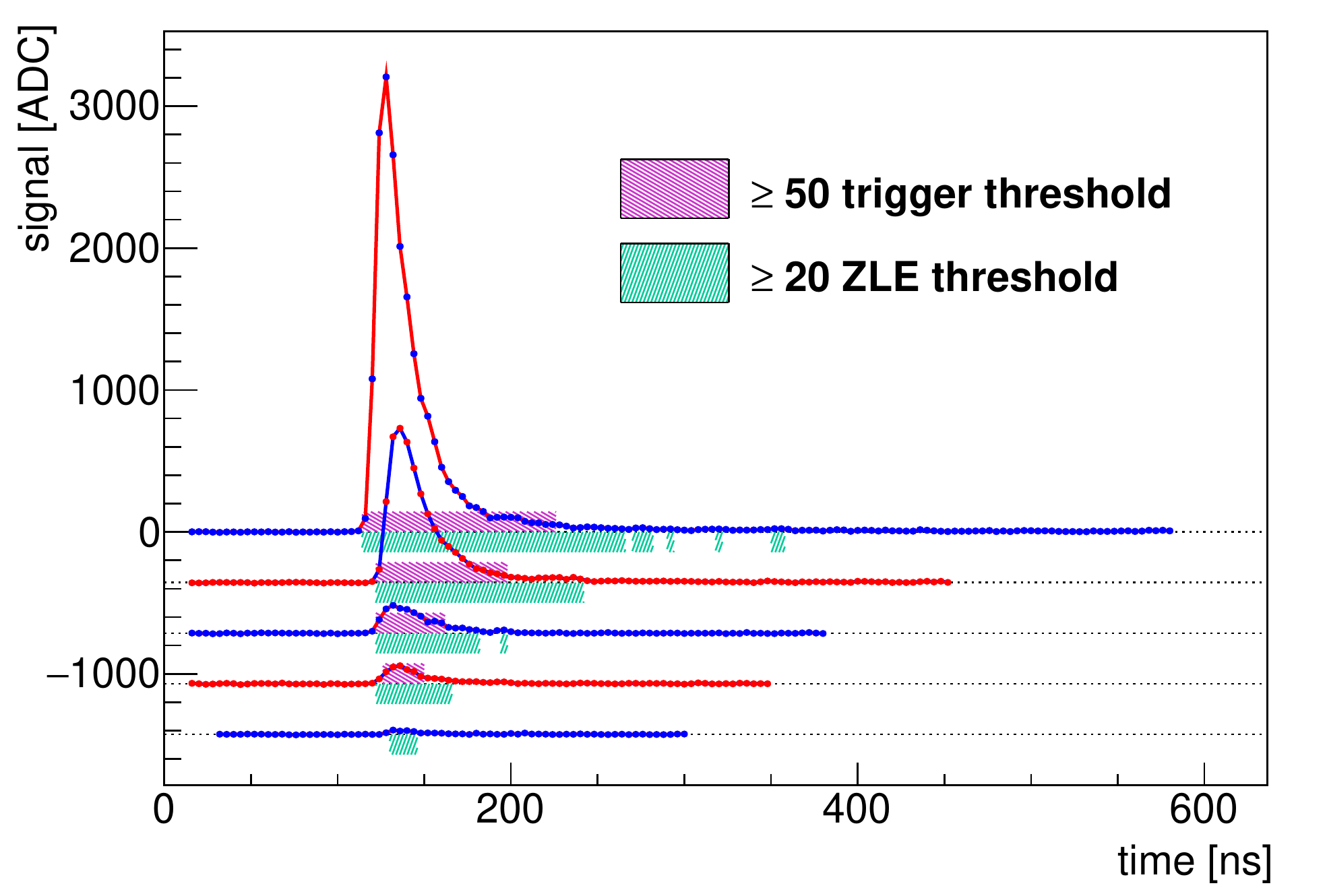}
\caption{Example DAQ trigger readout (y axis offset for clarity).  Pictured waveforms are inverted and baseline-subtracted with respect to the raw DAQ output; see Section~\ref{subsec:low} for  details on low-level waveform processing.  
Blue (red) waveform datapoints correspond to PMT channels at high (low) $z$.  
Pink and green highlighted waveform regions are those above the 50~ADC and 20~ADC trigger and ZLE thresholds, respectively.  The global trigger is generated by the first segment pair coincidence above trigger threshold (top two waveforms). All of the pictured PMT channels would have portions of their waveforms read out.}%; only the green-highlighted portions of 5 displayed example traces are retained after ZLE application.}
\label{fig:triggering}
\end{figure}

Events may produce light in one or more segments, with a typical multiplicity of $<$5\% of all segments.
To greatly reduce the data volume transferred, the WFDs' firmware applies ``Zero Length Encoding'' (ZLE) to suppress empty signals.
A ZLE threshold of 20 ADC channels above baseline (2 photoelectrons) is applied to the waveform data,
    and sections more than 24 samples before or 20 samples after the nearest above-threshold value are suppressed
    (possibly including all 148 samples).  
This reduced stored data volume is illustrated in Figure~\ref{fig:triggering}.  

The post-ZLE waveforms are transferred by CAEN CONET2 optical fiber from the WFDs to two four-link
    CAEN A3818 optical fiber cards in separate DAQ computers (with two or three WFDs daisy-chained for readout on each fiber link).
Each link is handled by an independent readout process for maximum parallel throughput.
Without ZLE, the 85~MB/s bandwidth of each fiber link would be the main data-rate-limiting bottleneck.
With ZLE, data rates are $\lesssim 10\%$ of the fiber capacity, with the limiting factor being a fast readout cycle
    before the maximum event buffer size of the WFDs (1023 triggers) overflows.
Testing indicates that the DAQ falls behind on readout (resulting in data loss) at trigger rates $\gtrsim 9\cdot 10^4$/s.

The DAQ computers send the data streams to a 20\,TB RAID-6 disk array over the local 10\,Gb/s ethernet network,
    and to a ``real-time'' analysis process generating monitoring plots for a detector status webpage.
Data are recorded in the binary format produced by the boards, slightly modified with extra header blocks
    and removal of fully-ZLE-suppressed waveform headers, with gzip compression.
The data are transferred from the RAID array for analysis and archiving on Oak Ridge and Livermore National Laboratory computer clusters,
    with the RAID array providing $\gtrsim 2$ weeks storage buffer capacity in the event of network outages to the remote facilities.

\subsection{Physics Datasets}
\label{subsec:physdata}

The analysis described in this paper uses data taken with the PROSPECT detector from March 5, 2018 to October 6, 2018.  
During this time period, which spanned five HFIR fuel cycles, the PROSPECT detector was in physics data-taking mode for 183 days; the HFIR reactor was on (off) for 105 (78) of these days.  
Calibration data-taking accounted for an additional eight calendar days of data-taking.  
Physics data for eight (five) of these reactor-on (-off) calendar days were not used for physics analysis due to PMT HV or other data quality issues not identified until after data-taking.  
In total, 95.6 (73.1) calendar-days of reactor-on (-off) data passing all quality checks were used for the physics analyses described in this paper.  

To provide improved background rejection a 106 segment inner fiducial volume is defined.  IBD events reconstructed in all outer segments and two inner segments in the bottom back corner of the detector (high $x$ and low $y$) are not included in the IBD candidate dataset.  PMTs in 64 of 154 detector segments ultimately exhibited current instabilities during physics data-taking.   These segments, comprising 42\% of the total active detector volume, were not used in the physics analyses described in this paper.   Of these, 36 were amoung the fiducial segments considered in the IBD selection criteria (described in Section~\ref{sec:select}).  This corresponds to 34\% of the fiducial volume.  
%A total of 70 fiducial and 20 non-fiducial cells are used for the oscillation and spectrum results.  
The impact on the data analysis is described in detail in~Section~\ref{subsec:eff}.  

%%%%%%%%%%%%%%%%%%%%%%%%%%%%%%
\section{Low-Level Data Processing, Calibration, and Event Reconstruction}
\label{sec:calib}

PROSPECT data is analyzed to select antineutrinos interacting via inverse beta decay in and around the inner detector volume.  
This selection involves analysis criteria on the reconstructed timing, position, energy, and pulse shapes of signals collected from the active segments of the detector.  
For the PROSPECT sterile neutrino oscillation analysis, establishing consistent energy scales between segments is essential.  
For the \uFive~spectrum measurement, accurate determination of the relationship between true antineutrino energy and reconstructed energy is important.  
The following section describes how raw digitized waveforms are processed to reconstruct and calibrate each of these quantities for PROSPECT physics analyses.  

%%%%%%%%%%%%%%%%%%%%%%%%%%%%%%%
\subsection{Pulse Definition and Low-Level Metrics}
\label{subsec:low}

Raw waveform data is initially stored to disk by the DAQ in the proprietary binary format
    produced by the CAEN V1725 digitizer cards, 
    slightly modified to include additional run header information 
    and strip out data blocks containing no waveforms.
This process produces eight separate files, written to disk in parallel, 
    each containing the output of two or three V1725 cards 
    sharing a common optical fiber link to the DAQ readout.
The parallel readout scheme facilitates uninterrupted data throughput, 
    necessary to prevent data loss from buffer overflows of the on-board memory of each digitizer card.
The separate raw readout files are later collated in time sequence into a single ROOT file~\cite{ROOT},
    with hardware board/channel numbers mapped to a 
    channel numbering scheme by segment number.

\begin{figure}[hptb]
\includegraphics[trim = 0.0cm 0.25cm 0.0cm 0.25cm, clip=true, 
width=0.49\textwidth]{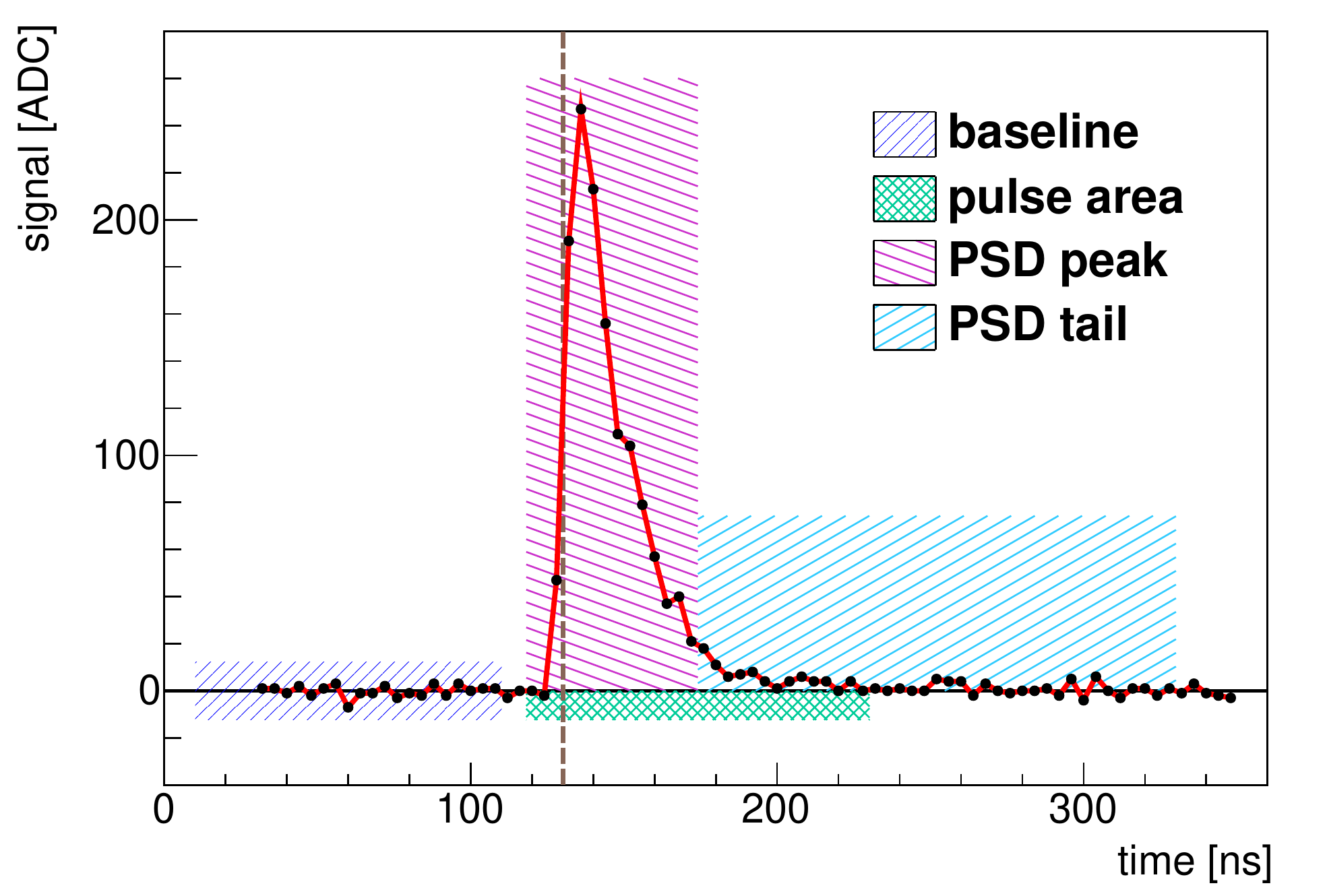}
\caption{Example analysis of a typical (smaller signal) electron pulse. The half-height leading edge timing (dashed vertical) determines windows for baseline subtraction, pulse area, and PSD.  The pictured waveform has been inverted and baseline-subtracted with respect to the raw DAQ output.}
\label{fig:pulseintegrals}
\end{figure}

The waveform file is then analyzed to locate and characterize pulses.
Each waveform is represented by a sequence of 14-bit integer ADC samples
    for contiguous 4\,ns digitization intervals.
The negative-polarity waveform is inverted so higher sample values
    indicate larger charge signals.
One global maximum sample and any number of local maxima 
    (separated by at least 20 samples from any higher point)
    are identified as initial pulse candidates.
The waveform baseline is calculated from the average of the median 8 samples
    in the range of 5 to 30 samples before the global maximum.
This baseline value is subtracted from all samples in the waveform
    for subsequent analysis.
The global maximum, and any local maxima at least 30 ADC units above the baseline,
    are considered as pulses for further analysis.  
    One such pulse is shown in Figure~\ref{fig:pulseintegrals} to visually illustrate the quantities of interest for each selected pulse.  

Each pulse's area $S$ is calculated from the sum of samples
    in the range from 3 samples before to 25 samples after the maximum location.
The pulse's arrival time $t$ is determined by scanning backwards
    from the pulse's maximum sample location to the first level-crossing
    at 50\% of the maximum.
The arrival time is linearly interpolated to the 50\% point
    between the two samples bracketing the level crossing.
    
A pulse shape discrimination (PSD) value is calculated for each pulse as the
    ``tail-over-total'' ratio of pulse areas
    between 11 and 50 samples after $t$ to the area
    between 3 samples before and 50 samples after $t$,
    integrated assuming trapezoidal interpolation between samples.
This choice of integration windows was selected to maximize the PSD
    figure-of-merit for discriminating neutron captures from
    similar-energy $\gamma$-ray interactions.

The time-ordered list of analyzed pulses found in each waveform --
    arrival time $t$, area $S$, PSD, along with baseline $b$ 
    and peak height $h$ -- 
    is written to an HDF5 table format file.

\subsection{Pulse Clustering and Pairing}
\label{subsec:ppairs}

The next stage of analysis uses the HDF5-format pulse data to extract
    low-level calibration constants from ambient background events.
These calibration constants are stored to a calibrations database,
    to be used in a later pass converting the pulse data
    to calibrated physics metrics involving ionization energy, time, and positions.

Prior to performing calibration procedures, however, pulse data are grouped into ``clusters'' of pulses nearby in time, defined as having arrival times between subsequent pulses separated by no more than 20\,ns.  
Within the cluster, pulses are paired between PMTs on opposite sides
    of the same segment.  
Segment pulses without a matching pair -- either because the other channel
    was turned off, or the signal fell below acquisition thresholds
    on the opposite side -- are retained by the data processing infrastructure,
    but are excluded from subsequent calibrated data analysis for results shown in this paper.  
Paired pulses are processed and combined to produce
    calibrated physics values describing the interaction producing the collected waveform in that segment:
    its time, position, visible energy deposition, and PSD.
    
%%%%%%%%%%%%%%%%%%%%%%%%%%%%%%%
\subsection{Timing Calibration}
\label{subsec:time}

The pulse arrival time variables $t_0^i$ and $t_1^i$
    for the two PMTs on segment $i$ are transformed into
    a conjugate pair of variables:
    a segment hit time $t^i = \frac{1}{2}(t_0^i + t_1^i)$,
    and a timing difference $\delta t^i \equiv t_1^i - t_0^i$.
The segment hit time is, to first order, independent
    of ionization position along the segment, as increased
    light transport time to one end cancels decreased
    transport time to the other.  
The differential time is independent of absolute event
    time in the run, and strongly correlated with
    hit position along the segment.  
    
Relative timing offsets between channels arising from electronics effects and cable length variations 
    are calibrated out using through-going cosmogenic muon tracks.
Candidate muon tracks are identified by a pulse ADC area ($S$) sum above $10^5$ and at least 4 paired segments.
Muon tracks crossing the full width of a segment produce signals which exceed
    the dynamic range of the digitizer, resulting in saturated waveforms
    with nonlinear degraded energy and timing information for energy depositions
    above $\sim15$\,MeV.
However, shorter ``corner-clipping'' track sections produce a range of
    well-formed waveform signals.
Muon statistics are sufficient to calibrate timing on a run-by-run basis: typically one hour, but sufficient even for five minute calibration source runs.  
    
Muon tracks provide signals across multiple segments with approximately simultaneous
    origin times, up to the muon transit speed through the detector.
Muon transit time is estimated from a Principal Components Analysis (PCA)
    trajectory fit to the pulse pair data.
For each pair $i,j$ of segments in the event with ``corner-clipping''-range
    signals, mean and variance of the segment-to-segment
    distributions $T^{ij} \equiv t^i - t^j - t_\mu^{ij}$ 
    and $\delta T^{ij} \equiv \delta t^i - \delta t^j$ are tallied,
    where $t_\mu^{ij}$ is the estimated muon transit time between segments.

The collection of averaged $\overline{T}^{ij}$ and $\overline{\delta T}^{ij}$ values
    defines an overdetermined linear system of equations for segment-to-segment timing offsets,
    up to a common-mode offset.
This system is solved using the least-squares method to determine average timing offsets $\overline{t}^i$
    (with common-mode constraint $\sum_i \overline{t}^i = 0$)
    and differential offsets $\overline{\delta t}^i$ for each segment.
These timing values, saved to the calibration database, are subtracted from
    the raw $t^i$ and $\delta t^i$ values for a pulse pair
    to yield the reconstructed event time $t_{rec}^i \equiv t^i - \overline{t}^i$
    and position-dependent $\Delta t^i \equiv \delta t^i - \overline{\delta t}^i$.

\begin{figure}[hptb!]
% figures from DocDB 2314
\includegraphics[trim = 0.0cm 0.15cm 0.0cm 0.4cm, clip=true, 
 width=0.49\textwidth]{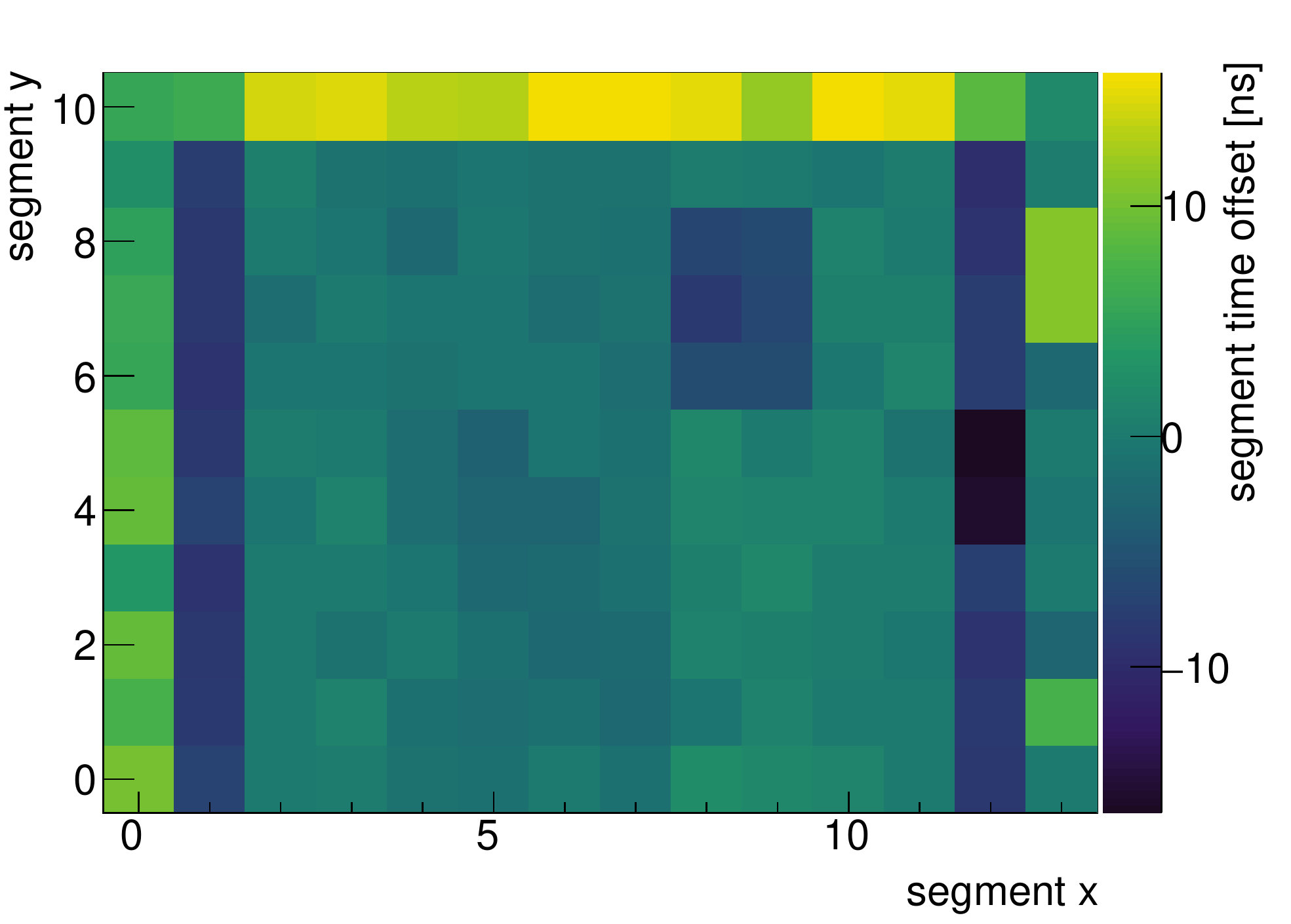}
\includegraphics[trim = 0.0cm 0.15cm 0.0cm 0.4cm, clip=true, 
width=0.49\textwidth]{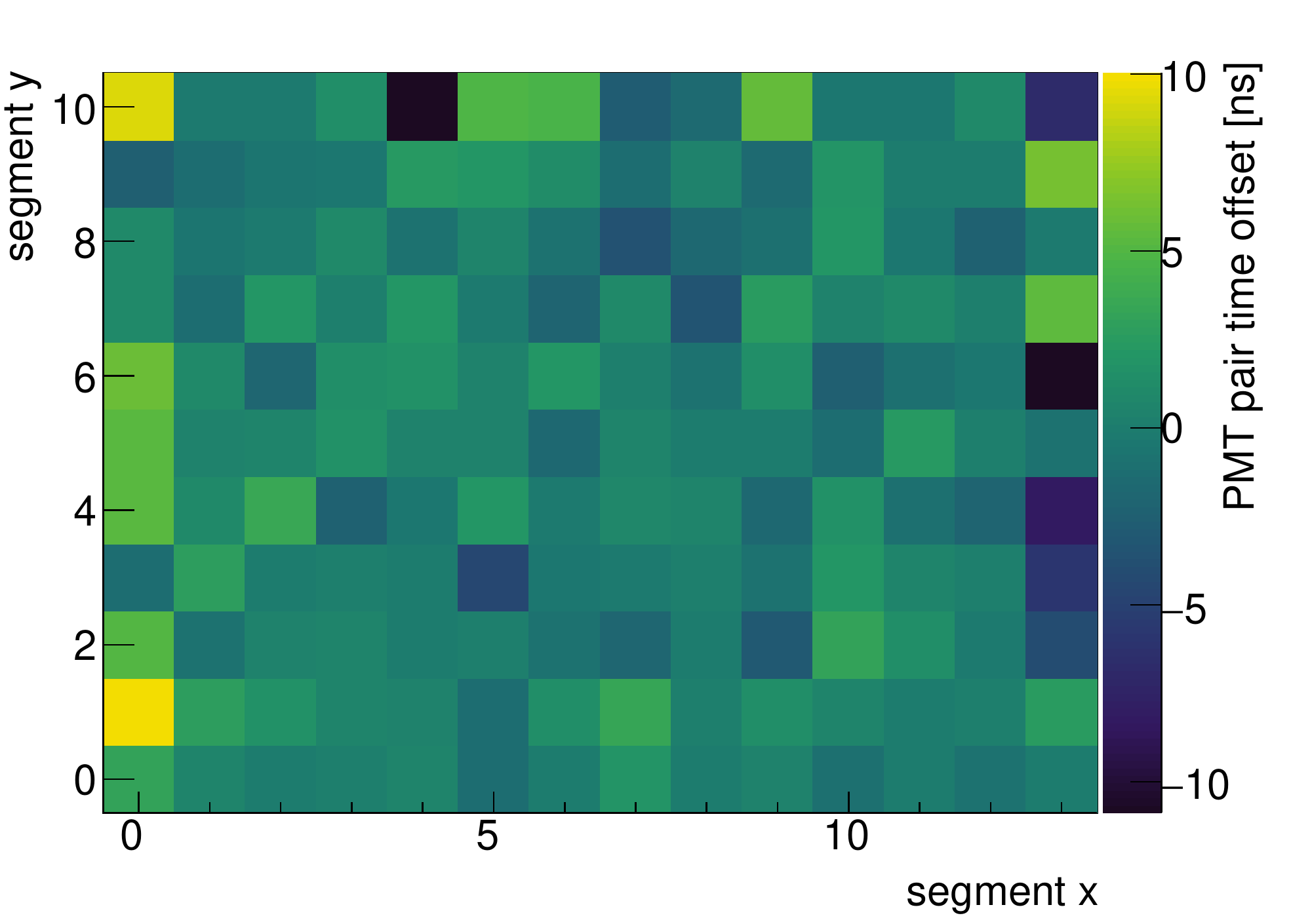}
\caption{Segment timing calibration constants extracted from one March 2018 physics run containing no inactive PMT channels; not all pictured channels are used in the final IBD selection.  Segment (0,0) is closest to the reactor core.  
Top: $\overline{t}^i$ segment time offsets; large-scale features visible from
    board clock offsets and systematic difference between ET and Hamamatsu PMT transit times.
Bottom: $\overline{\delta t}^i$ PMT pair offsets. More transit time variation is seen between ET PMTs.}
\label{fig:timing}
\end{figure}

Figure~\ref{fig:timing} shows the segment timing calibrations extracted from a typical one-hour run.
Timing differences $\lesssim 10\,$ns arise mainly from differences in PMT transit times,
    with systematic offsets between the ET and Hamamatsu PMT models,
    plus board-to-board clock $t_0$ offsets in discrete 8\,ns intervals.
Board-to-board $t_0$ offsets are prone to vary run to run;
    modulo this effect, the extracted timing calibration offsets have
    a run-to-run scatter $\lesssim 20$\,ps, and long-term drifts $< 1$\,ns over months.

%%%%%%%%%%%%%%%%%%%%%%%%%%%%%%%
\subsection{Combined PSD Parameter}
\label{subsec:psd}

To produce a single pulse PSD value, the PSD values from the two channels in a segment pair are corrected to remove residual position dependence and then statistically combined.  

Position variation of the PSD observed by each PMT for minimum-ionizing event tracks
    is mapped out using the corner-clipping muon hits also used for timing.
The PSD tail fraction is observed to increase with distance from the PMT,
    explicable by a wider spread in photon transit distances to the photocathode
    delaying light further from the shortest-path arrival edge.
The observed distribution is empirically fit as a function of $\Delta t$,
    for each segment in each run, with a three-parameter curve $p \cdot (1 + d \cdot [1-e^{k\Delta t}])$.
    
\begin{figure}[hptb!]
\includegraphics[trim = 0.0cm 0.05cm 0.0cm 0.0cm, clip=true, width=0.49\textwidth]{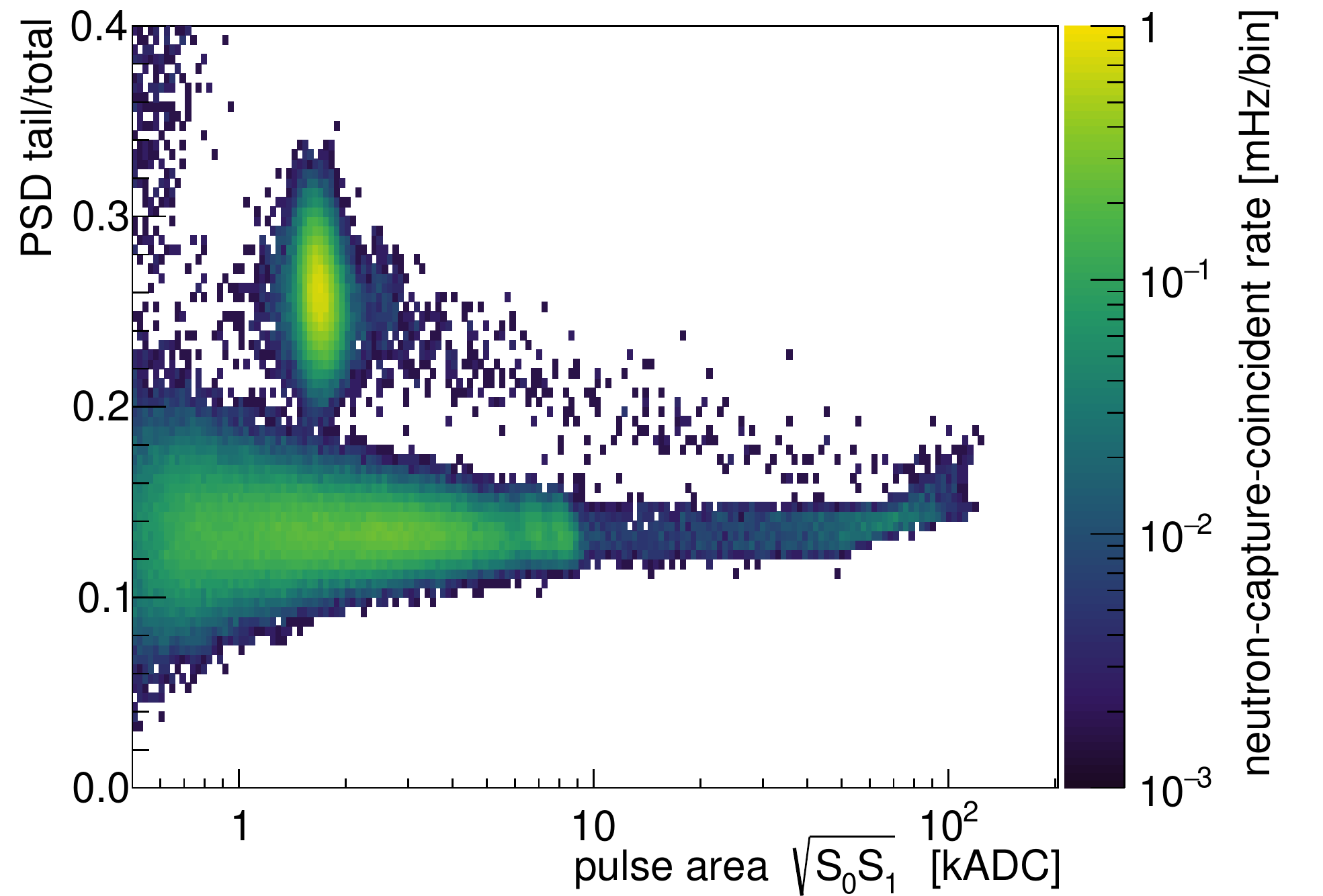}
\caption{
    Calibrated pulse PSD value versus signal amplitude for one calendar-hour of pulses occurring in time coincidence with cosmic muon signals.  
    Amplitude is defined as the product of pulse areas $S_0$ and $S_1$ for that segment.  
    Neutron capture signals on $^6$Li are clearly visible in a localized region of amplitude and high PSD, above a band of $\gamma$-produced signals of low PSD.  
}
\label{fig:sigpsd}
\end{figure}

The measured position dependent component $p \cdot d \cdot [1-e^{k\Delta t}]$
    is subtracted off of the PSD from each pulse,
    leaving a distribution centered around $p$ for electron-like events, and a higher but still position independent distribution of high-ionization-density events.
The two position-corrected PSD values for each pulse are averaged together, weighted
    by the estimated number of photoelectrons in each pulse,
    into a single PSD value.  
    Figure \ref{fig:sigpsd} shows a calibrated PSD distribution for pulses occurring after candidate muon tracks, which include a large population of ${^6}$Li-captured spallation neutrons.   
    Calibrated pulse PSD values are plotted versus uncalibrated signal amplitude -- defined as the product of pulse areas $S_0$ and $S_1$ for that segment's low-$z$ and high-$z$ PMT channel, respectively.  

The $p,d,k$ PSD values track the long-term changes in detector light transport.
While the position-dependent terms $d$,$k$ are calibrated out,
    the long-term variation in $p$, trending towards lower values as
    increased attenuation filters out longer
    light paths, remains.  
Rather than calibrating it out, the time-varying value of $p$
    is used for defining PSD cuts.
The center of the PSD distribution for n+$^{6}$Li capture events is also tracked
    for use in neutron capture identification cuts.  

%%%%%%%%%%%%%%%%%%%%%%%%%%%%%%%
\subsection{Position Calibration}
\label{subsec:position}

Both the relative timing and relative signal amplitude between PMTs
    provide information about the position of events along the segment length.
A position estimate is calculated both from timing $\Delta t$
    and from the log ratio of pulse areas $R \equiv \ln S_1/S_0$.  
    
\begin{figure}
\includegraphics[trim = 0.0cm 0.5cm 0.cm 0.0cm, clip=true, width=0.49\textwidth]{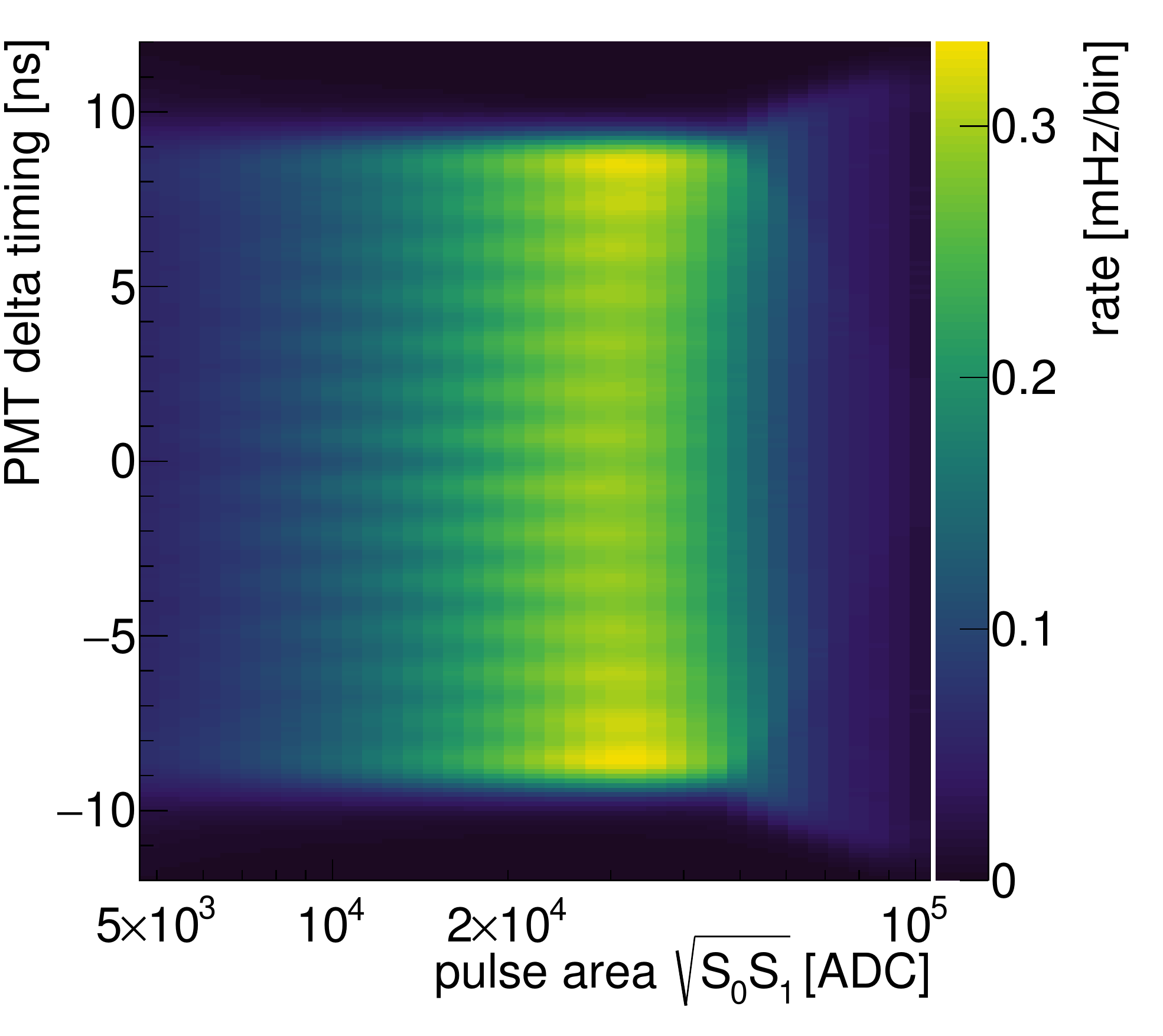}
\caption{$\Delta t$ versus signal amplitude for muon hits
    (summed over all PMTs. Fine structure variations are visible
    in the signal amplitude region corresponding to corner-clipping muon tracks; see the text for a detailed description. Increasing time spread at high amplitude is due to saturation of the ADC dynamic range.}
\label{fig:hobbes}
\end{figure}

The $\Delta t$ distribution for previously-described corner-clipping muon hits is recorded in each segment, and is plotted in Figure~\ref{fig:hobbes}.  
The distribution is not broadly uniform across the segment due to geometric selection efficiencies for this event type.  
However, the edges of the distribution provide well-defined markers for the ends of the active scintillator volume.  
Additional high-frequency variations are also present across the distribution, 
    corresponding to light transport perturbations caused by the diffusely-reflecting plastic support rod clips holding the edges of the specularly-reflecting optical grid panels (described in Section~\ref{sec:exp} and Ref.~\cite{prospect_grid}).  
The corner-clipping muon event class is more sensitive to these than events uniformly
    distributed over the detector bulk, since scintillation occurs in the segment corners near these clips.

The $\Delta t$ distribution shown in Figure~\ref{fig:hobbes} is fit to extract the distribution edges and the
    fine-structure wiggle positions across the segment.  
    A position model $z = a \Delta t + b (\Delta t)^3$ is used, 
    combined with empirical parameters for large-scale resolution and shape.  
This two-term position model (linear and cubic components) produces agreement
    to better than 1\,cm with dedicated calibration source position scans.  
%In addition, the modelled $z$ distribution exhibits periodicity matching at mm~scale that expected from the as-built spacing of diffulsely-reflecting support rod clips.  

%\begin{figure}
%\todo[MPM]{\textbf{FIGURE:} some kind of figure showing ADC log-ratio versus calibrated dt.}
%\caption{Log light ratio versus $\Delta t$ for corner-clipping muon hits,
%    providing connection between signal-ratio-based and timing-based position estimates.}
%\label{fig:lightratio}
%\end{figure}

To estimate position from relative light collection,
    the log signal ratio $R$ is fit against $\Delta t$,
    which is in turn linked to $z$ by the procedure above.  
For this step, a linear fit plus cubic correction $R = a + b \Delta t + c (\Delta t)^3$ is employed.  
Parameters for both timing-based and amplitude-based calibration curves $z(\Delta t)$ and $z(R)$ are stored 
    to the calibration database for later numerical evaluation.  
A final reconstructed position $z_{rec}$ for each pulse is formed from a statistically-weighted average of its timing- and amplitude-based $z$ estimates.  
It is found that removal of either the time or the amplitude based information from $z_{rec}$ produces noticeable degradation in reconstructed position resolution.  

\begin{figure}[t]
\includegraphics[trim = 0.5cm 0.1cm 0.25cm 0.0cm, clip=true, width=0.49\textwidth]{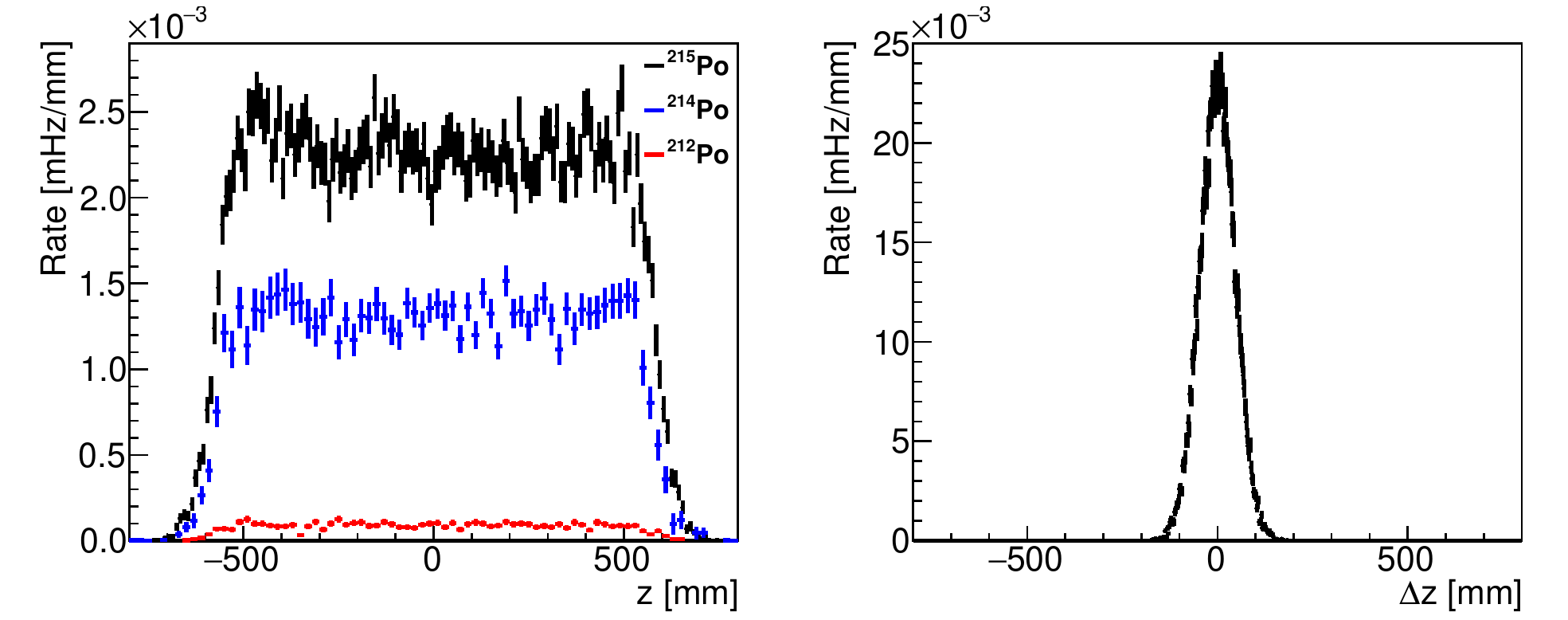}
\caption{Position distribution (left) of $^{215}$Po (black), $^{214}$Po (blue) and $^{212}$Po (red) $\alpha$-particles distributed uniformly throughout the detector and position coincidence distribution (right) of $^{219}$Rn-$^{215}$Po $\alpha$-$\alpha$ decays. The centered position of the absolute distributions relative to the segment center and the width of the relative $^{219}$Rn-$^{215}$Po position coincidence demonstrate the accuracy and precision of the z-reconstruction, respectively.}
\label{fig:position}
\end{figure}

The general features of reconstructed pulse positions $z_{rec}$ are illustrated for a single detector segment in Figure~\ref{fig:position} using a high-purity selected set of polonium $\alpha$ decay events in the PROSPECT scintillator, which arise from the presence of added $^{227}$Ac~\cite{prospect_nim}, and naturally occurring $^{238}$U, and $^{232}$Th decay chain isotopes.  
The energy, position, and time coincidence requirements for these datasets are described in Section~\ref{subsec:perform}.  
For all three polonium isotopes, uniform $z_{rec}$ distributions are centered on $z_{rec}=0$ with a width consistent with expectation based on the 117.6~cm active segment length.  
The resolution of $z_{rec}$ is illustrated by the the gradual reduction in rates at segment ends (high $|z_{rec}|$), and by the $z_{rec}$ coincidence observed between $^{215}$Po and its $\alpha$ decay parent $^{219}$Rn, which decays at an effectively identical location.  

%The most versatile
%790 event category is a high-purity sample of detector-intrinsic
%791 (219 Rn,215 Po) correlated alpha decays produced by 227 Ac de-
%792 liberately dissolved into the scintillator [69].  
%\todo{Text describing relevancce of fig:position}
%Figure~\ref{fig:lightratio} shows an example $R$ vs. $\Delta t$ fitted distribution.

As scintillator optical properties slowly evolve with time,
    so do both $z(R)$ and $z(\Delta t)$.
Collecting sufficient statistics to resolve the fine structure 
    in the $\Delta t$ distribution for each segment requires combining
    data over week-timescale periods.
The whole dataset is thus subdivided into 11 calibration periods
    for measuring and applying position calibrations.

%%%%%%%%%%%%%%%%%%%%%%%%%%%%%%%
\subsection{Energy Calibration}
\label{subsec:energy}

\newcommand{\Evis}{\ensuremath{E_\mathrm{vis}}}

The PROSPECT detector's segmented construction,
    coupled with scintillator nonlinearity (quenching) and trigger acquisition thresholds,
    complicates event-by-event extraction of interaction energies.
Rather than attempt reconstructing the initial energy of each interaction,
    the PROSPECT calibration effort is divided into two components:
    extracting a consistent measure of the visible energy, E$_{vis}$ 
    (light production after scintillator nonlinearity, but before
    light transport and PMT gain effects),
    and adjusting parameters in a Monte Carlo (MC) based detector response model to accurately reproduce
    data observables in E$_{vis}$ space.
This section describes the first component, calibration of position- and time-dependent variations in light collection.
Adjusting the response model to match absolute energy scale is discussed in Section~\ref{sec:ereco}.

Inputs for reconstructing the E$_{vis}$ of a segment interaction
    are the two pulse area signals $S_0, S_1$ from each segment end,
    and the reconstructed longitudinal position in the segment $z_{rec}$.  
The statistically optimal way to combine this information into a single
    E$_{vis}$ number, given the dominant uncertainty of photoelectron (PE)
    counting statistics fluctuations on the pulse area values,
    is to sum the estimated total number of PE counted by both PMTs,
    and divide out a position-dependent light collection factor,
    \begin{equation}\label{eq:evis}
\begin{aligned}
\Evis = \frac{S_0 n_0 / g_0 + S_1 n_1 / g_1}{n_0\eta_0(z_{rec}) + n_1\eta_1(z_{rec})},
\end{aligned}
\end{equation}
where $g_i$ is the pulse area signal per E$_{vis}$ deposited at segment center (combining effects of light production, light transport, and PMT/readout gain), $n_i$ is the estimated number of photoelectrons collected per E$_{vis}$ at segment center, and $\eta_i(z)$ is the position-dependent light transport efficiency to each PMT, normalized to 1 at segment center.

Neutron capture signals on $^6$Li provide a monoenergetic 
    reference continuously available from natural backgrounds throughout the scintillator volume,
    cleanly separable from dominant $\gamma$-ray backgrounds by both PSD and time correlations.
The high-ionization-density tracks of the $^4$He-$^3$H products are well into the
    scintillator's nonlinear quenching range, so the E$_{vis}$produced cannot be
    accurately predicted from first principles.
From the absolute energy scale calibration described in Section~\ref{sec:ereco},
    $\gamma$-ray calibration source spectra are reconstructed to the correct
    ({\em i.e.} MC-matching) E$_{vis}$ when the neutron capture peak is
    scaled to fall at $\Evis = 0.526$\,MeV.
    
The neutron capture signal is measured for each PMT for each run to
    determine the gain-stabilizing $g_i$ calibration constants.  
The neutron capture peak is also used to map out the light transport
    curves $\eta_i(z)$, summed over approximately two-week long periods 
    for sufficient statistics.  
The accuracies of $n_0$ and $n_1$ is not critical to the result,
    since these only define weightings that cancel out
    --- sub-optimal estimates of $n_i$ would inflate statistical
    scatter in the result, without shifting the mean,
    insofar as $g_i$ and $\eta_i$ are accurate.
The value of $n_i$ is determined from the width of the n+$^6$Li capture peak width.  
Gain-stabilizing constants are recorded into calibration databases and applied on a run-by-run basis, while light transport and photoelectron collection constants are recorded and applied in two-week intervals.  

\begin{figure}[hptb!]
\includegraphics[trim = 0.0cm 0.15cm 0.1cm 0.25cm, clip=true, 
width=0.49\textwidth]{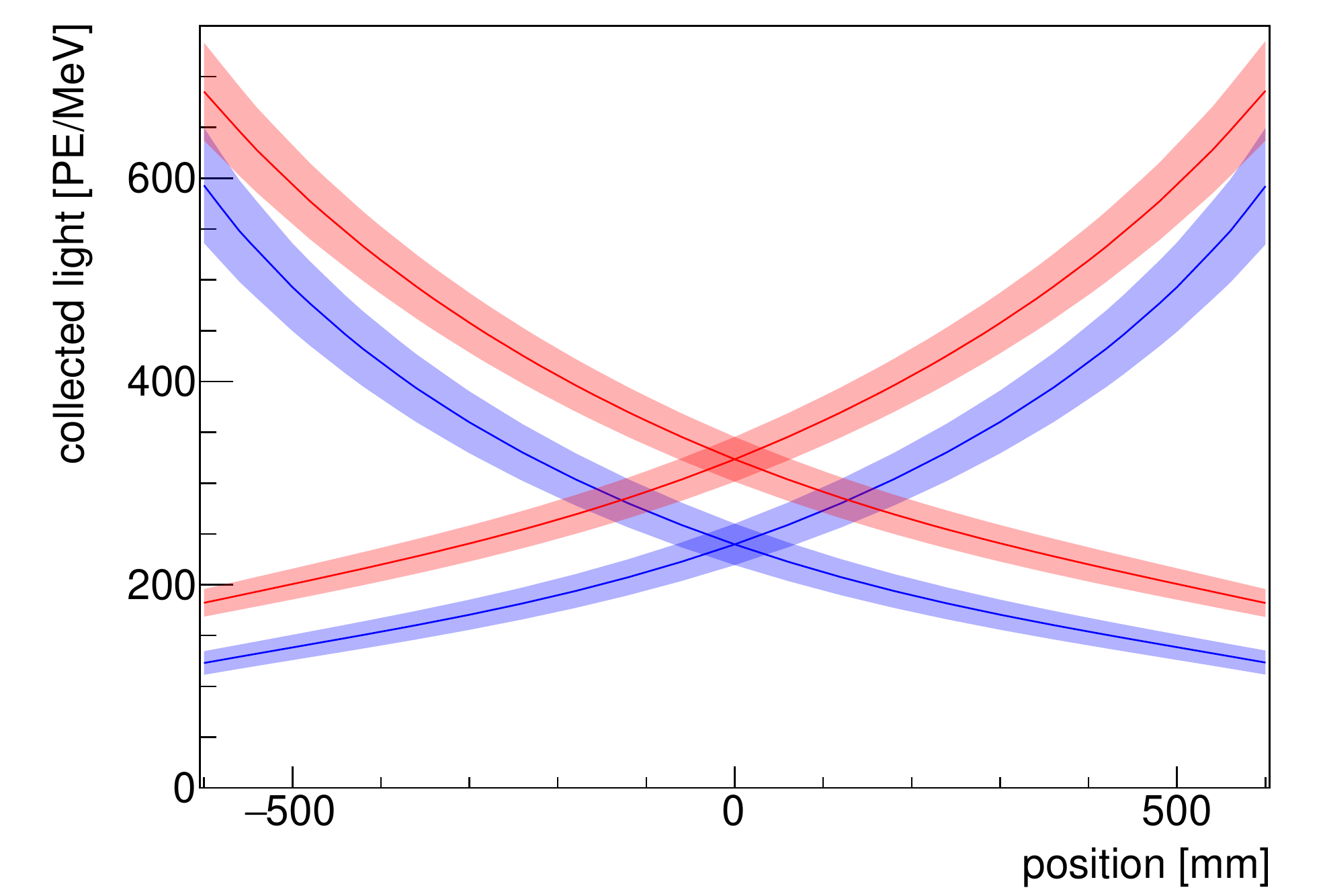}
\caption{Light collection $n_i\eta_i(z)$ averaged over channels, at beginning (upper red curves) and end (lower blue curves) of the data period used for the present analysis.  Bands indicate the RMS spread between  channels.}
\label{fig:lightcurves}
\end{figure}

%Using light collection calibration constants $n_i\eta_i(z_{rec})$ for two specific calibration time periods, 
Figure~\ref{fig:lightcurves} illustrates the magnitude of the time variation of the position-dependent light transport variation that must be taken into account to achieve stable E$_{vis}$ calibration.
For a single channel, the overall level of light collection varies by a factor of 3-5 along $z_{rec}$, with substantial variation between segments.  
Variation in light collection as a function of $z_{rec}$ is reduced to  roughly 50\% when information from both channels is combined.  
A substantial reduction in light collection is also clearly visible between the beginning and end of the dataset used for this analysis: at the segment center, a light reduction of 30\% is observed over the 7 calendar month data-taking period.  

\newcommand{\Esmear}{\ensuremath{E_\mathrm{smear}}}

\begin{figure}[hptb!]
\includegraphics[trim = 0.0cm 0.25cm 0.0cm 0.25cm, clip=true, width=0.49\textwidth]{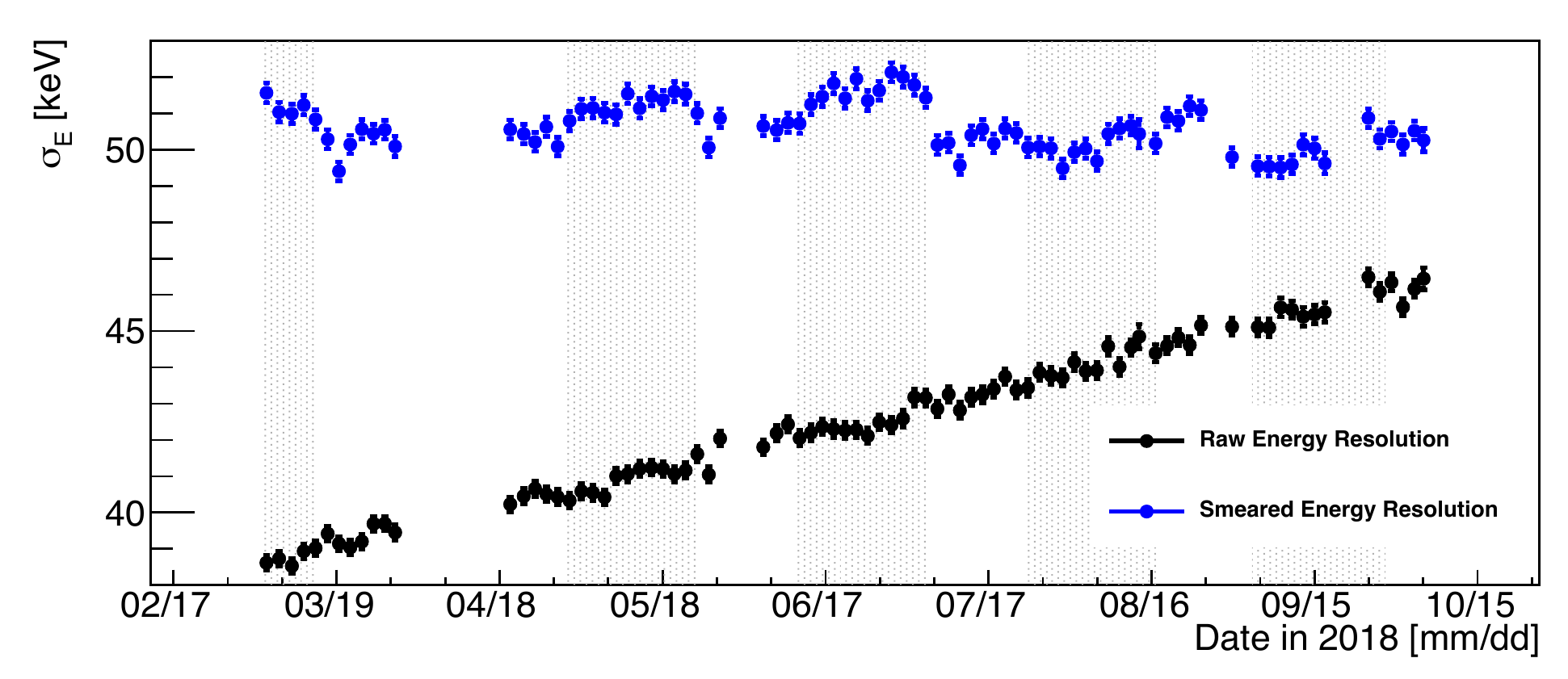}
\caption{E$_{rec}$ resolution of the $^{215}$Po peak from $^{219}$Rn-$^{215}$Po $\alpha$-$\alpha$ decays before (black) and after (blue) applying E$_{rec}$ energy smearing.}
\label{fig:Esmear}
\end{figure}

Degradation of scintillator optical properties with time
    causes a continuous gradual degradation of E$_{vis}$ resolution.
For constructing energy spectra in a uniform manner across different time periods, which permits straightforward reactor-off data subtraction and simpler interpretation of spectrum results, a ``smeared'' energy E$_{smear}$ is produced by
    adding random fluctuations to E$_{vis}$ to reduce the resolution
    to the equivalent of 325 photoelectrons/MeV in all segments at all times.
Figure~\ref{fig:Esmear} shows the long-term stability of E$_{smear}$ energy resolution 
    for the $^{215}$Po peak described in the previous section, 
    compared to the time-varying E$_{vis}$ resolution.

%Changes in light collection and DAQ trigger settings also cause
%    time-varying trigger efficiency for recording low energy interactions.
%To unify detector trigger response over time,
%    a 90\,keV software trigger threshold (slightly above acquisition thresholds)
%    is applied to all calibrated energy depositions
%    --- any segment hit below this threshold is removed from all subsequent processing.

\subsection{Event Reconstruction}
\label{subsec:reco}

As interactions of \nuebar and other particles in the PROSPECT inner detector will often produce pulses in multiple detector segments, it is necessary to analyze physics events at the cluster level.  
Thus, reconstructed cluster physics metrics are primary inputs to the higher-level PROSPECT oscillation and \uFive~physics analyses.  
Cluster formation was described previously in Section~\ref{subsec:ppairs}.  

To ensure consistency in cluster energy and multiplicity definitions despite variations in per-segment energy response with time coupling with hardware thresholds, only reconstructed pulses with E$_{smear} > $ 90~keV  are considered for analysis in reconstructed clusters.   
This threshold was estimated to be above the ZLE ADC threshold at all positions in all segments for the entire dataset by examining each segment's pulse energy spectrum shape in the vicinity of the trigger threshold at different times.  
To account for unexpected biases in the analysis method, the 90~keV energy cut threshold is treated with a $\pm$5~keV uncertainty when comparing predicted and measured IBD datasets.  
This uncertainty allows for small variations in the multiplicity of predicted events, which naturally propagates to an uncertainty in predicted reconstructed energy.  

Reconstructed physics quantities for individual clusters are formed using the reconstructed quantities of the included pulses.  
Cluster time, T$_{rec}$, is defined as the median $t_{rec}$ of the individual included pulse times.  
Cluster energy, E$_{rec}$, is defined as the sum of the reconstructed smeared energies \Esmear\ of all contained pulses.  
Cluster $z$-position and segment number, Z$_{rec}$ and S$_{rec}$, are defined as the $z_{rec}$ and segment number of the highest-energy contained pulse.  
Cluster segment multiplicity, as well as the energies, PSD values, and z-positions of each segment pulse, are also stored for use in later steps of the analysis.  
All of these cluster-related variables are used in the IBD signal selection process.  
Cluster E$_{rec}$ and S$_{rec}$ are used as primary inputs to the sterile neutrino oscillation analysis, while E$_{rec}$ is also a primary input to the \uFive~\nuebar spectrum analysis.  

%In the following section, we describe how clusters are formed and how reconstructed quantities related to cluster position and energy are defined for formed clusters.  
%Given the importance of \nuebar energies in PROSPECT's oscillation and \uFive~\nuebar spectrum measurement, we will also describe in detail how radioactive calibration sources and intrinsic detector backgrounds are used to define the relationship between the reconstructed energies of clusters, $E_{rec}$, and total true deposited energy.  

%\subsection{Cluster Building}
%\label{subsec:cluster}
%Clusters are identified by grouping together reconstructed pulses such that each item in the cluster is separated by no more than time 20~ns in $t_{rec}$ from the one before or after.  
%This window length is substantially larger than the estimated resolution associated with  absolute pulse times $t_{rec}$ ($<$1~ns), as well as the physics timescales of muon, electron, positron, gamma, and proton energy deposition in the detector.  
%Due to comparatively longer neutron propagation timescales, it is expected that multiple above-threshold proton recoils from the same neutron may occasionally not be combined into the same cluster.  

%%%%%%%
% N.B.: there is a cluster-pileup cut applied to IBD prompts, 
% but NOT a system-wide exclusion of closely-spaced clusters. 
% Removed commented-out text suggesting a generic close-cluster filter here. --MPM

%%%%%%%%%%%%%%%%%%%%%%%%%%%%%%%
\subsection{Calibration Performance}
\label{subsec:perform}

The stability of the energy, position and PSD metrics as a function of time and segment can be characterized using a variety of background categories present in normal physics data-taking runs, encompassing a range of particle types, energies, and spatial topologies.  
The most versatile event category is a high-purity sample of detector-intrinsic ($^{219}$Rn,$^{215}$Po) correlated $\alpha$ decays produced by $^{227}$Ac deliberately dissolved into the scintillator.   
The selection criteria and time-integrated rate for this signal are summarized in Table~\ref{tab:cutvals}.  
The total rate of this signal in the detector, 0.4~Hz, enables daily characterization of energy- and $z$-related performance metrics, as well as time-integrated comparisons between datasets from differing detector segments.  
Notably, the compact topology of these $\alpha$ coincidences also enables characterization of the stability of $z$-position reconstruction resolution with time.  
A similar high-purity sample of correlated ($^{214}$Bi,$^{214}$Po) and ($^{212}$Bi,$^{212}$Po) ($\beta+\gamma$,$\alpha$) decays from the $^{238}$U  and $^{232}$Th decay chains can also be found in the dataset due to natural radioactive contamination in the inner detector.  
Selection criteria and rates for these events are also summarized in Table~\ref{tab:cutvals}.  
Due to the presence of $\gamma$-rays in the prompt signal and the significant path length of the $\beta$-particles, they are not ideal for characterizing the $z$-resolution of the detector.  

\begin{table}[htbp!]
\centering
\begin{tabular}{|c|c|c|c|c|c|}
\hline 
\multirow{2}{*}{Decay} & \multicolumn{4}{|c|}{Selection Criteria} & Rate \\ \cline{2-5}
& E$_{rec}$ (MeV) & PSD & Pulses & $\delta$t$_{rec}$ ($\mu$s) & (mHz)\\ \hline 
$^{219}$Rn $\alpha$ & (0.57,1.15) & (0.19,0.36) & 1 & \multirow{2}{*}{(0,5000)} & \multirow{2}{*}{403}\\
$^{215}$Po $\alpha$ & (0.66,1.15) & (0.19,0.36) & 1 & & \\ \hline
$^{214}$Bi $\beta+\gamma$ & $<$4.00 & (0.05-0.22)  & Any & \multirow{2}{*}{(10,710)} & \multirow{2}{*}{165}\\
$^{214}$Po $\alpha$ & (0.72,1.00) & (0.17,0.34) & 1 & & \\ \hline
$^{212}$Bi $\beta+\gamma$ & $<$3.00 & (0.05-0.22) & Any & \multirow{2}{*}{(0.7,1.7)} & \multirow{2}{*}{55}\\
$^{212}$Po $\alpha$ & (0.95,1.27)& (0.17,0.34) & 1 & & \\ \hline
%\hline
\end{tabular} 
\caption{Selection criteria and rates for correlated decay signals in PROSPECT used for performance evaluations.  For bismuth decays, given PSD cut values are applied to the highest energy pulse in the cluster; relaxed time-dependent PSD cuts are also applied to other pulse clusters. Integrated rates include only segments used in the oscillation and spectrum analyses.}
\label{tab:cutvals}
\end{table}

Various clean $\gamma$-ray signals can also be identified for use in stability studies.  
A sample of mono-energetic 2.2~MeV $\gamma$-ray produced by $n$-H capture in the detector can be obtained from a 10-200~$\mu$s window following cosmogenic muon signals in the detector.  Cosmogenic muon signals are defined as events with summed pulse energies greater than 15~MeV, while the purity of the n-H sample can be further improved with tight cuts on the electron-like PSD band.  
Finally, prominent $\gamma$-ray peaks are visible in the low-PSD single trigger energy spectrum originating from intrinsic $^{208}$Tl contamination in the detector and from capture of reactor generated neutrons on metals in the HFIR complex and the PROSPECT shielding package

\begin{figure}
\includegraphics[trim = 0.cm 0.5cm 0.0cm 0.0cm, clip=true, width=0.48\textwidth]{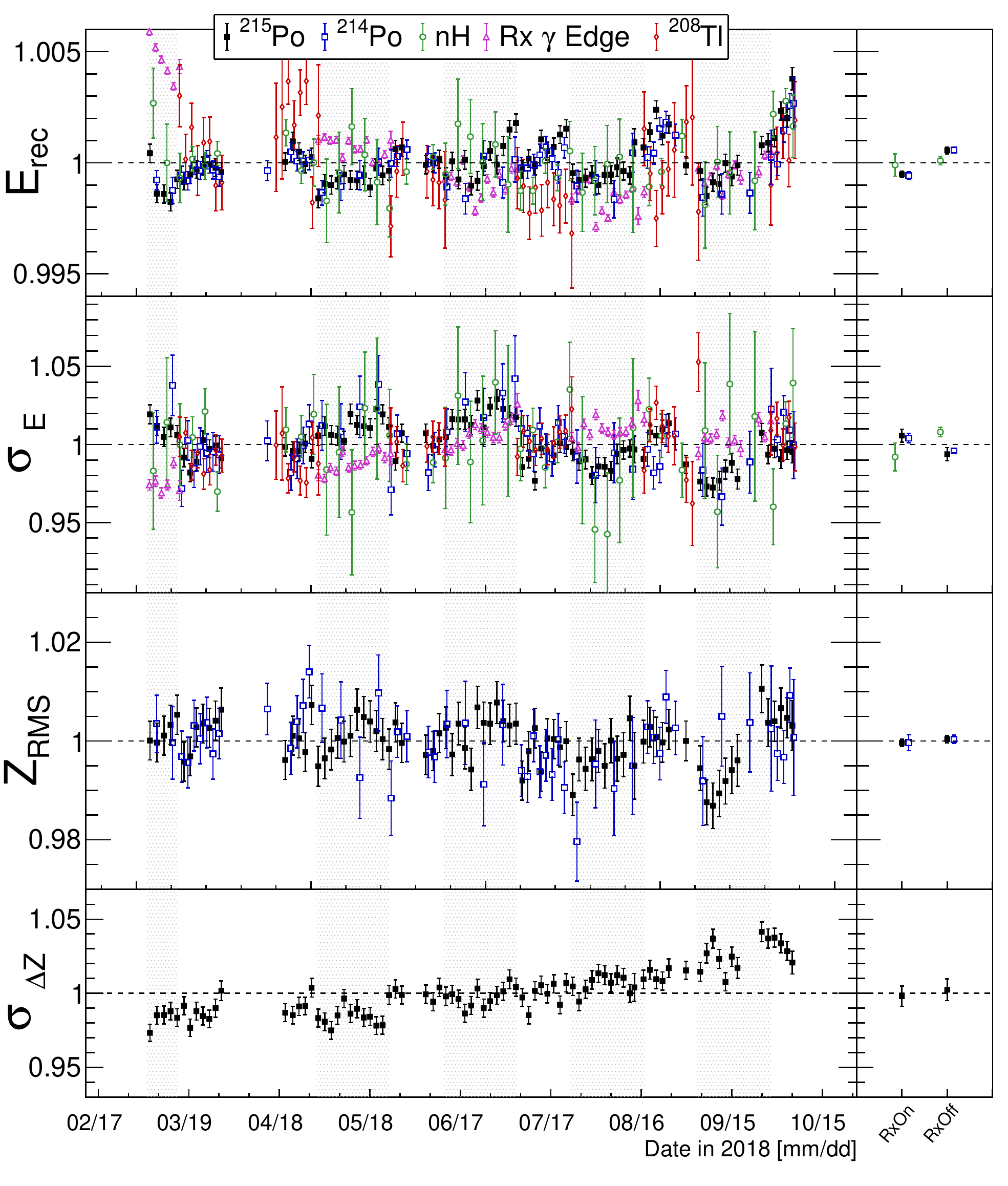}
\caption{Stability of pulse-level reconstructed physics metrics related to energy and longitudinal position ($z$).  
Stability is pictured over time, as well as between reactor-on and reactor-off periods.  
Metrics are calculated for $^{215}$Po (black) and $^{214}$Po (blue) $\alpha$ decays uniformly distributed throughout the detector, for nH captures (green), for $\gamma$-ray full-energy peaks from single $^{208}$Tl decay (red), and for and the highest-energy prominent reactor neutron capture peak edge (pink) during reactor-on and -off periods, respectively.  
Reconstructed metrics are described in more detail in the text.  
All quantities are shown relative to the average of all points in the dataset.  Light grey bands indicate reactor-on periods.  Right panel shows relative changes between reactor on and off datasets.  All error bars represent statistical uncertainties.} 
\label{fig:timeresponse}
\end{figure}

The time-stability of energy and $z$-related reconstruction metrics are summarized for these various sources in Figure~\ref{fig:timeresponse}.  
For each metric and event type, stability in time is expressed in reference to the mean value over the full dataset for that metric/event type; stability between reactor-on and reactor-off periods is expressed with respect to the mean of reactor-on and reactor-off values.  
E$_{rec}$ values for all sources are stable within $\pm$0.5\% over the full dataset, and to within 0.2\% between reactor-on and reactor-off periods.  
E$_{rec}$ resolutions are stable within $\pm$5\% over the full dataset, and within 2\% between reactor-on and reactor-off periods.  

Given the expected stability and uniformity of ($^{214}$Bi,$^{214}$Po) and ($^{219}$Rn,$^{215}$Po) distribution throughout the detector with time (Fig~\ref{fig:position}), the root mean square (RMS) of all coincidences' delayed reconstructed $z$ position, Z$_{RMS}$, should exhibit time-stability; any change in this quantity would indicate an alteration in the resolution of pulse $z$ reconstruction.  
This quantity is found to be time-stable within $\pm$1.5\%, corresponding to roughly 2~cm with respect to the $1.2$~m segment length.  
A more precise probe of $z$ resolution is provided by the distance between prompt and delayed  ($^{219}$Rn,$^{215}$Po) signals, $\sigma_{\Delta z}$.  
This metric exhibits a 7\% variation over time, corresponding to roughly 3.5~mm with respect to the 50~mm ($^{219}$Rn,$^{215}$Po) time-averaged $\sigma_{\Delta z}$.  
This variation in $z$ reconstruction capabilities is caused by the reduction in photon counting statistics due to decreased light collection over time, as described in the previous sections.  
Time variation in $z$-resolution for events with higher energies and larger spatial extent, such as IBD prompt positron signals, are likely to be less significant, due to higher photostatistics and to the finite cm-scale spatial extent of ionization tracks.  
Due to the interleaved nature of reactor-on and reactor-off datasets, this time variation results in $<$0.5\% difference in $\sigma_{\Delta z}$ between reactor-on and reactor-off periods.  
The minor impact of $z$-resolution time-dependence on the selection of IBD events will be discussed in more detail in Section~\ref{subsec:eff}.  

\begin{figure}[phtb!]
\includegraphics[trim = 0.2cm 0.0cm 0.0cm 1.0cm, clip=true, width=0.48\textwidth]{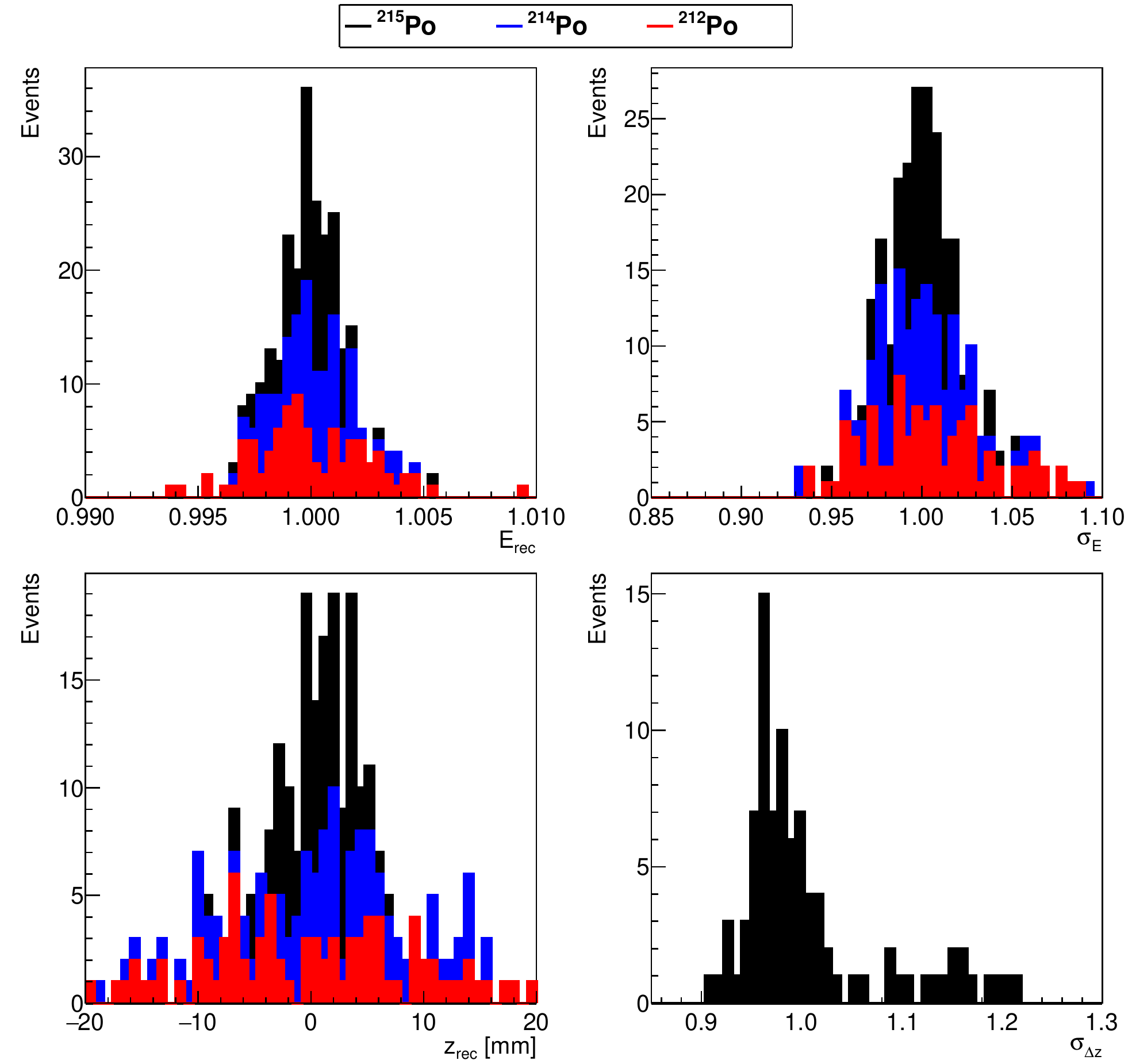}
\caption{Segment-to-segment stability of pulse-level reconstructed physics metrics related to energy and longitudinal position ($z$).  
Quantities are calculated for $^{215}$Po (black), $^{214}$Po (blue), and $^{212}$Po (red) $\alpha$ decays  uniformly distributed throughout the detector.  
Reconstructed quantities are described in more detail in the text.  All quantities are shown relative to the average of all points in the dataset with the exception of mean $z_{rec}$, which is plotted in~mm.} 
\label{fig:segresponse}
\end{figure}

Figure~\ref{fig:segresponse} provides similar reconstruction stability characterizations for the ensemble of detector segments.  
Reconstructed quantities for the $\gamma$-ray event classes are excluded because the segment multiplicity is greater than unity.  
Energy scales and resolutions are found to be identical to within $\pm$0.5\% and $\pm$7\% between all detector segments, respectively.  
To gauge the common alignment of $z$ between all segments, the mean -- rather than the RMS -- of the reconstructed $z$-position distribution for each segment is also plotted.  
The mean $z_{rec}$ for all segments are found to be aligned within $\pm$0.5~cm for $^{215}$Po events and within $\pm$2.0~cm for $^{212}$Po and $^{214}$Po events.  
Prompt-delayed position coincidence distributions for ($^{219}$Rn,$^{215}$Po) events are found to have variations in width ($\sigma_{\Delta z}$) of order 10\% or less, corresponding to a segment-to-segment variation of 0.5~mm or less.  

%To ensure that reconstructed $z$ positions are consistent between segments, Figure~\ref{fig:segresponse} also includes the average -- rather than the RMS -- of the reconstructed $z$ distribution. 
%These distributions for these events are found to be similar within BLAH\% for all segments.  

Variations in pulse-level reconstructed metrics with time and segment are propagated as systematic uncertainties in higher-level PROSPECT analyses.  
The treatment of these uncertainties are discussed in further detail in Sections~\ref{sec:osc} and~~\ref{sec:spec}.

%%%%%%%%%%%%%%%%%%%%%%%%%%%%%%%RECO AND ESCALE SECTION
%\input{sections/sec-reco.tex}

\section{Absolute Antineutrino Energy and Energy Resolution}
\label{sec:ereco}

For higher-level analyses, it is essential to define the relationship between reconstructed cluster energy, E$_{rec}$, and incoming antineutrino energy, E$_{\nu}$.  
This relationship is complex, given the presence of dead material throughout the antineutrino target, the segmented detector geometry, the small target size, and the non-linearity of light production in the scintillator.  
For \nuebar-related energy depositions, this relationship is defined using PG4, a GEANT-4 based~\cite{bib:Geant4} MC simulation of the PROSPECT detector, which is adjusted to reproduce the observed PROSPECT response to a wide variety of radioactive calibration source and intrinsic background energy depositions.  
This approach is in contrast to that recently presented by other reactor \nuebar experiments such as Daya~Bay, where geometric, scintillator, and electronics effects are independently modelled and parameterized, with energy non-linearities then matched to empirical fits of calibration and background energy spectra~\cite{bib:nim_escale}.  

\subsection{Monte Carlo Simulation Description}
\label{subsec:MC}
 
The PG4 MC simulation incorporates the essential aspects of the realized PROSPECT detector geometry described in Section~\ref{sec:exp}.  
The modelled dimensions of the scintillator  volume accurately reflect dimensions measured during detector assembly and scintillator preparation~\cite{prospect_ls}.  
The modelled optical grid features the as-measured average reflector and support rod dimensions, materials, and densities reported in Ref.~\cite{prospect_grid}.  
The most important aspects of both instrumented and un-instrumented segment support rods are also modeled, including radioactive source capsule materials and geometries as well as accurate air, acrylic, Teflon, PLA, and scintillator volumes.

The simulation includes the geometries and materials of the PMT housings, the acrylic support structure, the acrylic and aluminum tanks, and the inner and outer shielding packages.  
To simplify the simulation, all segments are given identical geometric and material properties.  
Modest simplifications are also applied to the support rod axis and calibration deployment system geometries. 
These simplifications are expected to have minimal impact on the PG4-determined relationship between true and reconstructed \nuebar energies.  

The non-linear optical response of the PROSPECT scintillator to energy depositions is not directly simulated via the computational-resource-heavy process of optical photon production and propagation.  
Instead, the fractional rate of conversion of true deposited energy to scintillation light is calculated step-by-step during Geant4 propagation of the particle using parameterizations of these physics processes:
\begin{equation}\label{eq:birks}
E_{MC} = A\sum_i (E_{scint,i}(k_{B2},k_{B2})+E_{c,i}(k_C)).
\end{equation}
The energy converted directly into scintillation light $dE_{scint}$ during simulation step $i$ is parameterized using Birks' law quenching~\cite{Birks}:
\begin{equation}\label{eq:ceren1}
\frac{dE_{scint}}{dx} = \frac{\frac{dE}{dx}}{1+k_{B1}\frac{dE}{dx}+k_{B2}(\frac{dE}{dx})^2},
\end{equation}
where $k_{B1}$ and $k_{B2}$ are first- and second-order Birks constants and $dE/dx$ is the true deposited energy in that step.  
Cerenkov light production and absorption and subsequent scintillation photon re-emission in simulation step $i$ is modelled as 
\begin{equation}\label{eq:ceren2}
E_{c} = k_{c}\sum_{\lambda}N_\lambda E_\lambda,
\end{equation}
where $N_\lambda$ is the number of Cerenkov photons emitted per unit wavelength, $E_\lambda$ is the energy of those Cerenkov photons, and $k_c$ is a normalization parameter that scales Cerenkov light production with respect to a default estimate based on simple scintillator refractive index assumptions.  
In Equation~\ref{eq:birks}, an overall scaling factor $A$ enables variation of the overall fractional rate of conversion of deposited energy into detected energy.  
We note that scintillation light from nuclear recoil signatures are modelled with two independent Birks parameters tuned to properly place the $n$-Li E$_{rec}$ peak location with respect to the $\gamma$-ray and $\beta$+$\gamma$ signatures used for calibration; recoil signatures in the energy range of interest for this analysis produce no Cerenkov light.

During the simulation, each step in deposited energy $E_{MC,i}$ is assigned to a running total for the appropriate segment.  
$E_{MC}$ for each segment following particle propagation is used to build synthetic waveforms based on measured shape templates and low-level detector calibration parameters.  
Waveform shape for each channel is assigned according to the magnitude of simulated scintillation light quenching for the relevant energy depositions, while waveform amplitude is determined by the magnitude of $E_{MC}$ and the position of deposited energy in $z$.  
Low-level pulse processing, cluster formation, and timing, PSD, position and energy reconstruction then proceed identically to that described above for real PROSPECT data.  

\subsection{Absolute Energy Response Determination}
\label{subsec:eresp}

PG4 MC simulations, run through PROSPECT's analysis infrastructure, can be used to generate simulated cluster E$_{rec}$ distributions and pulse multiplicities in response to any energy deposition given any combination of absolute energy response parameters ($A$,$k_{B1}$,$k_{B2}$,$k_c$). 
Data and PG4 cluster E$_{rec}$ and pulse multiplicity distributions can then be compared for a variety of radioactive sources, both deployed and intrinsic.  
For E$_{rec}$ spectra, datasets include  $\gamma$-ray sources $^{60}$Co (1.17+1.33~MeV), $^{137}$Cs (0.66~MeV), and $^{22}$Na (2$\times$0.511+1.27~MeV and 2$\times$0.511~MeV) deployed at the detector $z$ midpoint along a calibration axis near the ($x$,$y$) center of the detector, $n$-H  capture $\gamma$-rays from a similarly-deployed $^{252}$Cf spontaneous fission source (2.22~MeV), and $\beta$-dominated energy spectra from cosmogenically-produced $^{12}$B (3~MeV to 13.4~MeV).  
Pulse multiplicity distributions are included in the fit for all of the $\gamma$-ray sources listed above.  
All $\gamma$-ray datasets and the $^{252}$Cf dataset were obtained during special calibration campaigns in April and May 2018, respectively; the high-purity $^{12}$B dataset derives from special analysis cuts applied to the full physics dataset.  

%The segmentation of PROSPECT AD was non-negligible in energy reconstruction. 
%Particles' multi-segment-scattering in the detector introduced complicated nonlinear effects of reconstructed energy that dependent on segment multiplicity.
%The multiplicities of $\gamma$-ray events from $^{60}$Co, $^{137}$Cs, $^{22}$Na calibrations, and $n$-H capture interaction were also compared between data and MC, and the $\chi^2$ values were added in the summed $\chi^2$ of the spectral comparisons as

\begin{figure}[hptb!]
\includegraphics[trim = 0.5cm 0.6cm 1.8cm 0.6cm, clip=true, 
width=0.49\textwidth]{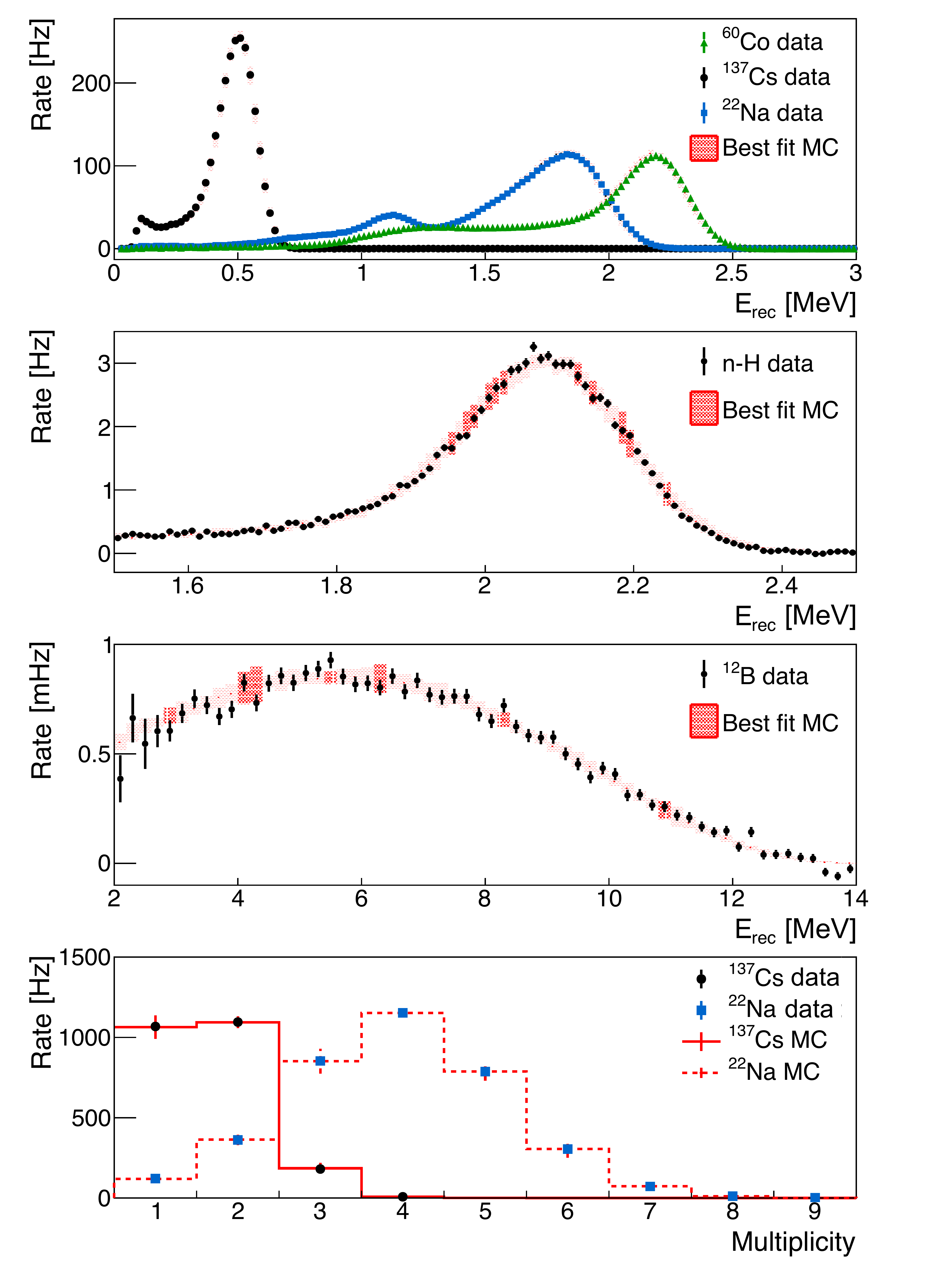}
\caption{Reconstructed distributions for calibration and best-fit PG4 MC datasets.  Top: E$_{rec}$ for detector-center $\gamma$-ray source deployments; Center top: E$_{rec}$ for $n$-H captures from a detector-center $^{252}$Cf source deployment; Center bottom: E$_{rec}$ for cosmogenically-produced $^{12}$B; Bottom: pulse multiplicity for detector-center $^{137}$Cs and $^{22}$Na source deployments, which represent the highest- and lowest-multiplicity calibration datasets. Error bands indicate statistical (data) and systematic (PG4) uncertainties.  Due to the presence of non-linearity and energy loss in dead materials, true and reconstructed energies should not be expected to align.}
\label{fig:calibE}
\end{figure}

To determine the nominal PROSPECT detector energy response model, cluster E$_{rec}$ and multiplicity distributions described above were simulated in PG4 for each grid point in a  4-dimensional detector response parameter space ($A$,$k_{B1}$,$k_{B2}$,$k_c$), and compared to the corresponding calibration datasets using the $\chi^2$ function: 
\begin{equation}\label{eq:escalechi2}
    \chi^2_{data-MC} = \sum_{\gamma} \chi^2_\gamma + \sum_{multi}\chi^2_{multi} + \chi^2_{^{12}\textrm{B}},
\end{equation}
In this comparison, $\chi^2_\gamma$ is the $\chi^2$ value for each $\gamma$-ray E$_{rec}$ distribution, $\chi^2_{multi}$ is the $\chi^2$ value for each of the two included $\gamma$-ray multiplicity distributions, and $\chi^2_{^{12}\textrm{B}}$ is the $\chi^2$ value of the $^{12}$B E$_{rec}$ distribution.  
The grid point providing the lowest $\chi^2$ value was chosen as the nominal energy model.  
Reconstructed energy and multiplicity distributions for the data and best-fit PG4  are shown in Figure~\ref{fig:calibE}.  
Both the shape and scale of these distributions show good agreement between data and the best-fit Monte Carlo.  
The best-fit parameters for this model are ($A$, $k_{B1}$, $k_{B2}$, $k_c$) = (1.0026$\pm$0.004, 0.132 $\pm$0.004~cm/MeV, 0.023$\pm$0.004~cm$^2$/MeV$^2$, 37$\pm$2\%), with a $\chi^2$/DOF (degrees of freedom) of 581.5/420.  
For the best-fit model, light is overwhelmingly contributed by direct scintillation from excitation and ionization: as an example, for the 2.22~MeV $n$-H capture de-excitation $\gamma$-ray, only 3.5\% of E$_{MC}$ is contributed by the Cerenkov process ($E_c$).  

Uncertainties on each of the four energy response parameters are assigned by identifying the maximum variation in each parameter value among all grid points with $\chi^2$ values within 1$\sigma$ of the best-fit model.  
For the $^{235}$U spectrum and oscillation physics analyses, an energy scale uncertainty covariance matrix reflecting these energy model parameters is then generated using these parameter variation ranges.  This scintillator-associated uncertainty is assumed to be correlated between all segments.  

To reduce the required parameter space dimension and computing time, the detector energy resolution smearing, per-pulse 90~keV analysis threshold, and PG4 geometry are held constant for all simulated grid points.  
These features and their uncertainties are determined using separate information, such as QA/QC studies and detector surveys, or data analyses that are unaffected by PG4 energy response parameters.  
The per-segment 5~keV threshold uncertainty is defined as given in Section~\ref{subsec:reco}, and is propagated as both a segment-correlated and a segment-uncorrelated uncertainty.   
Energy resolution uncertainty is described in the following section.  
Finally, PG4 studies indicate that the limited precision in measurements of the optical reflector panel masses can cause modest variations in detector energy response.  
The size of this uncertainty was estimated using PG4, along with the mass measurement precision of 1.7~kg reported in Ref.~\cite{prospect_grid}; this dead mass uncertainty is propagated in PG4 MC simulations as a 0.03~mm segment-correlated uncertainty in reflector thickness.  

\begin{figure}[hptb!]
\includegraphics[trim = 0.0cm 1.0cm 1.6cm 0.0cm, clip=true, 
width=0.49\textwidth]{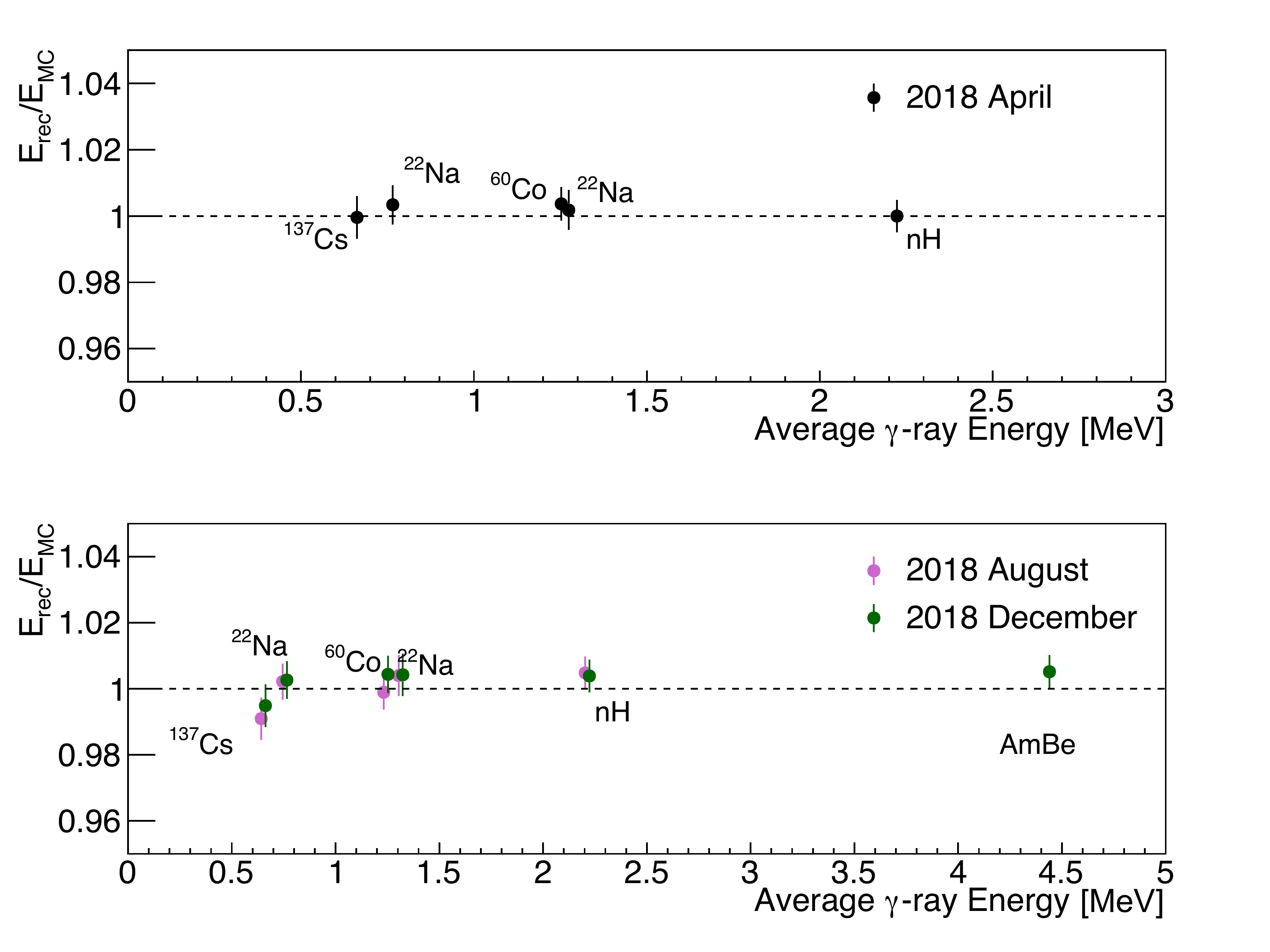}
\caption{Ratios of $\gamma$ calibration source reconstructed energy peak locations between data and PG4 MC simulations utilizing best-fit energy response modeling, plotted versus reconstructed $\gamma$-ray energy.  Error bars indicate statistical and systematic uncertainties.  Top: ratios for calibration source datasets used in the determination of the best-fit PG4 response model.  Bottom: ratios for calibration source datasets taken during different run periods.  Ratios for all datasets are within 1\% of unity.}
%, indicating excellent response modeling in PG4 for a wide variety of signal energies and multiplicities.} 
\label{fig:calibE_ratio}
\end{figure}

The overall agreement in measured and predicted response across the E$_{rec}$ energy distribution is further illustrated in Figure~\ref{fig:calibE_ratio}, which shows the ratio of the reconstructed $\gamma$-ray energy between data and the best-fit PG4 calibration dataset.  
This ratio is found to be unity within $\pm$1\% for all $\gamma$-ray calibration datasets used in the fit, with residues all lying within the error bands defined by the energy model and per-segment energy threshold uncertainties.  
For the $^{12}$B spectrum, the end point of the reconstructed (PG4-simulated) $\beta$ energy distribution, determined through Kurie plots, is $13.36\pm0.18$~MeV ($13.15\pm0.08$~MeV), indicating good agreements between data-PG4 $\beta$-particle E$_{rec}$ at higher energies.  
Data and PG4 $^{12}$B spectra are found to be most consistent when a relative shift of 0.38$\pm$0.41\% is applied; given the close correspondence between the properties of $^{12}$B electron and IBD positron kinetic energy depositions, this 0.41\% precision in verifying predicted and measured $^{12}$B energy scale agreement is also propagated as a segment-correlated energy scale uncertainty in the full detector response uncertainty covariance matrix.  

Similar data-PG4 comparisons are also shown in Figure~\ref{fig:calibE_ratio} for $\gamma$-ray and $^{252}$Cf calibration datasets acquired in August and December 2018, which were not used in the energy calibration procedure described above.  
Ratios are similarly statistically consistent with unity for these later datasets, demonstrating the stability of non-linearity effects and calibrated energy scales over time.  

\begin{figure}[hptb!]
\includegraphics[trim = 0.0cm 0.15cm 1.5cm 1.00cm, clip=true, 
width=0.49\textwidth]{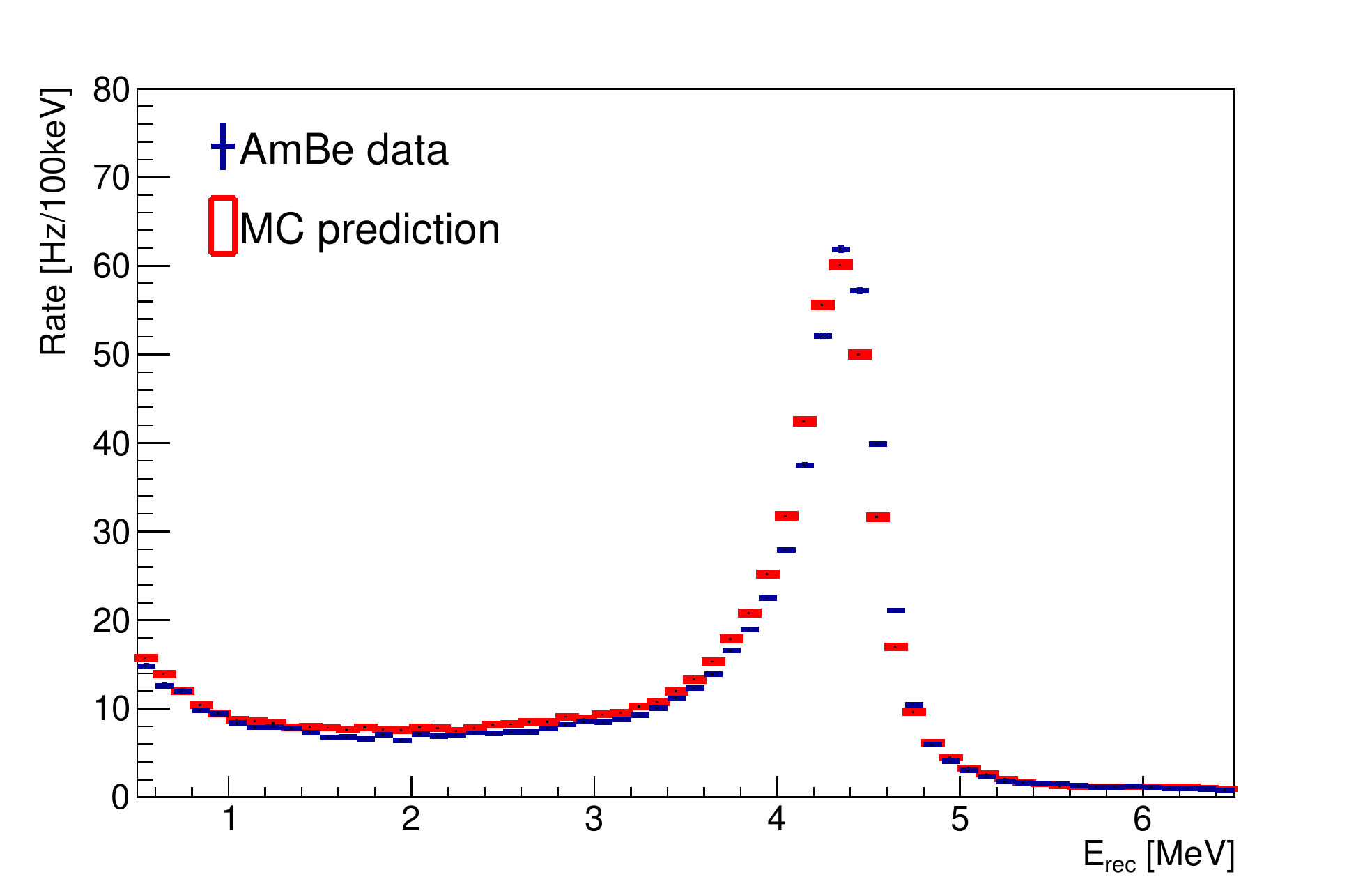}
\caption{Reconstructed and PG4-predicted energy spectrum of 4.43~MeV $\gamma$-rays from de-excitation of the first excited state of $^{12}$C following $\alpha$-particle capture on $^{9}$Be.  This signal was extracted from data featuring detector-center deployment of an $^{241}$Am-$^{9}$Be source.  Error bands indicate statistical (data) and systematic (PG4) uncertainties.} 
\label{fig:calibE_AmBe}
\end{figure}

While not included in the energy response model fitting, a special December 2018 detector-center deployment of an $^{241}$Am-$^{9}$Be source yielded a dataset containing 4.43~MeV $\gamma$-rays from de-excitation of the first excited state of $^{12}$C following $\alpha$-particle capture on $^{9}$Be.  
These signals were measured preceding neutron capture signals by requiring prompt-delayed time and position coincidence criteria identical to the IBD selection.  
Cuts rejecting high-PSD pulses within the prompt cluster enabled reduction of proton recoil contamination of the $^{12}$C de-excitation signature and more direct data-PG4 comparison of the monoenergetic $\gamma$-ray's energy deposition.  
As illustrated in Figure~\ref{fig:calibE_ratio}, and in more detail in Figure~\ref{fig:calibE_AmBe}, the energy scale of this feature is also accurately predicted by the best-fit PG4 MC to within 0.5\%, providing further confidence in PG4 modeling of response at high IBD positron energy.  

\subsection{Energy Resolution}
\label{subsec:eres}

\begin{figure}[hptb!]
\includegraphics[trim = 0.0cm 2.8cm 0.0cm 0.0cm, clip=true, 
width=0.49\textwidth]{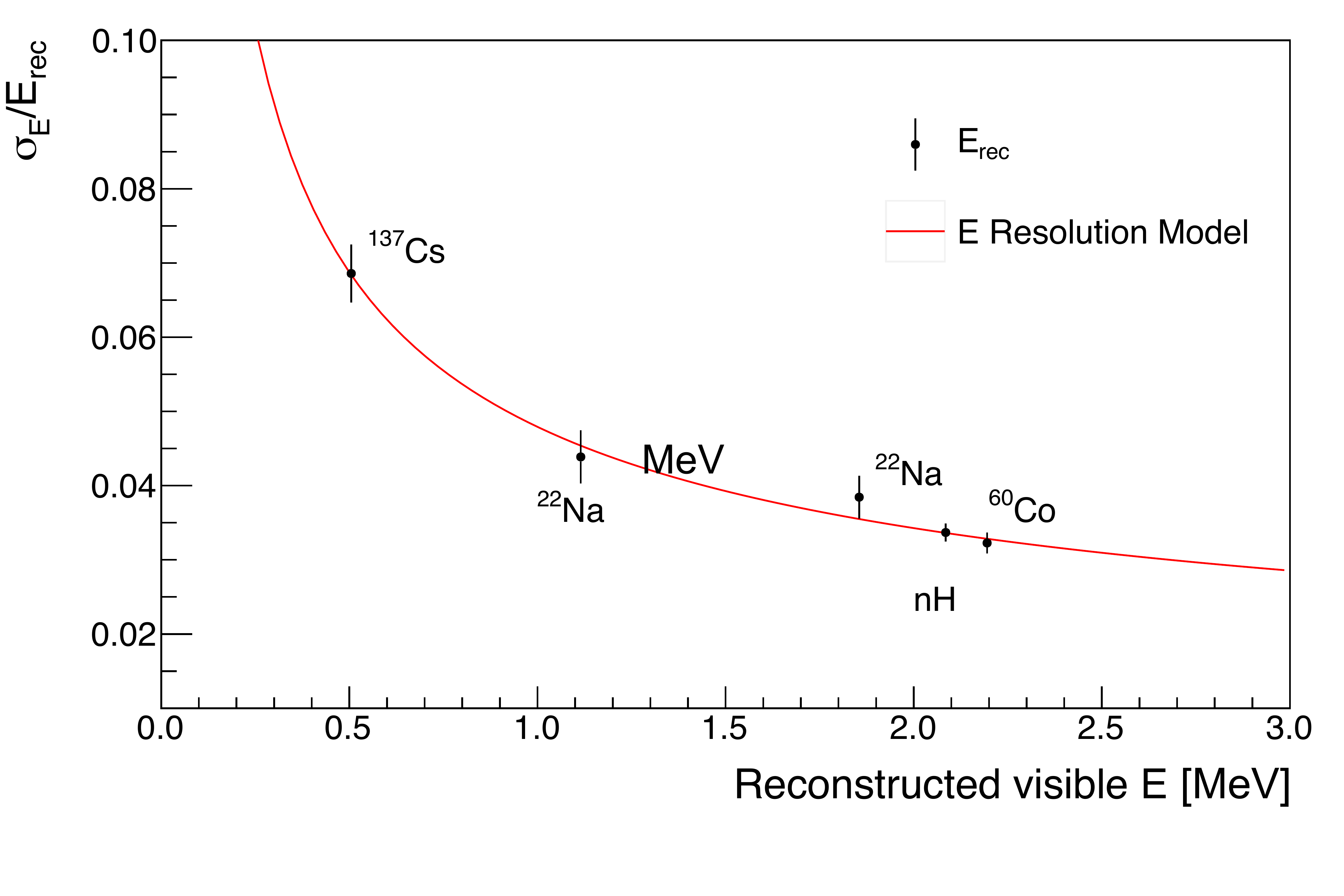}
\caption{Fractional PG4-modelled energy resolution (Eq.~\ref{eq:resolution}) versus reconstructed $\gamma$-ray energy.  Error bars indicate statistical and systematic uncertainties.  Good agreement is visible between the PG4 model and the displayed $\gamma$-ray calibration datasets.}
\label{fig:calibRes}
\end{figure}

The resolution in reconstructed energy distributions was also characterized for calibration $\gamma$-ray events.
The PG4 energy model was smeared with a Gaussian distribution whose $\sigma$ value was fitted with the resolution function
\begin{equation}\label{eq:resolution}
    \frac{\sigma_E}{E_{rec}} = \sqrt{a^2+\frac{b^2}{E_{rec}}+\frac{c^2}{E^2_{rec}}},
\end{equation}
where the first term is dependent on light collection inefficiency variations, the second term represents energy-dependent photostatistics, and the third term is related to PMT and electronics noise.
The best-fit energy resolution as a function of energy deposition is shown in Figure~\ref{fig:calibRes}; 
best-fit resolution parameters are found to be ($a$,$b$,$c$) = (1.15\%$\pm$0.47\%, 4.61\%$\pm$0.24\%, 0.0+1.3\%).  
The determined 1$\sigma$ spread in best-fit parameters is assigned as a correlated energy resolution uncertainty between all segments.  
%A re-fit of these parameters on the best-fit PG4 energy model is found to produce negligible change in resolution.  

We note that since both data and MC include inherent energy smearing due to loss of energy in non-scintillating regions, this contribution is not reflected in the fit parameters or in Figure~\ref{fig:calibRes}.  
This effective resolution contribution is characterized in Section~\ref{subsec:ibdmc}.  

\subsection{Determination of Position-Dependent Energy Response Variation}
\label{subsec:evar}

In addition to modeling absolute energy response effects in the PROSPECT detector center, PG4 MC simulations must also properly take into account position variations in IBD prompt E$_{rec}$ response due to proximity to the target boundary and to non-active segments.  
PG4 IBD MC simulations show that, to first order, variations in leakage of annihilation $\gamma$-ray energy into these regions results in a consistent shift in reconstructed IBD prompt E$_{rec}$.  
Proper modeling of these leakage effects was verified by performing segment-by-segment E$_{rec}$ comparisons between data and Monte Carlo for multiple $^{22}$Na source deployment locations.  
As a positron emitter, the $^{22}$Na source reflects the change in IBD energy scales resulting from variations in annihilation $\gamma$-ray energy leakage with detector position, as well as the distribution of IBD positron annihilation $\gamma$-ray energies among different detector segments.  

\begin{figure}[hptb!]
\includegraphics[trim = 1.3cm 0.1cm 0.0cm 0.0cm, clip=true, 
width=0.49\textwidth]{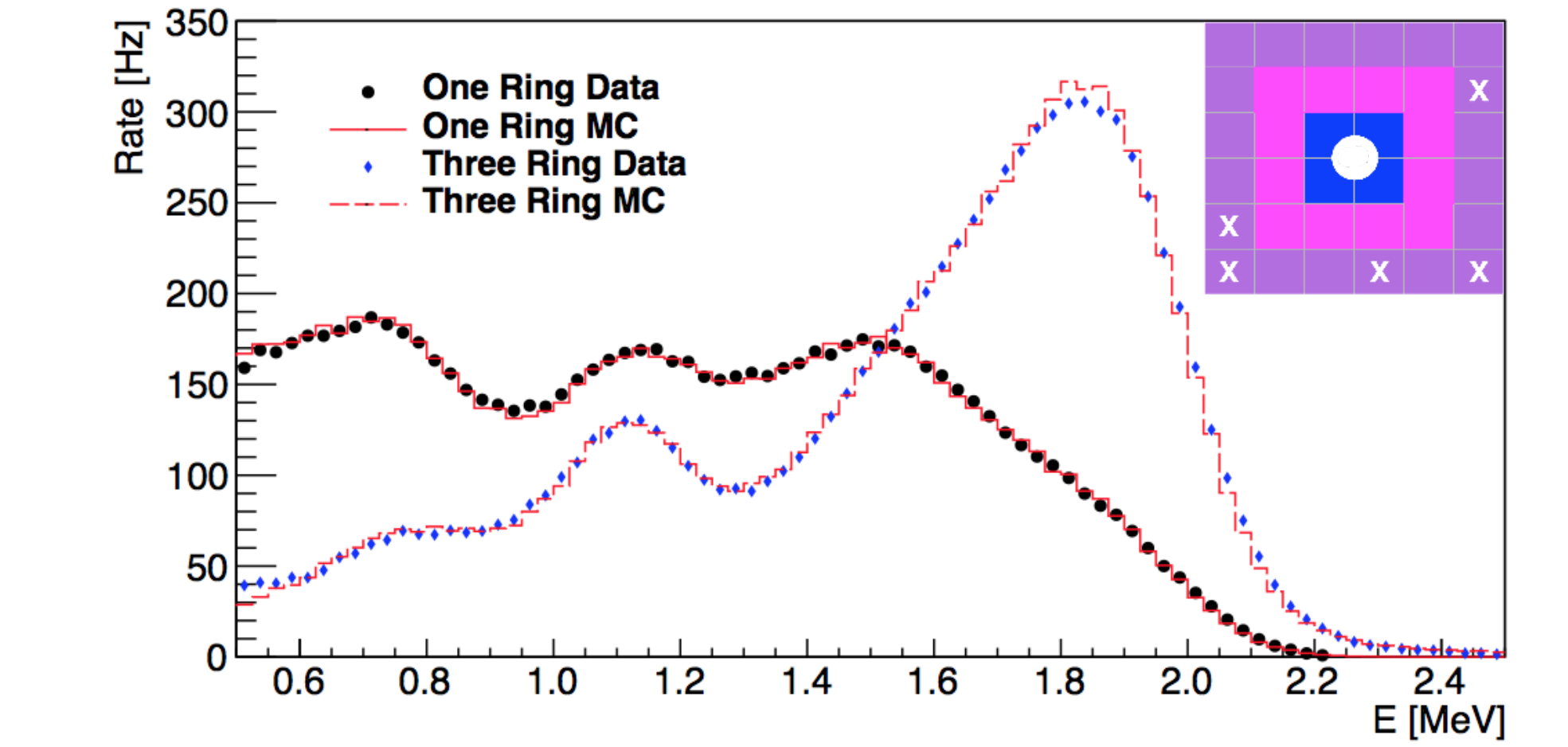}
\includegraphics[trim = 1.3cm 0.1cm 0.0cm 0.0cm, clip=true, 
width=0.49\textwidth]{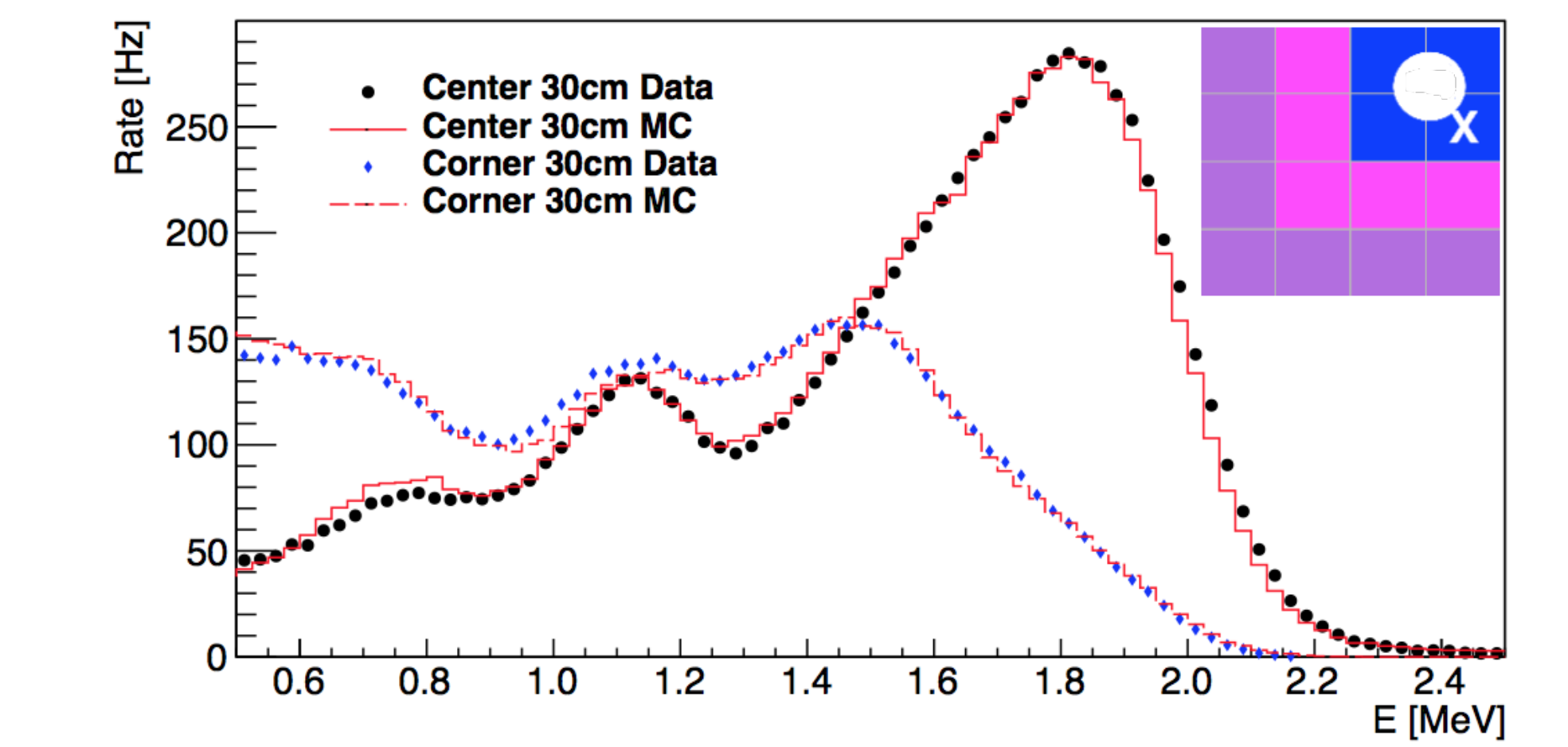}
\caption{Reconstructed energy distributions for calibration and best-fit PG4 MC $^{22}$Na source deployment datasets.  
Image insets depict the geometry of the source deployment axis and ring definitions.  `X' indicates an inactive segment; as this calibration run was taken earlier in the data-taking period, fewer dead segments are present in this analysis that in the IBD selection.  
Top: Detector-center source deployment segment-integrated energy distributions when either the nearest one or nearest three rings of detector segments are included in the integral.  Bottom: Three-ring energy distributions for source deployments at $z$=30~cm along detector center and detector corner calibration axes.  
}
\label{fig:gammaLeak}
\end{figure}

The latter effect is reflected in the top panel of Figure~\ref{fig:gammaLeak}, which shows the reconstructed spectrum from a $^{22}$Na source deployed at $z=0$ within a single ring of segments (four total segments) surrounding the $^{22}$Na source calibration axis, and within three rings of segments (36 total segments).  
The best-fit PG4 energy response model is also included for comparison.  
Incorrect modeling of the topology of annihilation $\gamma$-ray energy deposition would produce data-PG4 deviations that vary between one-ring and three-ring distributions.  
On the contrary, both the shape and scale of the PG4 and data distributions show good general agreement for both the one-ring and three-ring cases.  
By minimizing the $\chi^2$ between data and energy-shifted PG4 spectra, the relative data-PG4 scale shift for the one- and three-ring topologies is determined to be 4$\pm$1~keV and 5$\pm$1~keV respectively.  

Gamma energy leakage effects can also be demonstrated by comparing data and PG4 energy distributions for detector-center and detector-corner $^{22}$Na deployments. 
Figure~\ref{fig:gammaLeak} also shows 3-ring reconstructed energy distributions for a $^{22}$Na deployment at $z=$30~cm along the same calibration axis as above, and at $z=$30~cm along a calibration axis bordering the corner of the detector's fiducial volume.  
Again, good general agreement is found between the shape of data and PG4 distributions.  
Relative data-PG4 scale shifts are found to be 8$\pm$1~keV and 7$\pm$1~keV for these two detector positions respectively.  
This study indicates that PG4 IBD MC simulations reproduce variations in prompt energy scale arising from annihilation $\gamma$-ray energy leakage with keV-level precision.  
A conservative $\pm$8~keV uncertainty in prompt IBD energy scale due to modeling of annihilation $\gamma$-ray energy leakage is included as both a segment-correlated and segment-uncorrelated uncertainty in subsequent physics analyses.

%%%%%%%%%%%%%%%%%%%%%%%%%%%%%%%IBD SELECTION
\section{IBD Event Selection}
\label{sec:select}

Less than 1000 IBD positron-neutron pairs are expected to be produced per day in the PROSPECT inner detector by reactor antineutrinos during reactor-on data-taking periods.  
These IBD events exist amidst an intense background of reactor- and cosmogenically-produced $\gamma$-ray and neutron fluxes.  
To uncover this IBD signal, a highly effective selection based on pulse- and cluster-level reconstructed physics metrics must be performed.  
In the following section, we outline these selection criteria and discuss the expected stability of the resulting IBD detection efficiency.  
%Exact values for were ooptimized by parameter scans against effective stats metric?

%and describe how remaining backgrounds are estimated.  
%By applying the selection to the 2018 PROSPECT physics dataset described in \todo{Section}~\ref{WHAT}, we then describe the resultant signal and background datasets, as well as cross-checks performed to verify proper understanding and MC modelling of these selection cuts.  

\subsection{Antineutrino Selection}
\label{subsec:select}

The positron produced by a reactor antineutrino interaction in the PROSPECT scintillator will deposit up to about 8~MeV of kinetic energy in a small number (usually 1, 2, or 3) segments, with the highest energy deposition usually present in the segment in which the IBD interaction took place.  
The positron annihilates, almost always producing two 511 keV $\gamma$-rays, which will deposit energy in segments within a few tens of centimeters of the IBD interaction point.  
These positron-related low-density energy depositions occur on nanosecond timescales.  
Thus, the IBD selection requires an initial cluster with E$_{rec}$ between 0.8 and 7.2~MeV and individual reconstructed pulse PSD values  all within 2.0 standard deviations of the calibrated electron-like PSD mean.  
%PSD mean and variance, as described in Section~\ref{\todo{BLAH}},  are defined on a \todo{BLAH}-by-\todo{BLAH} bases as a part of low-level PSD calibrations procedures.  
No further cuts are made on the temporal or topological characteristics of the prompt cluster.
%This selection is $>$\todo{BLAH}\% efficient in its selection of IBD positrons.  

The IBD neutron is produced with less than a few tens of keV of kinetic energy and produces negligible scintillation light as it thermalizes.  
It then captures within a few tens of centimeters of the IBD interaction point with a roughly $50\,\mu$s time constant.  
Approximately 75\,\% of IBD neutrons capture on  $^6$Li, producing a $^3$H-$^4$He pair with 0.526~MeV of total visible energy.  
The high ionization density tracks of the capture products terminate within micrometers of the neutron capture point, producing scintillation light in a single segment.  
Thus, the IBD selection requires a single-pulse cluster within an (E$_{rec}$,PSD) phase space consistent with $n$-Li capture.  
That phase space is defined using the Gaussian-shaped feature corresponding to cosmogenic $n$-Li capture events in this space (Figure~\ref{fig:sigpsd}), with energy required to be within $\pm$3$\sigma$ of the mean value of 0.526~MeV and PSD within $\pm$2$\sigma$ of the PSD mean value.
This delayed cluster is required to occur within 120~$\mu$s of the prompt cluster; its segment S$_{rec}$ must be the same as or vertically/horizontally adjacent to that of the prompt cluster.  
If S$_{rec}$ are identical, the prompt-delayed Z$_{rec}$ difference is required to be less than 140~mm; if S$_{rec}$ are adjacent, Z$_{rec}$ spacing must be less than 100~mm.  

To remove activity associated with cosmogenic muons and other high-energy events capable of creating significant numbers of delayed secondaries, IBD candidates are rejected if their delayed capture times are within 200~$\mu$s of a preceding cluster with E$_{rec}>$ 15~MeV; this cut is referred to as a `muon veto.'
To similarly reject cosmogenic neutron-related activity, IBD candidates are rejected if their delayed capture occurs within 400~$\mu$s of another $n$-$^6$Li candidate, or within 250~$\mu$s of a preceding cluster with E$_{rec}>$ 0.25~MeV and at least one pulse with a PSD larger than 2$\sigma$ above peak of the electron-like PSD band.  
These cuts are referred to as the `neutron veto' and `recoil veto,' respectively.  
These three cuts are also referred to collectively as a `cosmic veto.'
IBD candidates are also rejected if either cluster occurs within 0.8~$\mu$s of a previous cluster; this cut, referred to as the `pile-up veto' reduces ambiguities in the calculation of trigger-related dead times.  

%To reject IBD candidates events associated with large neutron showers and subsequent combinations of inelastic neutron scatters and $n$-H and $n$-$^6$Li captures, IBD candidates are required to be separated from other IBD candidates by at least~800~$\mu$s.  

PG4 MC simulations of cosmogenic processes also indicate that neutron-related backgrounds are concentrated on the edges of the active region~\cite{prospect_nim}; for this reason, IBD candidates are rejected if their prompt or delayed S$_{rec}$ is within the outer-most layer of segments on the detector top and sides.  
Signals in two segments in the bottom back corner of the detector are similarly rejected due to high reactor-on trigger rates in these segments from reactor $\gamma$-ray backgrounds.  
IBD candidates are are rejected if prompt or delayed Z$_{rec}$ values are within 140~mm of the segment ends.  
These segment and $z$-end exclusion cuts are referred to as `fiducialization' in following sections.  

\begin{figure}[hptb]
\includegraphics[trim = 0.0cm 0.15cm 0.3cm 0.0cm, clip=true, 
width=0.48\textwidth]{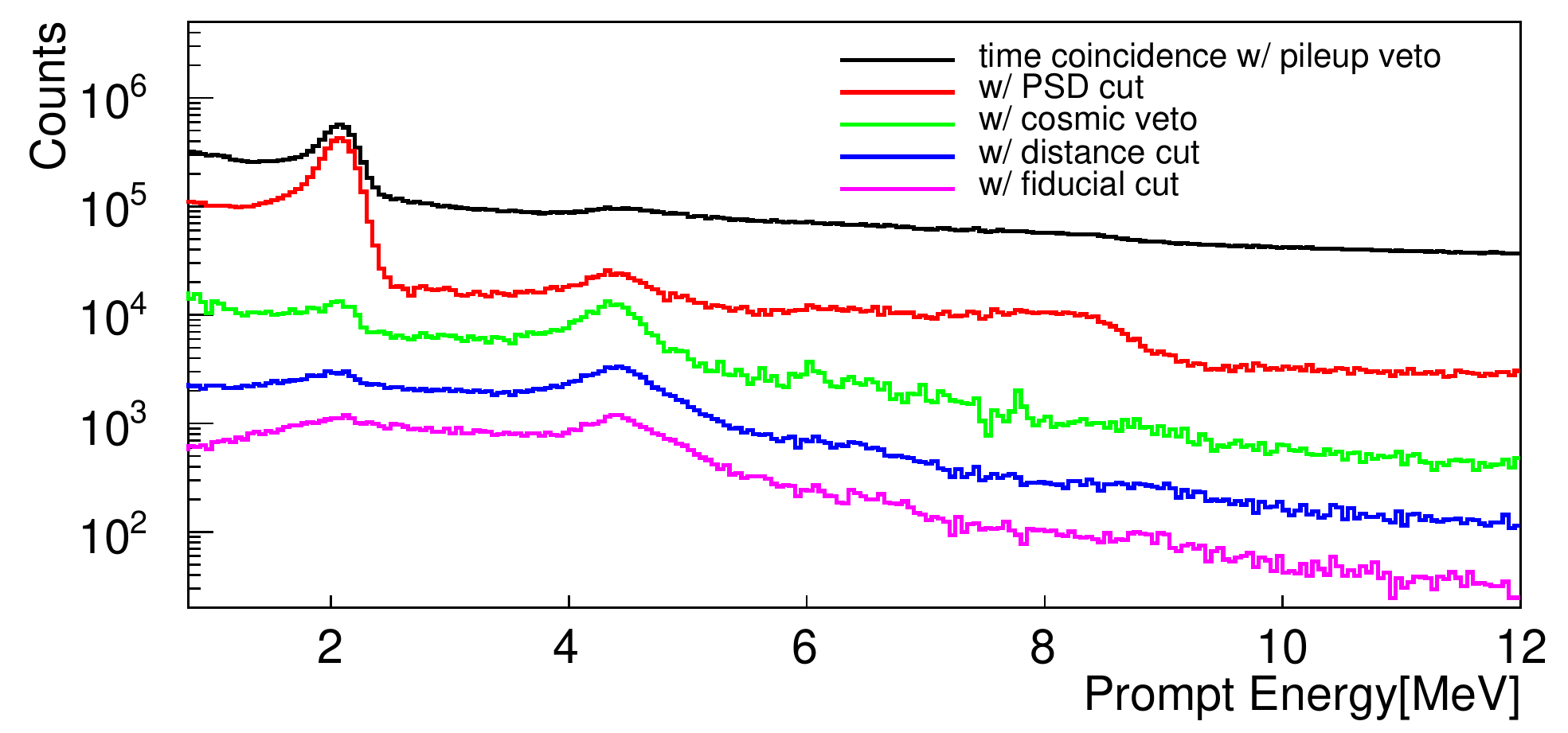}
\includegraphics[trim = 0.0cm 0.15cm 0.3cm 0.0cm, clip=true, 
width=0.48\textwidth]{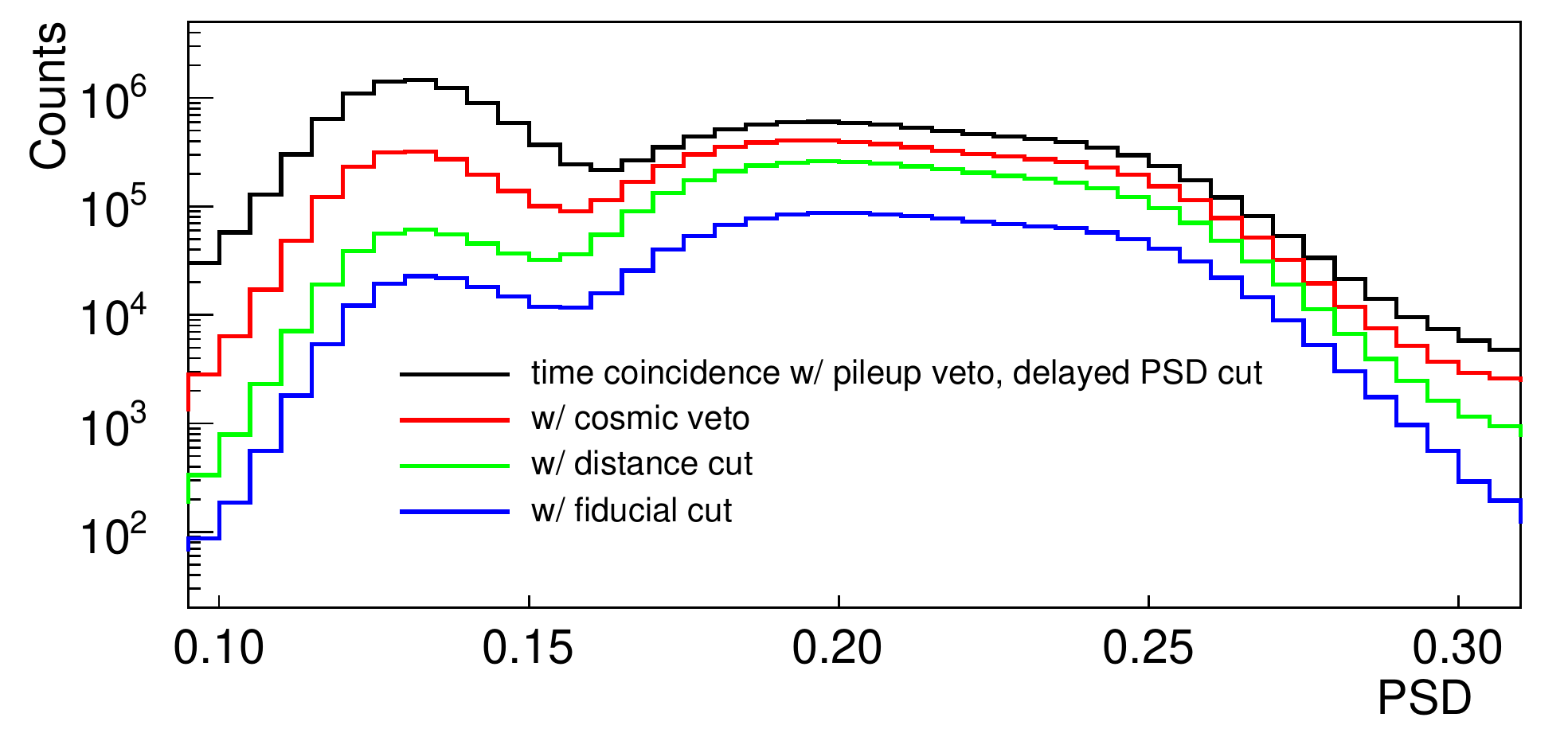}
\caption{Distributions of prompt $E_{rec}$ and reconstructed PSD of the highest-$E_{smear}$ prompt pulse (bottom) as IBD selection cuts are sequentially added to the PROSPECT reactor-on dataset.  Applied cuts are described in the text. 
Distributions include subtraction of accidentally time-coincident backgrounds, which is described in Section~\ref{sec:bkg}.}
\label{fig:select}
\end{figure}

Figure~\ref{fig:select} illustrates the reduction in IBD candidates upon sequential application of the IBD selection cuts described above during reactor-on data-taking; distributions include subtraction of accidentally time-coincident backgrounds, which is described in Section~\ref{sec:bkg}.  
A two to three order of magnitude reduction in IBD candidates is observed after all cuts are applied.  
The reactor-on prompt E$_{rec}$ distribution in Figure~\ref{fig:select} exhibits a smooth event distribution peaking between 2-3 MeV and falling at higher energies, consistent with the expected energy distribution of reactor $\nuebar$ IBD interactions; however, peak-like features also appear in this distribution, indicating the residual presence of background IBD candidates.  
The PSD distribution in Figure~\ref{fig:select} exhibits a double-humped structure matching that expected from prompt IBD positrons (low PSD) and prompt nuclear recoils (high PSD), gamma interactions from inelastic scatters (low PSD), and captures (high or low PSD for captures on $^6$Li and hydrogen, respectively) produced by cosmogenic neutrons.  
We note that due to integration over a broad energy and time range, the high and low PSD distributions observed in Figure~\ref{fig:select} are smeared out and provide an incomplete representation the detector's true PSD sepration capability.

\begin{figure}[hptb]
\includegraphics[trim = 0.0cm 0.15cm 0.0cm 1.5cm, clip=true, 
width=0.48\textwidth]{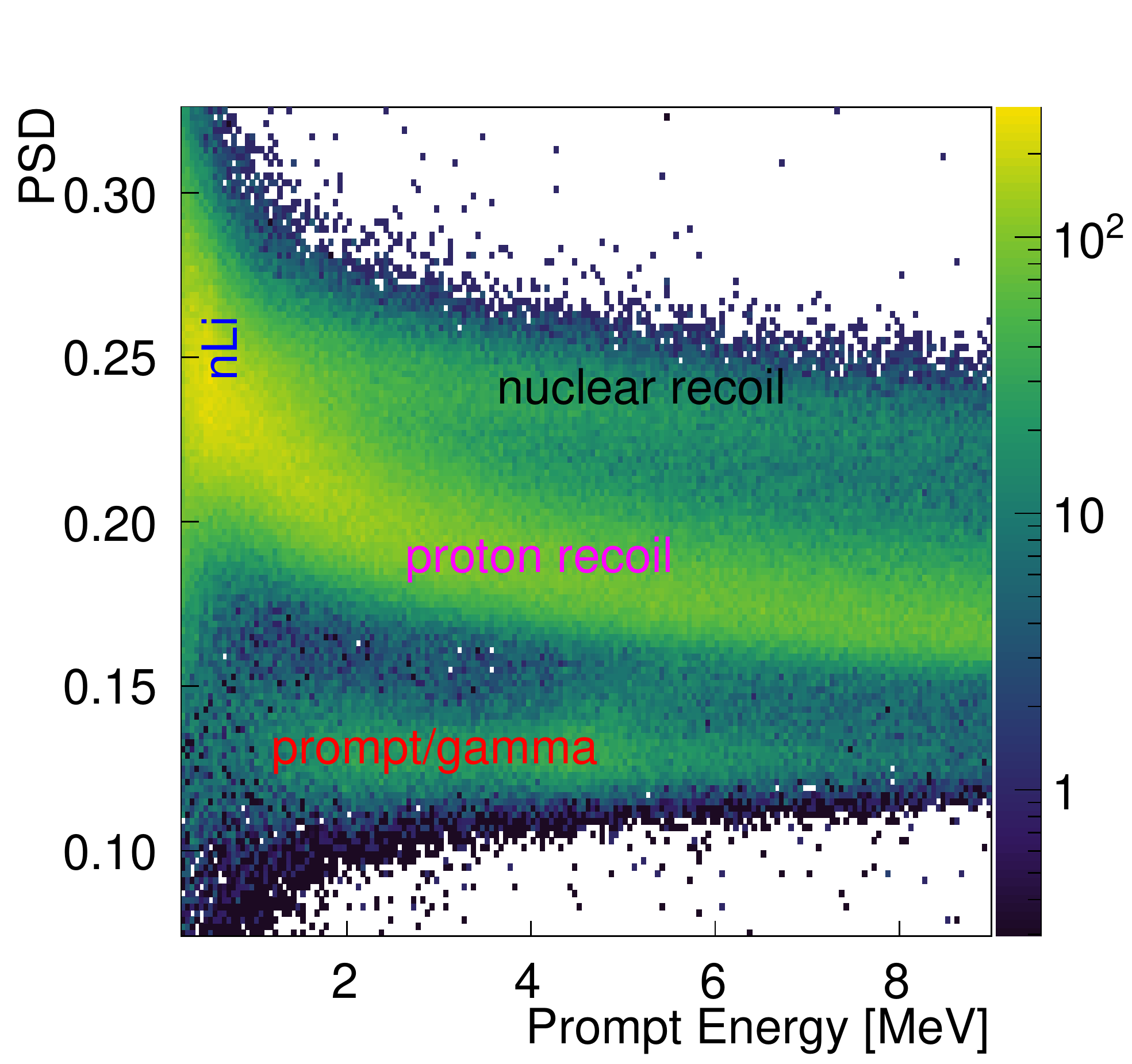}
\caption{Prompt energy/PSD distributions for selected IBD-like events from one reactor-on cycle.   Pictured are the total prompt E$_{rec}$ as well as the PSD value for the pulse of highest reconstructed energy within the prompt cluster.  The labelled regions contain IBD candidates (red), coincident ($n$-Li,$n$-Li) captures (blue), ($n$-$p$,$n$-Li) scattering and capturing fast neutrons on protons (magenta) and other heavier nuclei (black).  We note that a subset of prompt clusters inside the IBD candidate labelled band will also contain high-PSD pulses, and will not be selected as IBD candidates.}
\label{fig:select2D}
\end{figure}

IBD candidates are also investigated in Figure~\ref{fig:select2D} by simultaneously plotting the PSD and energy distributions for the most restrictive selection given in Figure~\ref{fig:select}.  
Pictured are the total summed prompt E$_{rec}$, as well as the PSD value for the pulse of highest reconstructed energy within the prompt cluster.  
The elongated band at low PSD value represents the area containing all selected IBD candidates, as well as a subset of non-IBD events containing sub-dominant prompt cluster pulses with high PSD values, e.g. due to the recoil from inelastic scattering.  
%The peak-like features visible in the top panel of~\ref{fig:select} are also visible within this band.  

Two other regions of potential IBD-like backgrounds are also highlighted in Figure~\ref{fig:select2D}.  
One isolated region at low energy and high PSD is produced by the time-coincident captures of two neutrons on $^6$Li, which are a signature of multi-neutron cosmogenic showers.  
Another region inhabiting a broad energy range at high PSD is produced by the scattering and subsequent $^6$Li capture of a single energetic cosmogenic neutron.   
These event classes, designated ($n$-Li,$n$-Li) and ($n$-p,$n$-Li), will be used to further investigate the impact of multi-neutron showers and high-energy cosmogenic neutrons on PROSPECT signals.  
In these investigations, the latter ($n$-p,$n$-Li) class will also include rejected events in the IBD-like band of Figure~\ref{fig:select2D} that contain a sub-dominant high-PSD prompt cluster pulse.  
The prompt parameter distribution in Figure~\ref{fig:select2D} clearly demonstrates the highly-effective reduction in copious multi-neutron and fast-neutron backgrounds made possible by PROSPECT's prompt PSD capabilities.  
Interestingly, an additional band visible at higher prompt PSD than the ($n$-p,$n$-Li) events is likely produced by fast neutron recoils on other heavier nuclei.

\subsection{IBD Monte Carlo Simulation}
\label{subsec:ibdmc}

After the parameter optimization described in the previous sections, PG4 IBD MC simulation datasets can be produced that include effects of energy response non-linearity, IBD positron energy loss and energy leakage, and energy resolution smearing.  
At the same time, the IBD MC must also accurately model the position distribution of IBD interactions within the PROSPECT detector, the behavior of IBD neutrons as they propagate through the detector, and detection efficiency variations associated with the IBD selection criteria.  
All of these aspects of the simulation are required to fully characterize the relationship between true \nuebar energy spectra and prompt IBD E$_{rec}$ spectra, which is essential when comparing predicted oscillated and unoscillated reactor \nuebar flux models to selected IBD candidates.

In the PG4 IBD MC simulation, an IBD vertex positioner module is first used to ensure proper placement of IBD interactions throughout the inner detector.  
To first order, IBD vertices are distributed according to a 1/r$^2$ distribution in the inner detector given the known reactor-detector center-to-center baseline reported in Section~\ref{sec:exp}.  
Vertices are generated in all detector materials, including the scintillator, optical grid components, PMT housing faces, and acrylic supports; vertex densities are varied to properly account for relative proton density differences between the materials in these different components.  
Vertex locations can be generated using either a point-like core geometry, or one incorporating the finite cylindrical shape of the reactor core as described in Section~\ref{sec:exp}.  
For the purpose of generating descriptions of detector IBD energy response, the point-like and cylindrical core geometry yield nearly identical results; given its quicker processing time, the point-like geometry is used.  
For the purpose of generating realistic distributions of $\nuebar$ production-interaction baselines for the oscillation analysis, the cylindrical reactor geometry is used.  

Final state positrons and neutrons are generated at each simulated IBD interaction vertex with kinetic energy and momentum distributions defined by the IBD cross-section~\cite{Vogel:1999zy} given the incoming neutrino direction and energy.
At reactor $\nuebar$ energies, this will produce IBD positrons (neutrons) with momenta preferentially directed back towards (away from) the reactor core.  
IBD neutrons, produced with $\mathcal{O}$(keV) energies, will thermalize and scatter prior to capture.  
The Geant4 libraries ``G4NeutronHPElastic" and ``G4NeutronHPThermalScattering" are implemented to model the propagation above and below 4~eV, respectively; the latter is modelled assuming thermal scattering by unbound hydrogen atoms.  
IBD positrons are propagated using the default Geant4 ``emstandard" libraries.  
The simulated detector geometry, translation from scintillator-deposited true energy to quenched energy, and PMT waveform simulation is as described in Section~\ref{subsec:MC}.  

\begin{figure}[hptb!]
\includegraphics[trim = 1.7cm 0.5cm 0.8cm 0.4cm, clip=true, width=0.49\textwidth]{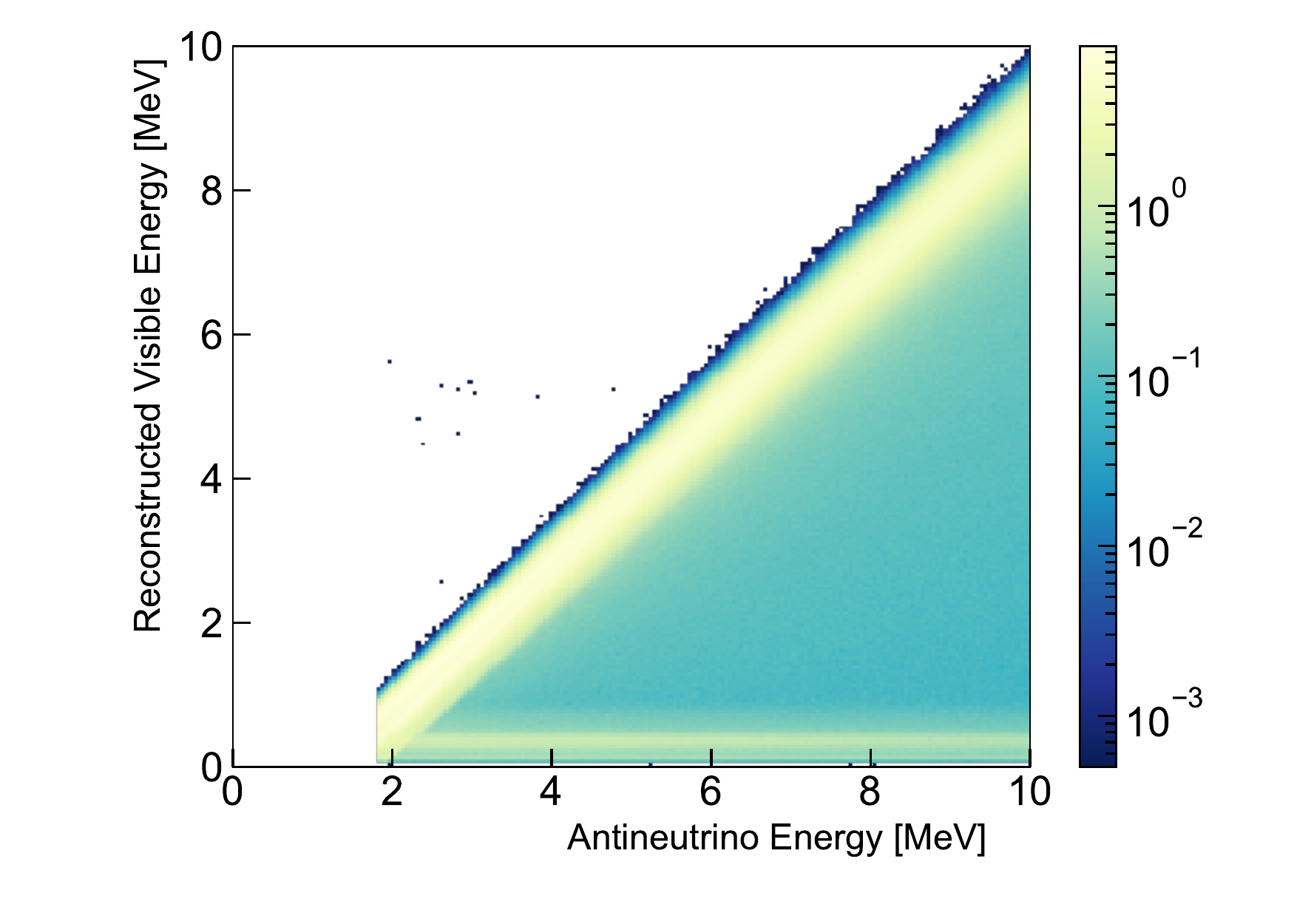}
\includegraphics[trim = 0.5cm 0.15cm 0.0cm 0.5cm, clip=true, width=0.49\textwidth]{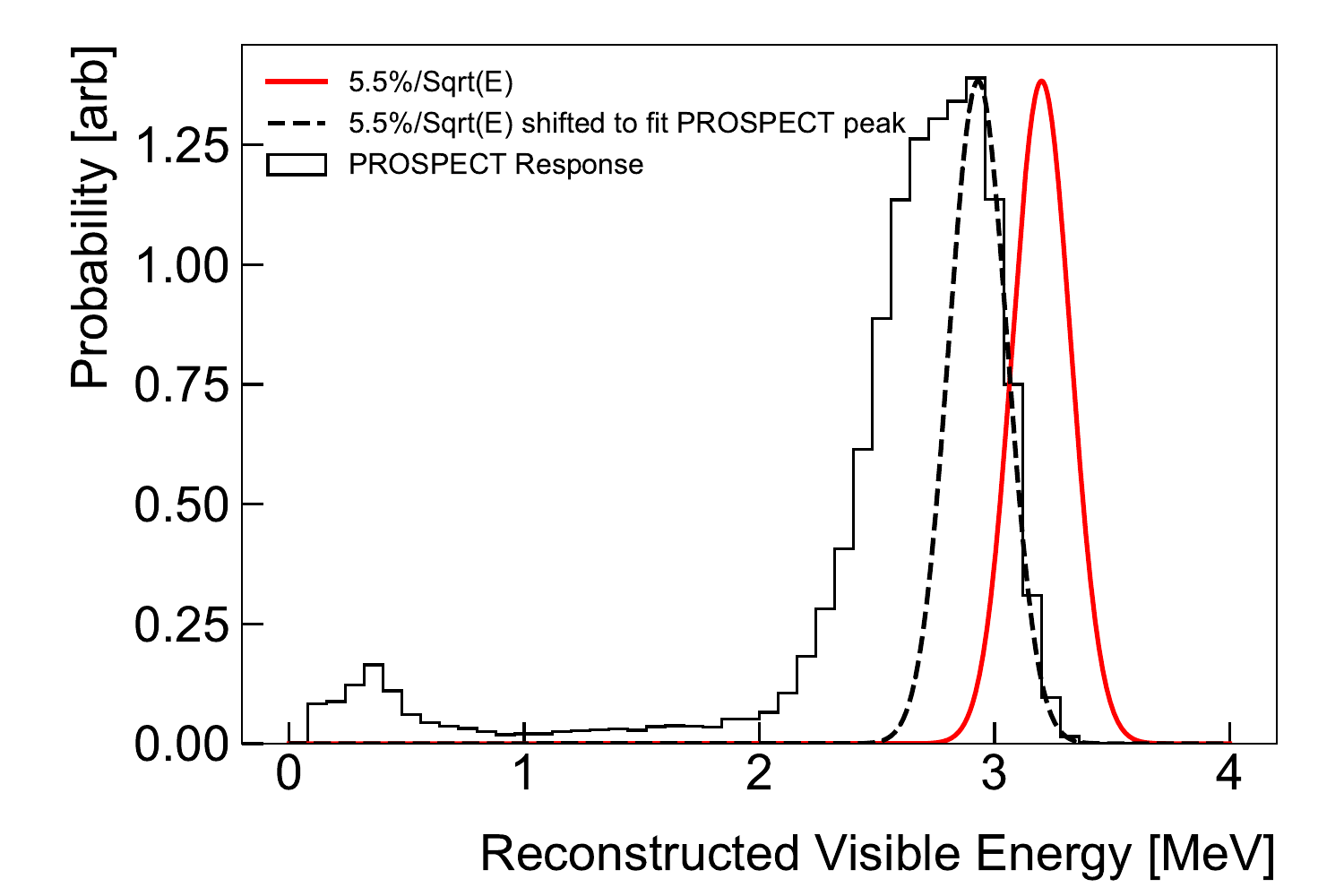}
\caption{Top: PROSPECT Detector response matrix describing the relationship between true \nuebar and reconstructed energies, as modelled by the best-fit PG4 detector simulation.  The matrix is generated using only output from the active detector segments used in the oscillation and spectrum analyses.  
Bottom: PG4-modelled E$_{rec}$ distribution in response to mono-energetic 4.0~MeV \nuebar evenly distributed throughout the detector.  A photostatistics-smeared, full-energy peak from this source is also plotted; see the text for detailed description of these distributions.} 
\label{fig:fullresponse}
\end{figure}

The position-integrated relationship between E$_{\nu}$ and E$_{rec}$ for the full IBD MC is illustrated in Figure~\ref{fig:fullresponse}; this detector response matrix is directly used in the PROSPECT \uFive~spectrum analysis, and is included in tabulated form in the attached supplementary materials.  
The matrix is generated using only output from the active detector segments used in these analyses.  
For the oscillation analysis, similar E$_{\nu}$ to E$_{rec}$ translation matrices are also generated separately for all individual PROSPECT segments.  
To simplify the generation of these per-segment matrices and address ambiguities related to true $\nuebar$ baselines, only MC IBD events with prompt S$_{rec}$ containing the true IBD vertex are considered.  
While this choice reduces the IBD MC sample by $3$\% for each active detector segment and ignores signal candidates from IBD interactions in inactive segments, these exclusions are found to produce negligible bias in the oscillation fit.  

Figure~\ref{fig:fullresponse} also includes an illustration of the E$_{\nu}$-E$_{rec}$ relationship for 4.0 MeV of monoenergetic \nuebar energy, corresponding to a vertical slice of the full detector response matrix.  
This distribution is accompanied by the true full-energy prompt positron signature expected from a 4.0~MeV neutrino as described by Equation~\ref{eq:IBD}, smeared by the 5.5\% photo-statistics energy resolution realized in the reconstructed IBD dataset.  
The added resolution smearing contributed by positron kinetic energy loss in non-active materials and annihilation $\gamma$-ray energy leakage is obvious here, and dominates the smaller photo-statistics resolution effect.  
A large off-diagonal contribution can be seen at low E$_{rec}$ arising largely from positron kinetic energy deposition in non-active detector regions.  
A relative offset between full and reconstructed energy peaks is also visible; this feature is a byproduct of both a mean per-event energy loss in non-active materials, as well as scintillator non-linearity effects which categorically reduce reconstructed energies below that of the true deposited energy.  
\subsection{IBD Detection Efficiency Variations}
\label{subsec:eff}

The efficiency of analysis cuts in selecting IBD interactions in active fiducial segments is estimated to be 30-40\% using PG4 IBD MC simulations.  
Some cuts are highly efficient: requirements on prompt E$_{rec}$ and PSD, prompt-delayed time coincidence, and segment and $z$ prompt-delayed spatial proximity cuts each remove less than 10\% of IBD events.  
Delayed cluster cuts are $\sim$70\% efficient, largely due to IBD neutron captures on nuclei other than $^6$Li.
Cosmogenic and closely-spaced cluster veto cuts remove $\sim$12\% (10\%) of the total detector live time during reactor-on (off)  periods.  
An additional $\sim$25\% inefficiency is introduced by $z$-fiducialization of each segment.  
%Detector fiducialization results in a sizable \todo{BLAH}\% reduction in IBD selection rates.  
 %variations in veto time during reactor-on periods arise from the time-varying reactor-related gamma rate, which will be described further in Section~\ref{\todo{BLAH}}.  
Precise quantitative demonstration of these detector-wide efficiencies is not elaborated upon further as this quantity is not a necessary input for the spectrum or oscillation  measurements presented in this paper.  

In contrast, relative variations in efficiency between segments, and between time periods, are important for both reported measurements, and must be characterized.  
Due to edge effects and non-active detector segments, the efficiency of IBD detection is expected to be position-dependent in PROSPECT.  
Relative efficiency variations between segments, if not correctly characterized, can mimic baseline-dependent \nuebar disappearance effects for low-$\Delta m^2$ scenarios.  
Segments with relatively high detection efficiencies also play an outsized role in determining baseline-integrated detector energy response; thus, an understanding of the fractional signal contribution of each segment is a necessary input to comparing predicted and detected \uFive~\nuebar spectra.  
Variations in detector performance exhibited by PROSPECT also result in time-varying IBD detection efficiency, which complicates the subtraction of backgrounds from the IBD signal.  
The remainder of this section will characterize IBD efficiency variations observed or expected in the PROSPECT detector, and describe any uncertainties or biases in the IBD signal associated with these variations.  
 
 \subsubsection{Position-Dependent Efficiency Variations}
\label{subsubsec:eff-pos}

\begin{figure}[hptb]
\includegraphics[trim = 2.7cm 14.3cm 2.2cm 2.6cm, clip=true, 
width=0.49\textwidth]{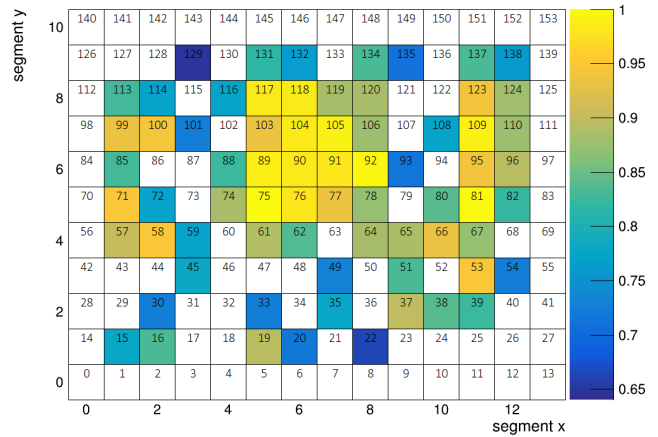}
\caption{Simulated IBD detection efficiency for each PROSPECT segment, given relative to the segment of highest efficiency.  The uncertainty for the relative efficiency of each segment is 0.5\%.  Also pictured is the default PROSPECT segment numbering scheme.}
\label{fig:releff}
\end{figure}

The primary driver of IBD selection efficiency non-uniformity with position is neutron mobility.   
Thermalizing IBD neutrons can migrate out of the active detector region, or into nearby inactive segments, where they are not detected.  
The magnitude of this effect is well-demonstrated in Figure~\ref{fig:releff}, which shows the simulated efficiency of detecting IBDs generated in each active fiducial segment, relative to the segment of highest efficiency.
Efficiencies are found to be as much as 25\% lower in segments adjacent to larger numbers of inactive or non-fiducial segments.

\begin{figure}[hptb!]
\includegraphics[trim = 0.0cm 0.75cm 0.3cm 1.15cm, clip=true, 
width=0.48\textwidth]{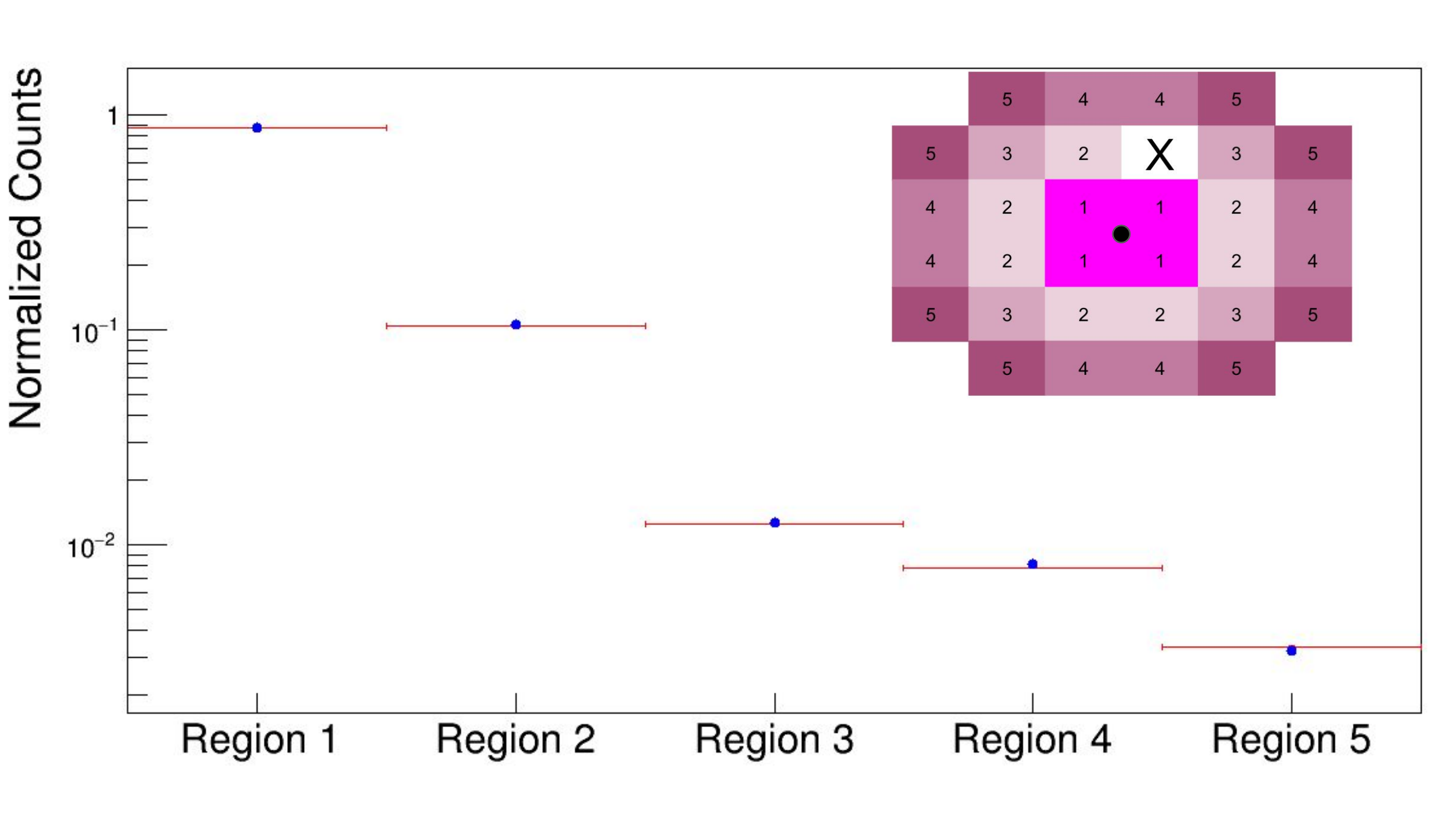}
\caption{Detected nLi capture rates for regions of differing distance to $^{252}$Cf source deployment location (black circle in region scheme). The inset image defines which segments are assigned to which region bin.  In this inset, `X' indicates an inactive segment; as this calibration run was taken earlier in the data-taking period, fewer dead segments are present in this analysis than in the IBD selection.  Blue dots represent data, while red lines represent PG4 MC simulations.}
\label{fig:mobility}
\end{figure}

Neutrons produced by a $^{252}$Cf source deployed for 1 hour in a calibration axis near the detector center were used to verify the modelling of neutron mobility by PG4.  
$^{252}$Cf neutron signals were selected by requiring time- and position-coincident clusters from prompt low-PSD fission $\gamma$-rays and delayed high-PSD fission neutron-$^6$Li captures.   Figure~\ref{fig:mobility} demonstrates the fractional contribution of $^{252}$Cf neutron captures in the different regions surrounding the source deployment axis.  
PG4-simulated fractional contributions using a custom $^{252}$Cf generator are also pictured.  
Good agreement is exhibited between predicted and measured capture locations.  

The mobility of the IBD positron and its annihilation $\gamma$-rays will also generate a segment-dependent variation in IBD cut selection efficiency.  
However, this effect is substantially smaller than that of neutron mobility: as an example, PG4 IBD MC predicts that the S$_{rec}$ for a selected IBD will differ from the IBD interaction segment only 3\% of the time, compared to a 25\% migration fraction for delayed neutrons.  
The small mobility effect, combined with relatively loose cuts  applied to prompt cluster energies and the absence of cuts on prompt signal topology, results in a percent-level efficiency variation associated with the prompt signal.  

Since prompt and delayed mobility effects are modelled in PG4, their impact on IBD are taken into account in the oscillation and spectrum analysis.   
%As demonstrated in the previous sub-section, prompt signal mobility impacts on IBD prompt energy response are also fully accounted for in PG4 simulations.  

Minor IBD segment-to-segment signal rate variations from a variety of other sources were also investigated.  
Given their small size, the following effects were not included in PG4 MC simulations.  
Instead, they were encapsulated in segment-uncorrelated signal rate systematic uncertainty estimates, along with uncertainties in the PG4-modelled efficiency variations.  

\begin{figure}[phtb!]
\includegraphics[trim = 0.3cm 0.15cm 0.5cm 8.0cm, clip=true, width=0.48\textwidth]{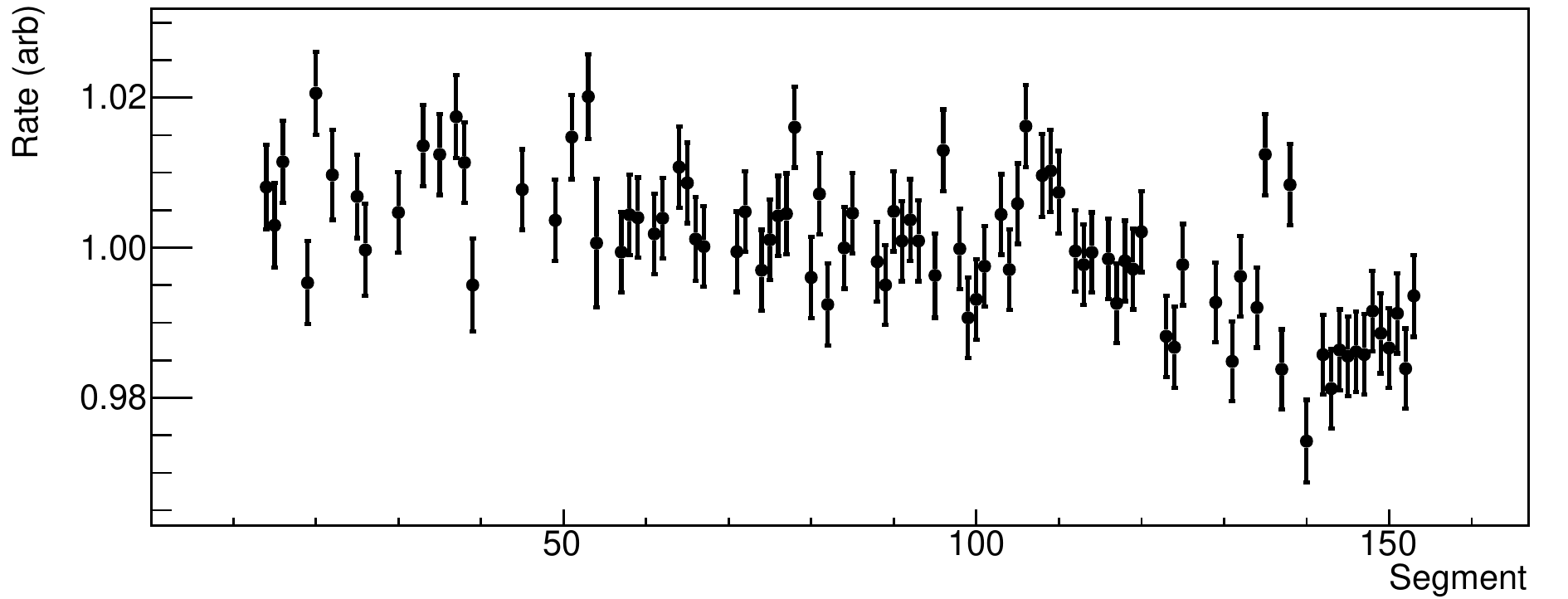}
\caption{Relative rate of detected correlated ($^{219}$Rn,$^{215}$Po) decays from $^{227}$Ac, as calculated for each fiducialized segment.  Segment numbers increase from detector bottom rows to top rows (increasing $y$), as illustrated in Figure~\ref{fig:releff}; within a row, segment numbers increase with increasing $x$.  All values are given relative to the mean of 3.3~mHz.  Error bars represent statistical uncertainties.}
\label{fig:rnporate}
\end{figure}

Detected IBD rate variations can arise from differing segment masses.   
Owing to the mm-level survey of the optical grid during detector assembly and the rigid optical grid mechanical structure, realized segment geometries are expected to have volumes identical to $<$1\%. 
Differences in fiducialization efficiencies can arise from inconsistent z$_{rec}$ between segments.   
As described in Section~\ref{subsec:perform}, $z$ offsets between segments are less than 1~cm, while $z$ resolutions for ($^{219}$Rn,$^{215}$Po) events vary between segments by less than 1~cm.  
Given the 89~cm fiducial segment length, this per-segment resolution variation corresponds to less than 2\% variation in $z$ fiducialization efficiency per segment.  
Characterization of the combined effects of variation in segment volumes and $z$-fiducialization can be performed by comparing rates of detection of uniformly-distributed correlated  ($^{219}$Rn,$^{215}$Po) decays between fiducial segments, which can be selected with extremely high efficiency ($>$99.9\%) and purity.  
As demonstrated in Figure~\ref{fig:rnporate}, rates are found to be similar in all fiducial segments to within $\pm$2\%.  

%For rates versus time, best-fit and true half-lives for $^{227}$Ac are also pictured.

Given the comparatively high PSD cut efficiencies and relatively consistent segment-to-segment PSD response, PSD cuts are expected to introduce negligible segment-to-segment variation in detected IBD signal rates.  
Cosmogenic and other IBD veto cuts are applied at the full-detector level and are also expected to have negligible impacts on relative IBD signal rates.  
Since none of the possible sources of IBD rate variation between segments for the oscillation analysis described above exceed 2\%, a conservative 5\% segment-uncorrelated efficiency uncertainty is applied.  

 \subsubsection{Time-Dependent Efficiency Variations}
\label{subsubsec:eff-time}

A variety of time-dependent aspects of the IBD selection have been identified.  
Many, such as variations in the optical and PSD performance of the detector, have been effectively mitigated during the process of low-level detector calibration, as described in Section~\ref{sec:calib}.   
Remaining time-dependent aspects of the selection after calibration must be quantified and either corrected or taken into account in uncertainty estimates.  

\begin{figure}[phtb!]
\includegraphics[trim = 0.0cm 0.25cm 0.0cm 0.0cm, clip=true, width=0.48\textwidth]{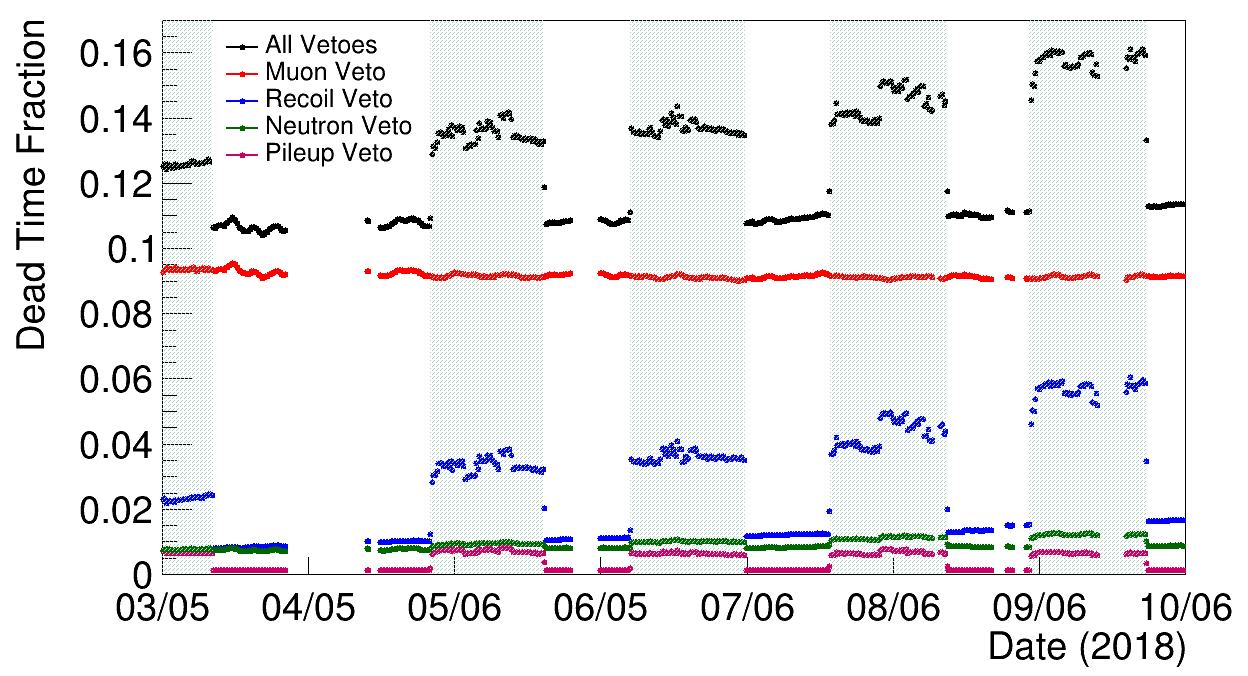}
\caption{Total dead time fraction associated with IBD selection veto cuts (black) as well as individual dead time fractions associated with the muon (red), recoil (blue), neutron capture (green) and pileup (pink) vetoes. Vertical green shaded bands indicate reactor-on periods.}
\label{fig:veto_eff}
\end{figure}

The primary source of time-dependence in detected IBD-like rates arises from changes in dead time fractions from muon, neutron, recoil, and pile-up veto cuts, which were described in Section~\ref{subsec:select}.  
These effects are illustrated in Figure~\ref{fig:veto_eff}, which shows, as a function of time, the fractional detector-wide dead time associated with these cuts.  
Veto fractions are systematically higher while the reactor is running.  
In addition, veto dead time fractions vary within individual reactor-on periods, while also increasing systematically with time during both reactor-on and reactor-off periods.  
Clearly, precisely correcting for dead time differences must be done in order to compare IBD-like rates between different time periods.  

\begin{table*}[htbp!]
\centering
\begin{tabular}{|c|c|c|c|c|c|}
\hline 
Event Type & Associated Veto & Reactor-Off Rate (Hz) & On-Off Offset (Hz) & Coefficient (\%/mbar) & On-Off Scaling (\%) \\ 
\hline 
single cluster & Pile-up & 1628 & 6708 & - & - \\ \hline
single $n$-$p$ & Recoil & 46.8 & 116 & - & - \\
single $n$-Li & Neutron & 11.5 & 2.85 & -0.57 $\pm$ 0.23 & 0.025 $\pm$ 0.015 \\
single muon & Muon & 497 & -2.3 & -0.16 $\pm$ $<$0.01 & 0.006 $\pm$ 0.000 \\ \hline
$n$-Li, $n$-Li & - & 0.012 & 8.5e-4 & -0.53 $\pm$ 0.01 & 0.022 $\pm$ 0.024 \\
$n$-$p$, $n$-Li & - & 0.33 & 4.2e-4 & -0.80 $\pm$ 0.02 & 0.033 $\pm$ 0.007 \\
IBD-like & - & 0.0052 & 7.1e-3 & -0.70 $\pm$ 0.01 & 0.028 $\pm$ 0.048 \\ \hline
%\hline
\end{tabular} 
\caption{Rates, barometric coefficients, and on-off scaling coefficients for different types of single (top) and correlated (bottom) event categories; barometric and scaling coefficients are used for cosmic background estimation in Section~\ref{subsec:bkg_corr}.  When relevant, the IBD veto cut associated with the listed event type is specified.  Time-integrated rates, as well as rate differences between reactor-on and -off periods, are provided.  Given the large non-atmospheric time-dependent changes in single $n$-$p$ and single cluster detection rates, atmospheric coefficients and reactor-off background scaling coefficients are not calculated for these signals.  }
\label{tab:atmscaling}
\end{table*}

To understand these veto fraction time variations, rates of the various vetoing event classes are investigated.  
Table~\ref{tab:atmscaling} overviews the rates of these and other event classes.  
Inclusive trigger rates, which determine the pileup veto dead time, are obviously the highest shown in Table~\ref{tab:atmscaling}, and exhibit substantial on-off differences.  
However, given the short veto window length for this veto (0.8~$\mu$s), this cut produces less than 1\% dead time during both reactor-on and reactor-off periods.  
Muons represent the second most common veto event class but exhibit relatively little variation between reactor-on and reactor-off periods.  
Thus, while the comparatively longer veto window length of this class (200~$\mu$s) produces the largest overall dead time contribution, it is relatively constant between reactor-on and reactor-off data periods.  
Recoil vetoes, the next most common class, exhibit relatively high rates as well as substantial on-off variation.  
This class largely arises not from true neutron-proton recoils, but from the small fraction of $\gamma$-ray flux in the high tail of the $\gamma$-like PSD distribution.  
Gamma fluxes vary substantially between reactor-on and -off periods and within individual reactor-on periods; see Section~\ref{subsec:bkg_acc} for an in-depth description of these variations.  
Moreover, the slow reduction in light yield described in Section~\ref{sec:calib} expands the overlap between high-PSD and low-PSD bands over time.  
For these reasons, this event class contributes the majority of time-dependence in total veto dead times.  
Neutron vetoes exhibit the lowest rate of any veto class, and contribute less than 1\% to total dead time.  

\begin{figure}[phbt!]
\includegraphics[trim = 0.3cm 0.8cm 0.0cm 0.4cm, clip=true, width=0.49\textwidth]{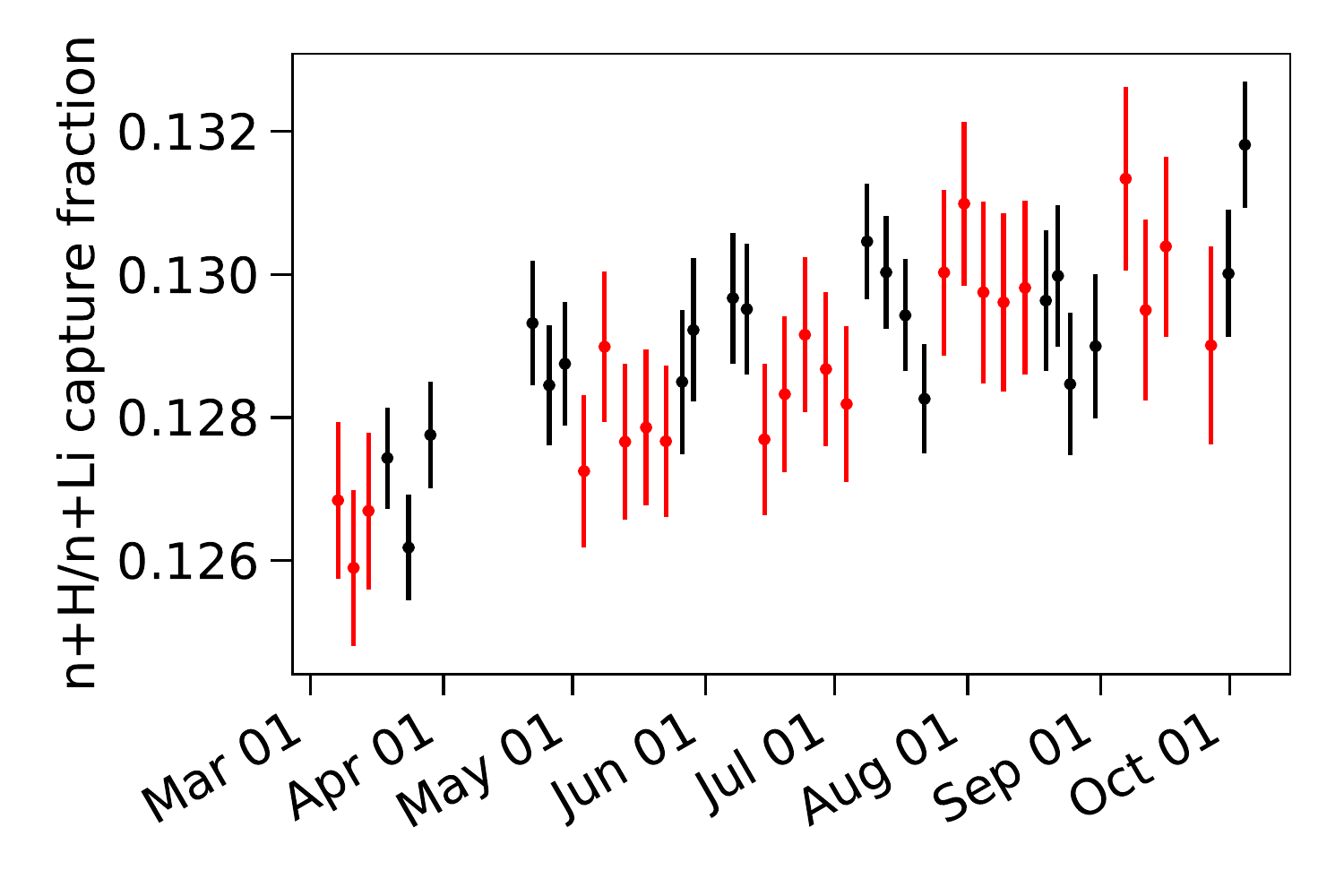}
\includegraphics[trim = 0.0cm 0.8cm 0.0cm 0.4cm, clip=true, width=0.49\textwidth]{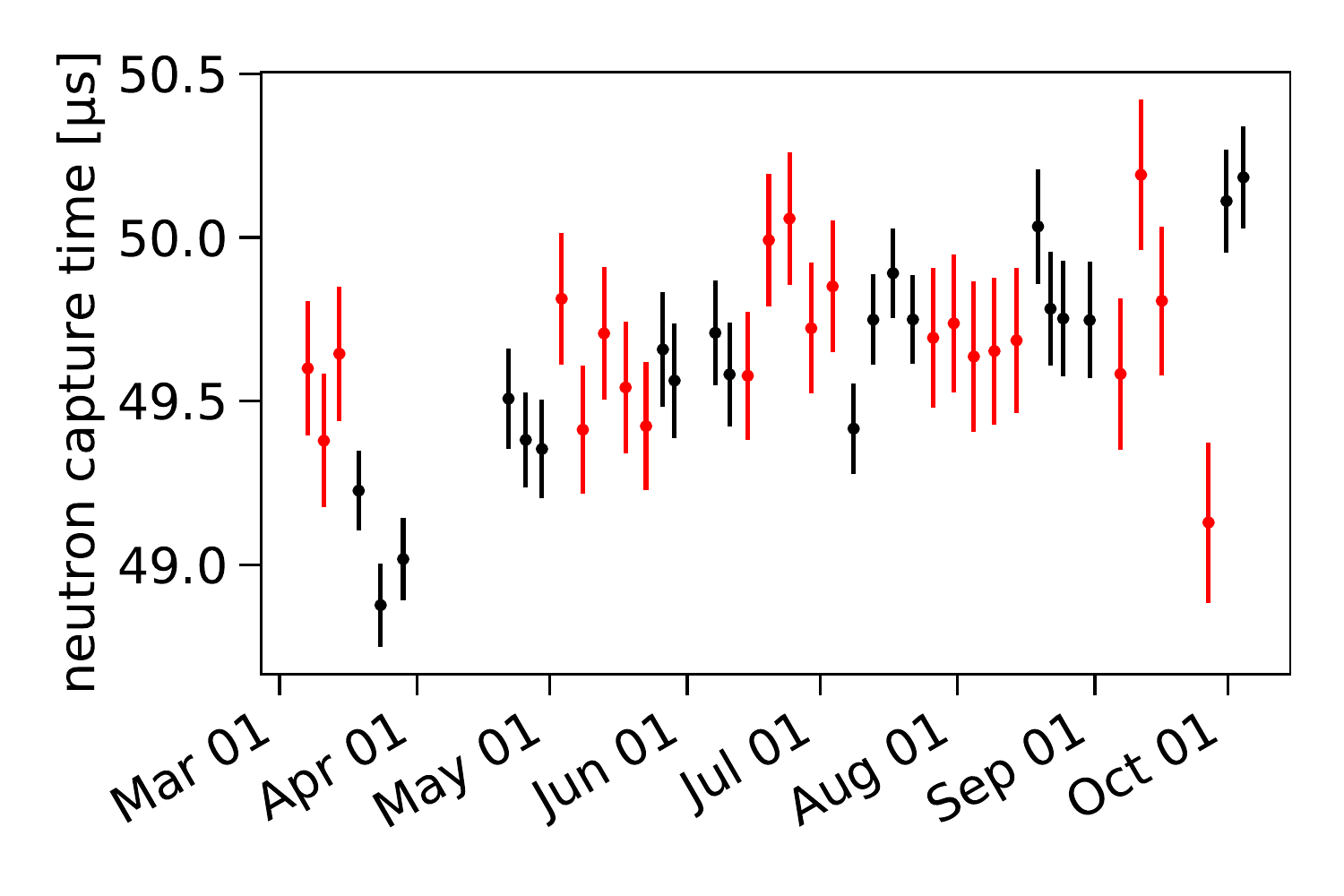}
\caption{Relative $n$-H capture fraction and capture times for cosmogenic neutron capture signals recorded in coincidence with a preceding $n$-$p$ scatter.  Red (black) points indicate reactor-on (-off) periods.  Small increasing trends can be observed in both variables, indicating a small reduction in $^6$Li concentration over the course the data-taking period.  Error bars represent statistical uncertainties.}
\label{fig:ncaptimedep}
\end{figure}

A sub-dominant additional source of time-dependence in the IBD selection is the reduction in the fraction of neutrons capturing on $^6$Li with time.  
Figure~\ref{fig:ncaptimedep} shows the increase in average capture time and the $n$-H capture fraction for cosmogenic neutrons.
Capture times are obtained by fitting coincidence time distributions between prompt recoil and delayed capture signals with the same coincidence and veto requirements as for IBD-like events; this event class, called ($n$-$p$, $n$-Li), was previously described in Section~\ref{subsec:select}.  
For $^6$Li capture fractions, $n$-Li cuts are identical to those applied in the IBD analysis, while $n$-H captures are delayed clusters with energies within 2.0$\sigma$ of the $\gamma$-like PSD band and 3.0$\sigma$ of the $n$-H peak energy.  
The ratio of $n$-H to  $n$-Li captures increases from 12.6\% to 13.2\% over the course of the physics dataset.  
As IBD cuts select only $^6$Li captures, this change will translate to a $\sim$0.7\% reduction in IBD detection efficiency over the course of the physics dataset.  

A decrease in $^6$Li capture rates resulting from a small reduction in $^6$Li concentration in the scintillator should be accompanied by an increase in capture times towards that expected in a pure hydrocarbon environment ($\sim$200~$\mu$s).  
Figure~\ref{fig:ncaptimedep} shows such an increase for the same ($n$-$p$, $n$-Li) dataset, from 49.1~$\mu$s to 50.2~$\mu$s.  
Using PG4, this 1~$\mu$s change in capture time for IBD events is found to produce a 1-2\% reduction in coincidence time cut efficiency.  
PG4 MC simulations also verify that both the change in capture time and $n$-H capture fraction are consistent with a fractional reduction of approximately 3\% in the scintillator's $^6$Li content.  
Capture time variations of generally similar magnitude appear to be present in all regions of the fiducial volume for this dataset within $\pm~1~\mu$s, with  lower (higher) increases observed in the bottom-most (top-most) row of fiducial detector segments.  
These changes are found to have negligible impact on PG4-predicted prompt energy spectra.  

If these two sub-dominant aspects of time dependence (reduction in capture time and increase in $n$-H capture fraction) observed in  various non-IBD event samples are combined, a position-integrated reduction in IBD detection efficiency of 2-3\% should be expected over the course of the physics dataset.  
%After correcting for veto dead time variations, the observed change in IBD-like rates during reactor-on periods is found to be consistent with this estimate, as will be discussed further in Section~\ref{sec:signal}.   
Interestingly, %, as shown in Figure~\ref{fig:rnporate} 
a reduction in ($^{219}$Rn,$^{215}$Po) event rates 3\% greater than that expected based on the 21.8~y $^{227}$Ac half life is also observed during the same physics dataset~\cite{berish_thesis}.  
The general correspondence between IBD and ($^{219}$Rn,$^{215}$Po) rate variations, common doping chemistry for $^{227}$Ac and $^{6}$Li, and neutron capture time and $n$-H fraction variations all appear to be consistent with a reduction in dopant concentration in the PROSPECT scintillator bulk; further dedicated chemical measurements of PROSPECT scintillator samples must be performed to verify this explanation.  

Finally, as mentioned in Section~\ref{subsec:perform}, modest degradation has been observed in the resolution of Z$_{rec}$ for ($^{219}$Rn,$^{215}$Po) events.  
Using PG4 IBD MC simulations, a similar broadening of the prompt-delayed $z$ coincidence distribution is estimated to produce less than 0.5\% reduction in IBD detection efficiency.  
%On the other hand, the width of prompt-delayed $z$ coincidence distributions for $^{12}$B events were found to improve over time by 2\% between the first and second half of the physics dataset, from 2.9 to 2.8~cm.  
%This resolution improvement results in a similarly small increase in IBD detection efficiency.  
This variation is also found to have no impact on PG4-predicted prompt energy spectra.  

The impact of these sub-dominant time-dependent IBD efficiency variations on high-level spectrum and oscillation analyses is expected to be negligible. 
For both analyses, variations in detection efficiency can complicate the scaling of reactor-off IBD candidate datasets to subtract cosmogenic backgrounds during reactor-on periods.  
This background-subtraction procedure, described in more detail in Section~\ref{sec:bkg}, is relatively insensitive to  monotonically decreasing efficiency due to the interleaved nature of reactor-on and reactor-off datasets.  
As demonstrated in Figure~\ref{fig:timeresponse}, linearly time-dependent quantities, such as the $z$-coincidence width for ($^{219}$Rn,$^{215}$Po) events, exhibit reactor on-off variations more than an order of magnitude smaller than variations between the beginning and end of the physics dataset.  
Any residual reactor on-off background scaling ambiguities or biases arising from detection efficiency variations are smaller than other sources of background scaling uncertainty; these additional uncertainties are discussed in more detail in the following section.  

For both the spectrum and oscillation analyses, any impact of efficiency time-dependence is minimized by the lack of substantial energy-dependence related to the effect.  
Regarding baseline dependence, which is most relevant to the oscillation analysis, statistical uncertainties on the baseline-uniformity of efficiency variations are smaller than the previously-defined 5\% per-segment normalization uncertainties  described above.

\section{Backgrounds}
\label{sec:bkg}

An array of backgrounds related to the reactor and cosmogenic activity accompany the IBD signal after the selection cuts in Section~\ref{sec:select} are applied.  The following section describes these various background sources.  

\subsection{Accidental Backgrounds}
\label{subsec:bkg_acc}

\begin{figure}[hptb!]
\includegraphics[trim = 0.0cm 0.0cm 0.0cm 0.0cm, clip=true, 
width=0.48\textwidth]{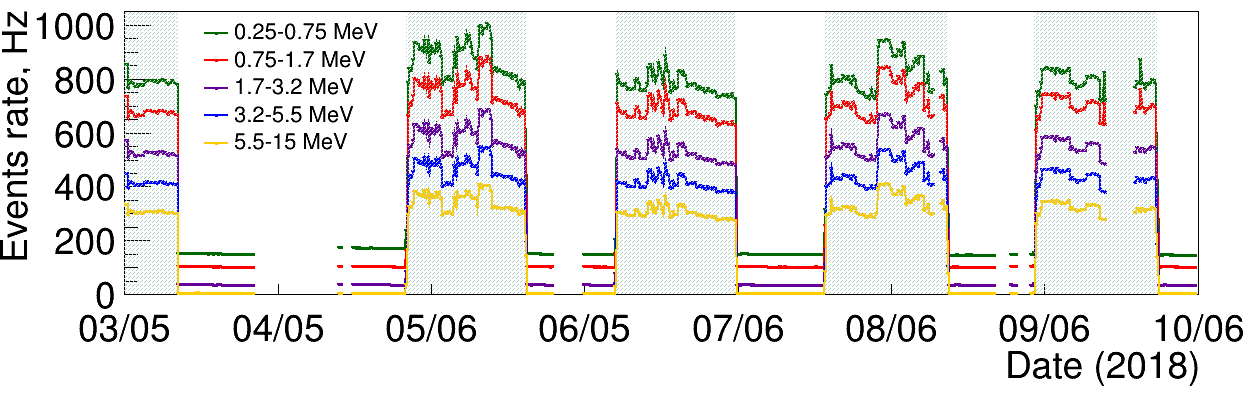}
\includegraphics[trim = 0.0cm 0.0cm 0.0cm 0.0cm, clip=true, 
width=0.48\textwidth]{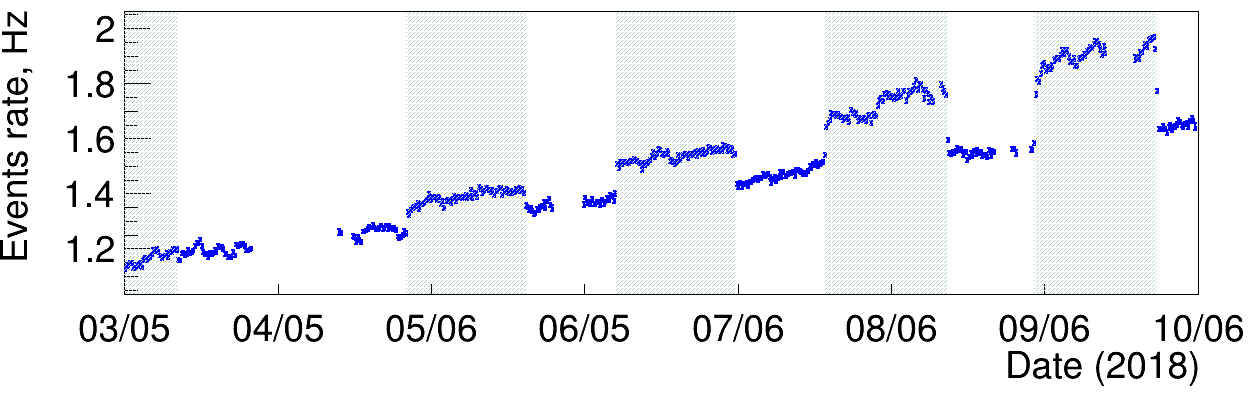}
\caption{Prompt-like (top) and delayed-like (bottom) cluster rates versus time.  For prompt-like clusters, rates are given for a variety of differing energy ranges.  Green vertical bands indicate reactor-on periods.} 
\label{fig:acc}
\end{figure}

Single $\gamma$-rays and single neutron captures from uncorrelated physics events can deposit energy in the PROSPECT detector in close enough spatial and temporal proximity to pass all IBD selection cuts.  
This category of background event is more common during reactor-on periods due to increased $\gamma$-ray fluxes due to the reactor and nearby neutron scattering experiments.  
This variation is illustrated in Figure~\ref{fig:acc}, which shows the rate versus time of clusters meeting the PSD, energy and topology requirements of either the prompt or the delayed IBD signal.  
Rate variations visible during individual reactor-on periods are caused by operations of nearby neutron scattering experiments.  

IBD prompt-like singles span a broad energy range during reactor-on periods, with a substantial high-energy contribution from reactor neutron capture on structural materials in the reactor building; prompt-like energy spectra soften substantially during reactor-off periods.  
Also visible in Figure~\ref{fig:acc} is an increasing rate of single IBD delayed-like events, with a noticeable difference in rates between reactor-on and reactor-off periods.  
This effect can be explained by noting, as discussed in Section~\ref{subsec:eff}, that a substantial proportion of high-PSD signals, including delayed-like events, are contributed by a small proportion of the plentiful $\gamma$-related activity with statistically high PSD values.  

\begin{figure}[hptb!]
\includegraphics[trim = 0.5cm 0.0cm 0.0cm 0.0cm, clip=true, 
width=0.48\textwidth]{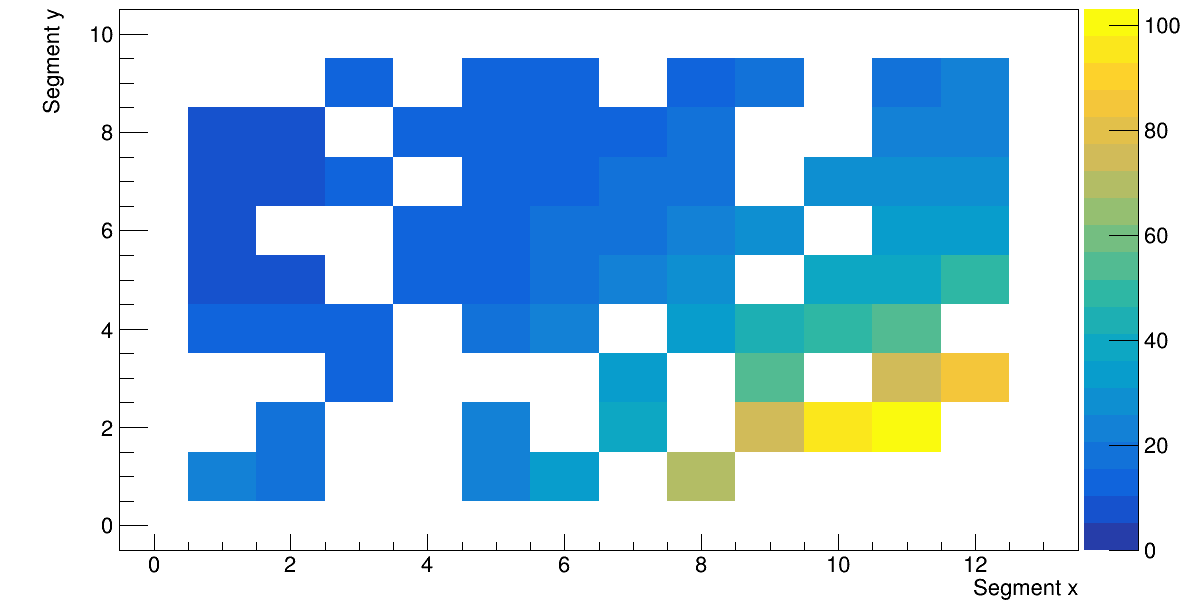}
\includegraphics[trim = 0.5cm 0.0cm 0.0cm 0.0cm, clip=true, 
width=0.48\textwidth]{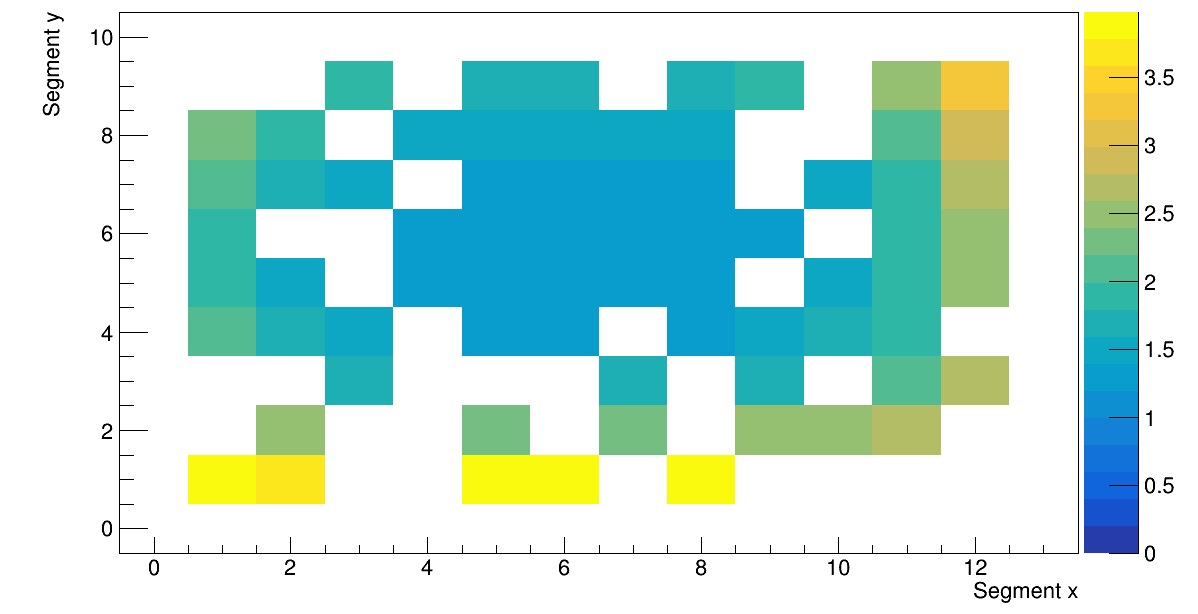}
\includegraphics[trim = 0.5cm 0.0cm 0.0cm 0.0cm, clip=true, 
width=0.48\textwidth]{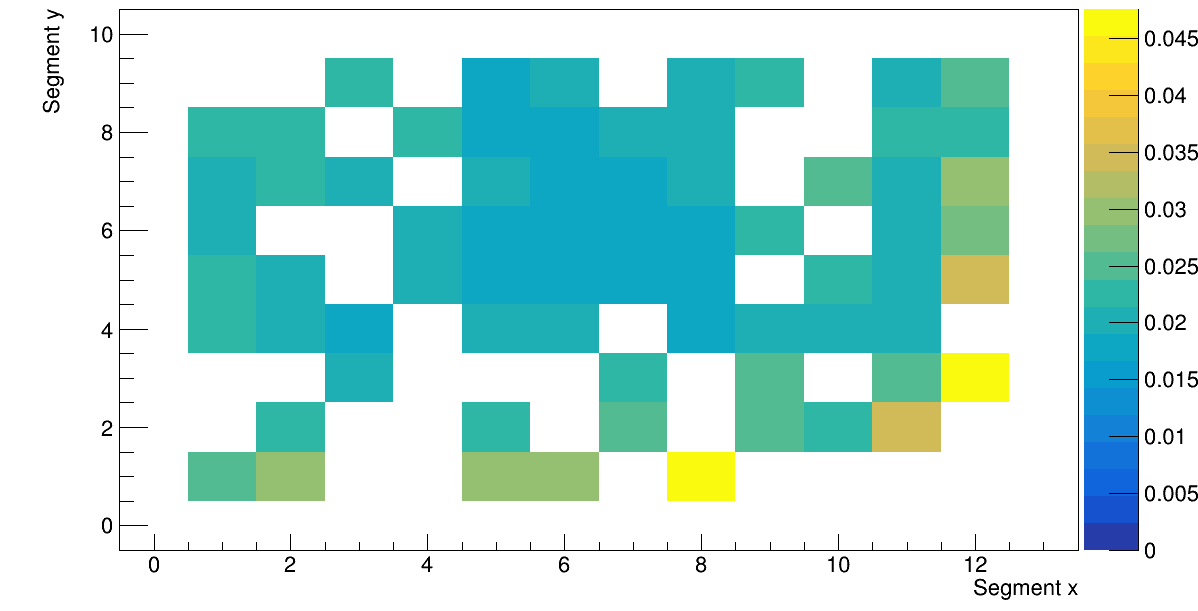}
\caption{Prompt-like (top, reactor-on; middle, reactor-off) and delayed-like (bottom, reactor-on) singles rates versus segment number.} 
\label{fig:acc_pos}
\end{figure}

The spatial distribution of prompt-like and delayed-like signal rates in the detector are shown in Figure~\ref{fig:acc_pos}.  
During reactor-on periods, prompt-like singles rates are found to be 2-10 times higher in segments in the bottom back (high-$x$, low-$y$) corner of the detector.  
This region of the detector receives comparatively less protection from the under-detector concrete monolith and from lead shielding lining the detector movement chassis.  
During reactor-off periods, prompt-like singles rates are found to exhibit substantially less segment dependence, with rates roughly two times lower in detector-interior segments.  
Delayed-like singles rates per segment are also found to be comparatively uniform, with roughly a factor of two variation across the detector during both reactor-on and reactor-off periods.  

%No reactor-dependent variation in the delayed-like singles rate is observed, indicating minimal intrusion of reactor-related neutron fluxes into the detector target region.  

The rate and physics properties of accidental backgrounds for this analysis were determined by collecting cluster pairs that pass all IBD cuts, with the exception of an altered (-12,-2)~ms requirement on prompt-like cluster time with respect to the delayed-like cluster.  
This time separation window excludes all relevant physics-correlated events, giving a pure, high-statistics accidental background sample identical to that in the IBD-like time coincidence window.  
After scaling this sample to account for the relative difference in coincidence time window lengths, it is directly subtracted from the IBD candidate sample with negligible associated uncertainty.  

\subsection{Cosmogenic Time-Correlated Backgrounds}
\label{subsec:bkg_corr}

As the PROSPECT detector is situated underneath minimal ($<$1 meter water equivalent) overburden, substantial contributions of time-correlated prompt-like and delayed-like cluster pairs are expected from cosmogenic muon and neutron fluxes.  
Some are included in the IBD candidate sample despite the dedicated cosmogenic veto cuts described in Section~\ref{subsec:select}.  
To estimate the contribution of these backgrounds to the IBD candidate sample collected during reactor-on data-taking, identical IBD selection cuts are also applied to the reactor-off dataset.  %, when negligible  \nuebar-related contributions are expected.  
Accidental backgrounds in the reactor-off IBD candidate dataset are similarly calculated and subtracted as described in the previous sub-section.  

\begin{figure}[hptb!]
\includegraphics[trim = 0.0cm 0.25cm 0.0cm 0.2cm, clip=true, 
width=0.48\textwidth]{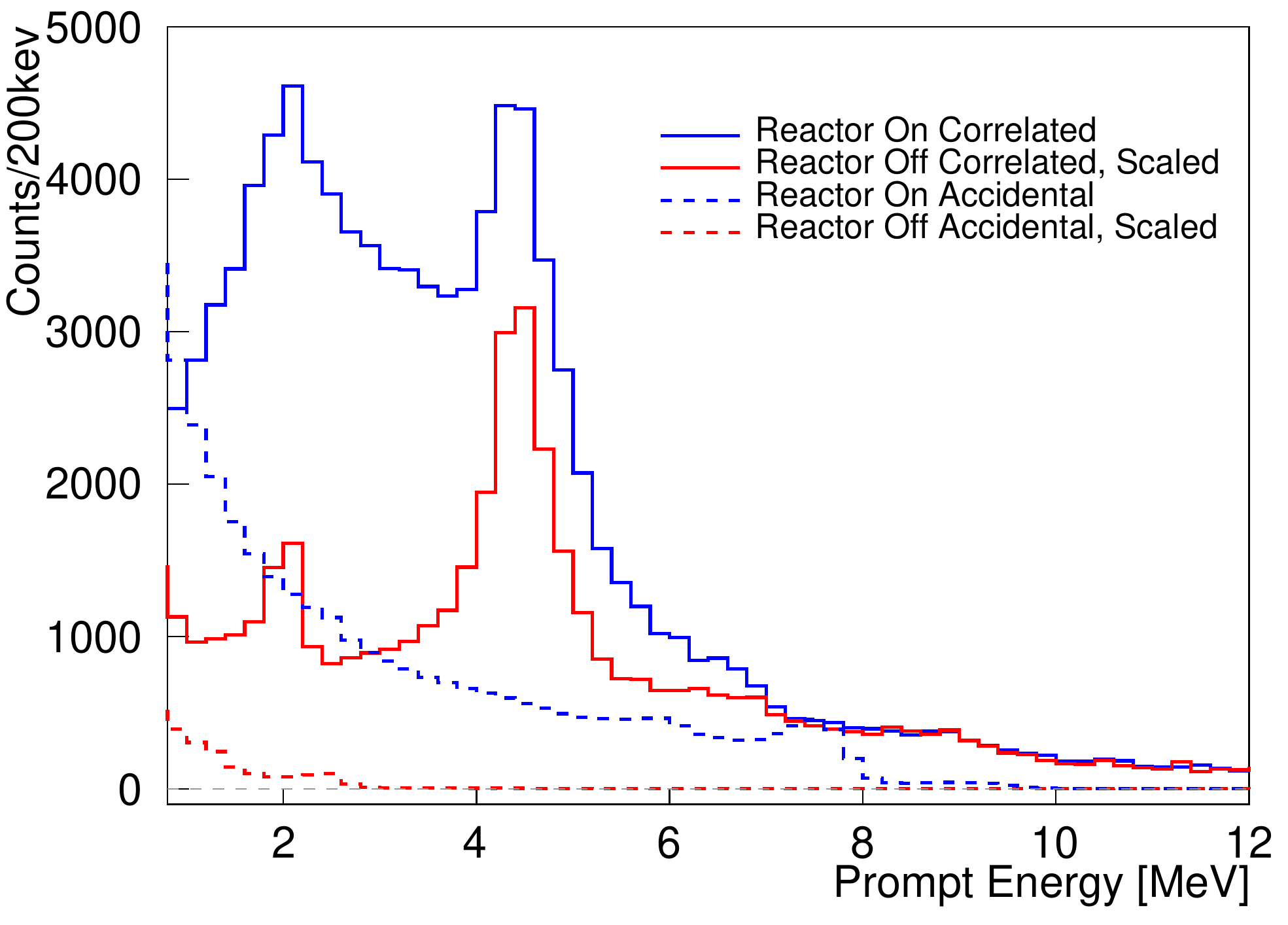}
\caption{The measured prompt E$_{rec}$ spectrum of correlated IBD-like candidates during reactor-on and reactor-off data periods.  Accidental backgrounds, described in the previous section, are drawn in dashed lines with corresponding colors.  Both reactor-off correlated and reactor-off accidental spectra are scaled to match reactor on exposure time.} 
\label{fig:cosmo}
\end{figure}

The prompt E$_{rec}$ spectrum of the reactor-off IBD candidate dataset, pictured in Figure~\ref{fig:cosmo}, exhibits contributions from three primary event categories.  
A peak in the spectrum centered at 2~MeV is characteristic of a $n$-H capture ; this feature can be caused by multi-neutron cosmogenic showers in which two neutrons of low incident energy capture within the inner detector.  
A peak in the spectrum centered at 4.5~MeV is characteristic of the 4.43~MeV $\gamma$-ray line of the first excited state of $^{12}$C; this feature is caused by the inelastic scatter and subsequent capture of one high-energy cosmogenic neutron in the detector.  
Finally, the continuum component of the spectrum encapsulates a combination of neutron-related processes, dominated by neutron-proton elastic scatters, inelastic neutron scatters, or a combination of these effects; both of these dominant continuum-contributing categories are produced by high-energy neutrons.  
PG4 MC simulations of pure cosmogenic neutron fluxes following the `Goldhagen' spectrum of Ref~\cite{goldhagen} are found to reproduce these primary features of the reactor-off IBD candidate spectrum.  
Simulations of primary cosmogenic neutrons and muons using the CRY cosmogenic simulator~\cite{CRY} indicate that neutrons are by far the dominant background source of these two.  
We note that these cosmogenic PG4 MC simulations are not used in any aspect of the cosmogenic background estimation and subtraction process for PROSPECT physics anlayses.  

%\begin{figure}[hptb!]
%\includegraphics[trim = 0.0cm 0.0cm 0.0cm 0.0cm, clip=true, 
%width=0.4\textwidth]{figures/PRD_Fig21_atm.png}
%\caption{IBD-like background rate during RxOff periods vs atmospheric pressure} 
%\label{fig:atmscaling}
%\end{figure}

\begin{figure}[hptb!]
\includegraphics[trim = 1.0cm 0.2cm 0.0cm 0.0cm, clip=true, 
width=0.49\textwidth]{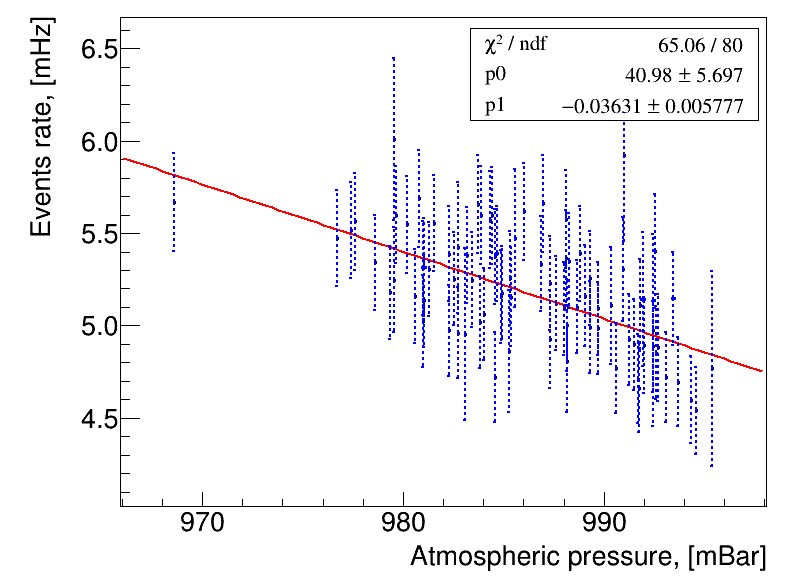}
\caption{Change in the rate of IBD-like events versus atmospheric pressure during reactor-off run periods.  Each point represents one day of reactor-off data.  The fitted trend is used to scale for the difference in average pressure between reactor-on and reactor-off data periods.  The average pressure difference between reactor-on and reactor-off periods is much smaller than the range of pressures depicted in this Figure.} 
\label{fig:atmscaling}
\end{figure}

Reactor-off cosmogenic backgrounds are subtracted from the reactor-on dataset after appropriately scaling the reactor-off dataset's normalization for relative differences in detector live-time and relative differences in absolute cosmogenic flux due to variations in atmospheric pressure.  
Rate corrections for atmospheric pressure are calculated using procedures similar to those documented in Refs.~\cite{bib:nucifer,prospect_osc}.  
Figure~\ref{fig:atmscaling} demonstrates the correlation between cosmogenically-produced IBD-like event rates and atmospheric pressure during reactor-off periods.  
%One can see that the relative vertical offset in rates between reactor-on and reactor-off periods is small, as reflected in Table~\ref{tab:atmscaling}.  
Using the fitted correlation coefficient also pictured in Table~\ref{tab:atmscaling}, (-0.70$\pm$0.01)~\%/mbar, along with the small average atmospheric pressure difference between interleaved reactor-on and reactor-off periods, a nominal reactor-off cosmogenic normalization scaling factor of 1.00$\pm$0.03\% is obtained.  
%Similar atmospheric correlation  coefficients and reactor-off scaling factors were also calculated  for a variety of cosmogenically-produced single cluster and coincident  cluster pair event classes, and are given in  Table~\ref{tab:atmscaling}.  
%All event classes show a reactor-off background scaling factor of $<$0.1\% from unity.
%A conservative 0.5\% uncertainty is assigned to this scaling factor for subsequent oscillation and spectrum analyses.  

Similar correlation coefficients were also determined for different cosmogenic physics event categories, including single muons, single $n$-Li captures, and time-coincident ($n$-$p$, $n$-Li) and ($n$-Li, $n$-Li).  
Associated correlation coefficients and on-off scaling factors for the various datasets are given in Table~\ref{tab:atmscaling}.  
The scaling factors for all event classes are found to be within $<$0.1\% of unity.  Nonetheless, a conservatively assigned 0.5\% uncertainty is used for the subsequent oscillation and spectrum analyses.  
We stress that atmospheric scaling factors are consistent between datasets in spite of relative offsets in absolute rates between reactor-on and reactor-off data periods, which were also given in Table~\ref{tab:atmscaling} and discussed in Section~\ref{subsec:eff}.  

During approximately 3 calendar days of reactor-off data-taking, the water level in the pool surrounding the reactor core was lowered from a nominal height of 3~m above the PROSPECT target volume $y$-center to 2~meters below it.  
Water level changes, performed to enable direct access to regions of the core vessel, were documented in paper logs taken by reactor operations staff, and shared with the PROSPECT collaboration.  
If this effective reduction in nearby shielding material has a substantial impact on the rate of cosmogenic IBD-like backgrounds in PROSPECT, a background scaling factor similar to that generated for atmospheric pressure variations must be calculated and applied to these data periods.  

The general accuracy of the water pool level documentation was verified with PROSPECT data by monitoring incident through-going muon rates at zenith angles corresponding to the location of displaced pool water.  
This analysis was enabled by a dedicated PROSPECT 3D muon tracking algorithm that exploits the relative charge and timing information from each PROSPECT segment.  
During periods of low pool water level, muon rates from these specific incident angles were found to increase by 2\% relative to adjacent data periods; a comparison of these same periods integrating over all zenith angles yielded negligible relative increases.  

Previously-discussed single $n$-$p$ and single $n$-Li cosmogenic neutron event classes, whose average rates are given in Table~\ref{tab:atmscaling}, were used to estimate variations in IBD-like backgrounds during low pool level periods.  
Comparing low pool level periods to nearby nominal pool level periods, rates of these two event classes are found to be unchanged within a conservative 2\% envelope.  
Rates of IBD-like events during these two time period groups are also found to be identical within 2\%.  
This 1.00 water pool scaling coefficient and its 2\% uncertainty applies only to the 5\% of reactor-off data experiencing low water pool levels.  
Thus, we apply no correction to account for this effect; this choice contributes negligibly (0.05\%) to the overall uncertainty in the previously-described correlated background atmospheric scaling uncertainty (0.5\%).  

\subsection{Other Time-Correlated Backgrounds}
\label{subsec:bkg_other}

A direct subtraction of the reactor-off backgrounds using the scale factor described above will not properly remove or account for any background component that scales differently in time than the cosmogenic flux.  
We have investigated three such background categories: neutrinos from spent nuclear fuel, time-coincident backgrounds from reactor $\gamma$-rays and neutrons, and time-correlated signals produced by radiogenic $\alpha$-particles in the PROSPECT detector.  

HFIR's spent nuclear fuel cores are stored in a pool directly adjacent to that housing the burning core, within 15~m of the PROSPECT detector.  
Due to the short cycle length for each HFIR core, the build-up of the long-lived fission products, such as $^{144}$Ce, $^{106}$Ru, and $^{90}$Y, is low compared to commercial reactor fuel.  
Using HFIR's mean cycle length and thermal power, the energy released per~\uFive~fission from Ref.~\cite{bib:kopeikin}, and standard nuclear databases~\cite{bib:JEFF,bib:ENSDF}, daily spent nuclear fuel $\nuebar$ contributions for each of the long-lived $^{235}$U fission products were individually calculated for one HFIR core~\cite{conant_thesis}.  
With conservative assumptions regarding spent core storage in the HFIR spent fuel pool, total spent fuel IBD contributions are found to be less than 0.1 per day, providing a negligible overall contribution to the IBD candidate dataset.  

Fast neutrons are produced in the matrix of the nuclear reactor core, but are very efficiently thermalized and attenuated by the light water pool surrounding the nuclear reactor core.  
Nonetheless, it is possible to generate reactor-produced, physics-correlated cluster pairs in the PROSPECT detector, either through travel of multiple neutrons from the same fission event to the inner detector, or through inelastic scattering of fast reactor neutrons or high-energy reactor-related $\gamma$-rays.  
The former process is highly unlikely: with 10$^{19}$ HFIR-produced neutrons per second at 85~MW$_{th}$, and a reactor-on trigger rate of 2~$\cdot$~10$^4$ Hz, the probability of PROSPECT detecting one (two) HFIR neutron(s) per fission is certainly less than 10$^{-15}$ (10$^{-30}$).  
This estimate is highly conservative, considering the limited dependence of  single $n$-Li and single $n$-$p$ rates on reactor status as shown in Figure~\ref{fig:acc} and Table~\ref{tab:atmscaling}.  
Nonetheless, such a probability indicates far less than 0.1 daily IBD candidates produced via this process.  
%As shown in Fig.~\ref{fig:atmscaling}, PROSPECT also detects a time-stable $\sim$0.17~Hz rate of time- and position-correlated clusters with high prompt and delayed PSD, a signature expected from fast neutron interactions in the detector.  
%The difference in rate between reactor-on and reactor off-periods, following correction for atmospheric pressure differences, is \todo[Olga]{BLAH$\pm$BLAH}~mHz, indicating no substantial excess of reactor-produced fast neutron signals.  

Given the high observed rate of single high-energy prompt-like clusters shown in Figure~\ref{fig:acc}, we also investigated the possibility of production of reactor-related correlated triggers from ($\gamma$,$n\gamma$) photo-neutron interactions in various PROSPECT detector materials, including lithium (scintillator), carbon (all components), deuterium (all components), boron (inner  shielding), oxygen (water shielding), aluminum (inner tank), and lead (shielding).  
Considering incident $\gamma$-ray energies, relevant cross-sections, and relative abundances within the detector, photo-nuclear interactions in PROSPECT are vastly more abundant in its lead shielding than in any other detector component.   
The contribution of photonuclear interactions in lead to IBD-like signatures in PROSPECT was estimated by performing PG4 MC simulations of high-energy $\gamma$-rays outside the detector shielding package with a flux normalization and spectrum tuned to reproduce rates of high E$_{rec}$ prompt-like clusters in PROSPECT during reactor-on periods (as in Figure~\ref{fig:acc}).  
These simulations estimate an IBD candidate rate of much less than one per day from this process.  

A similar consideration of reactor-produced neutron inelastic scattering processes in PROSPECT again reveals its lead shielding as the primary site of these interactions.  
With $\gamma$-ray fluxes expected to be significantly higher than reactor neutron fluxes in the lead shield, and comparable ($n$,$n\gamma$) and ($\gamma$,$n\gamma$) cross-sections in the relevant energy ranges, the former process is unlikely to dominate the latter in producing IBD backgrounds in the PROSPECT target.  
If inelastic reactor neutron interactions closer to the detector target are non-negligible, we would also expect an observed increase in detected ($n$-$p$,$n$-Li) events in PROSPECT during reactor-on periods; as noted in Table~\ref{tab:atmscaling}, we see no evidence of such an increase.  
%, we note that the increase in single proton recoil signatures during reactor-on periods (Table~\ref{tab:atmscaling}) is less than \todo{BLAH}~Hz, more than two orders of magnitude lower than the variation in high-energy $\gamma$-ray signatures.  
%Even this statement is highly conservative, as this on-off difference is almost certainly due to the leakage of reactor $\gamma$-ray signals into the proton recoil PSD band.  
%Even taking this difference at face value, since ($n$,$n\gamma$) cross-sections in lead are generally of similar magnitude to those of ($\gamma$,$n\gamma$), we  expect negligible contributions from both the former and the latter processes. 
%If fast reactor neutrons have penetrated the PROSPECT shielding, one would also expect an elevated rate of coincident 
%The lack of a statistically significant on-off difference between time-coincident $n-p$ $n-Li$ signatures in Table~\ref{tab:atmscaling} further indicates the absence of fast reactor neutrons penetrating the PROSPECT shielding.  \todo{check this statement at the end of all analysis}

Time-correlated IBD-like background contributions from radiogenically-produced ($\alpha$,$n$) interactions in organic scintillator detectors have been estimated by previous MeV-scale neutrino experiments~\cite{borexino_alpha, bib:prd_osc}.  
The primary process considered in these experiments is the $^{13}$C($\alpha,n$)$^{16}$O* interaction, which produces time-coincident signals from a prompt high-energy de-excitation $\gamma$-ray and the delayed neutron capture.  
Daya Bay estimates IBD-like signal rates of roughly 0.005 per ton of scintillator per day from $\alpha$-particle rates of roughly 0.5~Hz/ton~\cite{bib:prd_osc}.  
As described in Table~\ref{tab:cutvals}, $\alpha$-particles are primarily expected to be generated through decay products of $^{227}$Ac deliberately doped into the PROSPECT scintillator, which has an observed 0.3~Hz rate in the fiducial volume.  
Considering the Daya Bay $\alpha$-induced IBD backgrounds per ton given above, $^{227}$Ac chain products will generate much less than 0.1 IBD event per day in PROSPECT.  
Backgrounds from $\alpha$ processes on fluorine present in the PROSPECT optical grid's FEP linings were also considered and estimated to be negligible IBD contributors.  
It should also be noted than any time-stable radiogenic IBD background contribution would be identical between reactor-on and reactor-off periods, and would thus be properly removed during the subtraction of other correlated backgrounds.  

%Need to more generally show high-PSD triggers?
%Most prominent is the $n$-Li capture peak, with a detection rate of 5~Hz, or approximately 2~Hz/ton, as previously demonstrated in Fig~\ref{fig:acc}.  
%\todo{BLAH}: statement about any other peaks.  
%Scaling from Daya Bay's rate of IBD candidates per produced alpha, much less than 1 IBD candidate per day is expected from this process during both reactor-on and reactor-off periods.  

%Other time-correlated backgrounds, including detector-intrinsic (Rn,Po) and (Bi,Po) correlated decays, ($\alpha$,n) reactions from detector-intrinsic $\alpha$-particle emitters, and \nuebar emanating from spent nuclear fuel are expected to contribute less than 1\,\% of the reactor-off sample.  
%These backgrounds are expected to be constant in time, and are subtracted with negligible  systematic uncertainty using the method described above.  
%Fast neutron background surveys at HFIR~\cite{prospect_reactor} indicate that reactor-related IBD-like backgrounds are negligible.  

\subsection{Background Subtraction Validation}
\label{subsec:bkg_check}

In the following section we present analyses to demonstrate the reliability and accuracy of reactor-on background estimates.  

\begin{figure}[hptb!]
\includegraphics[trim = 0.0cm 0.5cm 0.0cm 0.75cm, clip=true, 
width=0.49\textwidth]{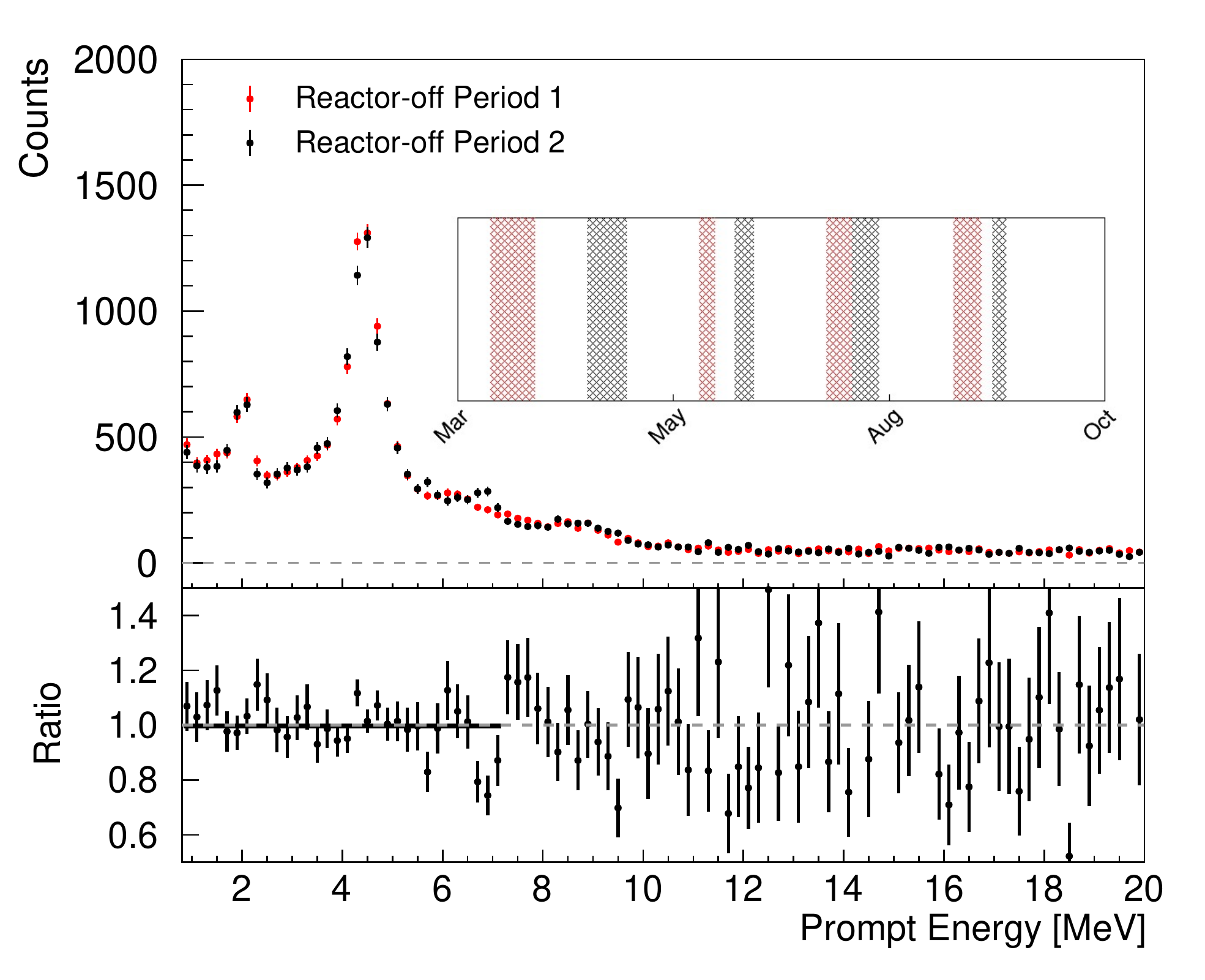}
\caption{The measured prompt energy spectrum of correlated candidates from  reactor-off data periods. Correlated candidates in period 2 are scaled to match period 1 exposure and corrected for relative atmosphere difference between two periods.  The figure inset indicates the breakdown of period 1 and 2 datasets within reactor-off periods.  The solid horizontal line in the bottom panel shows the best-fit normalization offset between datasets in the 0.8-7.2~MeV E$_{rec}$ range; see text for details.  Error bars represent statistical uncertainties.} 
\label{fig:offoff}
\end{figure}

Consistency between IBD-like datasets from different time periods demonstrates proper understanding of the level of time-stability of the detector's energy scale and IBD-like background contamination.  
This comparison for two different reactor-off time periods is shown in Figure~\ref{fig:offoff}.  
To more closely mimic the distribution of reactor-on and reactor-off periods in time due to the short HFIR cycle length, the two periods chosen for comparison in Figure~\ref{fig:offoff} are interleaved in time as shown in the figure inset; any systematic variation in efficiency or energy response occurring over extended timescales will have a reduced impact in this scenario.  
In addition, datasets are scaled to account for relative differences in atmospheric pressure between the two time period definitions; as in the comparison of reactor-on and reactor-off datasets, the scaling factor for this off-off comparison is much less than 1\%.  
The reactor-off datasets show consistency with one another:  comparison in the 0.8-7.2~MeV E$_{rec}$ range yields a $\chi^2$/DOF of 47.68/31.  
If the normalization is allowed to float between datasets, the best-fit offset in the 0.8-7.2~MeV energy range is found to be consistent with unity to 1\% statistical precision.  

PROSPECT IBD analyses rely on the correspondence of correlated IBD-like background rates and spectra between reactor-on and reactor-off periods.  
An explicit verification of this correspondence for IBD-like backgrounds is not possible, due to the presence of real IBD events during reactor-on periods.  
Instead, we have examined rates and spectra of correlated background event classes similar in appearance to IBD-like candidates in PROSPECT.

\begin{figure*}[hptb!]
% ~/Work/PROSPECT/objg/bin/OGManager -d psp.sql -e ../ObjectGraph/config/P2kAllbuts.cfg
% mv E_muon_veto.pdf PRD_Fig26_Muon_IBD.pdf
% mv Exclusive_cut_PSD.pdf PRD_Fig26_FN_IBD.pdf
\includegraphics[trim = 0.0cm 0.0cm 13.0cm 0.0cm, clip=true, 
width=0.48\textwidth]{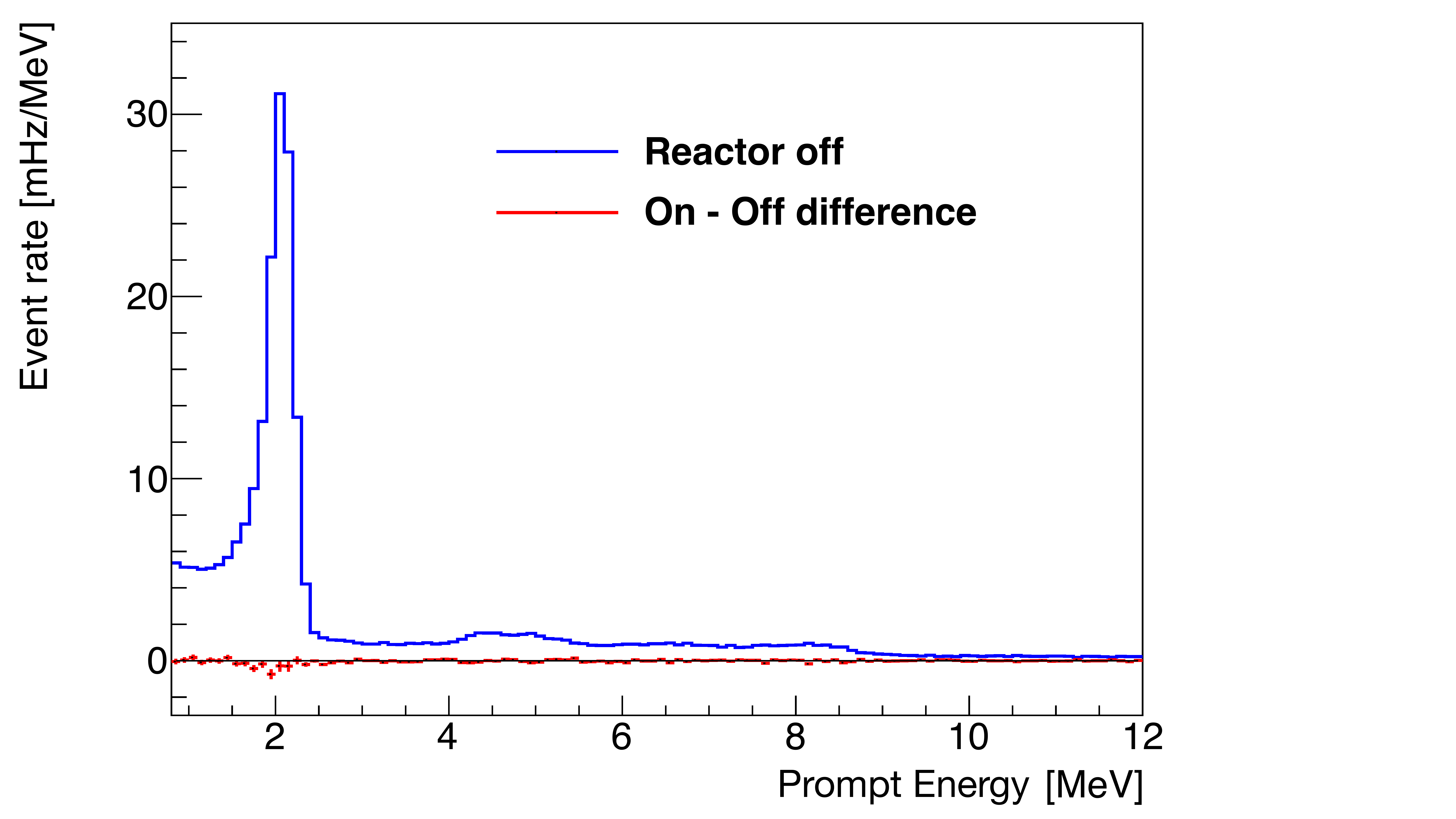}
\includegraphics[trim = 0.0cm 0.0cm 13.0cm 0.0cm, clip=true, 
width=0.48\textwidth]{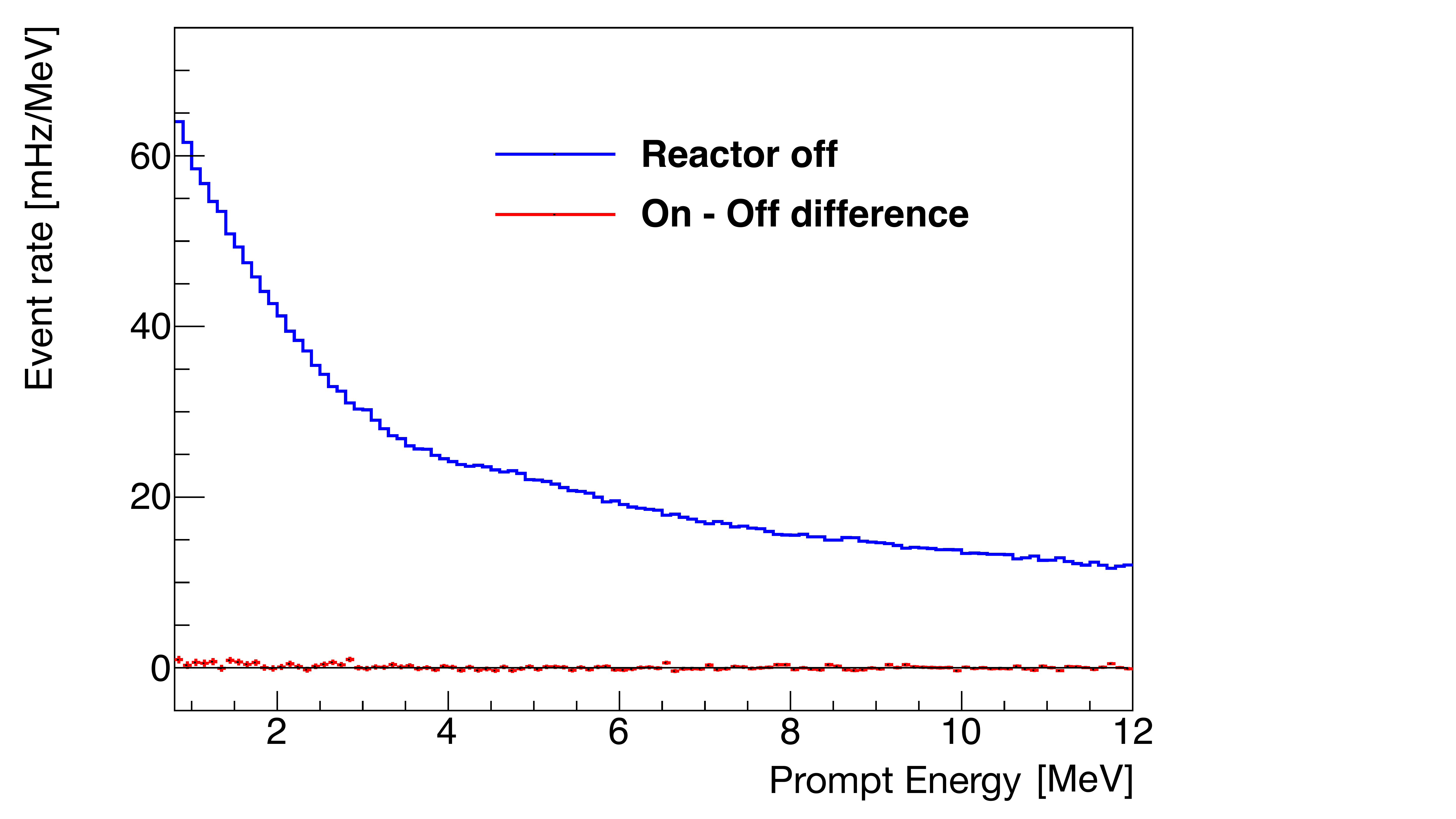}
\caption{Prompt E$_{rec}$ spectra for two classes of vetoed IBD-like events: IBD candidates rejected by a muon veto (left) and IBD candidates failing the prompt PSD requirement (right).  These IBD-like event classes are primarily produced by multi-neutron and fast neutron cosmogenic events, respectively.  Pictured curves represent reactor-off correlated vetoed IBD candidates (blue), and reactor-on correlated residuals following reactor-off background subtraction (red). Accidentals (between the IBD candidate pairs and with the veto-inducing event) have been subtracted out from all distributions. Due to the presence of  high-multiplicity showers in the dataset, substantial correlations are present between bins in the left-hand plot.} 
\label{fig:bkg_xcheck}
\end{figure*}

Figure~\ref{fig:bkg_xcheck} shows the correspondence for two specific event classes for reactor-on and reactor-off periods.
The first is IBD candidates rejected by a muon veto; muon cut definitions are outlined in Section~\ref{sec:select}.  
This class is overwhelmingly produced by neutronic signatures related to the initial vetoing cluster, particularly coincident captures of multiple neutrons.  
Similar events in which the vetoing particle does not traverse the PROSPECT target represent one source of expected IBD-like background.  
This event class contains a small expected contamination from true IBD events accidentally appearing in the muon veto window; this contribution is removed by appropriately scaling and subtracting the observed background-subtracted IBD signal described in Section~\ref{sec:signal}.  
The second event class is the ($n$-$p$,$n$-Li) dataset described in Table~\ref{tab:atmscaling} and Figure~\ref{fig:select2D}: IBD candidates failing the prompt PSD requirement.  
These events are overwhelmingly produced by interaction of fast cosmogenic neutrons in the PROSPECT target.  
In addition, many of the aforementioned potential sources of reactor-related correlated IBD-like backgrounds would also produce events in this category.  
This event class contains negligible contamination from true IBD interactions.  

Figure~\ref{fig:bkg_xcheck} demonstrates on-off correspondence by plotting the prompt energy spectrum of each event class during reactor-off periods, as well as the residual prompt spectrum during reactor-on periods after properly scaling and subtracting out this reactor-off signal.  
If reactor-off periods provide an accurate description of correlated backgrounds during reactor-on periods, the background-subtracted reactor-on signal for these event classes should be statistically consistent with no signal at all prompt energies.  
We note that in calculating statistical consistency for the muon-vetoed IBD-like event class, one must propagate statistical correlations between events and between prompt energy bins generated by the fact that many IBD-like candidates are often produced by the same cosmogenic interaction.  

For muon-vetoed IBDs, we find that reactor-on residuals are statistically inconsistent with zero at 2.9$\sigma$ confidence level in the vicinity of the $n$-H peak at 1.6-2.6~MeV prompt E$_{rec}$.  
The amplitude of this deficit in reactor-on signal is -2\% of the total reactor-off event class size in the $n$-H peak region, and shows no statistically significant variation with detector position.  
No statistically significant residual deficit or excess is observed in this event class outside the $n$-H peak region.  
IBD candidate events vetoed by preceding $n$-$p$ recoil signatures also exhibit a similar -3\% offset in the $n$-H peak region during reactor-on periods.  

Meanwhile, the ($n$-$p$,$n$-Li) event class pictured in Figure~\ref{fig:bkg_xcheck} showed a substantially smaller reactor-on residual excess: in the 0.8~MeV-7.2~MeV IBD prompt E$_{rec}$ range, the offset is +0.31\% $\pm$ 0.13\% the size of the reactor-off rate.  
This offset is similar in size to the current 0.5\% correlated cosmogenic background normalization uncertainty envelope described earlier in this section.  
A variety of other statistically independent non-signal cosmogenic event classes were also investigated and showed no meaningful excess in reactor-on data.  
Most notably, as will be described in the following Section, no residual reactor-on excess or deficit is observed within 2\% statistical uncertainty in IBD candidates above 8~MeV prompt E$_{rec}$, where negligible contributions from reactor \nuebar are expected.  

The observation of a residual reactor-on deficit for some event classes during PROSPECT reactor-on data periods is suggestive of unidentified time-variations in selection cut efficiencies, dead times, or accidental/cosmogenic background estimates, rather than the presence of unidentified reactor-produced correlated backgrounds~\cite{stereo_2019}.  
Issues related to detector response may also produce percent-level excesses in other event classes, depending on the cuts applied.  
Given the negligible estimated contributions from reactor-related correlated background in Section~\ref{subsec:bkg_other}, we suspect that response-related issues are responsible for both the deficits and excesses observed.  
As PROSPECT has been unable to precisely determine the underlying cause of this percent-level imperfection in its background subtraction procedure, the observed residuals are used to define additional uncertainty contributions to be applied to the \nuebar oscillation and spectrum analysis.  
First, an additional 1\% energy- and baseline-correlated reactor-off background normalization uncertainty is introduced to account for the small observed on-off excess in ($n$-$p$,$n$-Li) IBD-like events.  
An added 3\% uncertainty in the amplitude of the reactor-off nH peak in the 1.6-2.6 MeV region is also instituted to reflect the residual on-off deficit exhibited in muon- and recoil-vetoed IBD-like events; this uncertainty is treated as baseline-correlated, but is uncorrelated with respect to the reactor-off background normalization uncertainty.  
These additional uncertainties produce minimal degradation in oscillation and spectrum sensitivity; this conclusion remains unchanged when adjusting the level of assumed baseline correlation.

%IBD signal events above 
%The presence of a deficit, rather than an excess, in these event rates 
%BLAH: now describe uncertainty handling.

%of \todo{BLAH $\pm$ BLAH}~mHz (\todo{BLAH $\pm$ BLAH}\% of the reactor-off rate), and is also statistically consistent with zero above 0.8~MeV (p-value \todo{BLAH}).  
%This event class appears to exhibit a residual reactor-on contribution below the IBD signal prompt energy range that is 2-3\% the size of the total reactor-off signal in that region.  
%While this excess is not relevant to PROSPECT's IBD-based physics analyses, it could arise from small residual time-dependent variations in PROSPECT's calibrated PSD response or detector thresholding, or from a true population of reactor-related correlated IBD-like backgrounds.  
%%%%%%%%%%%%%%%%%%%%%%%%%%%%%%%SIGNAL and CROSS-CHECKS SUBSECTION
\section{Measured IBD Signal Sample}  
\label{sec:signal}

Following the application of cosmogenic and re-triggering vetoes to the 95.7 (73.1) calendar days of reactor-on (off) data described in Section~\ref{subsec:physdata}, IBD candidates were selected from 82.2 (65.2) days of reactor-on (off) live-time.  
During the reactor-on data-taking period, a total of 115852  IBD candidates are selected.  
Of these candidates 28358$\pm$18 are calculated to be contributed by accidental backgrounds, yielding a total of 87494$\pm$341 correlated IBD candidates.  
Following subtraction of 1309$\pm$4 accidental background events from the reactor-off IBD candidate dataset, a total of 29258$\pm$175 correlated IBD-like candidates are selected in the reactor-off dataset.  
Following application of live-time and atmospheric pressure scalings, this reactor-off IBD candidate tally corresponds to a total reactor-on cosmogenic background estimate of 36934$\pm$221.  
After subtraction of this background, a total of 50560$\pm$406 signal IBD events remain in the reactor-on dataset.  
The ratio of signal IBD to cosmogenic background (accidental background) events is estimated to be 1.37 (1.78).  
A summary of IBD candidate accounting is provided in Table~\ref{tab:rates}.  

\begin{table}[tb!]
\centering
\begin{tabular}{|c|c|c|c|}
\hline 
Category & Reactor-On & Reactor-Off \\ 
\hline 
Calendar Days & 95.65 & 73.09 \\
Live Days & 82.25 & 65.16 \\ \hline
IBD Candidates & 115852 & 30568  \\ \hline
Accidental Backgrounds & 28358$\pm$18  & 1309$\pm$4 \\ \hline
Correlated Candidates & 87494 $\pm$ 341 & 29258$\pm$175 \\ 
Rate Per Calendar Day & 915$\pm$ 4 & 400$\pm$2 \\ \hline
Cosmogenic Backgrounds & 36934$\pm$221 & N/A \\ \hline
Total IBD Signal & 50560$\pm$406 & N/A \\
Rate Per Calendar Day & 529 $\pm$4 & N/A \\
\hline
\end{tabular} 
\caption{Statistics of selected IBD candidates and accidental/cosmogenic backgrounds.  Errors, where included, represent statistical uncertainties in the relevant signal and background datasets.}
\label{tab:rates}
\end{table}

\begin{figure}[hptb!]
\includegraphics[trim = 0.0cm 0.0cm 0.0cm 0.0cm, clip=true, 
width=0.48\textwidth]{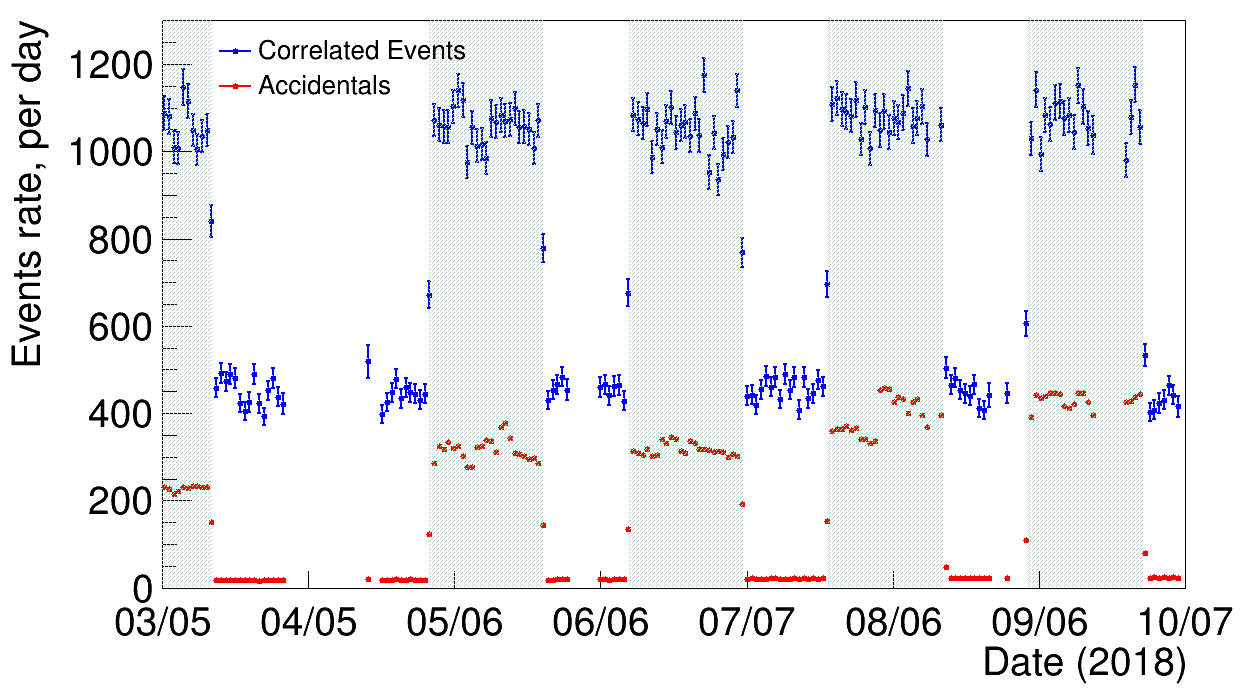}
\includegraphics[trim = 0.0cm 0.0cm 0.0cm 0.0cm, clip=true, 
width=0.48\textwidth]{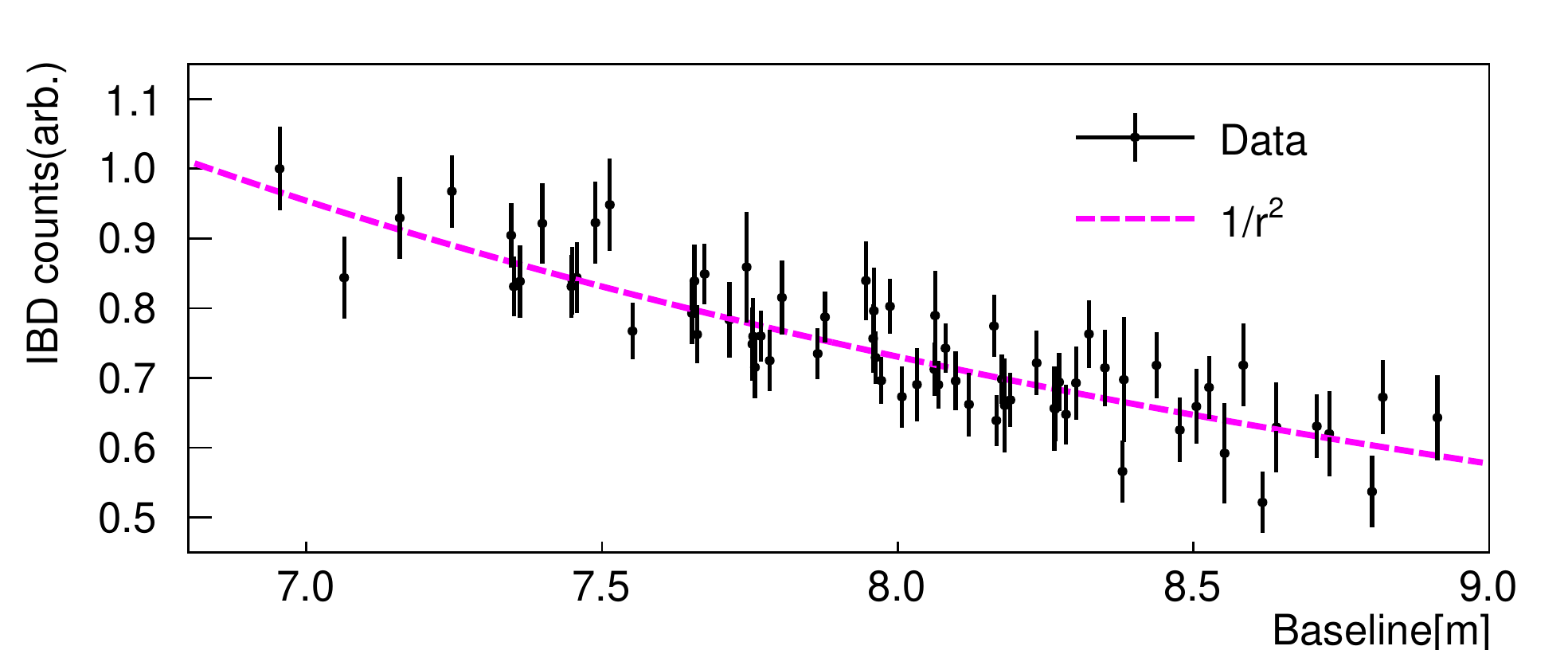}
\caption{Time dependence (top) and baseline dependence (bottom) of correlated IBD candidate rates.  For the bottom plot, the background-subtracted IBD signal is plotted.  Rates are integrated between 0.8 and 7.2~MeV prompt E$_{rec}$.  In the top plot, each point in most cases corresponds to one live-day, while in the bottom plot each point corresponds to one fiducial segment.  Error bars represent statistical uncertainties.} 
\label{fig:rates}
\end{figure}

The rate of correlated IBD candidates and accidental backgrounds per live-day is shown in Figure~\ref{fig:rates}.  
As described in Section~\ref{sec:bkg}, accidental backgrounds exhibit marked time-dependence, largely due to variations in prompt-like rates during reactor-on data-taking periods.  
The correlated IBD candidate rate is clearly dependent on reactor status; given the lack of reactor-correlated backgrounds (Section~\ref{subsec:bkg_other}), this dependence provides clear indication of observation of reactor antineutrinos.  
Smaller-amplitude deviations in these rates during reactor-on and reactor-off periods are caused by previously-described variations in cosmogenic fluxes, and thus cosmogenic IBD backgrounds, due to variations in atmospheric pressure.  

After applying subtraction of both accidental and correlated cosmogenic backgrounds, relative rates of IBD signals are shown in Figure~\ref{fig:rates} for each active fiducial segment; rates are normalized with respect to the shortest baseline, and are corrected for PG4-predicted relative variations in efficiency between segments.  
Efficiency-corrected IBD signal rates decrease with segment baseline, following the 1/r$^2$ distribution expected when sampling an isotropically-emitting compact $\nuebar$ source.  
The best-fit inverse-square function (with only amplitude parameter) pictured in Figure~\ref{fig:rates} provides a $\chi^2$/DOF of 72.4/69.  

\begin{figure}[hptb!]
\includegraphics[trim = 0.0cm 0.25cm 0.0cm 0.15cm, clip=true, 
width=0.48\textwidth]{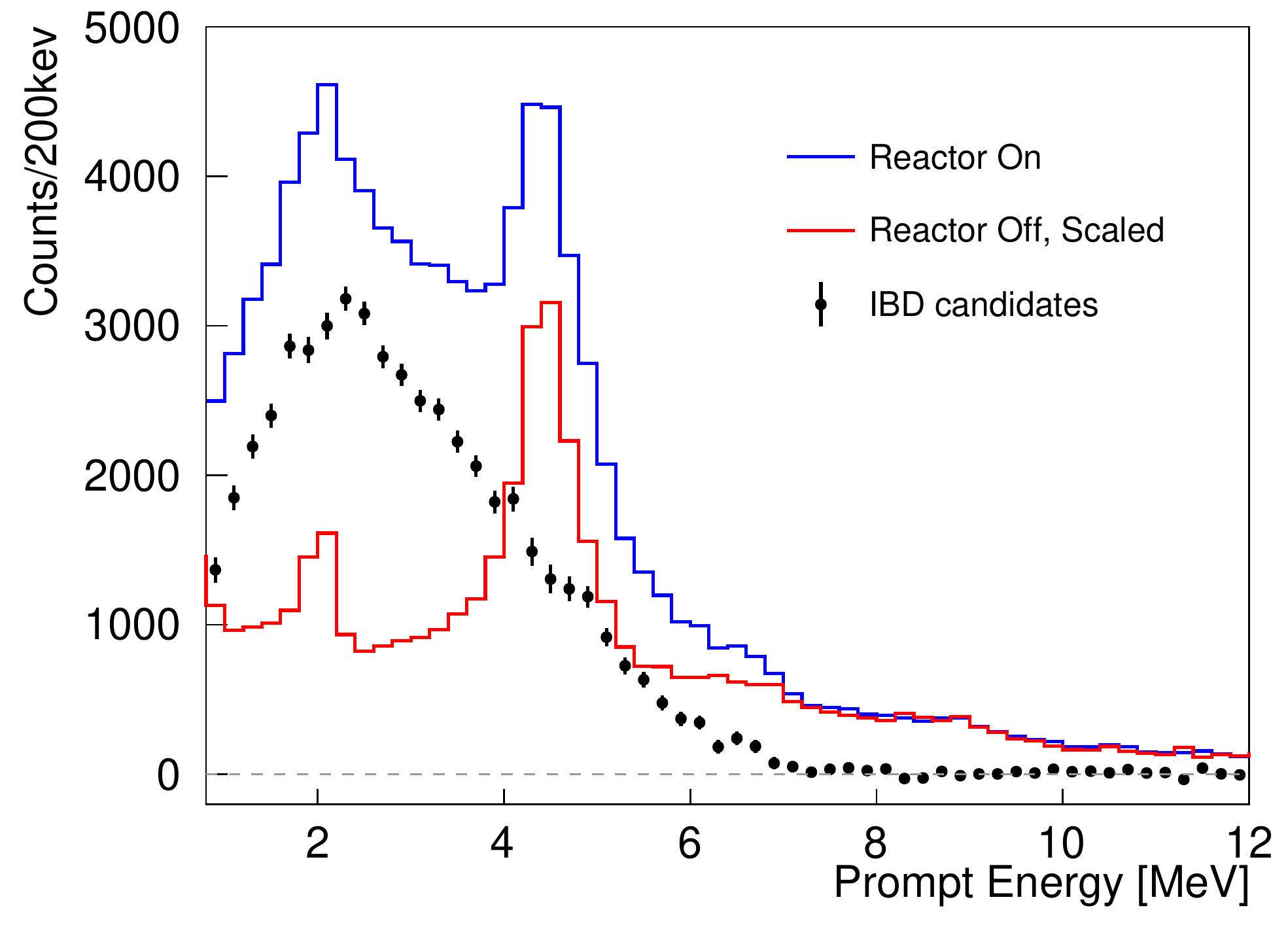}
\caption{The measured prompt visible energy spectrum of IBD events with both reactor-on and reactor-off correlated candidates displayed.  The reactor-off correlated candidates are scaled to match reactor-on exposure and corrected for atmosphere pressure difference between reactor on and off.   Only statistical errors are pictured for the background-subtracted IBD signal.} 
\label{fig:measuredspec}
\end{figure}

The prompt E$_{rec}$ spectrum of the accidentals-subtracted reactor-on IBD-like dataset is pictured in Figure~\ref{fig:measuredspec}, along with that of the cosmogenic background expected from the reactor-off dataset and the fully-background-subtracted IBD signal.  
After subtracting cosmogenic backgrounds, the IBD signal's prompt energy distribution matches the general expected shape of reactor $\nuebar$ interacting via IBD: 
count rates are highest in the 1-7~MeV range with a generally continuous appearance versus energy in this range despite the presence of peak-like features in the subtracted cosmogenic spectrum.  
Above 7~MeV, where reactor IBD signal contributions are expected to be minimal, background-subtracted IBD-like count rates are consistent with zero, indicating proper scaling of reactor-off data during reactor-on cosmogenic background subtraction.  
A quantitative comparison of the background-subtracted IBD signal distribution to zero from 8~MeV to 12~MeV yields a $\chi^2$/DOF of 20.9/20.  

\subsection{Signal Validation}  

To demonstrate a proper understanding of the background-subtracted IBD signal dataset, it is valuable to perform comparisons of IBD-like event distributions between different time periods and detector locations.  

\begin{figure}[hptb!]
\includegraphics[trim = 0.0cm 0.5cm 0.0cm 0.5cm, clip=true, 
width=0.48\textwidth]{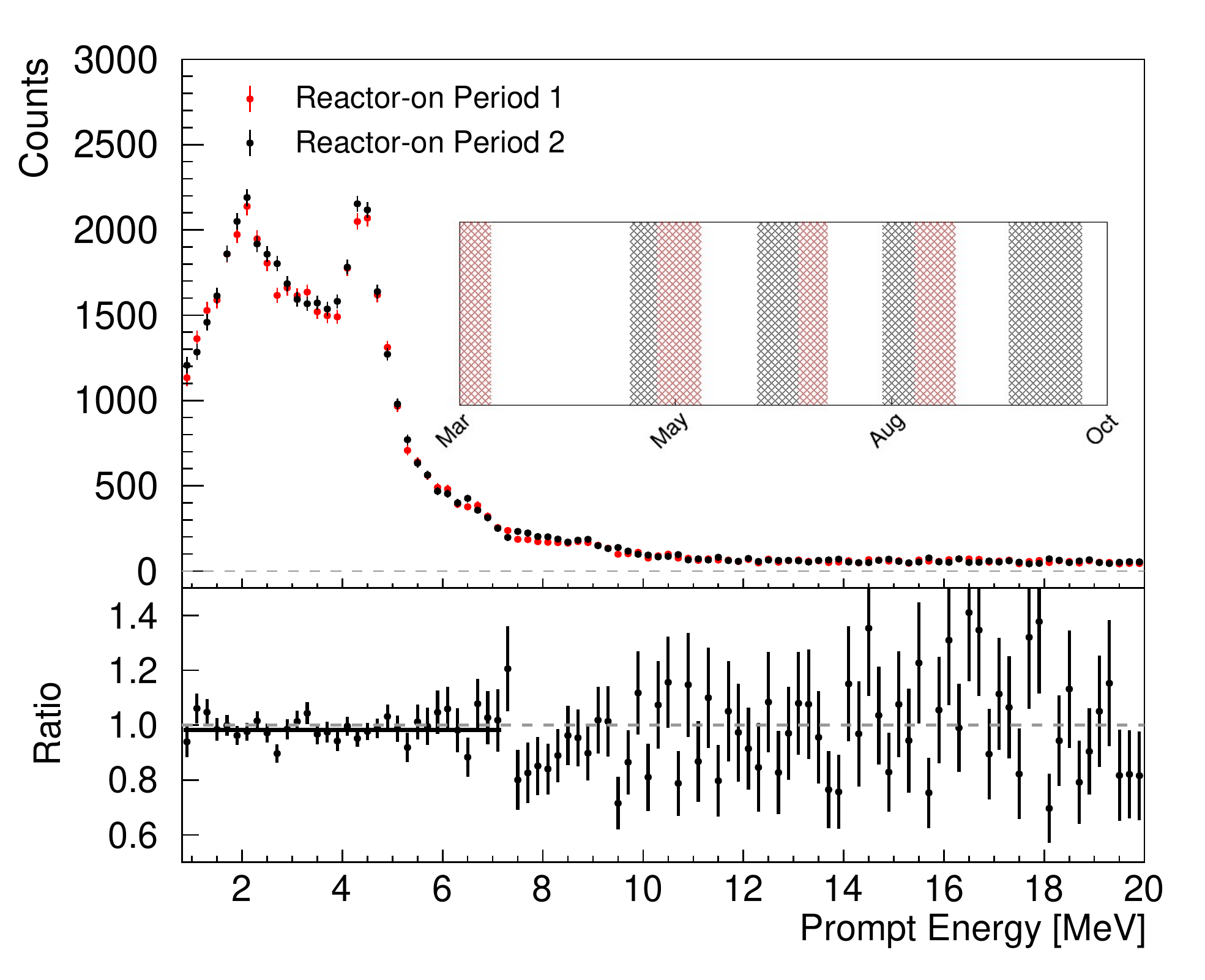}
\caption{The measured prompt energy spectrum of correlated candidates from  reactor-on data periods. Correlated candidates in period 2 are scaled to match period 1 exposure and corrected for relative atmosphere difference between two periods.  The figure inset indicates the breakdown of period 1 and 2 datasets within reactor-on periods.  The solid horizontal line in the bottom panel shows the best-fit normalization offset between datasets in the 0.8-7.2~MeV E$_{rec}$ range; see text for details.  Error bars represent statistical uncertainties.} 
\label{fig:onon}
\end{figure}

Given the stability in reactor thermal power during HFIR operation, a demonstration of time stability of the IBD selection can be provided by comparison of different reactor-on time periods.  
This comparison for two different reactor-on time periods is shown in Figure~\ref{fig:onon}.  
As in Figure~\ref{fig:offoff}, the two time periods are interleaved in time as shown in the figure inset.  
These datasets show consistency with one another: quantitative comparison between 0.8~MeV and 7.2~MeV yields a $\chi^2$/DOF of 26.2/31.  
If the normalization is allowed to float between datasets, the best-fit offset in the 0.8-7.2~MeV energy range is found to be less than 2\%, consistent with a hypothesis of equal normalizations within $\sim$2$\sigma$ statistical confidence level.

%As described in Section~\ref{subsec:eff}, a systematic reduction of efficiency in time is expected due to time-varying $z$-position resolutions and $^6$Li concentration in the PROSPECT active detector volume.  
%On the other hand, as described in Section~\ref{subsec:perform}, detector calibration procedures yield energy scales and resolutions are expected to be consistent across the PROSPECT dataset.  
%These characterizations of PROSPECT performance should be similarly reflected in obtained IBD-like candidate distributions.  
%In particular, if prompt $E_{rec}$ distributions are compared between early and late reactor-off and reactor-on periods, as opposed to the interleaved time periods shown in Figure~\ref{fig:onon}, PROSPECT's time-dependent detector response variations should be visible as a few-percent relative decrease in sample size of the late period with respect to the early period.  
%After applying \todo{BLAH}\% (\todo{BLAH}\%) rate corrections for relative differences in atmospheric pressure between early and late reactor-off (on) periods, best-fit normalization shifts of -\todo{BLAH}\% (-\todo{BLAH}\%) are found between early and late periods.  
%Thus, small time-dependent variations in IBD-like candidates match those expected from  characterizations of PROSPECT's detector response.  
%As with the time-interleaved datasets, prompt E$_{rec}$ distributions show good consistency between early and late reactor-off and reactor-on data periods.  

Due to the compact size of PROSPECT's inner detector, IBD interactions taking place in the inner-most and outer-most segments of its fiducial volume should exhibit differing levels of annihilation $\gamma$-ray energy leakage, leading to differences in prompt E$_{rec}$ spectra between these two regions.  
In addition, the presence of larger numbers of inactive segments near the detector bottom should lead to enhanced energy leakage for IBD interactions taking place in the bottom of the fiducial volume.  
These relative variations in response with position in the detector must be properly accounted for in predicted IBD signal distributions.  

\begin{figure}[hptb!]
\includegraphics[trim = 0.5cm 0.0cm 0.5cm 0.5cm, clip=true, 
width=0.49\textwidth]{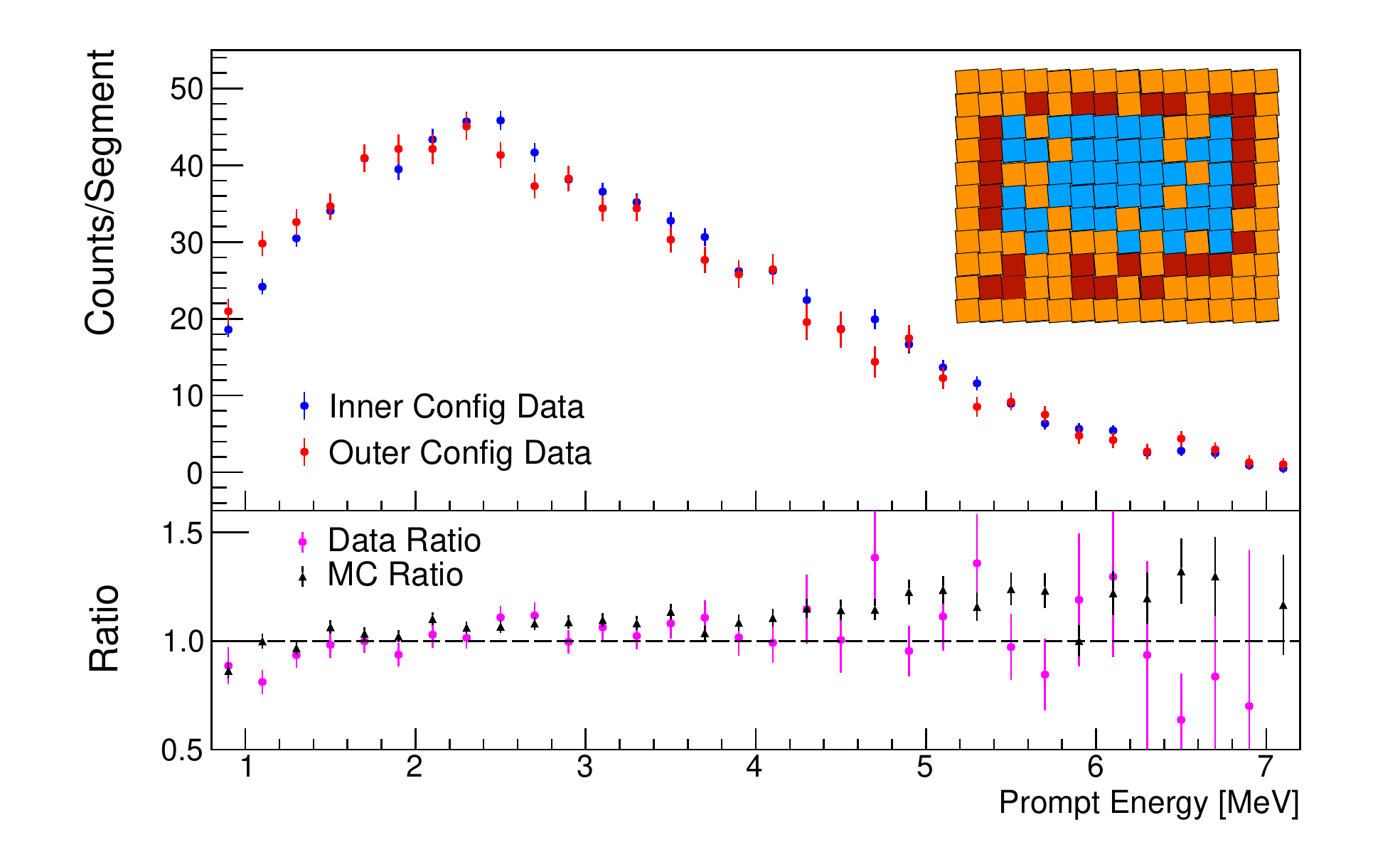}
\includegraphics[trim = 0.5cm 0.0cm 0.5cm 0.5cm, clip=true, 
width=0.49\textwidth]{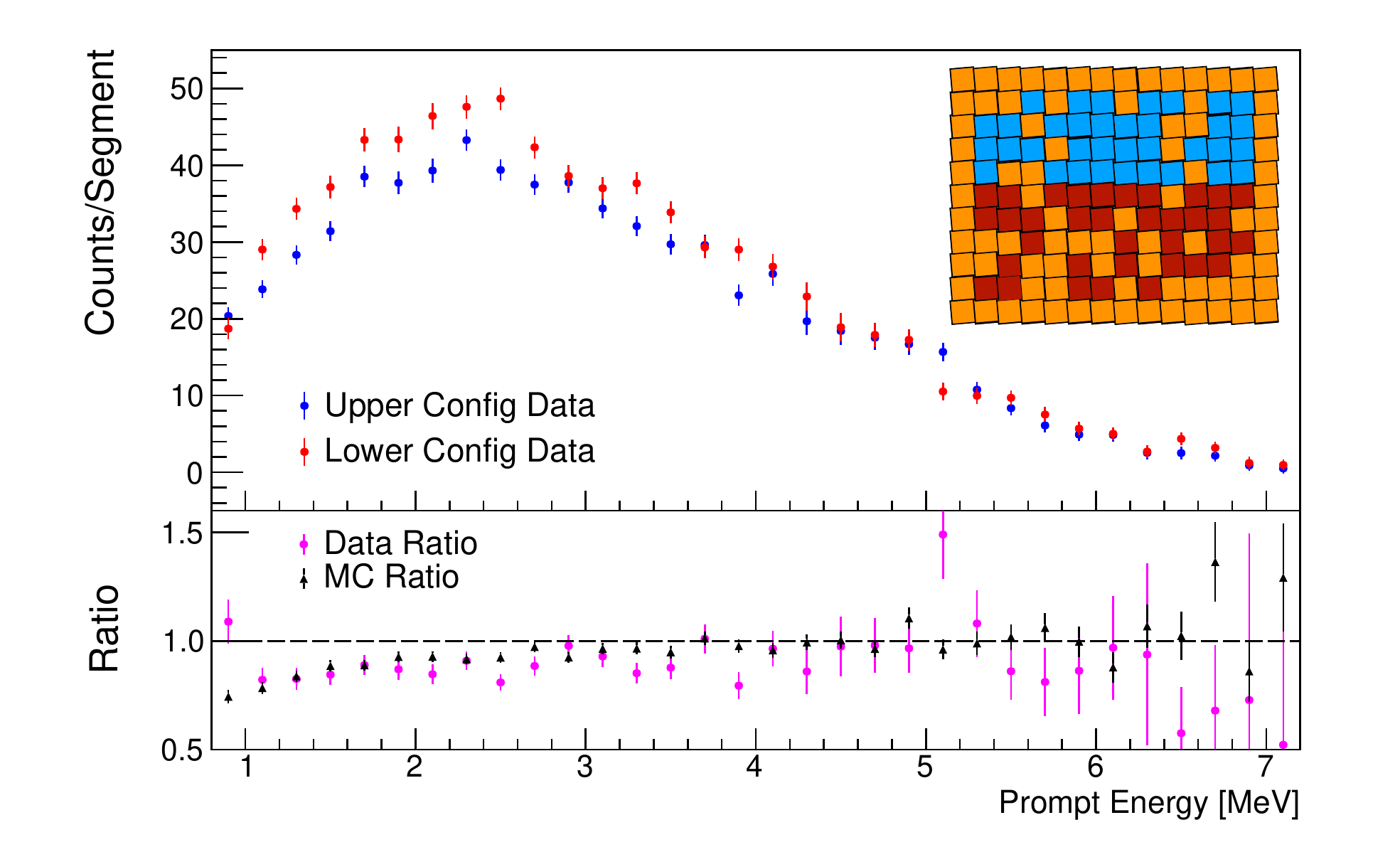}
\caption{Prompt E$_{rec}$ spectra and spectral ratios of IBD from different fiducial volume regions: detector inner-outer comparisons (top) and detector upper-lower comparisons (bottom).  Error bars represent statistical uncertainties.  The data-derived ratio between regions are compared to PG4-derived ratio predictions. See the text for detailed description.}
\label{fig:spatial}
\end{figure}

To verify proper modelling of these effects in the PROSPECT detector response model, background-subtracted IBD signal prompt E$_{rec}$ distributions are compared between these different detector regions in Figure~\ref{fig:spatial}.  
Figure insets illustrate which detector active segments are assigned to which category.  
Also pictured are the spectrum ratios between these two regions, in addition to that predicted by PG4 MC simulations of IBD interactions.  
Energy spectra and normalizations per segment should not be expected to be identical between regions due to the uneven distribution of dead and non-fiducial segments in the detector.  
However, deviations between regions should be correctly predicted by the PG4 IBD MC.  
Indeed, data-PG4 spectrum ratios between regions are generally consistent within the data's statistical limitations: a quantitative comparison of the data and PG4-predicted inner-outer (upper-lower) ratios give $\chi^2$/DOF of 56.6/31 (54.4/31).  
%Meanwhile, a fit of measured ratios to a flat value in E$_{rec}$ instead yield $\chi^2$/DOF of 47.0/31 (26.6/31); the PG4-based detector response model clearly provides a better description of the data than the simplistic assumption of position-independent energy response.  

\begin{figure}[hptb!]
\includegraphics[trim = 0.0cm 0.3cm 0.5cm 0.5cm, clip=true, 
width=0.49\textwidth]{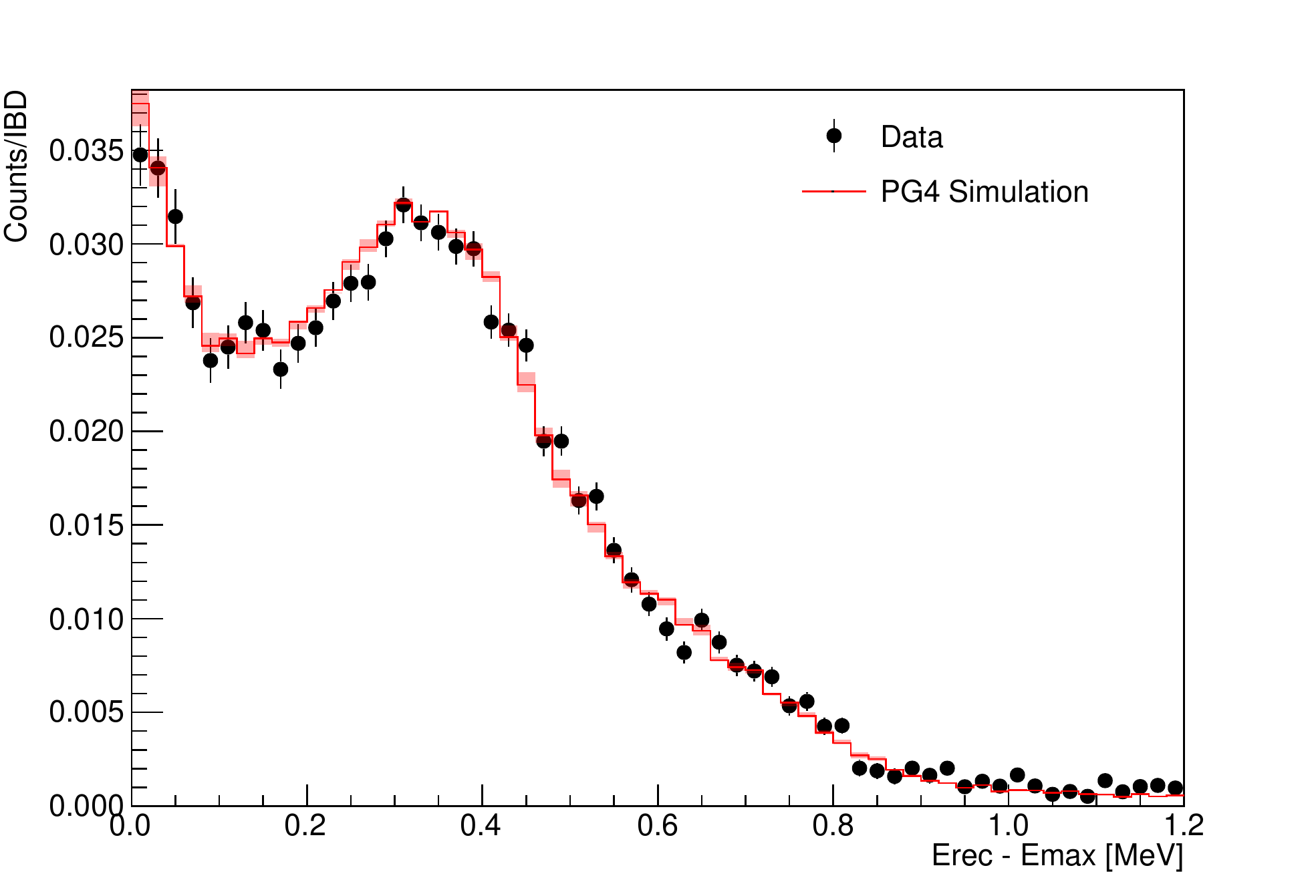}
\includegraphics[trim = 0.0cm 0.3cm 0.5cm 0.5cm, clip=true, 
width=0.49\textwidth]{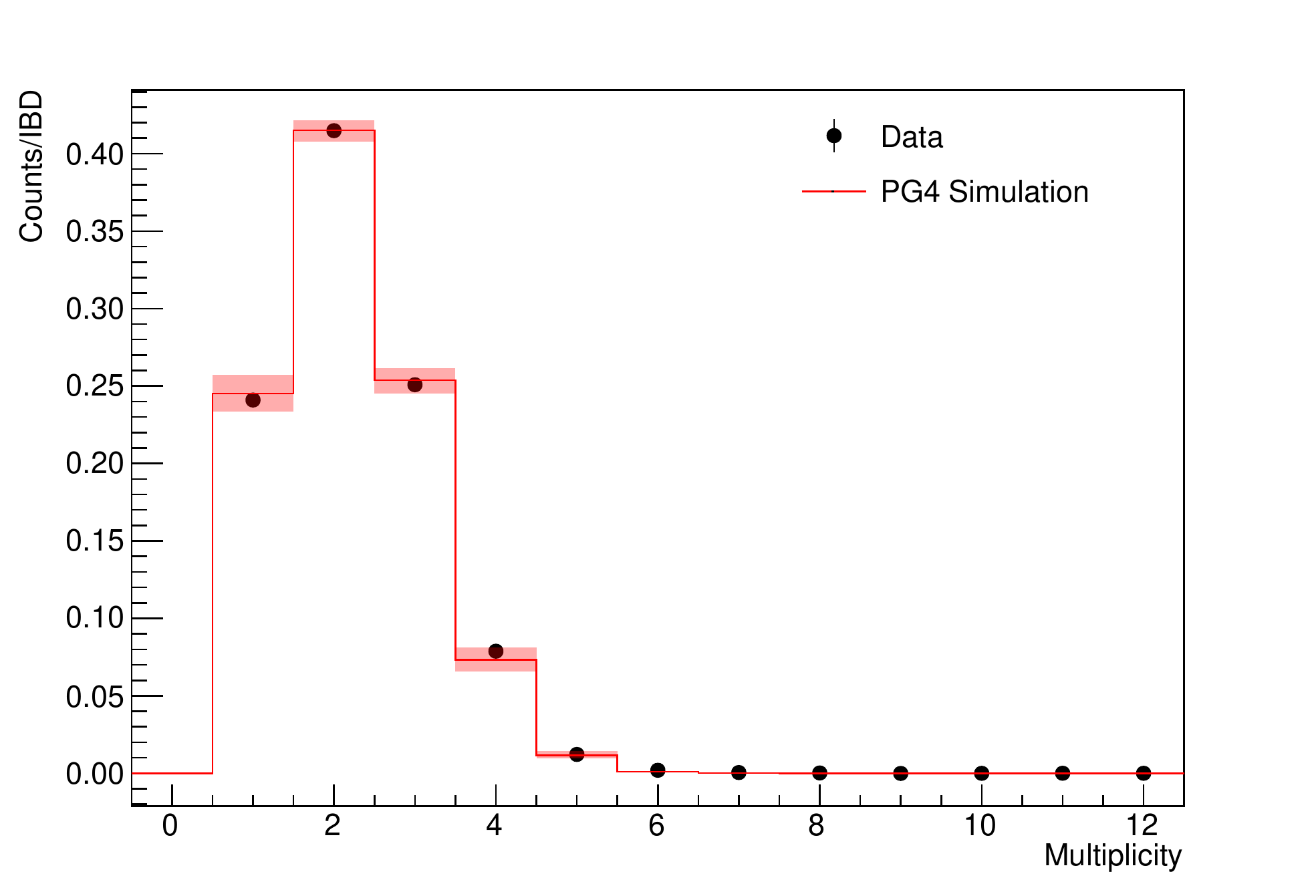}
\caption{Top: Energy deposition distribution outside the primary segment where IBD occurs. This is principally due to transport of annihilation $\gamma$-rays.   Bottom: Segment multiplicity of prompt cluster.  Error bands arising from consideration of the 5~keV thresholding uncertainty are also pictured in both panels for reference.} 
\label{fig:mult}
\end{figure}

The segmented nature of the PROSPECT target enables a variety of other cross-checks of the background-subtracted IBD dataset and modelling of these events.  
Whether due to IBD positrons traversing optical grid separators or migration of annihilation $\gamma$-rays, an IBD interaction in the PROSPECT detector more often than not produces reconstructed clusters spanning multiple segments.  
This effect is illustrated in Figure~\ref{fig:mult}, which shows the segment multiplicity of prompt clusters for the background-subtracted IBD signal.  
Both data and PG4 IBD MC simulations exhibit identical multiplicity distributions within systematic uncertainties, which are dominated by the $\pm$5~keV per-pulse analysis threshold uncertainty.  
This agreement is particularly reassuring, given the importance of pulse thresholding effects in determining event energy scales.  

Accurate modelling of IBD event topology is also demonstrated in Figure~\ref{fig:mult} by plotting the summed energy of all pulses (E$_{rec}$) excluding that with the highest reconstructed energy (E$_{max}$).  
This energy distribution is expected to be dominated by annihilation $\gamma$-ray energy depositions.  
Excellent agreement is found for this distribution between data and PG4 MC simulations, indicating accurate modelling of annihilation $\gamma$-ray energy depositions in the detector.  

\begin{figure}[hptb!]
\includegraphics[trim = 1.0cm 0.0cm 0.0cm 0.0cm, clip=true, 
width=0.49\textwidth]{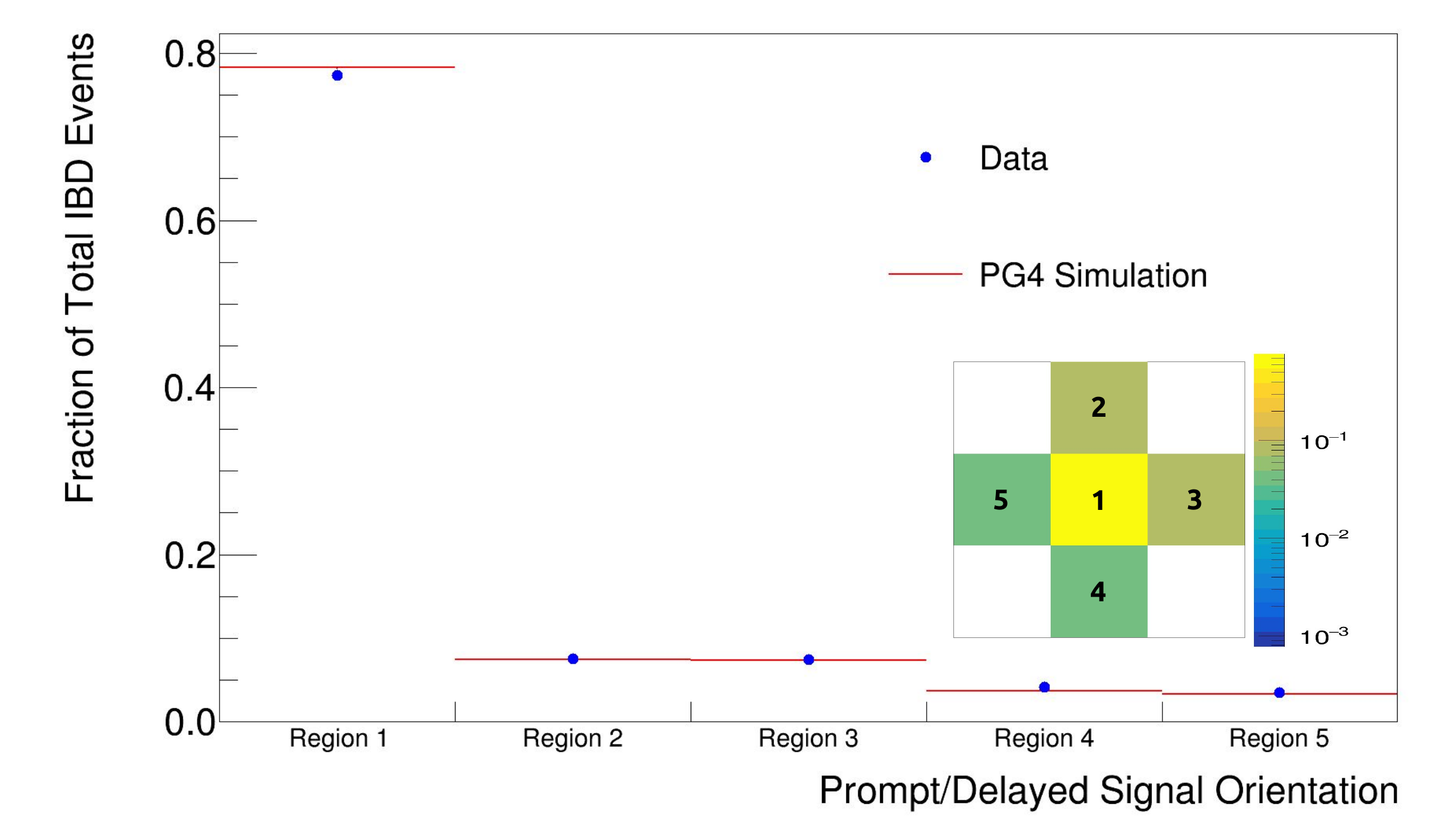}
\caption{Neutron capture segment relative to the segment with maximum prompt energy of data (blue) and simulation (red). The orientation convention is described in the graph's inset. Error bars for both data and simulation are comparable in size to the point marker width.}
\label{fig:IBDneutron}
\end{figure}

Detector segmentation also enables comparison of the relative positioning of prompt and delayed IBD signals with respect to one another in the detector target, as illustrated in Figure~\ref{fig:IBDneutron}.  
Approximately 77.4\%$\pm$0.5\% of IBD neutrons and positrons are found to have identical S$_{rec}$, indicating that IBD neutrons tend to capture in the same segment as their associated IBD interaction.  
This ratio is found to be 78.3\% for PG4 IBD MC, 0.9\%$\pm$0.5\% from the observed value.  
The data's marginally reduced IBD neutron mobility will result in smaller relative contributions from IBD interactions in inactive and non-fiducial segments.  
The impact of this added contribution on expected prompt E$_{rec}$ distributions is found to be small compared to those of other more dominant energy scale systematic uncertainties.  

When examining IBD signal events with different prompt and delayed S$_{rec}$, both data and PG4 show an outsized contribution from events with longer-baseline delayed S$_{rec}$.  
Events where the delayed $S_{rec}$ is `downstream' from the prompt $S_{rec}$ contribute 15.0\%$\pm$0.3\% of all IBD signal data events, while events with `upstream' neutrons contribute only 7.6\%$\pm$0.3\%.  
This difference in PG4 MC simulation is attributable to the non-negligible downstream kinetic energy of the final-state IBD neutron.  
The observation of this effect in PROSPECT provides an intriguing demonstration of the capabilities of segmented IBD detectors to statistically reconstruct the incoming direction of reactor $\nuebar$.

%%%%%%%%%%%%%%%%%%%%%%%%%%%%%%%OSC SECTION
\section{Sterile Neutrino Search Results}
\label{sec:osc}

Sterile neutrino oscillations are probed with the PROSPECT dataset by comparing prompt E$_{rec}$ spectra between different detector baselines.  
The following section will describe the appearance of the PROSPECT datasets in different baseline bins, introduce the statistical methods used to search for unexpected relative variations in E$_{rec}$ spectra between baselines, and present new sterile neutrino oscillation results based on the dataset described in Section~\ref{sec:signal}.  

%To perform a differential test of oscillation-induced spectral distortion, an IBD response model is generated for all detector positions using PG4, a GEANT4-based~\cite{bib:Geant4} Monte Carlo (MC) simulation package developed by the collaboration.  
\subsection{Datasets and Predictions}

\begin{figure}[hptb!]
\includegraphics[trim = 0.0cm 0.0cm 0.5cm 1.25cm, clip=true, 
width=0.49\textwidth]{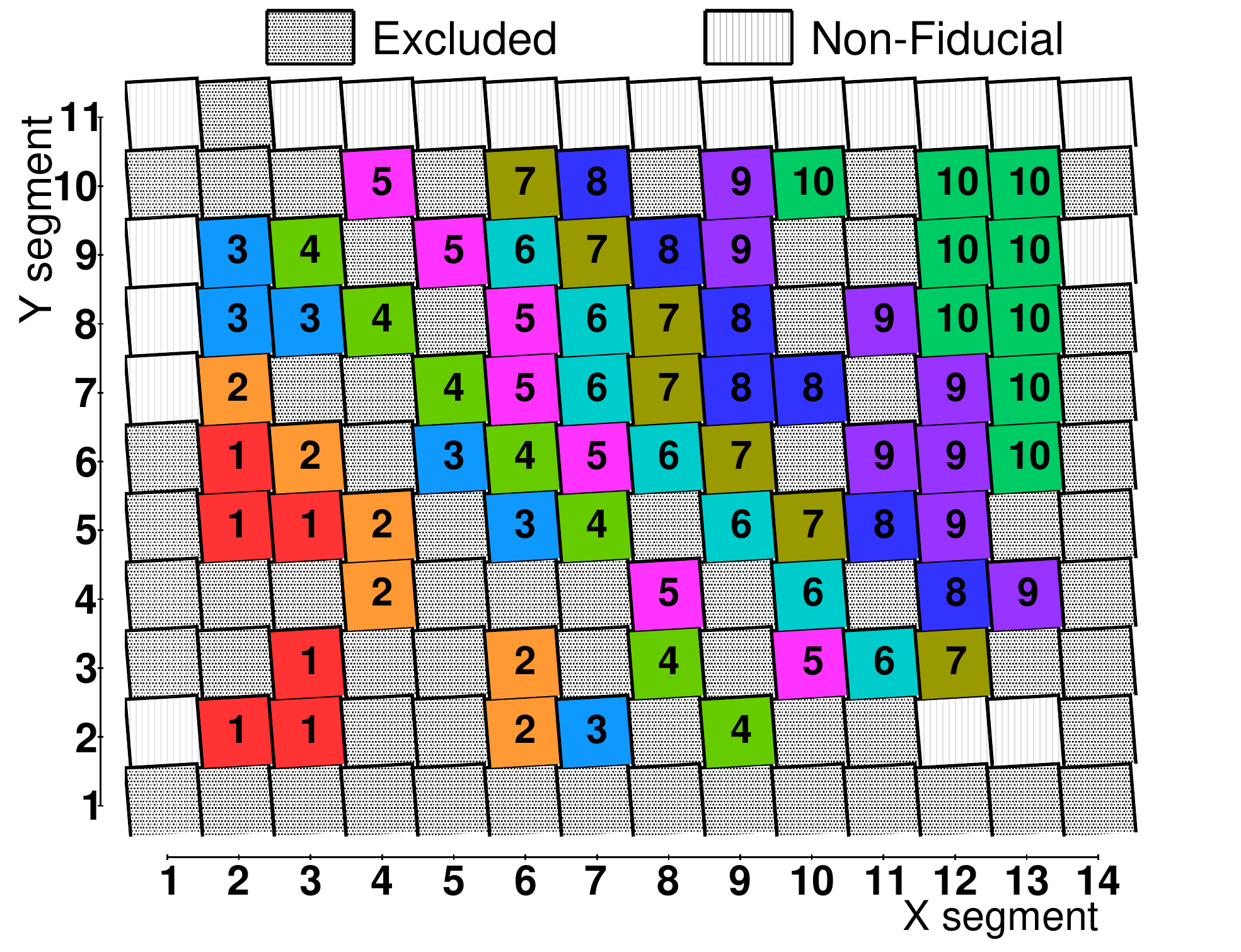}
\includegraphics[trim = 0.0cm 0.5cm 0.5cm 1.0cm, clip=true, 
width=0.49\textwidth]{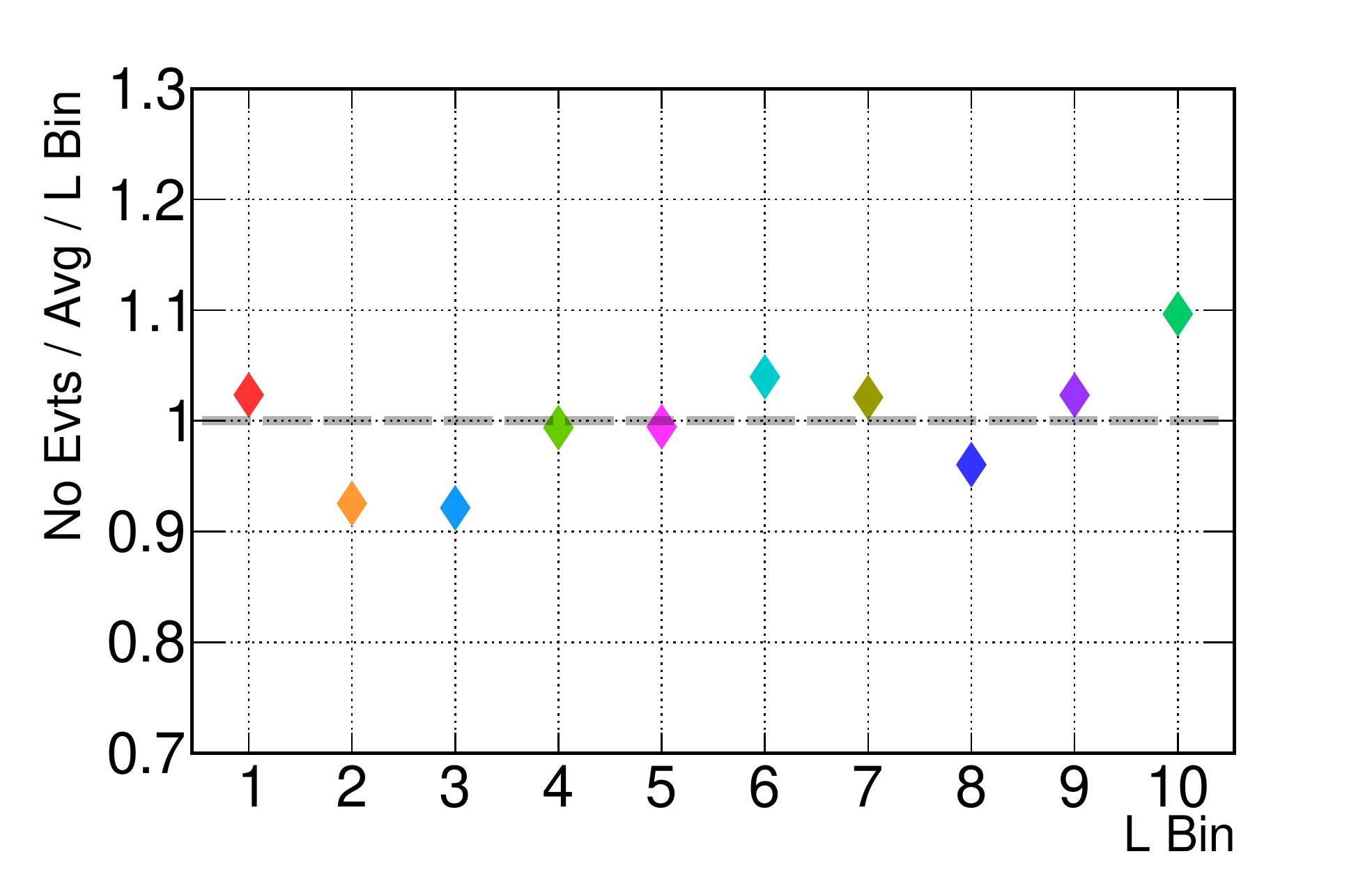}
\caption{Top: Baseline bin assignments for different active fiducial segments; excluded and non-fiducial segments are also designated.  Bottom: relative background-subtracted IBD signal counts per baseline bin; there are an average of roughly 5000 IBD signal counts per bin.} 
\label{fig:oscbinning}
\end{figure}

To perform the oscillation analysis, active detector segments are assigned to one of ten defined baseline ranges, or $l$, as illustrated in Figure~\ref{fig:oscbinning}.  
IBD events are then assigned to baseline bin $l$ according to their prompt S$_{rec}$.  
Segment $l$ assignments are chosen to produce roughly similar IBD signal statistics in each baseline bin.  
Given the 1/r$^2$ reduction of IBD signal events with baseline demonstrated in Fig~\ref{fig:rates}, this choice results in uneven baseline bin widths.  
This method differs from that described in the previous PROSPECT oscillation analysis~\cite{prospect_osc}, where the IBD dataset was separated into six bins of equal width; the new binning method provides better statistical coverage over a wider range of baselines and delivers better overall oscillation sensitivity.  
Roughly 5000~events are contained in each baseline bin $l$, with per-bin relative variations of 10\% illustrated in Figure~\ref{fig:oscbinning}.  

\begin{figure*}[hbtp!]
\includegraphics[trim = 0.0cm 0.0cm 0.0cm 0.0cm, clip=true, 
width=0.85\textwidth]{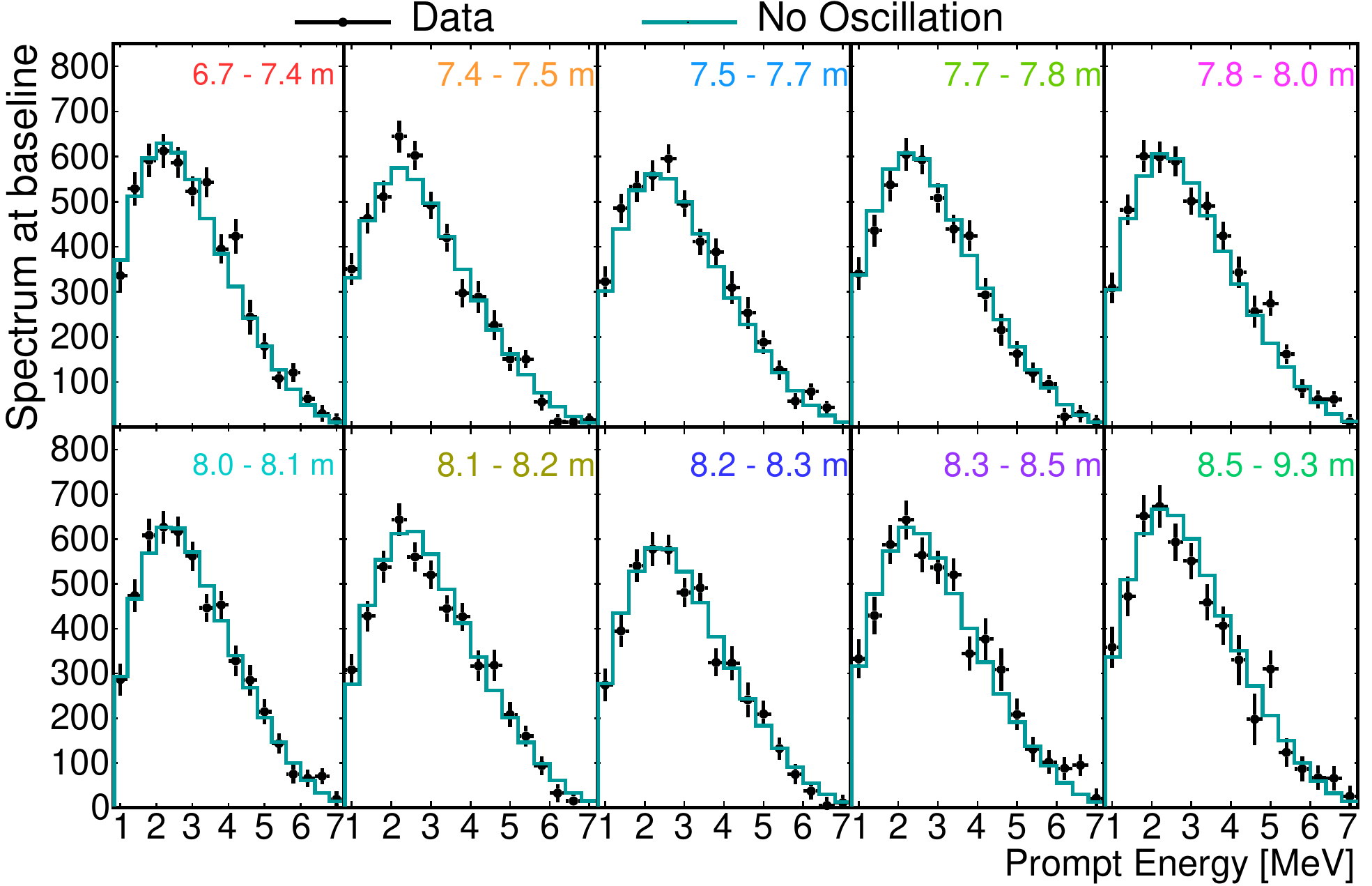}
\caption{Measured prompt E$_{rec}$ spectra for the ten baseline bins defined in Figure~\ref{fig:oscbinning}.  The PG4 no-oscillation prediction is also pictured as a solid line.  For the PG4 MC, a common normalization factor has been applied to match predicted and observed baseline-summed IBD signal counts.   Error bars represent statistical uncertainties.}
\label{fig:oscspec}
\end{figure*}

Prompt E$_{rec}$ spectra for background-subtracted IBD signal events in each $l$ bin, called $M_{l,e}$, are pictured in Figure~\ref{fig:oscspec}.  
Also pictured are the unoscillated IBD prompt E$_{rec}$ predictions $P_{l,e}$ for each baseline bin.  
$P_{l,e}$ are formed by applying the best-fit PG4-derived segment response matrices described in Sections~\ref{subsec:MC} and~\ref{subsec:ibdmc} to an IBD interaction generator following the $^{235}$U \nuebar~energy spectrum calculated by Huber~\cite{bib:huber} and the IBD cross-section of Ref.~\cite{Vogel:1999zy}.  
IBD vertex distributions for $P_{l,e}$ are generated assuming a finite cylindrical HFIR core geometry, as described in Section~\ref{subsec:ibdmc}.  
To remain consistent with procedures used for generating detector response matrices, the segment hosting each generated IBD's interaction is assigned as that reconstructed IBD event's prompt S$_{rec}$.  
While this choice effectively ignores a source of worsened resolution in knowledge of true \nuebar baselines, this contribution is negligible compared to the position resolution smearing related to the finite reactor core geometry and $l$ bin width.  

%Basically the baseline-correlated normalization of the MC is arbitrary.

Prior to application of detector response to produce IBD prompt $E_{rec}$ distributions, true $\nuebar$ energy distributions for each segment's IBD interactions can be distorted to account for the possible presence of sterile neutrino oscillations.  
This distortion is dictated by the parameters ($\Delta$m$^2_{41}$,sin$^2 2\theta_{14}$) as defined in Eq.~\ref{eq:osc}, as well as by the $\nuebar$ energies and true baselines corresponding to these IBD interactions.  
To accelerate the generation of oscillated predictions, each segment's $z$-center midpoint is used as the true generated $\nuebar$ interaction location for each IBD event.  
This choice serves to ignore the $\nuebar$ baseline (and oscillation) smearing provided by the $\mathcal{O}$(10~cm) range of \nuebar production-interaction baselines within a segment; however, this contribution is once again negligible compared to that of the finite HFIR core size.  

%\begin{equation}\label{eq:pred}
%P_{l,e} = \frac{N}{r^2}\mathrm{R}\bm{S_{HM}}
%\end{equation}
%Here, $r$ is the reactor-detector baseline

\begin{figure*}[hbtp!]
\includegraphics[trim = 0.0cm 0.0cm 0.0cm 0.0cm, clip=true, 
width=0.85\textwidth]{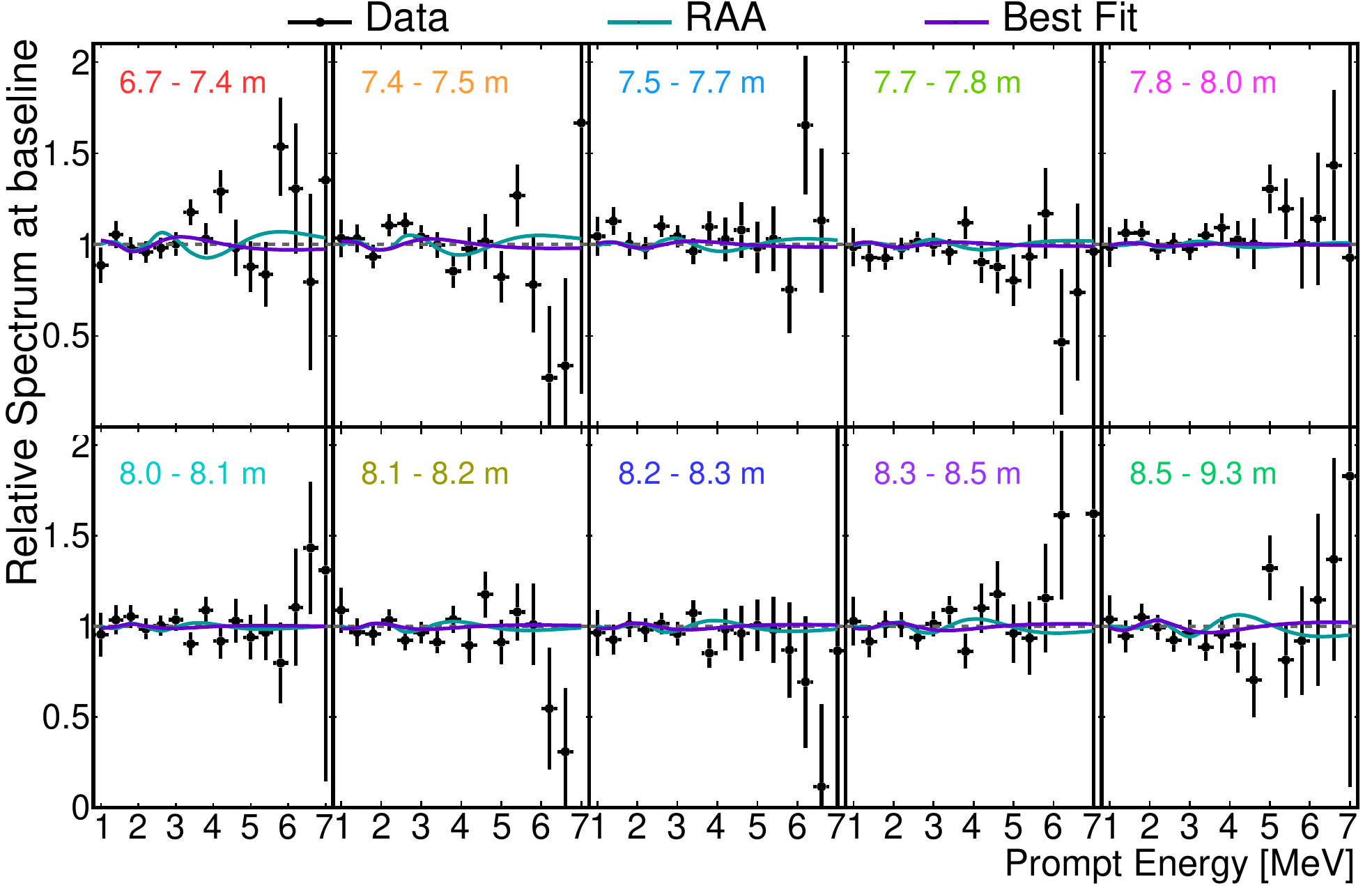}
\caption{Measured prompt E$_{rec}$ spectrum ratios ($\frac{M_{l,e}}{M_{e}}$, corrected by $\frac{P_{l,e}}{P_{e}}$) for the ten baseline bins defined in Figure~\ref{fig:oscbinning}.  PG4-predicted ratios in the presence of sterile neutrino oscillations matching those of the best-fit point of ($\sin^22\theta_{14}$,$\Delta m^2_{41}$) = (0.11,1.78~eV$^2$) and the `Reactor Antineutrino Anomaly' (RAA) best-fit point of Ref.~\cite{bib:mention2011} are also pictured as solid purple and blue lines, respectively.  In the absence of oscillations, the predicted ratio is unity for all energy-position bins.  Error bars represent statistical uncertainties.}
\label{fig:oscratio}
\end{figure*}

To ensure minimal dependence of the oscillation result on uncertainties in the shape and normalization associated with the Huber \uFive~reactor flux prediction, relative comparisons between measured prompt E$_{rec}$ are used to perform PROSPECT's oscillation measurement.  
These comparisons are based on the per-baseline measured and PG4-predicted content of each bin in baseline $l$ and energy $e$, $M_{l,e}$ and $P_{l,e}$, and on the detector-wide measured and predicted content of bin $e$, respectively:
\begin{equation}\label{eq:absSpectrum}
M_{e}=\sum_{l=1}^{10}M_{l,e}
\,\,\textrm{\,and\,}\,\,
P_{e}=\sum_{l=1}^{10}P_{l,e}.
\end{equation}
A detailed description of $M_{e}$ and $P_{e}$ will be given in Section~\ref{sec:spec}.  
For the oscillation analysis, $M_{l,e}$ are compared to the predicted per-baseline spectra $M_{e}\frac{P_{l,e}}{P_{e}}$.  
The latter quantity reduces the dependence on the underlying reactor $\nuebar$ model, while also correcting for relative energy response variations between baseline bins predicted by the PG4 simulation.  
The ratios between these two quantities for each baseline are shown in Figure~\ref{fig:oscratio}.  
An absence of short-baseline oscillation effects in $M$ will produce a flat ratio at unity; meanwhile, the presence of oscillation effects in $M_{l,e}$ and $M_{e}$ will alter this ratio in a manner also depicted in Figure~\ref{fig:oscratio}.  
Visual examination of the measured ratios in Figure~\ref{fig:oscratio} yields no immediate indication of non-flat trends similar to that produced by large-amplitude sterile neutrino oscillations.  

\begin{figure}[hptb!]
\includegraphics[trim = 0.0cm 0.0cm 0.0cm 0.0cm, clip=true, 
width=0.48\textwidth]{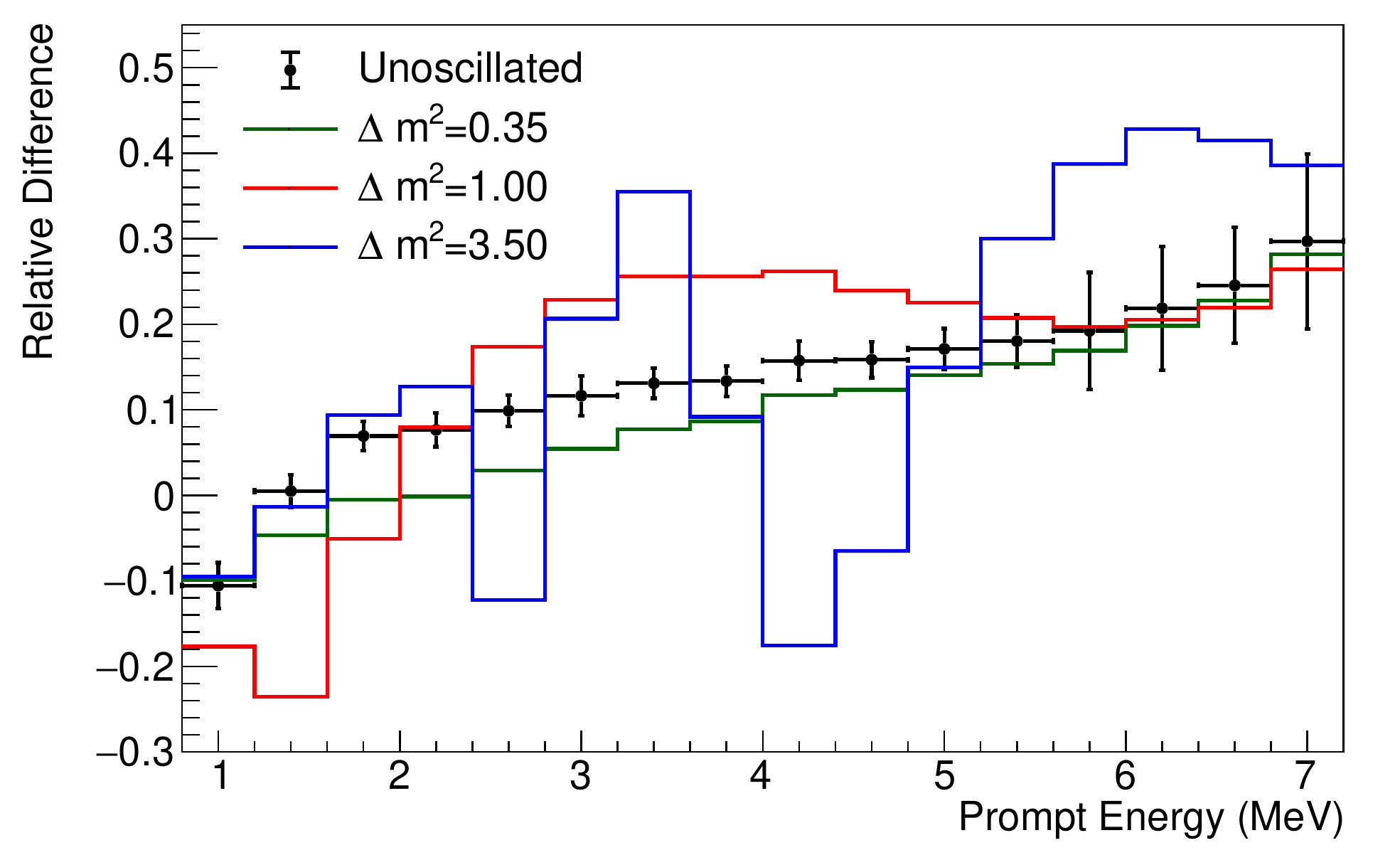}
%\includegraphics[trim = 0.0cm 0.0cm 0.0cm 0.0cm, clip=true, 
%width=0.48\textwidth]{figures/PRD_Fig30_110ratio.pdf}
\caption{PG4-predicted prompt E$_{rec}$ relative differences between baselines 5 and 1, $\frac{P_{5}-P_{1}}{0.5*(P_{5}+P_{1})}$, for the case of no oscillations (points), and in the case of sterile neutrino oscillations of 50\% amplitude and varying mass splittings.  Errors on the no-oscillation prediction represent the diagonals of the uncertainty covariance matrix between the different baselines.} 
\label{fig:oscrelvar}
\end{figure}

As each baseline bin $l$ is composed of segments of varying proximity to the detector edge and to inactive segments, some variations in $M_{l,e}$ are expected between different baseline bins even in the absence of oscillation effects.  
As mentioned above, PG4 is used to characterize these relative response variations, which are taken into account in $P_{l,e}$ predictions.  
To demonstrate the behavior of these relative response variations, Figure~\ref{fig:oscrelvar} shows the relative differences between un-oscillated predicted spectra $P_{1,e}$ and $P_{5,e}$ along with the impact of sterile neutrino oscillations on these ratios for differing mass-splitting values.  
High mass-splitting oscillations produce relative spectrum differences between baselines that are characteristically different than those produced by expected energy response variations.  
Thus, in the mass splitting region above 1~eV$^2$, statistical uncertainties are expected to dominate PROSPECT's sterile neutrino oscillation sensitivity.  
Below $\sim$0.5~eV$^2$, relative energy response variations and efficiency differences between baselines can mimic to an extent the behavior of oscillations; thus, uncertainties in these variations will also limit oscillation sensitivity in this mass-splitting range.  

%%No L/E plot, we now say.
%Given that the oscillation behavior described by Equation~\ref{eq:osc} follows an $L/E_{\nu}$ dependence, it is instructive to further aggregate all IBD signal candidates from Figure~\ref{fig:oscspec} into L/$E_{rec}$ space.  
%This visualization of the PROSPECT data is provided in Figure~\ref{fig:lovere}, along with the predicted distribution of IBD candidates in the presence of sterile neutrino oscillations.  
%BLAH say something about the y-axis that is plotted; BLAH mentione error bars are stats-only.
%Visual examination of the data again yields no unambiguous indication of the presence of sinusoidal variations in with L/$E_{rec}$.  
%We note that the distribution in Figure~\ref{fig:lovere} is not directly used in statistical determinations of allowed or excluded sterile neutrino parameter space; these operations are performed on the prompt E$_{rec}$ ratio distributions shown in Figure~\ref{fig:oscspec}.  

\subsection{Statistical Method}

To test for the possible existence of sterile neutrino oscillations, measured per-baseline prompt E$_{rec}$ spectra $M_{l,e}$ are quantitatively compared to predicted per-baseline prompt E$_{rec}$ spectra $M_{e}\frac{P_{l,e}}{P_{e}}$ in the presence of oscillation effects in $P_{l,e}$ and $P_{e}$ dictated by the parameters $\Delta m^2_{41}$ and $\sin^2 2\theta_{ee}$.  

For this purpose, a $\chi^2$ is defined as:
\begin{equation}
\label{eq:oscchi2}
\chi^2 = \bm{\Delta}^{\textrm{T}}\textrm{V}_{\textrm{tot}}^{-1}\bm{\Delta},
\end{equation}
where $\bm{\Delta}$ is a 160-element vector that represents the relative agreement between measurement and prediction in 10 $l$ bins and 16 $e$ bins: 
\begin{equation}\label{eq:delta}
\Delta_{l,e} = M_{l,e}- M_{e}\frac{P_{l,e}}{P_{e}}.  
\end{equation}
The 160 $\bm{\Delta}$ entries are grouped by baseline, running from shortest distance to highest distance.   
Within each baseline group, $\bm{\Delta}$ elements run from lowest to highest E$_{rec}$.  
%\ptsnote{Maybe we want to use $\Delta \chi^2_{min}$ instead of $\chi^2$}. 

\begin{figure}[hptb!]
%\includegraphics[trim = 0.0cm 0.0cm 0.0cm 0.0cm, clip=true, 
%width=0.42\textwidth]{figures/PRD_Fig31_osccov.png}
\includegraphics[trim = 0.25cm 0.4cm 0.5cm 1.25cm, clip=true, 
width=0.48\textwidth]{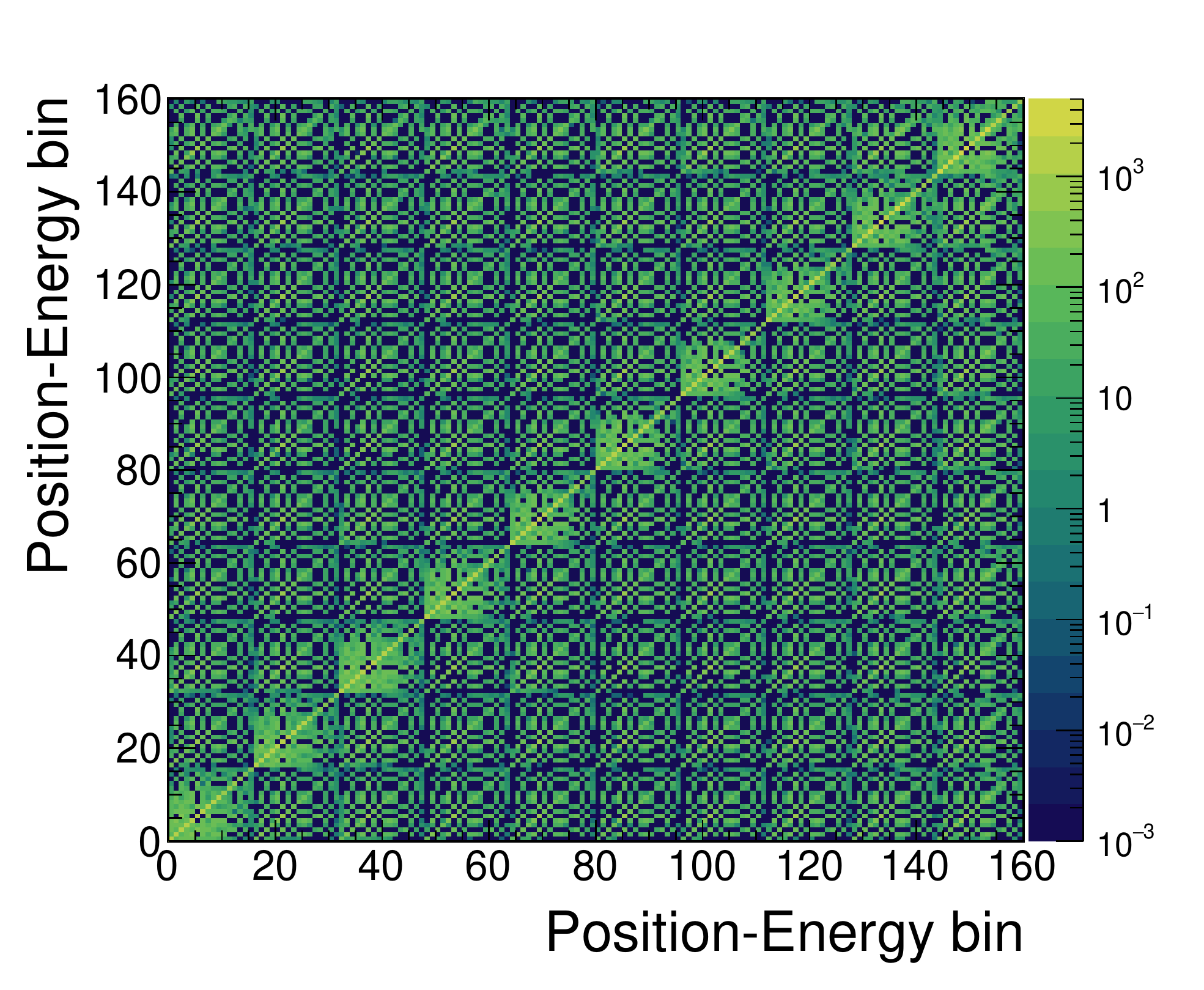}
\caption{Total uncertainty covariance matrix for the energy-baseline bins used for the PROSPECT oscillation analysis.  Full covariance matrix elements are computed by multiplying reduced covariance matrix elements by the relevant measured signal rates $M_i \cdot M_j$.  Sub-matrices of common baseline are visible within these covariance matrices, with baselines increasing with increasing $i$ and $j$.} 
\label{fig:osccov}
\end{figure}

Statistical and systematic uncertainties and their correlation between energy bins are incorporated into Eq.~\ref{eq:oscchi2} using the covariance matrix V$_{\textrm{tot}}$.  
This matrix is composed of the sum of individual statistical and systematic matrices V$_{\textrm{stat}}$ and V$_{\textrm{sys}}$.  
To highlight the relative magnitude of uncertainty contribution of different elements, the total uncertainty reduced covariance matrix is pictured in Figure~\ref{fig:osccov}. 
Each entry V$_{\textrm{tot}}^{i,j}$ is obtained by multiplying the corresponding reduced covariance matrix entry by $M_i \cdot M_j$.  
As mentioned above, the 160 $i$ and $j$ values in V$_{\textrm{tot}}$ are grouped by baseline, running from lowest to highest baseline with increasing $i$ and $j$.  
For example, the 10 sub-matrices appearing along the diagonal of V$_{\textrm{tot}}$ represent uncorrelated statistical uncertainties for each individual baseline.

Statistical uncertainties V$_{\textrm{stat}}$ are dominated by reactor-on IBD candidates and the subtracted cosmogenic background estimate based on the reactor-off IBD candidate dataset.  
Subtracted accidental backgrounds during reactor-on and reactor-off periods contribute little statistical uncertainty, owing to the large offset time window used to determine them.  
Uncorrelated statistical uncertainties from each dataset, which compose the diagonal of V$_{\textrm{stat}}$, are primarily determined by the Poisson error of each $l,e$ bin after properly scaling for relative live time and environmental differences between datasets and data periods.  
As each $M_{l,e}$ is a subset of the detector-integrated spectrum $M_{e}$,  correlations in $\Delta_{l,e}$ statistical uncertainties will exist between different $l$, resulting in off-diagonal contributions to V$_{\textrm{stat}}$.  

\begin{table*}[tb!]
\begin{tabular}{|l|c|c|c|l|}
         \hline
         Parameter & Section & Nominal Value & Uncertainty & Correlations \\
         \hline
         Absolute background normalization & \ref{subsec:bkg_corr}, \ref{subsec:bkg_check} & - & 1.0\% & Correlated between energies and baselines \\
         Absolute $n$-H peak normalization & \ref{subsec:bkg_check} &  - & 3.0\% & Correlated between energies and baselines \\
         Relative signal normalization & \ref{subsec:eff} & - & 5\% & Correlated between energies \\ 
         Baseline uncertainty & \ref{sec:exp} & - & 10~cm & Correlated between energies and baselines \\
         \hline
         First-order Birks constant &  \ref{subsec:eresp} & 0.132 MeV/cm & 0.004 MeV/cm & Correlated between baselines \\
         Second-order Birks constant &  \ref{subsec:eresp} & 0.023 MeV/cm & 0.004 MeV/cm & Correlated between baselines \\
         Cherenkov contribution &  \ref{subsec:eresp} & 37\% & 2\% & Correlated between baselines \\
         Absolute energy scale &  \ref{subsec:eresp} & - & 0.6\% & Correlated between baselines \\ 
         Absolute photostatistics resolution &  \ref{subsec:eres} & - & 5\% & Correlated between baselines \\ 
         Absolute energy leakage &  \ref{subsec:evar} & - & 8~keV & Correlated between baselines \\ 
         Absolute energy threshold & \ref{subsec:eresp}, \ref{subsec:reco} & & 5~keV & Correlated between baselines \\ 
         \hline
         Relative energy scale & \ref{subsec:perform}, \ref{subsec:eresp} & - & 0.6\% & Uncorrelated between baselines \\
         Relative photostatistics resolution & \ref{subsec:perform}, \ref{subsec:eres} & - & 5\% & Uncorrelated between baselines \\ 
         Relative energy leakage & \ref{subsec:evar} & - & 8~keV & Uncorrelated between baselines \\ 
         Relative energy threshold & \ref{subsec:eresp}, \ref{subsec:reco} & - & 5~keV & Uncorrelated between baselines \\ 
         Reflector panel thickness & \ref{subsec:eresp} & 1.18~mm & 0.03~mm & Uncorrelated between baselines \\ 
        \hline
         %\\ \hline
         %Energy Loss & 8 keV & Uncertainty in the energy lost by escaping 511 gammas
         %\\ \hline
         %Aluminum 28 Activation & 100\% & Uncertainty in the amount of aluminum 28 contributing to the spectrum
         %\\ \hline
         %Non-equilibrium Contributions & 100\% & Uncertainty in extrapolating the antineutrino spectrum from Mueller et al \cite{bib:mueller2011}
         %\\ \hline
         %Panel Thickness & 0.05 mm & Combined systematic and statistical uncertainty for the thickness of the panels separating segments
         %\\ \hline
         %Z Fiducial Cut & 25 mm & Uncertainty in the position of events near the edge of the fiducial volume
         %\\ \hline
         %Energy Threshold & 5 keV &  Uncertainty in the segment-by-segment energy threshold cut
         %\\ \hline
    \end{tabular}
\caption{Summary of systematic uncertainties taken into account in the oscillation systematic covariance matrix V$_{\textrm{sys}}$.  Where applicable, nominal parameter values are provided.  References to relevant sections where uncertainties are described are also given.}
\label{tab:osc_uncertainties}
\end{table*}

Systematic uncertainties in $\Delta_{l,e}$, as well as systematic correlations between different $l$ and $e$, are taken into account in the covariance matrix V$_{\textrm{sys}}$.  
Various sources of systematic uncertainty related to detector response, response stability with time and with detector position, and background estimates, have been described throughout previous sections in this paper.  
These sources of systematic uncertainty are overviewed in Table~\ref{tab:osc_uncertainties}, as well as being described briefly below:
\begin{itemize}
\item{\emph{Absolute background normalization and $n$-H peak uncertainty:} accounts for unexpected background variations between reactor-off and reactor-on periods, and for uncertainty in the atmospheric scaling factor.  Each is included as a baseline- and energy-correlated uncertainty within its relevant energy range; the two effects are treated as uncorrelated.}
\item{\emph{Relative signal normalization:} accounts for relative volume and efficiency variations between different baseline bins.  Included as an energy-correlated uncertainty.}
\item{\emph{Baseline:} accounts for uncertainty in the detector-reactor baseline, as described in Section~\ref{sec:exp}.  Included as an energy-correlated and baseline-correlated uncertainty.}

\item{\emph{Energy non-linearity model uncertainties:} accounts for uncertainty in best-fit Birks scintillator non-linearity parameters k$_{b1}$ and k$_{b2}$ and the Cerenkov light contribution k$_c$.  As all segments contain the same scintillator, these uncertainties are treated as baseline-correlated.} 
\item{\emph{Energy scale uncertainties:} accounts for linear energy scale uncertainties.  These are included as both a baseline-correlated and a baseline-uncorrelated uncertainty, reflecting the validations provided in Sections~\ref{subsec:eresp} and ~\ref{subsec:perform}, respectively. }
\item{\emph{Energy loss and leakage uncertainties:} accounts for uncertainties in PG4 MC modelling of energy scale offsets between different detector regions/locations, which arise from loss of energy in inactive detector materials.  Energy loss in optical grid reflectors is treated separately from energy losses due to leakage of $\gamma$-ray energy out of active detector regions.  These are included as both baseline-correlated and baseline-uncorrelated uncertainties.} 
\item{\emph{Energy threshold uncertainties:} accounts for uncertainties in reconstructed pulse energy thresholds, which play a key role in equalizing pulse multiplicities between different segments and different time periods.  These are included as both a baseline-correlated and a baseline-uncorrelated uncertainty.} 
\item{\emph{Photostatistics resolution uncertainties:} accounts for uncertainties in photostatistics resolution in Eq.~\ref{eq:resolution}.  These are included as both a baseline-correlated and a baseline-uncorrelated uncertainty,  reflecting the validations provided in Sections~\ref{subsec:eres} and ~\ref{subsec:perform}, respectively. } 
\end{itemize}

For each systematic uncertainty parameter described in Table~\ref{tab:osc_uncertainties}, a covariance matrix V$_x$ is produced through generation and characterization of systematically fluctuated MC datasets.  
This process proceeds by first generating 10$^3$ MC datasets and unoscillated $P_{l,e}$ datasets including variations of a single systematic uncertainty parameter following a Gaussian distribution with a 1$\sigma$ width as indicated in Table~\ref{tab:osc_uncertainties}.  
Toy MC $P_{l,e}$ distributions for baseline, signal normalization, and energy resolution, leakage and linear scale variations are generated via analytical adjustment of the default null oscillation PG4 IBD dataset; for the background normalization uncertainty, similar analytical adjustment is applied to the reactor-on cosmogenic background prediction. 
$P_{l,e}$ distributions for energy threshold systematic variations also use this default PG4 dataset, while applying a variety of reconstructed pulse energy threshold requirements in the analysis chain.  
For reflector panel thickness and scintillator non-linearity parameter uncertainties, $P_{l,e}$ distributions are obtained via generation of PG4 IBD MC datasets containing adjusted input simulation parameters; sample sizes are sufficiently large to ensure negligible MC-related stastical uncertainty contribution.
For the purposes of covariance matrix generation, we subsequently refer to systematically fluctuated $P_{l,e}$ distributions as $P_i$ and the un-fluctuated $P_{l,e}$ as $\overline{P}_i$.   

With systematically fluctuated datasets $P_i$ in hand, covariance matrix elements for each uncertainty parameter can be calculated as the average difference in fluctuated and un-fluctuated datasets,
\begin{equation}\label{eq:covmatrix}
\textrm{V}_{ij} = \left\langle(P_i-\overline{P}_i)(P_j-\overline{P}_j)\right\rangle,
\end{equation}
for any two entries $i$ and $j$ in $P$.
%The final systematic uncertainty reduced covariance matrix, obtained from summing reduced covariance matrices for each individual parameter, is also pictured in Figure~\ref{fig:osccov}.  
%Off-diagonal contributions are clearly more dominant in V$_{sys}$ than in V$_{stat}$.  
It is clear from the large size of on-diagonal elements in V$_{\textrm{tot}}$~from Figure~\ref{fig:osccov} that uncorrelated statistical uncertainty contributions are of substantially larger size than systematic uncertainty contributions.  

\subsection{Oscillation Results}

\begin{figure}[hptb!]
\includegraphics[trim = 0.25cm 0.25cm 0.5cm 0.5cm, clip=true, 
width=0.49\textwidth]{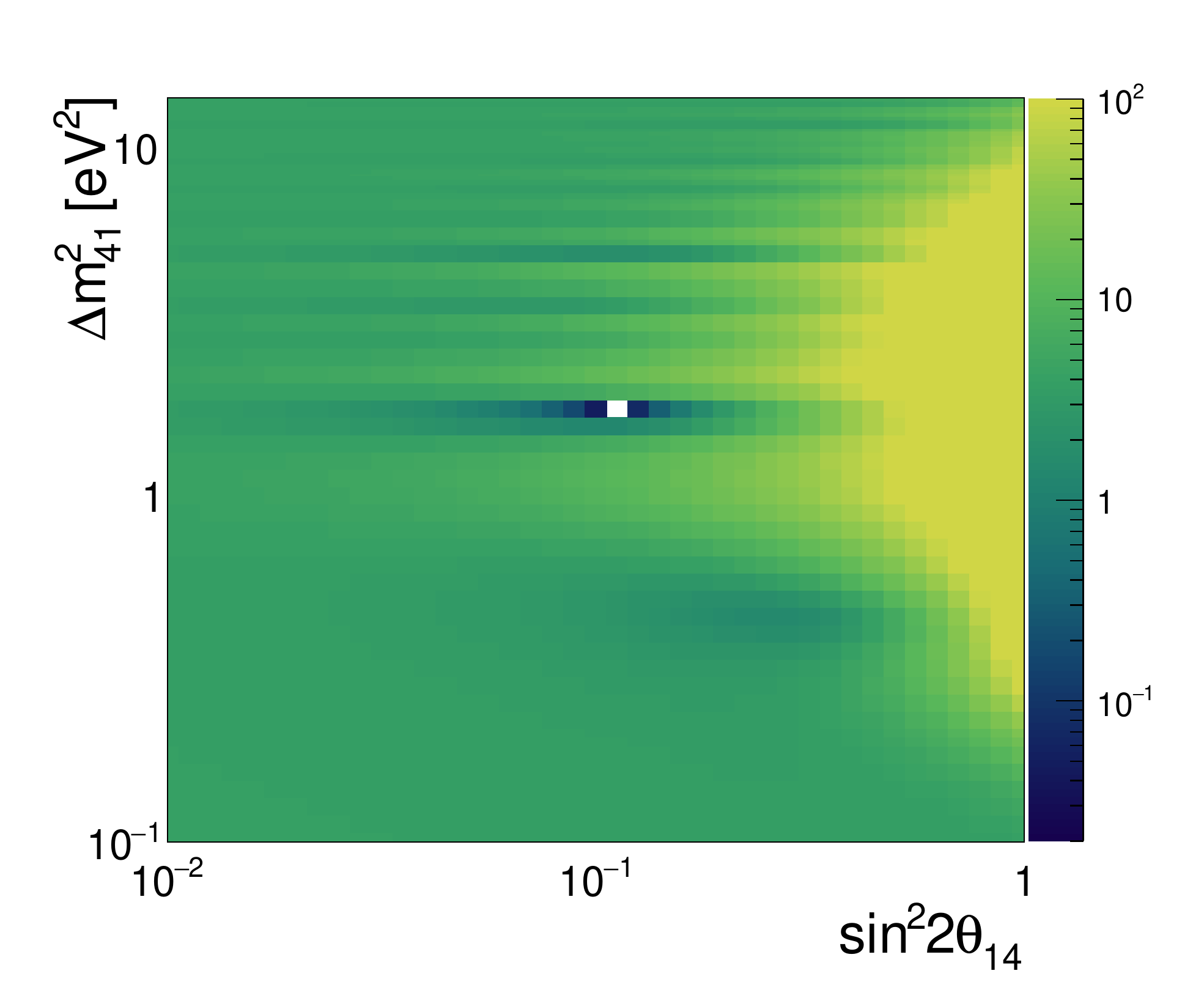}
\caption{The value of $\Delta \chi^2$ obtained for each (sin$^2 2\theta_{14}$,$\Delta$m$^2_{41}$) grid point, relative to the best-fit point (white square) at (0.11,1.78~eV$^2$).  The $\chi^2$ definition is provided in Eq.~\ref{eq:oscchi2}.  The white spot corresponds to the location of the best-fit point ($\Delta\chi^2=0$).} 
\label{fig:chi2map}
\end{figure}

Using the PROSPECT IBD candidate E$_{rec}$ spectra $M_{l,e}$ described in Section~\ref{sec:signal}, the covariance matrices V$_{sys}$ and V$_{stat}$ described in the previous section, and PG4-generated oscillated $P_{l,e}$ spectra, the $\chi^2$ of Equation~\ref{eq:oscchi2} can be calculated for each point in the tested sterile neutrino parameter space.  
Calculated $\Delta\chi^2$ with respect to the best-fit point in phase space are pictured in Figure~\ref{fig:chi2map}.  
The minimum value ($\chi^2_{min}$/DOF) of 119.3/142 was identified at the grid point ($\sin^22\theta_{14}$,$\Delta m^2_{41}$) = (0.11,1.78~eV$^2$).  
This $\chi^2_{min}$/DOF of 0.84 is slightly higher with respect to the previous minimum, 0.74, reported at ($\sin^22\theta_{14}$,$\Delta m^2_{41}$) = (0.35,0.5~eV$^2$) by PROSPECT in Ref~\cite{prospect_osc}.  
This new $\chi^2_{min}$ value should also be contrasted with that obtained in the case of null oscillations ($\theta_{14}$=0), where the $\chi^2$/DOF is 123.3/144; while this $\Delta \chi^2$ of 4.0 indicates that the null oscillation case does not provide the best match to the data, further statistical analysis must be done to quantify the level of preference for non-zero oscillations.  
These two $\chi^2$ can also be compared to 135.1, the $\chi^2$ value obtained at the `Reactor Antineutrino Anomaly' (RAA) best-fit point of Ref.~\cite{bib:mention2011},  ($\sin^22\theta_{14}$,$\Delta m^2_{41}$) = (0.165, 2.39  eV$^2$). 
This emphasizes that the dataset also contains a preference for the null oscillation hypothesis over this suggested region of oscillation parameter space.  

Based on the $\chi^2$ values in Figure~\ref{fig:chi2map}, two distinct statistical approaches were used to define oscillation parameter space regions allowed and excluded by the data.  
The first method, called the Gaussian CL$_s$ method~\cite{cls}, is based on testing multiple pairs of hypotheses. 
To assign the exclusion confidence level, for each point in (sin$^2 2\theta_{14}$,$\Delta$m$^2_{41}$) parameter space three values are needed:
\begin{align}\label{eq:cls_teststat}
\Delta T = \Delta \chi^2_{min}(x)_{1}-\Delta \chi^2_{min}(x)_{0} \\
\overline{\Delta T_{0}} = \Delta \chi^2_{min}(x^{Asimov}_{0})_{1} \\
\overline{\Delta T_{1}} = -\Delta \chi^2_{min}(x^{Asimov}_{1})_{0}, 
\end{align}
where the $\Delta \chi^2$ in all cases are calculated using Equation~\ref{eq:oscchi2}.  
$\Delta \chi^2_{min}(x)_{0}$ and $\Delta \chi^2_{min}(x)_{1}$ are calculated with the PROSPECT data against the null oscillation hypothesis and oscillation hypothesis with parameters $(\Delta m^2_{41}$,$\sin^2 2\theta_{14}$) respectively.  
$\Delta \chi^2_{min}(x^{Asimov}_{0})_{1}$ is calculated with the unoscillated Asimov dataset~\cite{cls} tested against the oscillation hypothesis given by the parameters ($\Delta m^2_{41}$,$\sin^2 2\theta_{14}$).  
$\Delta \chi^2_{min}(x^{Asimov}_{1})_{0}$ is its converse, calculated for oscillated Asimov dataset with parameters (sin$^2 2\theta_{14}$,$\Delta$m$^2_{41}$) tested against the null oscillation hypothesis.  

Once the values from Equation~\ref{eq:cls_teststat} are known, the value of CL$_s$ can be computed using:
\begin{equation}\label{eq:cls_detail}
\textrm{CL}_{s}(x) = \frac{1-p_1}{1-p_0} \approx 
\frac{1+\textrm{Erf}\large(\frac{\overline{\Delta T_{1}}- \Delta T(x)}{\sqrt{8|\overline{\Delta T_{1}}|}}\large)}{1+\textrm{Erf}\large(\frac{\overline{\Delta T_{0}}- \Delta T(x)}{\sqrt{8|\overline{\Delta T_{0}}|}}\large)}.
\end{equation}

\begin{figure}[hptb!]
\includegraphics[trim = 0.25cm 0.25cm 0.5cm 0.5cm, clip=true, 
width=0.49\textwidth]{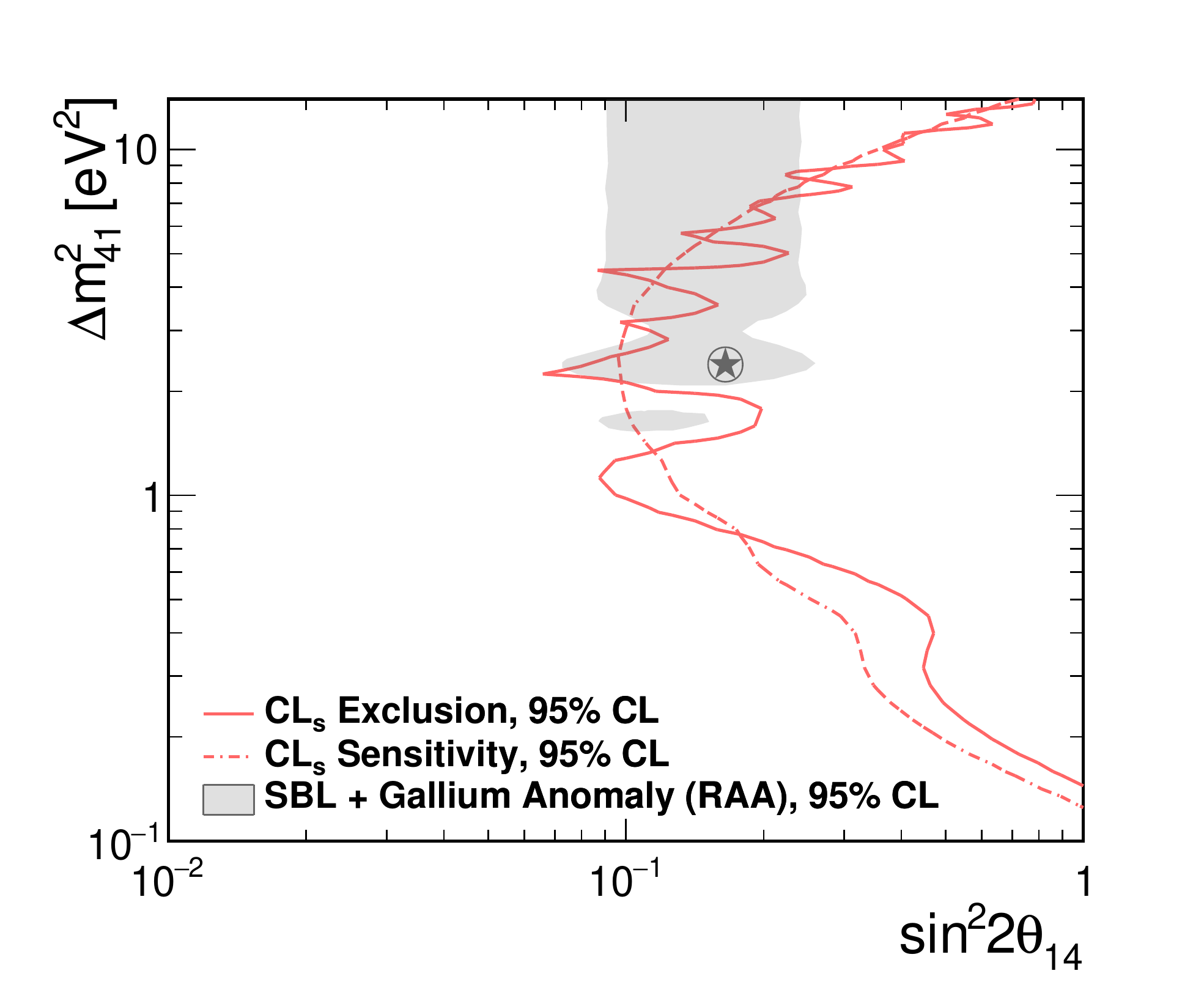}
\caption{Expected PROSPECT sterile neutrino oscillation sensitivity contour, as well as the exclusion contour corresponding to the $\Delta \chi^2$ distribution in Figure~\ref{fig:chi2map}.  Both contours are obtained using the Gaussian CL$_s$ method.  Also pictured is the RAA preferred parameter space and best-fit point from Ref.~\cite{bib:mueller2011}; the best-fit point is excluded at $>$95\% confidence level.} 
\label{fig:osc_cls}
\end{figure}

The point ($\sin^2 2\theta_{14}$,$\Delta m^2_{41}$) is said to be excluded by the given data at 2$\sigma$ confidence level if CL$_{s}<0.05$. 
The resulting 95\% confidence level CL$_s$ exclusion contour is shown in Figure~\ref{fig:osc_cls}.  
The RAA best fit is clearly excluded at better than 95\% CL.  
%To enable combination of PROSPECT's sterile oscillation parameter exclusion regions with those of other experiments, the quantities BLAH, BLAH, BLAH, and BLAH are included in supplementary materials accompanying this publication.  

\begin{figure}[hptb!]
\includegraphics[trim = 0.0cm 0.5cm 0.0cm 1.5cm, clip=true, 
width=0.49\textwidth]{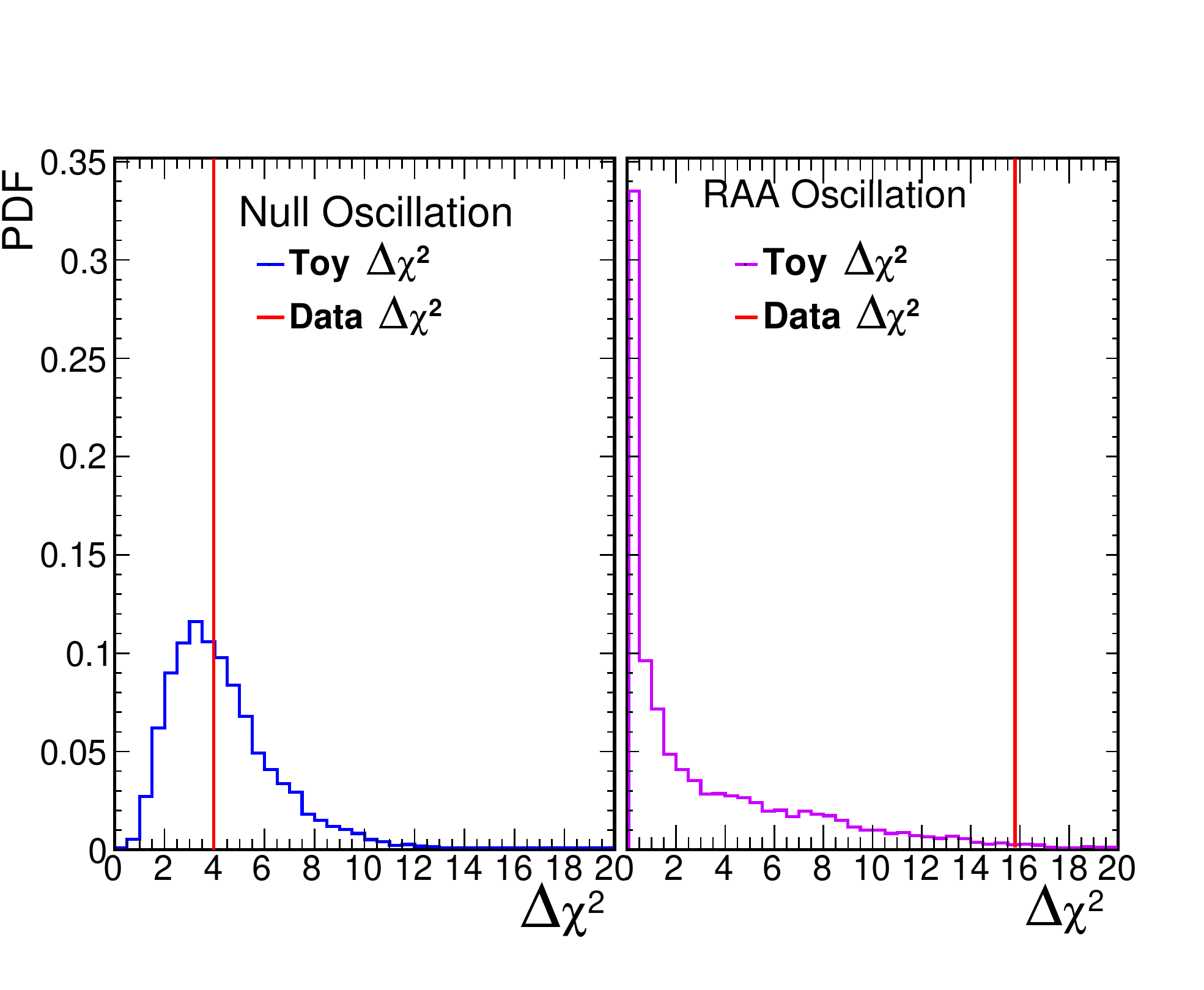}
\caption{Distributions of $\Delta \chi^2$ for toy MC datasets generated for the null oscillation (left, blue) and RAA best-fit point (right, magenta); $\Delta \chi^2$ are calculated between true and best-fit grid points individually for each toy.  Red vertical lines indicate the observed $\Delta \chi^2$ value from PROSPECT's data.  The observed value sits in the middle (higher end) of the distribution for the null (RAA) grid point, indicating good (poor) compatibility of the data with representative toy datasets from that grid point.  } 
\label{fig:osc_1dFC}
\end{figure}

The Gaussian CL$_s$ method provides a conservative excluded region that allows for easy combination with other experimental results, but it does not address the consistency of the data with respect to the null oscillation hypothesis.  
To remedy this, an examination of excluded sterile neutrino oscillation parameter space based on the the input $\chi^2$ map in Figure~\ref{fig:chi2map} was performed using a Feldman-Cousins frequentist approach~\cite{fc}, similar to that described in Ref.~\cite{prospect_osc}.  
This approach was first used to determine the level of preference observed in PROSPECT data for the best-fit point described above with respect to the null hypothesis, and with respect to the RAA best-fit point.  
For the null hypothesis, 10$^3$ individual toy datasets were generated by taking an unoscillated model spectrum at each baseline and adding a vector of independent random variables multiplied by a Cholesky decomposition of the full covariance matrix. 
This ensures that all toy results include the proper correlated and uncorrelated variations across baselines and energies.
These toy PROSPECT datasets represent the range of expected measurements likely to be delivered by PROSPECT in the absence of sterile neutrino oscillations given the range of expected statistical and systematic variations described above.  
Each toy PROSPECT dataset was then fit in a manner similar to that described above for the observed PROSPECT data.  
The $\Delta \chi^2 = \chi^2_{\textrm{null}} - \chi^2_{\textrm{min}}$ values calculated for all toys then form a distribution of expected $\Delta \chi^2$ values, as shown in Figure~\ref{fig:osc_1dFC}.  
The $\Delta \chi^2$ value obtained by a fit to the PROSPECT dataset was then compared to this distribution; the observed $\Delta \chi^2$ value, 123.3 - 119.3 = 4.0, is found to be smaller than 57\% of $\Delta \chi^2$ generated by the toy null oscillation datasets, indicating little incompatibility with the no-oscillation hypothesis.  

The same test was performed on the RAA best-fit point using 10$^3$ oscillated toy MC datasets.  
For the measured data, the best-fit $\chi^2$ mentioned above forms a $\Delta \chi^2$ value of 16.1 with respect to the $\chi^2$ obtained at the RAA best-fit point.  
When compared to the distribution of $\Delta \chi^2$ values from the RAA-oscillated toy datasets described above, we find that the observed $\Delta \chi^2$ value corresponds to a p-value of 1.5\%, as shown in Figure~\ref{fig:osc_1dFC}.  
This indicates that the RAA best-fit point is excluded by the PROSPECT data at the 2.5$\sigma$ confidence level.  

\begin{figure}[hptb!]
\includegraphics[trim = 0.25cm 0.25cm 0.5cm 0.5cm, clip=true, 
width=0.49\textwidth]{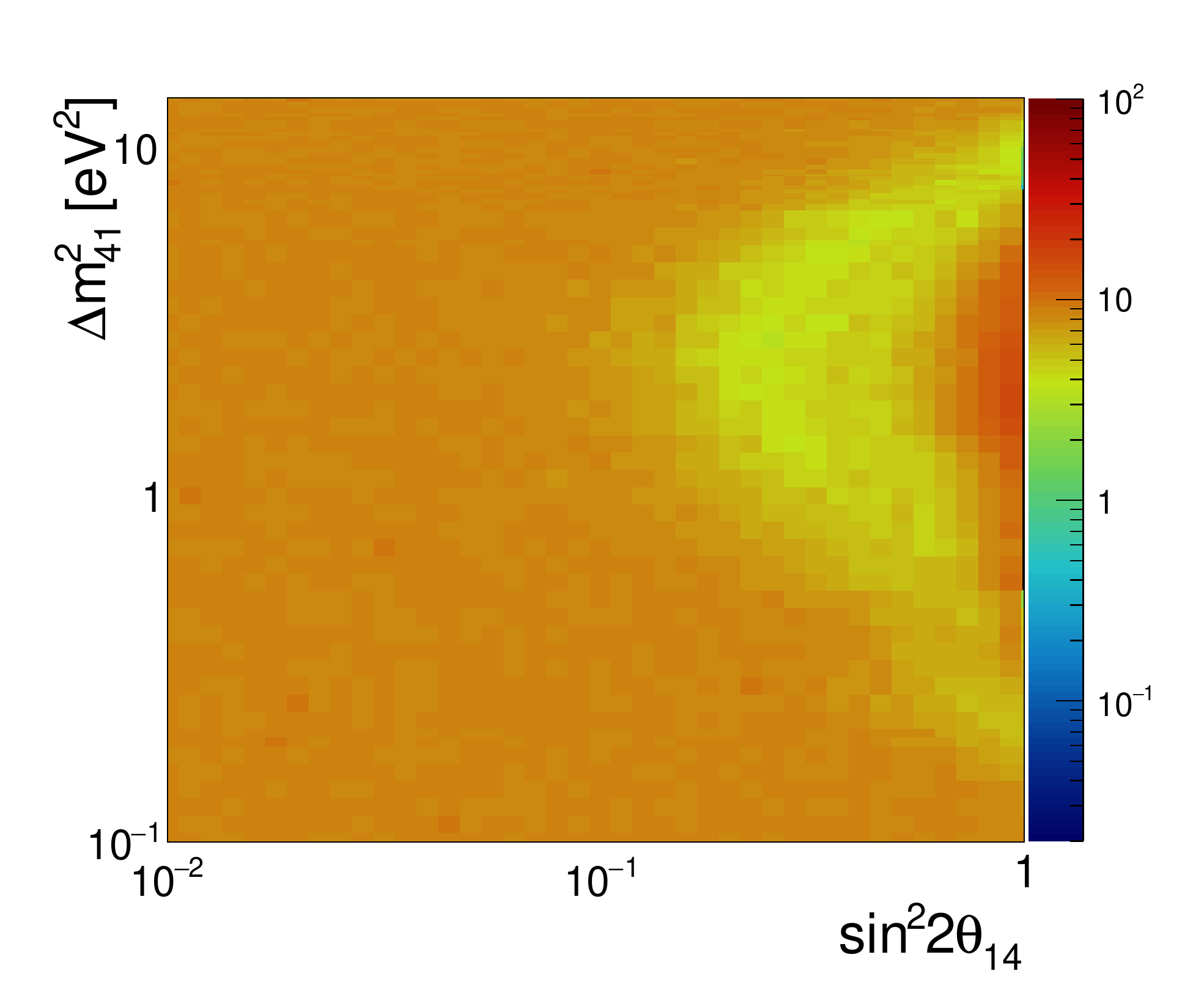}
\caption{Map of critical $\Delta \chi^2$ values indicating 95\% CL incompatibility with that grid point's predicted oscillatory behavior; generated using the Feldman-Cousins (FC) frequentist approach.  For reference, the incorrect assumption of an $\chi^2$ distribution with two degrees of freedom yields a flat map with $\Delta \chi^2$ = 5.99.} 
\label{fig:osc_chi2crit}
\end{figure}

Similar $\Delta \chi^2$ profiles were generated for each point in an examined grid of ($\Delta$m$^2_{41}$, sin$^2 2\theta_{14}$) values.  
At each grid point, a critical value, $\Delta \chi^2_{crit}$, is identified below which 95\,\% (2\,$\sigma$) of all 10$^3$ toy dataset-derived $\Delta \chi^2$ fall.  
The map of $\Delta \chi^2_{crit}$ values for each grid point in oscillation parameter space is shown in Figure~\ref{fig:osc_chi2crit}.  

It is worth noting that assuming these $\Delta \chi^2$ distributions follow a $\chi^2$ distribution with two degrees of freedom, as might be naively done when fitting two oscillation parameters, $\Delta$m$^2$ and sin$^2$2$\theta$, would yield a common $\chi^2_{crit}$ value of 5.99 across the pictured oscillation parameter space.  
This outcome is clearly at odds with the confidence level definitions of Figure~\ref{fig:osc_chi2crit} derived via the Feldman-Cousins approach.  
In particular, the incorrect $\chi^2_{crit}$ value associated with this inappropriate statistical treatment, for the case of the null hypothesis, would yield a p-value of 0.17, smaller than the p-value of 0.57 reported by the Feldman-Cousins approach.  
For the RAA best-fit point, this treatment leads to a p-value of 0.0004, smaller than the correct 0.015 p-value.  
Thus, it appears that this incorrect statistical interpretation of observed $\Delta \chi^2$ values will lead to over-statement of levels of statistical disagreement between data and the no-oscillation hypothesis, as well as under-statement of the level of compatibility between the data and some regions of non-zero oscillation parameter space.  
This observation is consistent with discussions in a variety of other publications~\cite{fc, stat_sterile, conrad_sterile}, and underscores the importance of using correct statistical treatments, such as the Gaussian CL$_{s}$ or Feldman-Cousins approaches.  

\begin{figure}[hptb!]
\includegraphics[trim = 0.25cm 0.25cm 0.5cm 0.5cm, clip=true, 
width=0.49\textwidth]{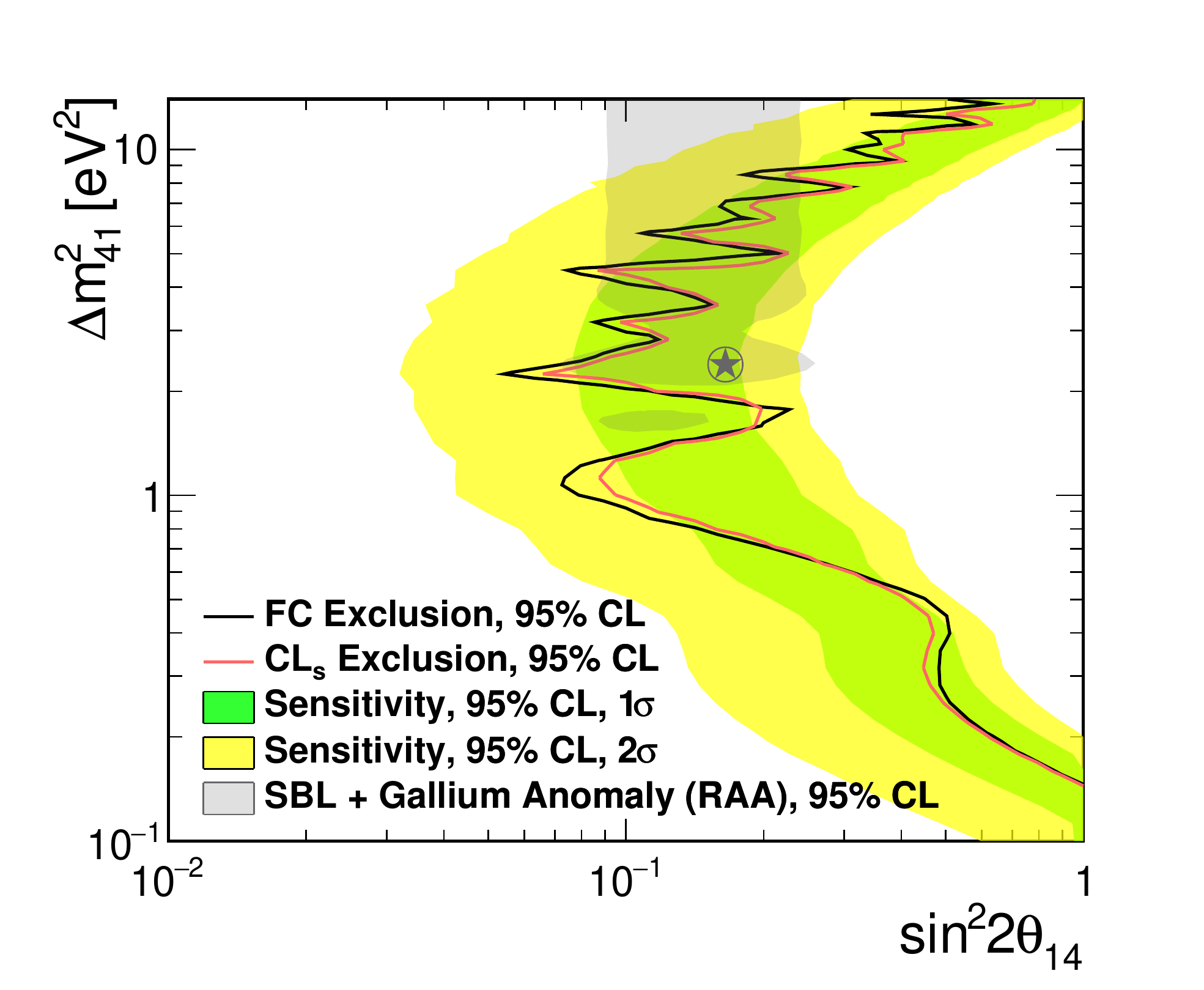}
\caption{Oscillation exclusion contours derived using the Gaussian CL$_s$ and Feldman-Cousins (FC) methods.  Also pictured are the 1$\sigma$ and 2$\sigma$ (green and yellow) exclusion ranges produced by PROSPECT toy MC datasets, as well as the RAA preferred parameter space and best-fit point from Ref.~\cite{bib:mueller2011}.} 
\label{fig:osc_xcheck}
\end{figure}

Using the Feldman-Cousins approach, an oscillation parameter space exclusion contour was assigned in (sin$^2 2\theta_{14}$,$\Delta$m$^2_{41}$) space to the observed $\chi^2$ values pictured in Figure~\ref{fig:chi2map}. 
A 95\,\% confidence level exclusion contour, shown in Figure~\ref{fig:osc_xcheck}, can be drawn by identifying all oscillation parameter space grid points whose data-derived $\Delta \chi^2$ between that grid point and the best-fit exceeds the $\chi^2_{crit}$ value given in Figure~\ref{fig:osc_chi2crit}.  
The present dataset excludes significant portions of the Reactor Antineutrino Anomaly allowed region~\cite{bib:mention2011}.  
This exclusion shows good agreement with that derived using the Gaussian CL$_s$ method.   

The colored bands included in Figure~\ref{fig:osc_xcheck} indicate, for each $\Delta$m$^2$ value, the range of sin$^2$2$\theta_{14}$ values at which the 95\% CL exclusion boundary appears for unoscillated toy MC datasets; green and yellow ranges contain 1$\sigma$ and 2$\sigma$ of all toys' 95\% CL exclusion boundaries.  
By comparing the observed exclusion region to these bands, one can assess the compatibility of the spectral ratio data in Figure~\ref{fig:oscratio} with the range of expected unoscillated PROSPECT spectral ratios.  
The exclusion region formed by the PROSPECT data sits within the green 1$\sigma$ region for most $\Delta m^2$ values, indicating that the observed spectral ratios are typical of those expected based on the systematic and statistical variations described in the previous section.  

\section{Spectrum Analysis}
\label{sec:spec}

Using the data and detector response model described the previous sections, the detected E$_{rec}$ spectrum of IBD interactions can be compared to theoretical predictions.
A total of 50560~$\pm$~406(stat) IBD events have been detected, with a cosmogenic (accidental) signal to background of 1.4 (1.8). 
This is the highest statistics measurement of the \uFive{}~\nuebar spectrum to date. 

Since \uFive{} is the only primary fissile isotope that can be studied in isolation, this measurement enables improved interpretation of measurements from low-enriched uranium (LEU) power reactors such as those used by the $\theta_{13}$ experiments.
These experiments have observed discrepancies between predicted and detected \nuebar energy spectra~\cite{bib:prl_reactor,bib:reno_shape,dc_bump}. 
In this section we present an updated PROSPECT measurement of the \uFive~\nuebar spectrum from HFIR, compare it to theoretical predictions, and perform further analysis to gauge the source of the deviation from predictions at high energy observed by LEU experiments.  

\subsection{Modelling the HFIR \nuebar Spectrum}
\label{subsec:nonfuel}
\begin{figure}
    \centering
    \includegraphics[trim = 0.25cm 0.0cm 0.0cm 0.0cm, clip=true, 
width=0.49\textwidth]{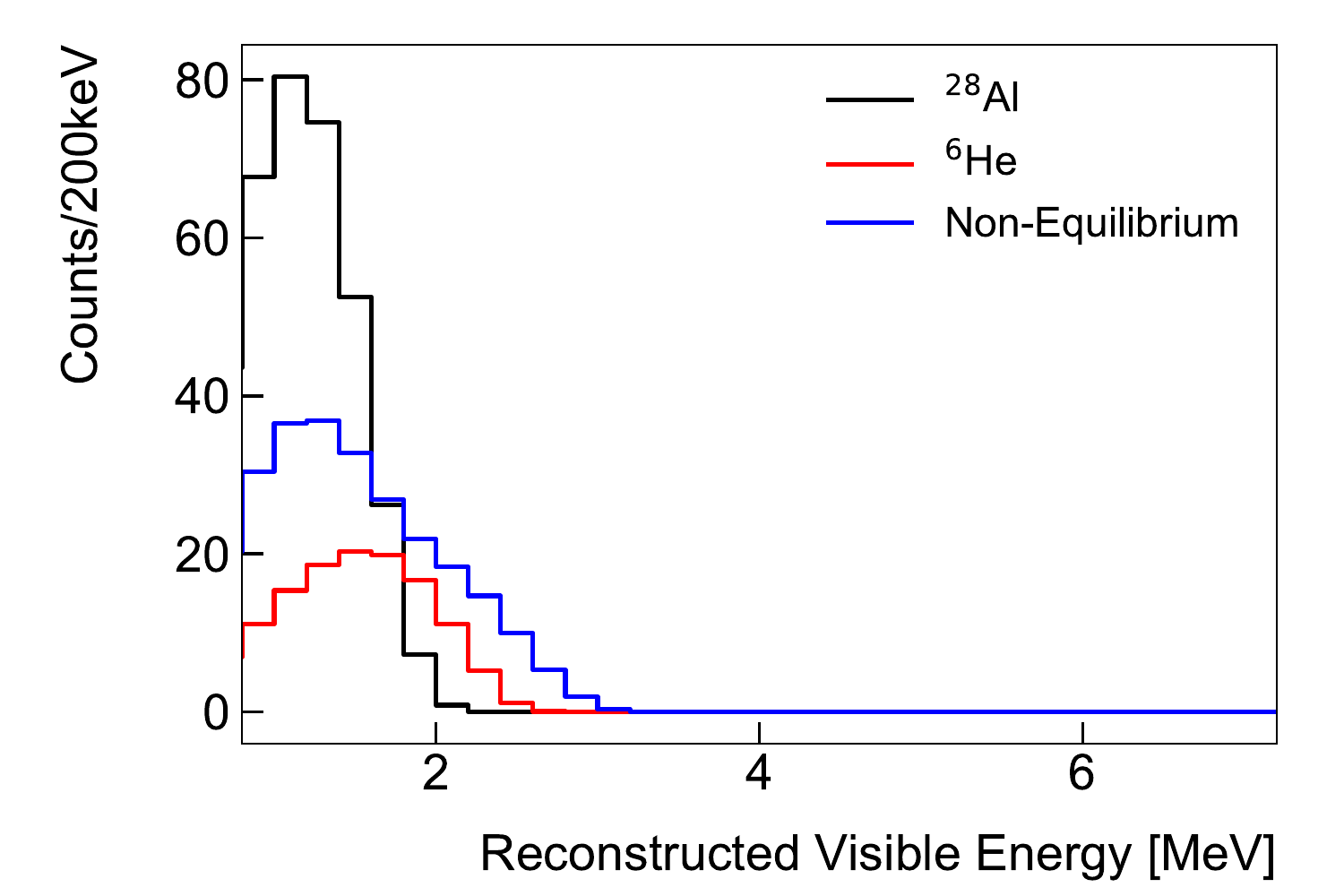}
    \caption{Corrections added to the predicted \uFive{} spectrum to account for non-equilibrium isotopes and neutrinos from $^{28}$Al and $^{6}$He.}
    \label{fig:hfir_corrections}
\end{figure}

More than 99\% of the \nuebar produced by High Flux Isotope Reactor are due to U-235 fission.
However, small fluxes of neutrinos are produced from neutron activation of the surrounding material. 
The two dominant non-\uFive{} sources of \nuebar are $^{28}$Al from the fuel cladding and $^6$He generated in the beryllium neutron reflector that surrounds the core~\cite{nonfuel}.
Each of these contribute less than 1\% of the total observed \nuebar flux and they are limited to the low-energy region of the spectrum ($<$4~MeV true neutrino energy). 
The predicted contribution to the detected spectrum for each of these is shown in Fig~\ref{fig:hfir_corrections}.

The leading theoretical model of \uFive{} \nuebar emission the Huber beta conversion model from Ref.~\cite{bib:huber}. 
This model converts a measured electron spectrum from neutron irradiation of fissile material into an \nuebar energy spectrum using `virtual beta-branches'.
Since the irradiation time in these measurements is relatively short compared to HFIR's 24-day cycle, corrections are needed to account for the production of non-equilibrium isotopes. 
The procedure laid out in Ref~\cite{bib:mueller2011} is followed to determine the correction needed to match the exposure in this measurement.
This correction is also shown in Fig~\ref{fig:hfir_corrections}.
The inverse beta decay cross-section from Ref~\cite{bib:ibd_xs} is used to convert the \nuebar flux to a predicted spectrum. 

These components are summed to produce the model of the HFIR \nuebar spectrum that the PROSPECT detector is exposed to. 
The total \nuebar{} spectrum is passed through the detector response model to produce a predicted E$_{rec}$ spectrum which can be compared to the PROSPECT measurement.
Further details of the HFIR prediction can be found in the Supplemental Material. 

%\subsection{Uncertainty Treatment - BEN!}
%As in the previous sections, this analysis uses covariance matrices to quantify the impact of each uncertainty.
%Table~\ref{tab:spec_uncertainties} details the individual uncertainties considered in this analysis.
%Statistical uncertainties make up the majority of the data effects.
%An additional 0.5\% uncertainty is added to the background, due to variations in reactor off periods.
%Model uncertainties are related to differences between HFIR operation and the Huber model~\cite{nonfuel}\todo[Ben]{Update this to PRC citation}.
%This includes non-equilibrium isotope corrections, as well as $^6$He and $^{28}$Al activation.
%The remaining covariance matrices are related to detector uncertainties.
%4123These include the physical properties of the detector, such as nonlinearity, energy loss, Cherenkov contributions, and wall thickness, as well as analysis cuts such as muon veto variations, fiducial volume, and energy thresholds.

\subsection{Statistical Treatment}

A $\chi^2$ metric is used to quantify the comparison between the measured spectrum and the beta conversion \uFive~model prediction:

\begin{align}
\chi^2_{min} = \Delta^{T} {V}^{-1} \Delta, \\
\Delta_i \equiv N_i^{obs} - N_i^{pred}\times (1 + \eta),
\label{eqn:min_chi2}
\end{align}

\noindent where $\Delta_i$ is the difference between the measured and predicted events in the $i$th E$_{rec}$ bin including a free-floating nuisance parameter $\eta$ to account for the normalization.  

The total uncertainty covariance matrix ($\mathrm{V}$) is used to determine the minimum $\chi^2$ for the measurement, including all uncertainties from signal and background statistics, detector, background, and reactor-related systematics, and from the theoretical model for the \uFive~\nuebar spectrum.  
Statistical uncertainties from signal and background datasets are determined using methods similar to those for the oscillation analysis.  
For reactor-related spectrum uncertainties, a 100\% uncertainty is assumed for all non-\uFive{} corrections and for the non-equilibrium correction.  
For theoretical model uncertainties, the Huber model's published covariance matrix is converted into PROSPECT E$_{rec}$ space via Cholesky decomposition.  

\begin{table*}[tb!]
\centering
\begin{center}
    \begin{tabular}{|l|c|c|l|}
         \hline
         Parameter & Section & Uncertainty & Description \\
         \hline
         Background Normalization & \ref{subsec:bkg_corr}, \ref{subsec:bkg_check}& 1\% & Accounts for variation between reactor-off periods
         \\ \hline
         $n$-H Peak & \ref{subsec:bkg_check} & 3\% & Accounts for uncertainty on background subtraction in the $n$-H peak region
         \\ \hline
         Detector Non-linearity & \ref{subsec:eresp} & 0.002 & Uncertainty for Birks non-linearity in energy deposition
         \\ \hline
         Cherenkov Contribution & \ref{subsec:eresp}& 0.41 & Uncertainty on Cherenkov contributions to collected photons
         \\ \hline
         Energy Scale & \ref{subsec:eresp}& 0.004 & Uncertainty on linear energy scale
         \\ \hline
         Energy Resolution & \ref{subsec:eres} & 5\% & Uncertainty in photostatistics contribution to energy-dependent resolution
         \\ \hline
         Energy Loss & \ref{subsec:evar}& 8 keV & Uncertainty in energy lost by escaping 511~keV $\gamma$-rays
         \\ \hline
         $^{28}$Al Activation & \ref{subsec:nonfuel} & 100\% & Uncertainty in the amount of $^{28}$Al contributing to the spectrum
         \\ \hline
         Non-equilibrium Correction & \ref{subsec:nonfuel}& 100\% & Uncertainty in extrapolating \nuebar{} contribution from long-lived fission daughters
         \\ \hline
         Panel Thickness & \ref{subsec:eresp} & 0.03 mm & Uncertainty in mass of the panels separating segments
         \\ \hline
         Z Fiducial Cut & \ref{subsec:eff} & 25 mm & Uncertainty in the position of events near the edge of the fiducial volume
         \\ \hline
         Energy Threshold & \ref{subsec:eresp}, \ref{subsec:reco} & 5 keV &  Uncertainty in the segment-by-segment energy threshold cut
         \\ \hline
    \end{tabular}
    \caption{Descriptions and values of the individual uncertainties combined to produce the final covariance matrix.}
    \label{tab:spec_uncertainties}
\end{center}
\end{table*}

For detector and background systematic uncertainties, 
a covariance matrix was generated for each contribution by either varying parameters in simulated data, or by analytically varying the Huber spectrum \cite{bib:huber} passed through the full detector response.  
Values used for each uncertainty were chosen as the result of a dedicated study of each effect.  
These effects include the physical properties of the detector, such as nonlinearity, energy loss, Cherenkov contributions, and wall thickness, as well as components of analysis cuts or signal definition, such as fiducial volume, energy threshold, or background subtraction.  
Table~\ref{tab:spec_uncertainties} details the individual uncertainties considered in this analysis.   
Detailed description of the origin of each of these systematic uncertainties has been provided throughout the previous sections of this paper.   

\begin{figure}
    \centering
    \includegraphics[trim = 0.25cm 0.0cm 0.0cm 0.0cm, clip=true, width=0.49\textwidth]{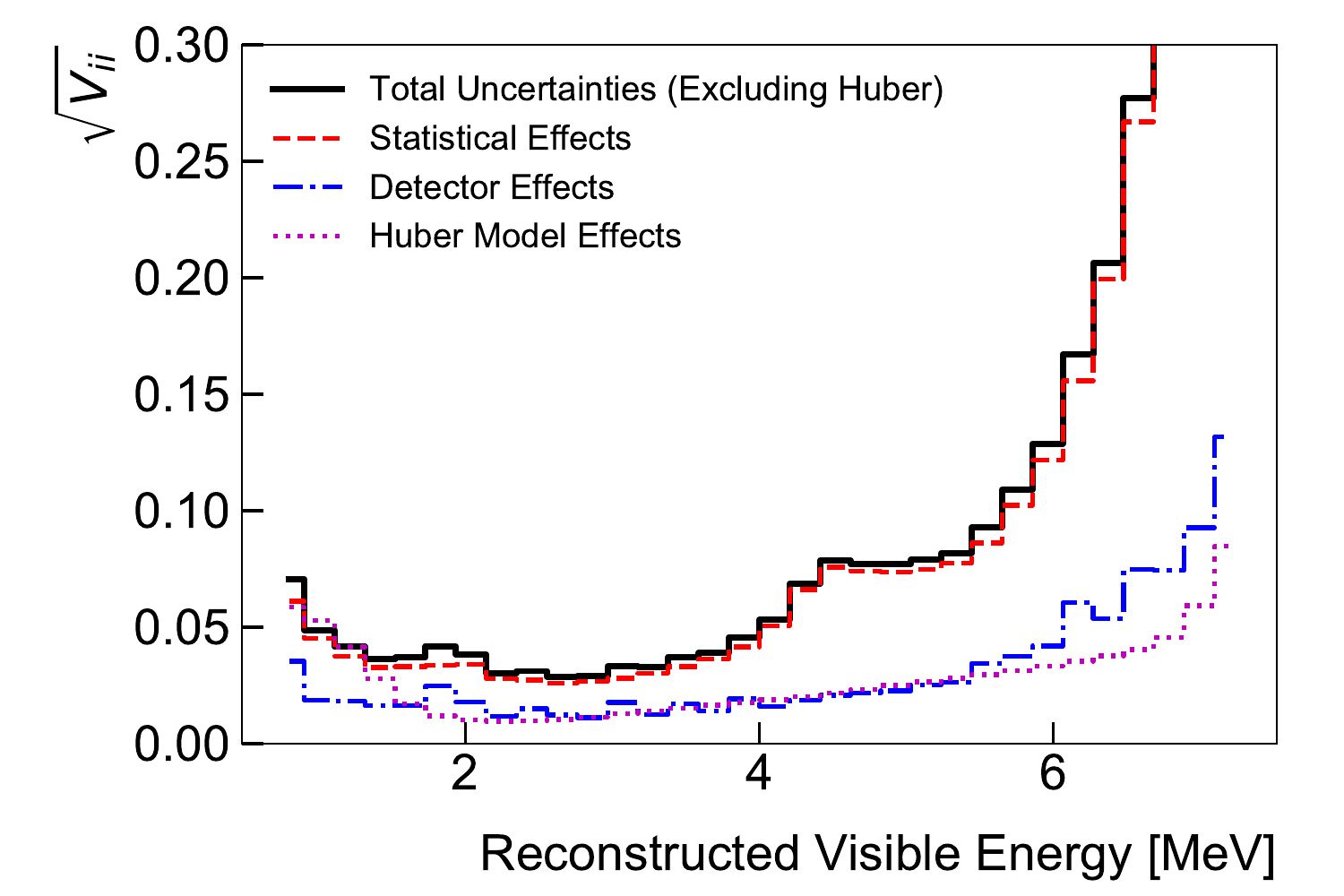}
\includegraphics[trim = 0.25cm 0.0cm 0.0cm 0.0cm, clip=true, 
width=0.49\textwidth]{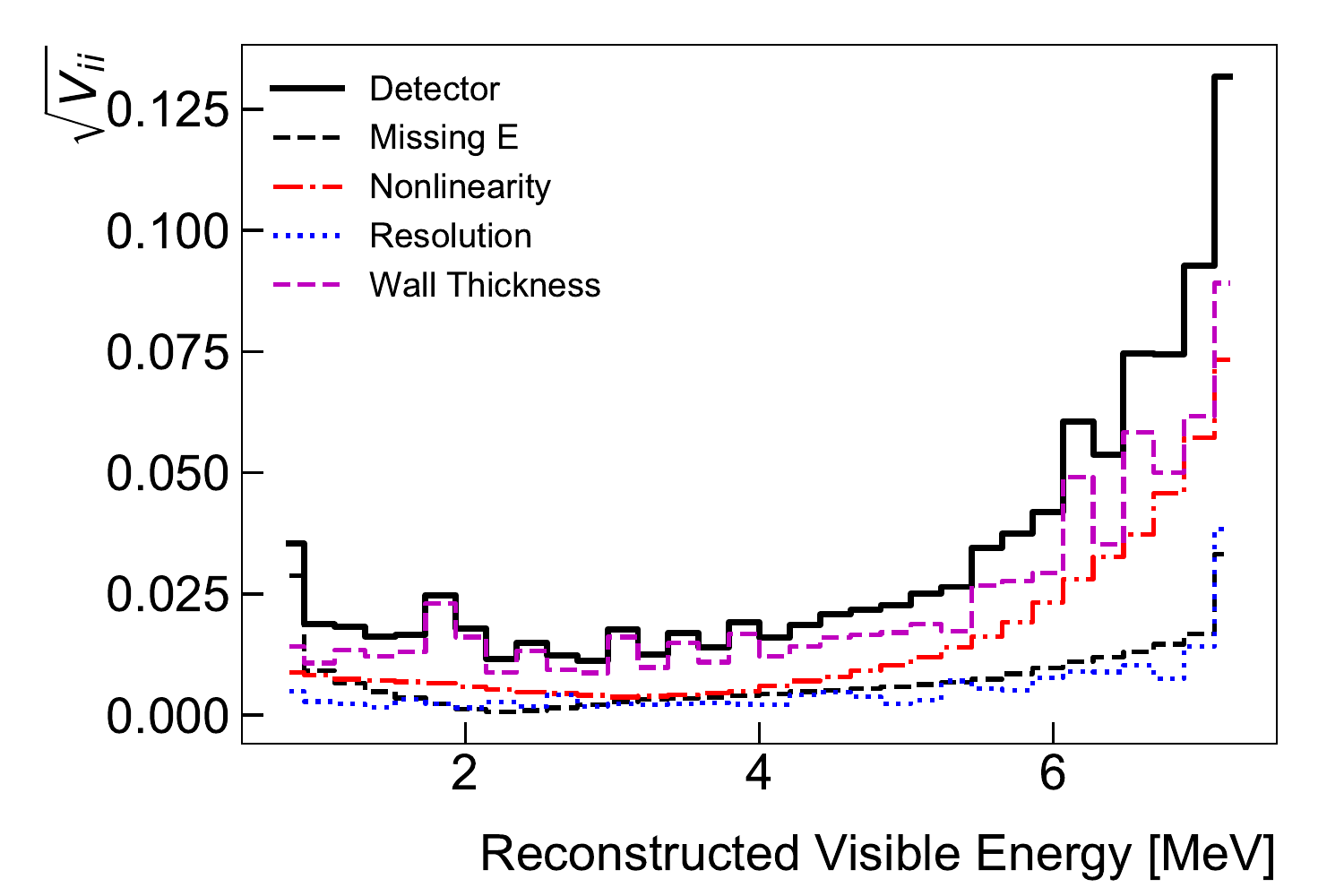}
    \caption{Uncertainties for the PROSPECT \uFive \nuebar spectrum measurement, represented by the square root of the uncertainty covariance matrix diagonal elements. Top: comparison of the three categories of uncertainties: statistics, detector, and model. Bottom: comparison of the individual contributions to the detector uncertainty.}
    \label{fig:spec_cov_mat}
\end{figure}

To provide an illustration of the relative contribution from different uncertainty sources for the spectrum analysis, 
Fig~\ref{fig:spec_cov_mat} shows the diagonal elements of the various categories included in the full uncertainty covariance matrix.  
Statistics clearly serve as the dominant source of uncertainty for the current \uFive~spectrum measurement, with detector-related systematic uncertainties as the largest sub-dominant uncertainty contributor.  
Reactor and model-related uncertainties provide the smallest overall uncertainty contribution.  
Figure~\ref{fig:spec_cov_mat} also provides a breakdown of the largest detector-related contributors.  
The dominant sources of detector systematic uncertainty are the limitations of understanding of the detector's $E_{rec}$ scale and non-linearity, as well as the uncertainty in the total dead mass contributed by the reflecting walls of the optical grid.  

\subsection{Results}

\begin{figure}
    \centering
    \includegraphics[trim = 0.25cm 0.0cm 0.0cm 0.0cm, clip=true, 
width=0.49\textwidth]{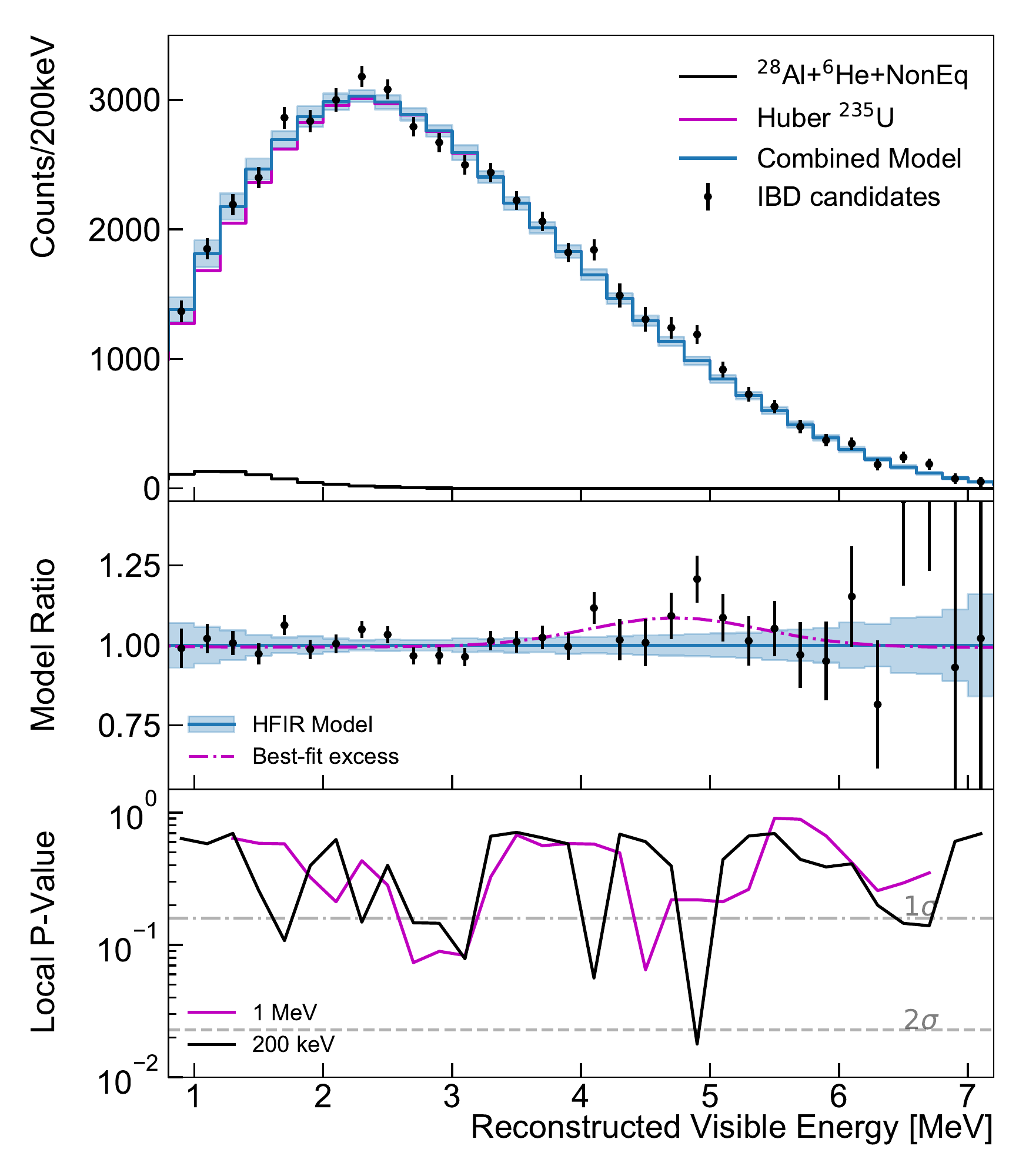}
    \caption{Top: Comparison of the \uFive{} model to the measured PROSPECT E$_{rec}$ spectrum. Middle: Ratio of the measurement to the HFIR prediction based on the Huber model. Bottom: The local p-value from 1~MeV- and 200~keV-wide sliding windows, quantifying any local deviations from the model prediction.  Error bars on data points represent statistical uncertainties, while error bands on the model represent systematic uncertainty contributions as represented in Figure~\ref{fig:spec_cov_mat}.}
    \label{fig:spec_huber_comp}
\end{figure}

The comparison of the Huber model to the measured spectrum is shown in Fig~\ref{fig:spec_huber_comp}.
The normalization of the model is determined by a minimization of the $\chi^2$ in the [0.8,7.2]~MeV region.
A $\chi^2$/DOF of 30.79/31 is observed, corresponding to a one-sided p-value of 0.48. 
To further quantify if any specific region of the spectrum is contributing significantly to this total $\chi^2$, additional nuisance parameters are added in 200~keV and 1~MeV-wide windows and a new $\chi^2_{min}$ determined for each. 
This $\Delta\chi^2$ can be interpreted as the local contribution to the total $\chi^2$.
The corresponding single-sided p-values are determined from the $\Delta\chi^2$ and plotted in Fig~\ref{fig:spec_huber_comp}.
Small excursions are observed in the 2.5\;MeV and 5\;MeV regions using this method.  
However, no region shows more than 2$\sigma$ deviation within the 1~MeV model prediction windows used.  

\begin{figure}
    \centering
    \includegraphics[trim = 0.25cm 0.0cm 0.0cm 0.0cm, clip=true, 
width=0.49\textwidth]{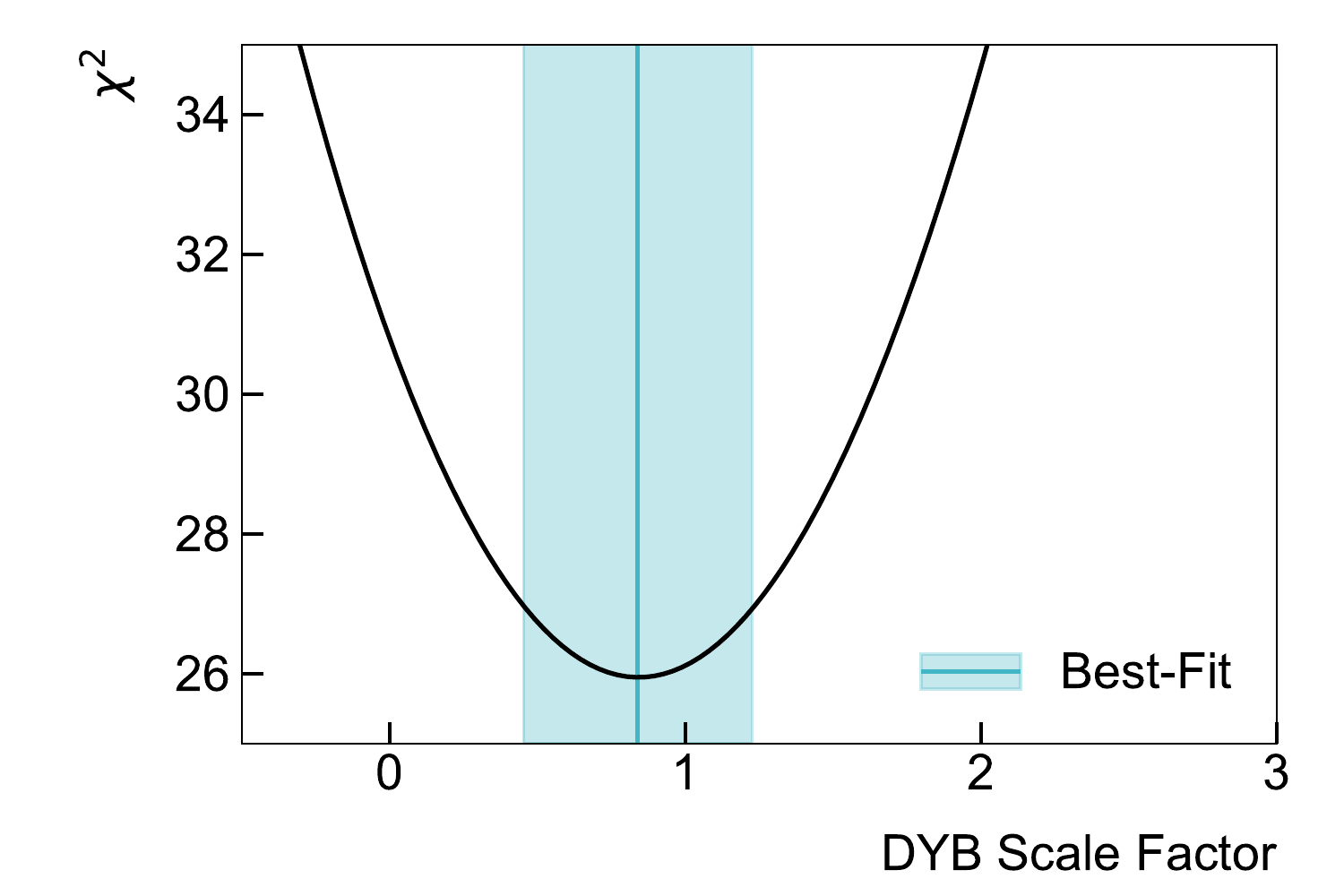}
    \caption{Resulting $\chi^2$ curve for the fit of a single parameter quantifying the amplitude of a Gaussian, whose sigma and mean are fixed by the unfolded Daya Bay spectrum, added to the HFIR model prediction. A best-fit value of 0.84$\pm$0.39 is found with a $\Delta \chi^2$ of 4.84 with respect to $A$=0.}
    \label{fig:dial_a_bump_chi2}
\end{figure}

Precision measurements at nuclear power reactors have observed discrepancies between predicted and detected \nuebar energy spectra. 
Most notably, a wide excess of events between 4-6~MeV E$_{rec}$ has generated much interest in the community. 
As these LEU reactors burn a time-evolving mixture of fuel, it is difficult to disentangle the isotopic origin of this distortion. 
To test whether PROSPECT observes such a feature, a Gaussian with mean 5.678 MeV and sigma 0.562 MeV is added to the HFIR model in true neutrino energy prior to applying the detector response.
This mean and sigma of the Gaussian are obtained from fitting the unfolded Daya Bay spectrum~\cite{bib:cpc_reactor}.  
%whose mean and sigma are fit to the unfolded Daya Bay spectrum is added to the HFIR model prior to applying the detector response.
The amplitude ($A$) of this addition, in units where a Daya Bay-sized distortion is equal to one, is varied yielding the single parameter $\chi^2$ curve shown in Fig~\ref{fig:dial_a_bump_chi2}.
A best-fit distortion of 0.84$\pm$0.39 is observed.
Fig~\ref{fig:spec_huber_comp}b shows a comparison of the data to both the best-fit distortion and the unmodified HFIR predicted spectrum. 

The data are consistent with a distortion of equal size to that observed by the $\theta_{13}$ experiments ($A=1$).
However, the data disfavor a null-hypothesis of no distortion in the \uFive{} spectrum ($A=0$) at 2.17$\sigma$, as well as a \uFive{} spectral distortion of the size ($A=1.78$) required to be the sole source of the $\theta_{13}$ measurements at 2.44$\sigma$.

\section{Summary}
\label{sec:summary}

During 96 calendar days of reactor-on data-taking between March and October 2018, the PROSPECT experiment observed over 50,000 inverse beta decay interactions of \nuebar produced by \uFive{} fission product decays by the highly-enriched 85~MW HFIR reactor.  
Despite deployment on the earth's surface in a high-background reactor facility environment, the PROSPECT IBD analysis is capable of selecting more signal IBD events than either cosmogenic-induced backgrounds (signal-to-background ratio of 1.4) or accidental backgrounds (signal-to-background ratio of 1.8).  
In overviewing the signal and background modelling, estimation, and validation processes, a number of unexpected but useful PROSPECT capabilities were also demonstrated, such as its performance of cosmic muon tomography of the HFIR water pool, and its ability to determine the direction of propagation of an observed flux of reactor \nuebar.  

In order to probe short-baseline reactor antineutrino disappearance with PROSPECT, reconstructed prompt energy spectra at ten different reactor-detector baselines were compared.  
In particular, baseline-dependent variations in detected energy spectra would indicate disappearance produced by oscillation between active and sterile neutrino sectors.  
In this paper, it was shown using two different statistical techniques that these relative baseline comparisons indicated no significant indication of sterile neutrino oscillations.  
While a best fit to the data in the sterile neutrino parameter space is found at (sin$^2$2$\theta_{14}$,$\Delta$m$^2$) = (0.11,1.78~eV$^2$), this preference is very mild with respect to the no-oscillation hypothesis, which is disfavored with a p-value of only 0.57.  
However, the canonical Reactor Antineutrino Anomaly best-fit point given in~\cite{bib:mention2011} is substantially disfavored at the 2.5$\sigma$ confidence level.  
Other regions of parameter space in the $\sim$0.1-15~eV$^2$ mass splitting range are disfavored at more than 95\% confidence level by PROSPECT’s data.  

By integrating the measured prompt energy spectra over all baseline ranges, PROSPECT has also reported on a new measurement of the \uFive~\nuebar energy spectrum.   
PROSPECT's updated \uFive{} spectrum result shows good agreement with the beta conversion \nuebar prediction of Huber~\cite{bib:huber}, with a $\chi^2$/DOF of 30.79/31.  
By measuring a nearly pure sample of \nuebar resulting from \uFive~fission, PROSPECT is  able to assess hypotheses regarding the origin of differences between modelled and measured energy spectra from \nuebar experiments at LEU commercial reactor cores, specifically in the high-energy 5-7~MeV \nuebar energy regime.  
The energy spectrum measured by PROSPECT is consistent with a scenario in which the spectral data-model discrepancy observed by Daya Bay is present in all fissioning isotopes.  
Conversely, PROSPECT's data disfavor at 2.4$\sigma$ confidence level a scenario in which \uFive~\nuebar{} are solely responsible for the Daya Bay high-energy data-model discrepancy.  
A scenario in which which no discrepancy exists in the \uFive~\nuebar spectrum is similarly disfavored at 2.2$\sigma$ confidence level.  

\section{Acknowledgements}

This material is based upon work supported by the following sources: US Department of Energy (DOE) Office of Science, Office of High Energy Physics under Award No. DE-SC0016357 and DE-SC0017660 to Yale University, under Award No. DE-SC0017815 to Drexel University, under Award No. DE-SC0008347 to Illinois Institute of Technology, under Award No. DE-SC0016060 to Temple University, under Contract No. DE-SC0012704 to Brookhaven National Laboratory, and under Work Proposal Number  SCW1504 to Lawrence Livermore National Laboratory. This work was performed under the auspices of the U.S. Department of Energy by Lawrence Livermore National Laboratory under Contract DE-AC52-07NA27344 and by Oak Ridge National Laboratory under Contract DE-AC05-00OR22725. Additional funding for the experiment was provided by the Heising-Simons Foundation under Award No. \#2016-117 to Yale University. 

J.G. is supported through the NSF Graduate Research Fellowship Program and A.C. performed work under appointment to the Nuclear Nonproliferation International Safeguards Fellowship Program sponsored by the National Nuclear Security Administration’s Office of International Nuclear Safeguards (NA-241). This work was also supported by the Canada  First  Research  Excellence  Fund  (CFREF), and the Natural Sciences and Engineering Research Council of Canada (NSERC) Discovery  program under grant \#RGPIN-418579, and Province of Ontario.

We further acknowledge support from Yale University, the Illinois Institute of Technology, Temple University, Brookhaven National Laboratory, the Lawrence Livermore National Laboratory LDRD program, the National Institute of Standards and Technology, and Oak Ridge National Laboratory. We gratefully acknowledge the support and hospitality of the High Flux Isotope Reactor and Oak Ridge National Laboratory, managed by UT-Battelle for the U.S. Department of Energy.

\bibliographystyle{apsrev4-1}
\bibliography{refs}{}

\end{document}